\documentclass[12pt]{article}
\usepackage{epsf}
\begin{document}

\def\lesssim{\;\raise0.3ex\hbox{$<$\kern-0.75em\raise-1.1ex\hbox{$\sim$}}\;}
\def\gtrsim{\;\raise0.3ex\hbox{$>$\kern-0.75em\raise-1.1ex\hbox{$\sim$}}\;}

\begin{center}
{\Large \bf Neutrino Emission from Neutron Stars}\\[0.5ex]
\end{center}

\begin{center}
         D.G.\ Yakovlev$^{1,5}$, A.D.\ Kaminker$^1$ \\
         O.Y.\ Gnedin$^2$, and P.\ Haensel$^{3,4}$
\end{center}
\noindent
$^1$Ioffe Physical Technical Institute,
Politekhnicheskaya 26, 194021, St.-Petersburg, Russia \\
{\it e-mail: yak@astro.ioffe.rssi.ru} and {\it kam@astro.ioffe.rssi.ru}\\
$^2$Institute of Astronomy,
Madingley Road, Cambridge CB3 0HA, England \\
{\it e-mail: ognedin@ast.cam.ac.uk}\\
$^3$Copernicus Astronomical Center,
Bartycka 18, PL-00-716, Warsaw, Poland \\
{\it e-mail: haensel@camk.edu.pl}\\
$^4$ D\'epartment of d'Astrophysique Relativiste et de Cosmologie -- \\
UMR 8629du CNRS, Observatoire de Paris, F-92195 Meudon Cedex, France\\
$^5$ Institute for Theoretical Physics,
University of California, Santa Barbara, CA 93106, USA\\ 

\begin{abstract}
We review the main neutrino emission mechanisms
in neutron star crusts and cores.
Among them are the well-known reactions
such as the electron-positron annihilation,
plasmon decay, neutrino bremsstrahlung of electrons
colliding with atomic nuclei in the crust, as well as
the Urca processes and neutrino bremsstrahlung in nucleon-nucleon
collisions in the core.  We emphasize recent
theoretical achievements, for instance, band structure effects
in neutrino emission due to scattering of electrons  
in Coulomb crystals of atomic nuclei.
We consider the 
standard composition of matter (neutrons, protons,
electrons, muons, hyperons) in the core,
and also the case of exotic constituents such as
the pion or kaon condensates and quark matter.
We discuss the reduction of the neutrino emissivities
by nucleon superfluidity, as well as the
specific neutrino emission produced by Cooper pairing
of the superfluid particles.
We also analyze the effects of strong magnetic fields on some reactions,
such as the direct Urca process and the neutrino
synchrotron emission of electrons.
The results are presented in the form convenient
for practical use.
We illustrate the effects of various neutrino reactions
on the cooling of neutron stars.
In particular, the neutrino emission
in the crust is critical in setting
the initial thermal relaxation between the core and the crust.
Finally, we discuss the prospects of exploring the 
properties of supernuclear matter
by confronting cooling simulations with
observations of the thermal radiation from isolated
neutron stars.
\end{abstract}

\newpage

\tableofcontents

\newpage

\section{Overview}
\label{chapt-overview}

\subsection{Introduction}
\label{sect-overview-introduct}

Neutron stars are the most fascinating stars
known in the Universe.
Indeed, measured masses of these tiny stars are around 
$M \simeq 1.4 \, M_\odot$,
but they are expected to have very small radii, $R \simeq 10$ km.
Accordingly, they
possess an enormous gravitational energy,
$G M^2/R \sim 5 \times 10^{53}$ erg $\sim 0.2 \, M c^2$, and
surface gravity, $G M/R^2 \sim 2 \times 10^{14}$ cm~s$^{-2}$.
Since the gravitational energy constitutes a large fraction
of the rest-mass energy, neutron stars are relativistic
objects; space-time is essentially curved within
and around them, implying they can behave as strong gravitational
lenses (e.g., Zavlin et al.\ 1995).
%
%
Neutron stars are
very dense. Their mean density
$\bar{\rho} \simeq 3M/(4 \pi R^3) \simeq 7 \times 10^{14}$
g~cm$^{-3}$ is several times the standard nuclear density, 
$\rho_0 = 2.8 \times 10^{14}$ g~cm$^{-3}$.
The central density is larger, reaching up to
(10--20) $\rho_0$.  Thus, the cores of neutron stars
are composed of a strongly compressed nuclear matter.
This compression,
which cannot be reproduced in laboratory,
is provided by the very strong gravitational force.
Accordingly, neutron stars can be treated as unique
astrophysical laboratories of superdense matter.
The main constituents of neutron star cores
are thought to be strongly degenerate neutrons
with the admixture of protons
and electrons, although other particles may also be available.

Observationally, neutron stars manifest themselves
in different ways, for instance, as
radio pulsars,
X-ray pulsars,
anomalous X-ray pulsars (e.g., Mereghetti et al.\ 1998),
%
%
X-ray bursters (e.g., Lewin et al.\ 1995),
%
%
X-ray transients (e.g., Campana et al.\ 1998),
%
%
sources of quasiperiodic X-ray oscillations
(e.g., Psaltis et al.\ 1998, Kluzniak 1998),
%
%
%
and soft-gamma repeaters
(e.g., Cline et al.\ 1999 and references therein).
%
%
They are the sources of electromagnetic radiation in all wavelength
bands, from radio to hardest gamma-rays, and they are known to
work as efficient accelerators of highly energetic particles.
Their birth in supernova explosions is accompanied by
the most powerful neutrino outburst.  One of them was detected
by the underground neutrino detectors as a signal of the
1987A supernova
explosion in the Large Magellanic Cloud
(e.g., Imshennik and Nadezhin 1988, Burrows 1990).
%
%
%
Evolution of orbital parameters of the double neutron
star binary containing PSR 1913+16, the Hulse--Taylor pulsar,
indicates that the binary system emits gravitational waves.
Merging neutron stars are thought to be among the
best candidates
for direct detection of gravitational waves.

In other words, neutron stars are extremely interesting to observe
and study theoretically. Their main ``mystery''
is the equation of state of dense matter in their
cores. It cannot be derived from first principles
since the exact calculation would require
the exact theory of nuclear interactions
and the exact many-body theory
to account for in-medium effects.  While lacking both of these
theories, many model equations of state
have been constructed in the last decades (e.g., Shapiro and Teukolsky 1983, 
Glendenning 1996, Weber 1999, Lattimer and Prakash 2000).
They vary widely
from the soft to the moderate and stiff ones
and produce very different
structure of neutron stars.  This opens a possibility
to constrain the equation of state by comparing
theory with observations in many ways.
The majority of methods are based on the determination (constraint)
of the stellar mass and/or radius and the comparison
with the mass-radius diagrams for different
equations of state.
In particular, one can use the data
obtained from the mass measurements in close binaries containing
neutron stars,
the minimum spin periods of the
millisecond pulsars, the identification of the kilohertz
quasi-periodic oscillations with the orbital frequency
in the last stable orbit, and from other related methods.
Unfortunately, no decisive argument
has been given so far in favor
of the stiff, moderate or soft equations of state.
One can definitely rule out
only the ultra-soft equations of state which give the
maximum stellar mass lower than $1.44 \, M_\odot$,
the accurately measured mass of the Hulse--Taylor pulsar
(Thorsett and Chakrabarty 1999).
%
%

In this review we focus on another method to explore
the internal structure of neutron stars --- by
comparing the theory of neutron star cooling with
observations of the thermal radiation from the surfaces
of isolated neutron stars.  For about $10^5$--$10^6$ years
after their birth in supernova explosions, neutron
stars cool mainly via neutrino emission from their
interiors. Although their matter is very dense,
they become fully transparent for neutrinos about
20 seconds after the birth
(e.g., Burrows and Lattimer 1986,
Prakash et al.\ 1997).
%
%
Thus, the neutrinos
produced in numerous reactions leave neutron stars freely,
providing a powerful source of cooling.
Therefore, it is important to study the whole variety of
neutrino reactions in different
neutron star layers in order to analyze the cooling.
Some mechanisms have been
reviewed not long ago by
Bisnovatyi-Kogan (1989), Pethick (1992),
Itoh et al.\ (1996), and Yakovlev et al.\ (1999b).
%
%
%
More references to
the original and review papers are given in subsequent chapters.

The present review differs from other papers
in several respects. First, we discuss the neutrino
mechanisms in all neutron star layers, from the
crust to the inner core,
paying attention to open theoretical problems. Second, we consider
the neutrino emission in the non-superfluid and non-magnetized matter
as well as in the presence of superfluidity
of baryons and the possible strong magnetic field.
In particular, we analyze the
neutrino emission due to Cooper pairing of baryons,
which is forbidden in non-superfluid matter, and the
synchrotron neutrino emission of electrons
forbidden without the magnetic field.
Third, we present the results in a unified form,
convenient for use in neutron star cooling codes.
Fourth, we discuss the relative importance of all
these neutrino reactions for the various stages of the cooling.

The review is organized as follows.
In the next section, we describe briefly
the structure of neutron stars.  In Chapt.\ \ref{chapt-nucrust}
we analyze various neutrino reactions in neutron star
crusts. In Chapt.\ \ref{chapt-nucore} we describe
neutrino processes in non-superfluid and non-magnetized
neutron star cores. In Chapt.\ \ref{chapt-nusup}
we consider the reactions in superfluid
and magnetized cores.
Finally, in Chapt.\ \ref{chapt-cool}
we discuss the efficiency of various reactions
for neutron star cooling, present new results
of cooling simulations and compare them with observations.

\subsection{Structure of neutron stars}
\label{sect-overview-struct}

A neutron star can be subdivided into the {\it atmosphere}
and four internal regions: the {\it outer crust},
the {\it inner crust}, the {\it outer core}, and the {\it inner core}.

{\bf The atmosphere} is a thin layer of plasma which determines the spectrum
of thermal electromagnetic radiation of the star.
In principle, this radiation contains valuable information
on the stellar parameters (temperature, gravitational acceleration
and chemical composition of the surface, magnetic field, etc.)
and, as a result, on the internal structure.
Geometrical depth of the atmosphere varies from some ten centimeters
in a hot star down to some millimeters in a cold one.

Neutron star atmospheres have been studied theoretically by many
authors (see, e.g., review papers by Pavlov et al.\ 1995,
Pavlov and Zavlin 1998, and references therein).
%
%
%
Construction of the atmosphere models, especially for cold stars
with the effective surface temperature $T_s \lesssim 10^6$~K
and strong magnetic fields 10$^{11}$--10$^{14}$~G,
is incomplete owing to serious theoretical problems
concerned with the
calculation of the equation of state and spectral opacity of
the atmospheric plasma.

{\bf The outer crust (outer envelope)} extends from the
bottom of the atmosphere to the layer of density
$\rho_{\rm d} \approx
4 \times 10^{11}$ g~cm$^{-3}$ and has a depth of
a few hundred meters
(Shapiro and Teukolsky 1983). It is composed of ions and
electrons. A very thin (no more than several meters in a hot star)
surface layer contains non-degenerate electron gas
but in the deeper layers the electrons are degenerate.
At $\rho \lesssim 10^4$ g~cm$^{-3}$ the electron plasma
may be non-ideal and the ionization may be incomplete.
At higher densities the electron gas is almost ideal and
the atoms are fully ionized, being actually bare atomic nuclei.
As the electron Fermi energy grows with $\rho$,
the nuclei capture electrons and become neutron-rich
(Sect.\ \ref{sect-nucrust-beta}). At the base of the outer crust
($\rho = \rho_{\rm d}$) the neutrons start to drip from the nuclei
and form a free neutron gas.

The state of the degenerate electrons is characterized by
the Fermi momentum $p_{{\rm F}e}$
or the relativistic parameter $x_r$:
\begin{equation}
      p_{{\rm F}e} = \hbar (3 \pi^2 n_e)^{1/3}, \quad
      x_r={p_{{\rm F}e} \over m_e c} \approx
      100.9 \left( {\rho_{12} Y_e} \right)^{1/3},
\label{pF}
\end{equation}
where $Y_e=n_e/n_b$ is the number of electrons per baryon,
$n_e$ is the number density of electrons, $n_b$ the number density
of baryons, and $\rho_{12}$
is the mass density in units of $10^{12}$ g cm$^{-3}$.
The electron degeneracy temperature is
\begin{equation}
       T_{\rm F}= (\sqrt{1+x_r^2} -1)\,T_0, \quad
       T_0 \equiv {m_e c^2 \over k_{\rm B}} \approx
       5.930 \times 10^9~~{\rm K},
\label{TF}
\end{equation}
where $k_{\rm B}$ is the Boltzmann constant. The chemical
potential of strongly degenerate electrons
equals $\mu_e=m_e \,(1+x_r^2)^{1/2}$. The electrons
become relativistic at $x_r \gg 1$, i.e., at $\rho \gg 10^6$ g cm$^3$.

The state of the ions (nuclei) in a one-component ion plasma
is determined by the ion-coupling parameter
\begin{equation}
     \Gamma = {Z^2 e^2 \over a k_{\rm B} T}
     \,  \approx \, 0.225 \, x_r \, {Z^{5/3} \over T_8},
\label{Gamma}
\end{equation}
where $Ze$ is the nuclear charge,
$a=[3/(4 \pi n_i)]^{1/3}$
is  ion--sphere (Wigner--Seitz cell) radius,
$n_i$ is the number density of
ions, and $T_8$ is the temperature in units of $10^8$ K.
For the densities and temperatures of interest
for neutrino reactions (Chapt.\ \ref{chapt-nucrust})
the ions constitute either a strongly coupled
Coulomb liquid ($1 < \Gamma < \Gamma_m$),
or a Coulomb crystal ($\Gamma > \Gamma_m$), where
$\Gamma_m \approx 172$ corresponds
to the solidification of a classical one-component Coulomb liquid
into the body centered cubic (bcc) lattice (Nagara et al.\ 1987).
%
%
The corresponding melting temperature is
\begin{equation}
     T_m= {Z^2 e^2 \over
     a k_{\rm B} \Gamma_m}
     \approx 1.32 \times 10^7 Z^{5/3}
     \left( {\rho_{12} Y_e} \right)^{1/3}~~
     {\rm K}.
\label{Tm}
\end{equation}
%

An important parameter of a strongly coupled ion system
is the ion plasma temperature
\begin{equation}
     T_p = {\hbar \omega_p \over k_{\rm B}}
     \approx 7.83 \times 10^9
     \left({Z Y_e \rho_{12} \over A_i } \right)^{1/2}~{\rm K},
\label{Tp}
\end{equation}
determined by the ion plasma frequency
$\omega_p=\left( {4\pi Z^2 e^2 n_i / m_i }  \right)^{1/2}$,
where $m_i \approx A_i m_u$ is the ion mass and
$m_u = 1.66055 \times 10^{-24}$~g is the atomic mass unit.
The plasma temperature characterizes
thermal vibrations of the ions.  If $T \gtrsim T_p/8$,
the vibrations can be treated classically,
while at $T \ll T_p$ they are essentially quantum.
In the layers of interest for neutrino reactions,
the transition from the classical to the quantum-mechanical regime
takes place at temperatures lower than $T_m$.

Vibrational properties of the bcc Coulomb crystal of ions
immersed in the uniform electron gas
are well known (e.g., Pollock and Hansen 1973).
%
%
The spectrum contains three phonon branches.
Generally, the phonon frequencies are
$\omega_s(k) \sim \omega_p$ ($s=$ 1, 2, 3).
At large wavelengths, $k \to 0$, two branches
behave as transverse acoustic modes
($\omega_s \propto k$ for $s=1$ or 2),
while the third is longitudinal and optical ($\omega_s \to \omega_p$).
However, because of the compressibility of the electron gas,
the optical mode turns into the acoustic
one at very small $k$.

{\bf The inner crust (inner envelope)}
extends from the density
$\rho_{\rm d}$ at the upper boundary
to $\sim 0.5 \rho_0$ at the base.
Its thickness can be as large as several kilometers.
The inner crust is composed of
the electrons, free neutrons and neutron-rich atomic nuclei
(Negele and Vautherin 1973,
Pethick and Ravenhall 1995).
%
%
%
The fraction of free neutrons increases with density.
At the bottom of the crust, in the density range from
$10^{14}$ to $1.5 \times 10^{14}$ g~cm$^{-3}$,
the nuclei can be non-spherical and form clusters
(Lorenz et al.\ 1993;
Pethick and Ravenhall 1995).
%
%
The nuclei disappear completely
at the crust-core interface.

The free neutrons $n$ in the inner crust may
be superfluid.  The superfluidity
is thought to be
produced by Cooper pairing via
the attractive part of the singlet-state neutron-neutron
interaction. Superfluidity occurs when temperature
falls below the critical temperature $T_{cn}$.
References to the numerous microscopic calculations
of $T_{cn}$ can be found, for instance, in Yakovlev et al.\ (1999b).
The results are very sensitive to the model of $nn$-interaction
and the many-body theory employed.  The most important common feature
is that $T_{cn}$ increases with density at $\rho \gtrsim \rho_{\rm d}$,
reaches maximum (from $10^8$ to $10^{11}$ K,
depending on the microscopic model) at subnuclear densities,
and decreases to zero at $\rho \sim \rho_0$.
The initial increase is associated with the growth of
the effective $nn$ interaction strength with increasing density. The decrease
of $T_{cn}$ occurs where
the effective singlet-state $nn$ attraction
turns into repulsion. In the models which take into account
in-medium (polarization) effects $T_{cn}$ is typically
several times lower than in the models which ignore these effects.
The critical temperature is sensitive to the values of the
effective neutron mass which determines the density of states
of neutrons near the Fermi surface: the lower the effective mass
the smaller $T_{cn}$. Note, that
the bound nucleons within atomic nuclei can also be in superfluid state.

{\bf The outer core} occupies the density range
$0.5 \rho_0 \lesssim \rho \lesssim 2 \rho_0$
and can be several kilometers deep.
It is composed mainly of neutrons with some admixture
(several percent by number)
of protons $p$ and electrons $e$
(the so called {\it standard} nuclear composition).
For $\rho \gtrsim \rho_0$, where 
the electron chemical potential $\mu_e > m_\mu c^2 = 105.7$ MeV,
a small fraction of muons ($\mu$) appear.
The composition of $npe$-matter at densities below the muon threshold
is determined by the conditions
of electric neutrality and beta-equilibrium with respect to the
reactions
$ n \to p + e + \bar{\nu}_e$,
$\; p + e \to n + \nu_e $, where $\nu_e$ and $\bar{\nu}_e$ stand for
electron neutrino and antineutrino, respectively.
The electric neutrality requires the
electron and proton number densities be equal,
$n_p = n_e$.
Beta equilibrium implies then the following relationship
between chemical potentials of the particles:
$ \mu_n = \mu_p + \mu_e$.
The neutrino chemical potential is ignored here
since neutron stars are transparent for neutrinos.
In the presence of muons
the condition of electric neutrality reads
$n_p=n_e+n_\mu$, and the equilibrium with respect
to the weak processes involving muons implies
$\mu_e=\mu_\mu$, in addition to the 
beta-equilibrium condition $\mu_n=\mu_p+\mu_e$.
All $npe\mu$-plasma components are strongly degenerate.
The electrons and muons form almost ideal Fermi-gases.
The electrons are ultrarelativistic, 
while the muons are nonrelativistic near the threshold
of their appearance and become moderately relativistic
at higher densities.
The neutrons and protons, which interact via nuclear forces,
constitute a strongly non-ideal, non-relativistic
Fermi-liquid.

With increasing density,
the particle Fermi energies grow, so that new particles
can be created.  
The appearance of new particles in the inner core
is discussed below.

Let us emphasize
that a self-consistent quantum theory
of matter of supernuclear density has not been constructed yet.
Many theoretical equations of state have been proposed 
(e.g., Lattimer and Prakash 2000) which
can be divided conventionally into the
{\it soft}, {\it moderate} and {\it stiff} with respect
to the compressibility of matter.
These equations of state vary considerably
from one another.

Almost all microscopic theories
predict the appearance of neutron and proton superfluids
in neutron star cores
(see Yakovlev et al.\ 1999b for references).  These superfluids
are believed to be produced by the $nn$ and $pp$ Cooper pairing
due to the attractive part of the nuclear potential and can be
characterized by the critical temperatures
$T_{cn}$ and $T_{cp}$.
The proton superfluidity is
accompanied by superconductivity (most likely, of second type)
and affects the evolution of the internal magnetic field
(see, e.g., Ruderman 1991).
%
%

Owing to a sufficiently low concentration of protons,
the proton pairing is
produced by the singlet-state attractive part of the $pp$ interaction.
Although the singlet-state $nn$ interaction in neutron star cores
is repulsive, as discussed above, 
some part of the triplet-state
$nn$ interaction is attractive leading to neutron pairing.
As in the crust, the microscopic theories give very
model-dependent values of $T_{cn}$ and $T_{cp}$.
According to these theories,
the density dependence of $T_{cn}$
has a maximum at supranuclear
density. The maximum values
vary from $10^8$~K to about $10^{10}$~K in different models.
Usually,
the dependence $T_{cp}(\rho)$ also has a maximum at
some supernuclear density. The maximum values fall in the same range as
$T_{cn}$.

The critical temperatures $T_{cn}$ and $T_{cp}$
are much more sensitive to the microscopic models of dense matter
than the pressure, the main ingredient of the equation of
state.  There is no direct link between
these quantities: the pressure is the bulk property
determined by the entire Fermi-seas of the particles,
while the superfluidity is the phenomenon occurring near the
Fermi surfaces.  However, both the superfluidity and pressure
depend on the repulsive core of the nucleon-nucleon interaction.
For a less repulsive (more attractive) core, one has a softer equation of
state and higher $T_c$.  Therefore, the superfluid state of
neutron star cores is related in some way to the equation of state.
As in the crust,  $T_c$ in the stellar core
is sensitive to in-medium effects.

The outer core of a low-mass neutron star extends all the way
to the center.
More massive stars possess also {\bf the inner core}.
It can be several kilometers in radius
and have a central density as high as $(10-15) \rho_0$.
The composition and equation of state of the inner core
are poorly known.
Several hypotheses have been
discussed
in the literature and it is impossible to reject any of them
at present:

(1) Large proton fraction ($>11\%$) and/or hyperonization
of matter ---  the appearance of
$\Sigma$, $\Lambda$ and other hyperons
(see, e.g.,
Shapiro and Teukolsky 1983, Balberg and Barnea 1998,
Balberg et al.\ 1999). The fractions of
$p$ and $e$ may become so high that the powerful $npe$ direct Urca
process of neutrino emission becomes allowed
(Lattimer et al.\ 1991) as well as the similar reactions
involving hyperons
(Prakash et al.\ 1992).
%
%
%
In these cases the neutrino luminosity
of the star is enhanced by 5--6 orders of magnitude
(Sect.\ \ref{sect-nucore-Durca}) compared to the standard neutrino
luminosity produced mainly by the
modified Urca processes (Sect.\ \ref{sect-nucore-Murca}).
This accelerates considerably
the cooling of neutron stars.

(2) The second hypothesis,
proposed in different forms by
Bahcall and Wolf (1965a, b), Migdal (1971), Sawyer (1972),
and Scalapino (1972),
%
%
%
%
%
%
assumes the appearance of 
pion condensation. The condensates soften the equation of state
and enhance the neutrino luminosity by allowing
the reactions of the direct Urca type
(Sect.\ \ref{sect-nucore-exotica}).
However, many modern microscopic
theories of dense matter predict weakly polarized pionic degrees
of freedom which are not in favor of pion condensation
(Pethick, 1992).

(3) The third hypothesis predicts a
phase transition to the strange quark matter composed
of almost free $u$, $d$ and $s$ quarks with small admixture of
electrons 
(see Weber 1999 for review).
%
%
%
%
%
In these cases the neutrino
luminosity is thought to be considerably higher than the standard
luminosity due to switching on the direct Urca processes
involving quarks (Sect.\ \ref{sect-nucore-exotica}).

(4) Following
Kaplan and Nelson (1986), Nelson and Kaplan (1987)
and Brown et al.\ (1988),
%
%
%
%
several authors considered the hypothesis of kaon condensation.
Similarly to pion condensates,
kaon condensates may also enhance the neutrino
luminosity by several orders of magnitude
(Sect.\ \ref{sect-nucore-exotica}).
A critical analysis of the theories of kaon condensation
was given by Pandharipande et al.\ (1995).
%
%
The present state of the theories has been reviewed,
for instance, by Ramos et al.\ (2000).

We see that the composition of dense matter
affects the neutrino emission and, hence, neutron star cooling.
These effects are discussed in the next chapters.
Notice that in all cases the matter can be in superfluid state.
For instance,
Takatsuka and Tamagaki (1993, 1997a, b, c)
calculated the neutron and proton superfluid gaps
in the matter with pion condensates.
%
%
%
%
%
%
The same authors
(Takatsuka and Tamagaki 1995)
studied nucleon superfluidity in the presence of kaon
condensation. The pion or kaon condensates mix strongly
the neutron and proton states and may induce the triplet-state
pairing of quasi-protons. 
If hyperons appear in the
$npe$-matter,
they can also be superfluid
(Balberg and Barnea 1998).
%
%
In all these cases the critical temperatures are of the same
order of magnitude as $T_{cn}$ and $T_{cp}$ in ordinary $npe$ matter.
Some authors discussed
the superfluidity of quark matter
(e.g., Bailin and Love 1984, Iwasaki 1995,
Schaab et al.\ 1996, 1997a,
%
%
%
%
%
Scha\"afer and Wilczek 1999, Rajagopal 1999).
%
%
In this case the situation is more intriguing since 
the theoretical models
predict two possibilities. The first is pairing
of like quarks leading to the superfluid gaps of a few MeV
and critical temperatures of a few times $10^{10}$ K,
not much higher than $T_c$ in $npe$ matter.
The second is pairing of different quarks 
($ud$, $us$, $ds$) which could be much more
efficient producing $k_{\rm B}T_c$ up to 50 MeV, about one-tenth
of the quark Fermi energy ($\sim 500$ MeV).
This pairing requires 
the Fermi momenta of different quarks to be sufficiently
close; it weakens with increasing the difference of
these Fermi momenta. If realized the pairing leads to
the strongest superfluidity and superconductivity
of dense matter (often referred to as color superconductivity
since Cooper pairs of quarks possess color).
Cooling of neutron stars containing
quark matter with superfluidity of two
types has been analyzed recently 
in several papers (e.g., Page et al.\ 2000).

Another complication comes from the possible
existence of strong magnetic fields in neutron star
interiors. It is widely accepted that the field
on the surfaces of
radio pulsars can be as large as $10^{13}$ G. Thompson and Duncan (1996)
predicted theoretically the existence of the so-called magnetars,
neutron stars with the magnetic fields
several orders of magnitude stronger than
in ordinary radio pulsars.
%
%
Some of the soft-gamma repeaters
or anomalous X-ray pulsars may be magnetars
(e.g., Colpi et al.\ 2000),
%
%
although this interpretation requires further confirmation
(e.g., Harding et al.\ 1999).
%
%
It is natural to assume that the internal magnetic
field can be even higher than that on the surface. The internal field 
can be
confined in the crust or be distributed over the entire star.
If protons or hyperons
are superfluid (superconducting) in the core,
the magnetic
field would likely exist in the form of fluxoids (the quantized
magnetic flux tubes, Sect. \ref{sect-nusup-fluxa}). In the absence
of superconductivity, the magnetic field
is thought to be microscopically
uniform. A strong magnetic field can affect the
neutrino reactions and stellar cooling.
In particular, a quasi-uniform magnetic field
opens a new neutrino reaction, the synchrotron
neutrino emission by electrons (Sect.\ \ref{sect-nucrust-syn}),
which is forbidden without the magnetic field.  The effects
of magnetic fields on the neutrino reactions in the
crust and core are studied in Chapts.\ \ref{chapt-nucrust},
and \ref{chapt-nusup}, respectively.
Some of the magnetic field effects on neutron star cooling
are analyzed in Chapt.\ \ref{chapt-cool}.

\newpage

\section{Neutrino emission from neutron star crusts}
\label{chapt-nucrust}

\subsection{Main neutrino reactions}
\label{sect-nucrust-introduc}

{\bf (a) Reactions and their properties}

In this chapter we consider various neutrino reactions
in neutron star crusts.
These reactions are important sources of energy loss in
a cooling neutron star at initial stages of the thermal
relaxation in the stellar interior
(first 10-100 years of the neutron star life,
Chapt.\ \ref{chapt-cool}).
In addition, studying neutrino emission from the stellar crusts
is an excellent introduction to the problem of neutrino
reactions in the stellar cores
(Chapts.\ \ref{chapt-nucore} and \ref{chapt-nusup}).

While considering neutron star crusts,
we deal with an extremely rich spectrum of physical
conditions ranging from an ``ordinary" electron-ion plasma
in the outermost surface layers to the strongly
coupled plasma of atomic nuclei (or their clusters),
electrons and free neutrons near the crust-core
interface (Sect.\ \ref{sect-overview-struct}). The problem is
complicated further by the possible presence of very strong magnetic
fields and nucleon superfluidity (of free neutrons
and nucleons within atomic nuclei) in the inner crust.
Accordingly, there is a variety of neutrino emission
mechanisms that may be important in different
crust layers for certain temperature intervals.
In order to simplify our consideration,
we mainly restrict ourselves only to those mechanisms
which may affect the thermal evolution of
neutron stars (Chapt.\ \ref{chapt-cool}). Therefore
we will {\it not} consider in detail the neutrino emission
from the outer layers of density $\rho \lesssim 10^{10}$ g cm$^{-3}$.
These layers contain a negligible fraction of the neutron
star mass and cannot be the sources of significant neutrino energy
losses. Moreover, we will consider
temperatures $T \lesssim 10^{10}$ K
which are expected in the crusts of cooling neutron
stars older than, say, one minute.
At such temperatures, neutron stars become fully transparent to neutrinos.
In addition, there is no need to study too low
temperatures, $T \lesssim 3 \times 10^6$ K, because
the neutrino emission becomes too weak
to affect the stellar evolution.

Therefore, under the conditions
of study, the electrons constitute strongly degenerate,
ultrarelativistic, almost ideal electron gas. This
greatly simplifies our treatment of the neutrino emission.
The atomic nuclei mainly
form strongly coupled Coulomb liquid or crystal, and may
form a liquid crystal near the crust base. Free neutrons
which appear in the inner crust constitute a strongly
interacting 
 Fermi liquid. All of these properties of the  matter
affect the neutrino emission.

Our primary goal is to obtain the {\it neutrino
emissivity} $Q$ (energy carried away by neutrinos and antineutrinos
per second per unit volume) in various reactions.
It is the total emissivity which is the most important
for the thermal evolution of neutron stars
(Chapt.\ \ref{chapt-cool}). Thus, we will study
neither the spectrum, nor the angular 
distribution of the emitted neutrinos,
nor the neutrino
scattering and propagation in the plasma (which would be
most important for proto-neutron stars, see, e.g., 
Prakash et el.\ 1997).

The calculations of the emissivities are based on the
Weinberg-Salam-Glashow theory of electroweak interactions
(e.g., Okun' 1984). 
Typical energies of the emitted neutrinos are 
of the order of the thermal energy,
$k_{\rm B}T$, or higher. For temperatures 
of practical interest, $T \gtrsim 3 \times 10^6$ K,
these energies are higher than the possible neutrino masses.
Accordingly, one can employ the approximation of massless neutrinos
while calculating the neutrino emissivity.
In practical expressions
we will take into account the generation of neutrinos and
antineutrinos of three flavors ($\nu_e$, $\nu_\mu$, and
$\nu_\tau$) although the expressions
will be presented in the form ready to incorporate any number
of neutrino flavors.
On the other hand, the thermal energies of interest
are much smaller than the intermediate boson mass,
$\sim$ 80 GeV. This enables us to use the reduced
4-tail Feynman weak interaction diagrams instead of the more
complicated pairs of three-tail diagrams tied by the intermediate-boson
exchange line.

The neutrino processes in the crusts of cooling neutron stars
have much in common with those in normal stars at the late
stages of their evolution (particularly, in presupernovae)
or in the cores of white dwarfs. These processes have been
studied since the beginning of the 1960s in a number of
seminal papers (see, e.g., Pinaev 1963,
%
%
Fowler and Hoyle 1964,
%
%
Beaudet et al.\ 1967,
and references therein, as well as references in subsequent
sections of this chapter). The early studies were conducted
in the framework of an old, simplified Fermi theory of weak interactions
and required revision in the 1970s after the Weinberg-Salam-Glashow
theory was widely accepted. However, the revision consisted
only of simple replacements of the weak interaction normalization
constants associated with the inclusion of neutral currents.
It did not destroy the earlier results. Thus, we will not
put special emphasis on the type of the weak interaction
theory, used in the references, assuming the reader
will not be confused by the somewhat different
normalization constants.
The general neutrino processes in stellar interior have been reviewed
by several authors, for instance, by Imshennik and Nadyozhin
(1982), Bisnovatyi-Kogan (1989), Itoh et al.\ (1996).
%
%
%

The main neutrino reactions in neutron star
crusts
are listed in Table \ref{tab-nucrust-list}.
The electron-positron pair annihilation, plasmon decay,
electron synchrotron neutrino emission,
and photoneutrino emission involve
electrons (positrons) and collective electromagnetic modes
(associated mainly with electrons). 
The appropriate neutrino emissivities can be calculated
precisely as a function of {\it two parameters}, the temperature and electron
number density. 
We can precisely determine these emissivities
for any given model of dense matter.

The next reaction in Table \ref{tab-nucrust-list} is
the neutrino bremsstrahlung due to scattering
of electrons off atomic nuclei. It is the process
based solely on weak and electromagnetic interactions
but its emissivity depends on the correlations between the nuclei
and on the proton charge distribution within the nuclei.
Thereby, it is linked to
a specific model of dense matter,
although it does not vary
strongly from one model to another.

Other processes in Table \ref{tab-nucrust-list}
involve weak nucleon reactions.
They are subdivided into two
groups: the beta decay reactions (including Urca processes)
and the processes connected with strong interaction
of free neutrons in the inner neutron star crust
(neutrino emission in neutron-neutron collisions,
in neutron-nucleus collisions,
and due to Cooper pairing of free neutrons). The
emissivities of these processes depend on
a microscopic model of the matter (e.g., on
the critical temperature of superfluidity of free neutrons).\\

\begin{table}[t]
\caption{Main neutrino processes in a neutron star crust$^{\ast)}$}
\begin{center}
  \begin{tabular}{||llll||}
  \hline \hline
  No.   & Process   & {} & Sect.  \\
  \hline
 1 & $e^-e^+$ pair annihilation      & $e e^+ \to \nu \bar{\nu}$
                                     & \protect{\ref{sect-nucrust-annih}}  \\
 2 & plasmon decay                   & $ \gamma \to \nu \bar{\nu}$
                                     & \protect{\ref{sect-nucrust-plasmon}}  \\
 3 & electron synchrotron            & $ e \to  e \nu \bar{\nu}$
                                     & \protect{\ref{sect-nucrust-syn}}  \\
 4 & photoneutrino emission          &
                                       $ e + \gamma \to  e \nu \bar{\nu}$
                                     & \protect{\ref{sect-nucrust-egamma}} \\
 5 & electron-nucleus bremsstrahlung & $ e (A,Z) \to  e (A,Z) \nu \bar{\nu}$
                                     & \protect{\ref{sect-nucrust-ebrems}} \\
 6 & beta processes (including Urca) & $ e (A,Z) \to (A,Z-1) \nu_e  \quad
                                         (A,Z-1) \to (A,Z) e \bar{\nu}_e$
                                     & \protect{\ref{sect-nucrust-beta}}\\
 7 & Cooper pairing of neutrons      & $ nn \to \nu \bar{\nu}$
                                     & \protect{\ref{sect-nucrust-nn}} \\
 8 & neutron-neutron bremsstrahlung  & $ nn \to nn \nu \bar{\nu}$
                                     & \protect{\ref{sect-nucrust-nn}} \\
 9 & neutron-nucleus bremsstrahlung  & $ n(A,Z) \to n(A,Z) \nu \bar{\nu}$
                                     & \protect{\ref{sect-nucrust-nn}} \\
  \hline \hline
  \end{tabular}
\begin{tabular}{l}
  $^{\ast)}$  $\gamma$ means a plasmon or photon; $(A,Z)$ stands for
  an atomic nucleus
\end{tabular}
\label{tab-nucrust-list}
\end{center}
\end{table}

{\bf (b) Illustrative model: ground-state matter}

For illustration purposes, we will mainly use
the model of {\it ground-state matter} in the neutron star crust.
This state of matter is energetically most favorable.
Its properties
will be described using the results of Haensel and Pichon (1994)
at densities below the neutron drip density and
the results of Negele and Vautherin (1973) and Oyamatsu (1993)
at higher densities.
The model assumes that only one nuclear species
is present at a given density (or pressure),
which leads to discontinuous variations of the nuclear composition with
density (pressure).
Thermal effects on the composition of dense matter are small
and can be ignored.

The model of ground-state matter 
(discussed in Sect.\ \ref{sect-nucrust-beta} in more details)
is based on the properties of atomic nuclei 
The nuclei can be treated as spherical
almost everywhere in the crust although
they might be non-spherical in the deepest layer of the inner crust
($10^{14}$ g cm$^{-3} \leq \rho \leq 1.5 \times 10^{14}$ g cm$^{-3}$).
We describe this layer
adopting model I of Oyamatsu (1993).
The layer consists of four sublayers.
The first sublayer below the crust of spherical nuclei
contains the rod-like nuclei. It is followed by a sublayer of
the slab-like nuclei,
and by the two sublayers with
the roles of the nuclear matter and
neutron matter reversed, the rod-like one
(neutron gas tubes in nuclear matter)
and the ``Swiss cheese" one
(spherical neutron gas bubbles in nuclear matter).
The latter is an analog of the phase of
spherical nuclei and is the last phase in the neutron-star crust.
At higher density the nuclei dissolve into the
uniform matter of the neutron star core.

The properties of atomic nuclei in all layers of neutron
star crusts are conveniently described
by introducing the Wigner--Seitz cells of
different geometries (e.g., spheres for the spherical nuclei,
or cylinders for the rod-like ones).
Following Oyamatsu (1993), we employ
the parameterization of the local
neutron and proton number density distributions
within a Wigner--Seitz cell of the form
\begin{equation}
  n_j(r)= \left\{
     \begin{array}{ll}
        (n_j^{\rm in} - n_j^{\rm out})
        \left[ 1 - \left(  r \over R_j
        \right)^{t_j} \right]^3 + n_j^{\rm out}, &  r< R_j, \\
        n_j^{\rm out},           & r \geq R_j,
     \end{array}
        \right.
\label{nucrust-Oya}
\end{equation}
where $r$ is the distance from the cell center
(e.g., from the center of the sphere or from the
axis of the cylinder),
$j=n$ or $p$, and $n_j^{\rm in}$, $n_j^{\rm out}$,
$t_j$ and $R_j$ are the adjustable parameters.
These parameters are presented by Oyamatsu (1993)
at several values of the density $\rho$ 
for the spherical and nonspherical nuclei.
For spherical nuclei, they are consistent with those
following from the results of Negele and Vautherin (1973).

Kaminker et al.\ (1999a) interpolated the fit parameters
(separately within each phase).
The interpolation smears out jumps in the nuclear composition
with increasing $\rho$, but they have little effect
on the neutrino emission.
Notice that $n_n^{\rm out}=0$ at $\rho$ below the neutron drip density.
In the phases with
spheres, rods and slabs, $n_p^{\rm out}=0$,
and $n_n^{\rm out}$ describes the number density
of free neutrons, while the region $r < R_n$ is occupied
by the nucleus itself (with $n_n^{\rm in} > n_n^{\rm out}$).
In the two last ``bubble" phases with the roles of nuclear matter and
neutron matter reversed,
$n_p^{\rm out} \neq 0$, and $n_j^{\rm out} > n_j^{\rm in}$,
i.e., the local number density of neutrons and protons
increases with distance $r$ from the center of the Wigner--Seitz
cell. With increasing $\rho$, the nucleon density
profiles become smoother, resembling uniform matter.
Thus, Kaminker et al.\ (1999a) obtained a simple analytic description
of the local density profiles of 
neutrons and protons for the
ground-state matter throughout
the neutron star crust. This description
will be referred to as the {\it smooth
composition} model of ground-state matter.\\

{\bf (c) Illustrative figures}

\begin{figure}[ht!]
\begin{center}
\leavevmode
\vskip 0.8cm
\hskip -1cm
\epsfxsize=8.0cm
\epsfbox{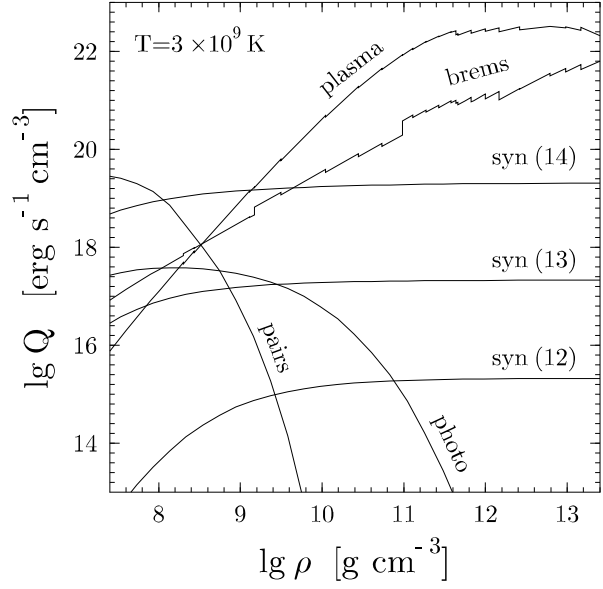}
\vskip -1cm
\end{center}
\caption[ ]{
Density dependence of neutrino emissivities
from ground-state matter of the neutron-star crust
due to various mechanisms at $T= 3 \times 10^9$~K.
Curves `syn (12)', `(13)', and `(14)'
refer to the synchrotron mechanism
(Sect.\ \protect{\ref{sect-nucrust-syn}}) at
$B=$ 10$^{12}$, 10$^{13}$ and 10$^{14}$~G, respectively.
Curve `pairs' corresponds to the neutrino emission due
to the annihilation of electron-positron pairs; it is almost
independent of $B$ at given $T$. Other curves are for $B=0$:
`brems' --- electron--nucleus bremsstrahlung
(Sect.\ \protect{\ref{sect-nucrust-ebrems}});
`plasma' --- plasmon decay (Sect.\ \protect{\ref{sect-nucrust-plasmon}});
`photo' --- photoneutrino process
(Sect.\ \protect{\ref{sect-nucrust-egamma}}).
}
\label{nucrust-fig93}
\end{figure}

\begin{figure}[ht!]
\begin{center}
\leavevmode
\vskip 0.8cm
\hskip -1cm
\epsfxsize=8.0cm
\epsfbox{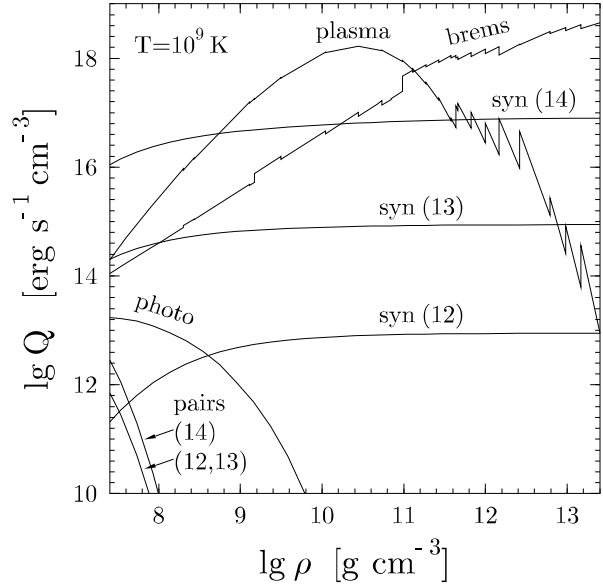}
\vskip -1cm
\end{center}
\caption[ ]{
Same as Fig.\ \protect{\ref{nucrust-fig93}}, but
at $T= 10^9$~K.
Pair annihilation depends noticeably on $B$.
}
\label{nucrust-fig91}
\end{figure}

\begin{figure}[ht!]
\begin{center}
\leavevmode
\vskip 0.8cm
\hskip -1cm
\epsfxsize=8.0cm
\epsfbox{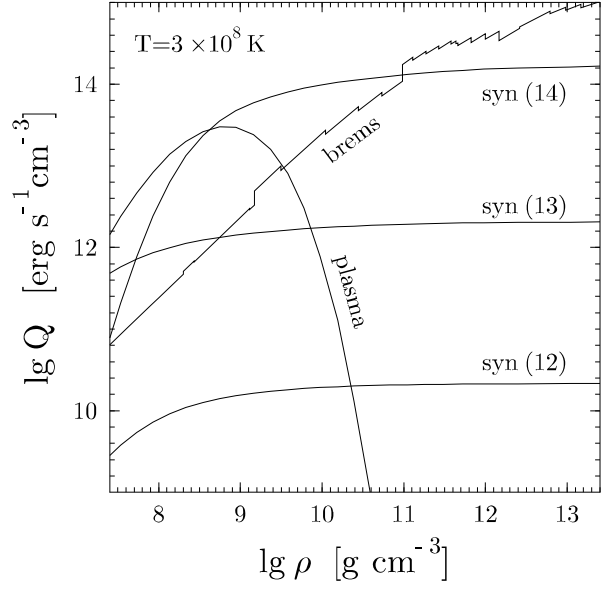}
\vskip -1cm
\end{center}
\caption[ ]{
Same as Fig.\ \protect{\ref{nucrust-fig93}}, 
but at $T= 3 \times 10^8$~K.
Pair annihilation and photoneutrino
processes become negligible.
}
\label{nucrust-fig83}
\end{figure}

\begin{figure}[ht!]
\begin{center}
\leavevmode
\vskip 0.8cm
\hskip -1cm
\epsfxsize=8.0cm
\epsfbox{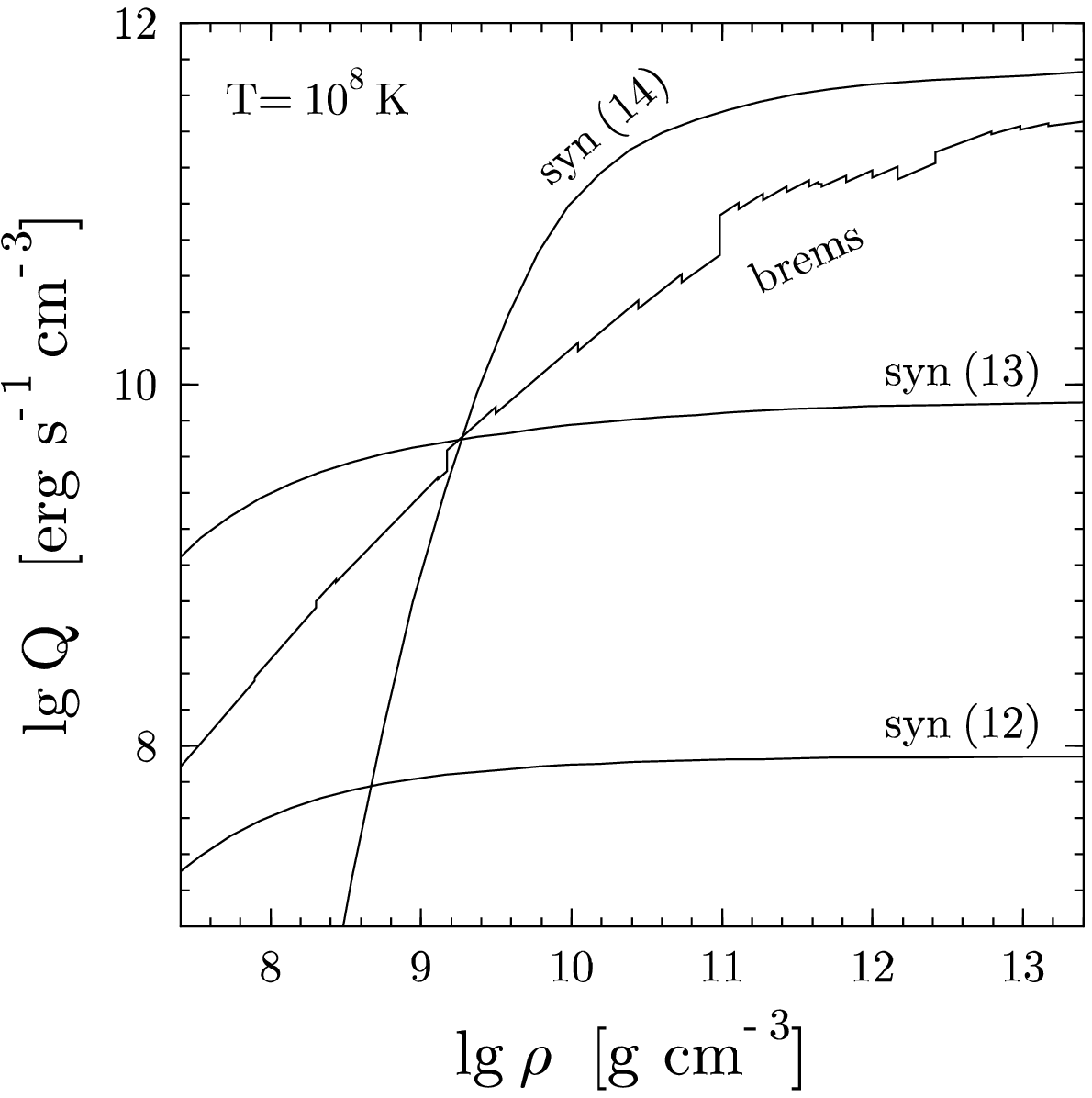}
\vskip -1cm
\end{center}
\caption[ ]{
Same as Fig.\ \protect{\ref{nucrust-fig93}}, 
but at $T= 10^8$~K.
Plasmon decay becomes negligible.
}
\label{nucrust-fig81}
\end{figure}

To make our analysis less abstract, in
Figs.\ \ref{nucrust-fig93}--\ref{nucrust-fig81} we display
the density dependence of the neutrino emissivity of the
main neutrino processes
for the four values of the temperature,
$T=3 \times 10^9$, $10^9$, $3 \times 10^8$,
and  $10^8$~K, respectively.
We adopt the ground--state model of matter
described above
and extend the density range
to somewhat lower values
in order to show clearly the efficiency of different
neutrino mechanisms. We exclude
densities higher than $3 \times 10^{13}$
g cm$^{-3}$
(near the bottom of the neutron star
crust); they will be considered in
Sects.\ \ref{sect-nucrust-ebrems}--\ref{sect-nucrust-overlook}.

The neutrino emissivities due to the pair annihilation
(`pairs', Sect.\ \ref{sect-nucrust-annih})
and synchrotron radiation of electrons
(`syn', Sect.\ \ref{sect-nucrust-syn})
are presented for the three values of the
magnetic field $B=10^{12}$, $10^{13}$ and $10^{14}$ G
in a magnetized neutron-star crust.
At the highest temperature,
$T=3 \times 10^9$ K, the pair-annihilation
emissivity is actually independent
of $B$ as long as $B \lesssim 10^{14}$ G.
Other curves are
calculated neglecting the effect of the magnetic fields
(for $B=0$). The curve `plasma' refers to the neutrino plasmon decay
process (Sect.\ \ref{sect-nucrust-plasmon}),
`photo' corresponds to the photoneutrino
reaction (Sect.\ \ref{sect-nucrust-egamma}).
The emissivity of the latter process is taken from
Itoh et al.\ (1989, 1996). Finally, `brems' is
the neutrino bremsstrahlung due to the electron-nucleus scattering.
Its emissivity is calculated using the formalism
described in Sect.\ \ref{sect-nucrust-ebrems}.
The jumps of the 
curves in Figs.\ \ref{nucrust-fig93}--\ref{nucrust-fig81} are 
mainly associated 
with the changes of nuclear composition.
These changes affect each curve in a different way.
Neutrino bremsstrahlung due to the electron-nucleus scattering
is affected in the entire density range displayed, since the jumps
of the atomic charge number influence directly Coulombic electron-nucleus
interaction. 
The neutrino emissivities of other
processes are determined by the electron and positron number
densities which are almost continuous at low 
densities, below the neutron drip density, but have
quite pronounced jumps at higher densities.
Thus, the appropriate curves
are nearly continuous below the neutron drip density.
At higher densities, the emissivity of the synchrotron radiation is almost
independent of the electron number density
and thus remains smooth, while the emissivity of
the plasmon decay reflects the jumps of
the electron number density (see Sect.\ \ref{sect-nucrust-plasmon}).

We return to Figs.\ \ref{nucrust-fig93}--\ref{nucrust-fig81}
while analyzing various neutrino processes in
subsequent sections.
We adopt the units in which $\hbar = c = k_{\rm B} =1$,
but give the final expressions in the standard physical
units.

%

%

%

%

\subsection{Annihilation of electron-positron pairs}
\label{sect-nucrust-annih}

{\bf (a) Emissivity}

We start with the process of neutrino-pair
emission due to the annihilation of electron-positron
pairs,
\begin{equation}
   e + e^+ \to \nu + \bar{\nu}.
\label{nucrust-annih}
\end{equation}
The process was proposed by Chiu and Morrison (1960)
%
%
and independently the same year by M.\ Levine (in a private communication
mentioned by Fowler and Hoyle 1964).

Although we consider a strongly degenerate electron
gas (Sect.\ \ref{sect-nucrust-introduc}), it contains 
a tiny fraction
of positrons determined by the condition of thermodynamic equilibrium.
The positrons and electrons annihilate
producing neutrino emission. This
process is most efficient in the low-density and high-temperature
plasma, where the positron fraction is the highest.
Indeed, calculations show that the process is extremely
efficient in a non-degenerate plasma of temperature
$T \gtrsim 10^{10}$ K, which we do not study
in detail. In a strongly degenerate electron plasma,
the process is suppressed because of the negligibly small
positron fraction. Unlike more complicated neutrino
processes, this one can be calculated
very accurately as a function of only two parameters, 
the temperature and electron number density. 

The pair annihilation is the simplest neutrino process
described by one four-tail Feynman diagram.
The interaction Hamiltonian is
\begin{equation}
    \hat{H}= {G_{\rm F} \over \sqrt{2}}\, J_\alpha l^\alpha,
\label{nucrust-annih-H}
\end{equation}
where $G_{\rm F}=1.436 \times 10^{-49}$ erg cm$^3$ is the Fermi
weak interaction constant,
\begin{equation}
     l^\alpha = \bar{\psi}'_\nu \gamma^\alpha (1 + \gamma^5) \psi_\nu
\label{nucrust-neutrino-current}
\end{equation}
is the neutrino weak 4-current ($\alpha$ runs the values 0, 1, 2, and 3,
and the signature ($+1,-1,-1,-1$) is adopted
to define 4-vectors), $\gamma^\alpha$ and $\gamma^5$ are the Dirac
matrices, $\psi_\nu$ is the neutrino wave function,
$\psi'_\nu$ is the antineutrino wave function, and $\bar{\psi}'_\nu$
denotes the Dirac conjugate. 
We define $\gamma^5=-i\gamma^0 \gamma^1 \gamma^2 \gamma^3$
in accordance with Berestetskii et al.\ (1982)
although the definition with opposite sign is 
also used in the literature.
The wave functions are taken
in the form
\begin{equation}
  \psi_\nu =  {u_\nu \over \sqrt{2 \epsilon_\nu}} \;
              {\rm e}^{i p_\nu x},
  \quad
  \psi'_\nu = {u'_\nu \over \sqrt{2 \epsilon'_\nu}} \;
              {\rm e}^{-i p'_\nu x},
\label{nucrust-psi-neutrino}
\end{equation}
where $x=(t,{\bf r})$ is the 4-vector of time-space coordinate,
$p_\nu=(\epsilon_\nu,{\bf p}_\nu)$ is the 4-momentum of the neutrino,
$p'_\nu=(\epsilon'_\nu,{\bf p}'_\nu)$ the 4-momentum of the antineutrino,
while $u_\nu$ and $u'_\nu$ are the standard bispinors.
Finally, $J=(J^0,{\bf J})$ in Eq.\ (\ref{nucrust-annih-H})
is the weak electron-positron 4-current:
\begin{equation}
     J^\alpha = \bar{\psi}'_e \gamma^\alpha (C_V + C_A \gamma^5) \psi_e.
\label{nucrust-electron-current}
\end{equation}
Here
\begin{equation}
   \psi_e = {u_e \over \sqrt{2 \epsilon_e}} \; {\rm e}^{-i p_e x},
   \quad
   \psi'_e = {u'_e \over \sqrt{2 \epsilon'_e}} \;
               {\rm e}^{i p'_e x}
\label{nucrust-psi-electron}
\end{equation}
are the wave functions
of the electron and positron, respectively;
$p_e=(\epsilon_e,{\bf p}_e)$ and $p'_e=(\epsilon'_e,{\bf p}'_e)$
are particle 4-momenta, $\bar{u}_e u_e = \bar{u}'_e u'_e = 2 m_e$,
$m_e$ being the electron rest-mass.
All the wave functions are
normalized in the ordinary quantum-mechanical manner,
per particle per unit volume.

According to Eq.\ (\ref{nucrust-electron-current}) the weak
electron current consists of the vector and axial-vector terms
containing, respectively, the vector and axial-vector
constants, $C_V$ and $C_A$. The process can produce
neutrino pairs of any flavor, $\nu_e \bar{\nu}_e$,
$\nu_\mu \bar{\nu}_\mu$ and  $\nu_\tau \bar{\nu}_\tau$.
Emission of the electron neutrinos goes via either 
charged or neutral electroweak currents. Under the conditions
of study (in which four-tail Feynman diagrams are
applicable, Sect.\ \ref{sect-nucrust-introduc}) the amplitudes
of both reaction channels are summed coherently,
and one has $C_{Ve}=2 \sin^2 \theta_{\rm W} + 0.5$,
$C_{Ae}=0.5$, where $\theta_{\rm W}$ is the Weinberg angle,
$\sin^2 \theta_{\rm W}=0.23$.
Emission of the muon or tau neutrinos can only go
through neutral electroweak currents, so that
$C_{V\mu}=C_{V\tau}= 2 \sin^2 \theta_{\rm W} - 0.5$ and
$C_{A\mu}=C_{A\tau}= -0.5$.

The general expression for the neutrino emissivity $Q_{\rm pair}$
follows from the Fermi Golden Rule:
\begin{equation}
     Q_{\rm pair} = (2\pi)^4 \; { G_{\rm F}^2 \over 2} 
     \sum_\nu \int
     { {\rm d}{\bf p}_e \over (2\pi)^3} \,
     { {\rm d}{\bf p}'_e \over (2\pi)^3} \,
     { {\rm d}{\bf p}_\nu \over (2\pi)^3} \,
     { {\rm d}{\bf p}'_\nu \over (2\pi)^3} \,
     (\epsilon_\nu+\epsilon'_\nu)\, f_e f'_e \;
     \delta^{(4)}(p_e + p'_e - p_\nu - p'_\nu) \,
     {\cal L}^{\alpha \beta} \, {\cal J}_{\alpha \beta},
\label{nucrust-annih-Q}
\end{equation}
where $f_e$ and $f'_e$ are the Fermi-Dirac distributions
of the electron and positron, and the summation
is over all neutrino flavors.
The expression contains 4-tensors
${\cal L}^{\alpha \beta}$ and ${\cal J}^{\alpha \beta}$
composed of bilinear combinations of
the matrix elements
$l^\alpha_{fi}$ and $J^\alpha_{fi}$ of the neutrino
and electron currents, respectively (with appropriate
initial $i$ and final $f$ states):
\begin{equation}
     {\cal J}^{\alpha \beta}  =
     \sum_{\rm spins} (J_{fi})^{\alpha \ast}\,(J_{fi})^{\beta},\quad
     {\cal L}^{\alpha \beta}  =
     (l_{fi})^{\alpha \ast}\,(l_{fi})^{\beta},
\label{nucrust-Jab}
\end{equation}
where the summation extends over the electron and positron spin states.

Using Eqs.\ (\ref{nucrust-neutrino-current}) and
(\ref{nucrust-psi-neutrino}), and the standard
prescriptions of quantum electrodynamics,
we have
\begin{eqnarray}
    {\cal L}^{\alpha \beta} & = &
    { 1 \over 4 \epsilon_\nu \epsilon'_\nu} \,
    \left[ \bar{u}'_\nu \gamma^\alpha (1+\gamma^5) u_\nu \right]^\ast
    \left[ \bar{u}'_\nu \gamma^\beta (1+\gamma^5) u_\nu \right]
\nonumber \\
    & = &
    { 1 \over 4 \epsilon_\nu \epsilon'_\nu} \,
    {\rm Tr} \left[ 
    (\gamma p_\nu) \, \gamma^\alpha (1+\gamma^5) \, (\gamma p'_\nu)\,
    \gamma^\beta (1 + \gamma^5) \right]
\nonumber \\
    & = &
    { 2 \over \epsilon_\nu \epsilon'_\nu} \,
    \left[ p_\nu^\alpha p_\nu^{\prime \beta}
          +p_\nu^\beta  p_\nu^{\prime \alpha}
          - g^{\alpha \beta} (p_\nu p'_\nu)
          - i \, e^{\alpha \beta \lambda \rho}
           p_{\nu \lambda} p'_{\nu \rho} \right].
\label{nucrust-ll}
\end{eqnarray}
Here, the bilinear combinations of neutrino bispinors and
antineutrino bispinors are replaced by the
neutrino and antineutrino polarization density matrices,
$(\gamma p_\nu)$ and $(\gamma p'_\nu)$;
$g^{\alpha \beta}$ is the metric tensor,
$e^{\alpha \beta \lambda \rho}$ is the antisymmetric
unit tensor, $(pp') \equiv p^\alpha p'_\alpha$ and
$(\gamma p) \equiv \gamma^\alpha p_\alpha$.

Tensor ${\cal J}^{\alpha \beta}$, associated with
the electron-positron current, is calculated in a similar fashion.
Now the bilinear combination of the electron bispinors must be replaced
by the electron polarization density matrix, $[(\gamma p_e)+m_e]/2$,
and the bilinear combination of the positron bispinors
by the positron density matrix, $[(\gamma p'_e)-m_e]/2$.
This gives
\begin{eqnarray}
    {\cal J}^{\alpha \beta} & = &
    { 1 \over 4 \epsilon_e \epsilon'_e } \, 
    {\rm Tr} \left\{
    [(\gamma p_e)+ m_e] \, \gamma^\alpha (C_V+ C_A \,\gamma^5) \,
    [(\gamma p'_e)-m_e] \, \gamma^\beta (C_V+ C_A \,\gamma^5)
    \right\}
\nonumber \\
    &=& { 1 \over \epsilon_e \epsilon'_e} \,
    \left\{ (C_V^2 + C_A^2) [p_e^\alpha p_e^{\prime \beta}
          +p_e^\beta  p_e^{\prime \alpha}
          -  g^{\alpha \beta} (p_e p'_e)]
    \right.
\nonumber \\
    & & \left. - m_e^2 (C_V^2-C_A^2) g^{\alpha \beta}
        - 2 i \, C_V C_A \, e^{\alpha \beta \lambda \rho}
           p_{e \lambda} p'_{e \rho} \right\} .
\label{nucrust-JJ}
\end{eqnarray}

From Eqs.\ (\ref{nucrust-ll}) and (\ref{nucrust-JJ}),
we have
\begin{eqnarray}
  {\cal L}^{\alpha \beta} {\cal J}_{\alpha \beta}
  & = & {4 \over \epsilon_e \epsilon'_e \epsilon_\nu \epsilon'_\nu}
    \;  \left[ (p_e p_\nu)(p'_e p'_\nu) (C_V+C_A)^2 \right.
\nonumber \\
   &  & + \left. 
    (p_e p'_\nu)(p'_e p_\nu) (C_V-C_A)^2
    + m_e^2 (C_V^2-C_A^2)(p_\nu p'_\nu) \right].
\label{nucrust-llJJ}
\end{eqnarray}
Integrating  Eq.\ (\ref{nucrust-annih-Q})
over ${\rm d}{\bf p}_\nu$ and ${\rm d}{\bf p}'_\nu$
with Lenard's identity (e.g., Berestetskii et al.\ 1982)
%
%
\begin{equation}
  \int {{\rm d}{\bf p} \; {\rm d}{\bf p}'
        \over \epsilon \epsilon'} \,
        p^\alpha p^{\prime \beta} \, \delta^{(4)}(k-p-p')=
        {\pi \over 6} \,
        \left(k^2 g^{\alpha \beta} + 2 k^\alpha k^\beta \right),
\label{nucrust-Lenard}
\end{equation}
where $k=p_e+p'_e$
is the 4-vector of the neutrino-pair momentum, we find
the emissivity 
\begin{eqnarray}
  Q_{\rm pair} & = & { 2 \, G^2_{\rm F} \over 3 \, (2 \pi)^7} \,
      \int {\rm d}{\bf p}_e \; {\rm d}{\bf p}'_e \,
      f_e f'_e \;
      { \epsilon_e + \epsilon'_e \over \epsilon_e \epsilon'_e } \,
\nonumber \\
     & & \times
      \left\{ C_+^2 \left[m_e^4+3m_e^2(p_e p'_e) + 2 (p_e p'_e)^2 \right]
            + 3 m_e^2 C_-^2 \left[m_e^2+(p_e p'_e) \right] \right\},
\label{nucrust-annih-Q1}
\end{eqnarray}
where $C_+^2= \sum_\nu (C_V^2+ C_A^2)=1.678$ and
$C_-^2 = \sum_\nu (C_V^2 - C_A^2)= 0.1748$. The numerical values
of $C_+^2$ and $C_-^2$ are obtained for the three neutrino flavors
with the values of $C_V$ and $C_A$ given above.
One can easily recalculate $C_\pm^2$ for any selected neutrino flavors.
Thus, we arrive at the equation for the
neutrino emissivity valid for any electron and positron
distributions $f_e$ and $f'_e$.

Generally, the 4-vector of the neutrino-pair momentum must be time-like,
\begin{equation}
   k^2 = \omega^2 - {\bf k}\cdot{\bf k} \geq 0.
\label{nucrust-annih-kk}
\end{equation}
This condition imposes the {\it kinematic restriction}
on the integration domain in momentum space in Eq.\ (\ref{nucrust-annih-Q1}).
Such restrictions can
be strong for some neutrino reactions, but not
for the pair annihilation: one always has
$(p_e+p'_e)^2>0$, and
the reaction is always allowed.
Since $\omega=\epsilon_e + \epsilon'_e$, the energy
of the emitted neutrino pair is always higher than $2 m_e c^2$.

In our case
\begin{equation}
   f_e=\left[ \exp \left( \epsilon_e - \mu_e \over k_{\rm B} T \right)
             + 1 \right]^{-1}, \quad
   f'_e=\left[ \exp \left( \epsilon'_e + \mu_e \over k_{\rm B} T \right)
             + 1 \right]^{-1}
\label{nucrust-annih-f}
\end{equation}
are the equilibrium Fermi-Dirac distributions 
independent of the orientations ${\bf p}_e$ and ${\bf p}'_e$
($\mu_e$ and $\mu'_e=-\mu_e$ are the electron and positron chemical potentials,
respectively). Therefore,
we can simplify the expression for $Q_{\rm pair}$
by integrating it
over the angles and writing it down 
as a sum of the decoupled
integrals over $p_e$ and $p'_e$.
In standard physical units 
\begin{eqnarray}
    Q_{\rm pair} & = & {Q_c \over 36 \pi} \,
    \left\{ C_+^2 \left[ 8(\Phi_1 U_2 + \Phi_2 U_1)
           - 2(\Phi_{-1} U_2 + \Phi_2 U_{-1})
	   + 7(\Phi_0 U_1 + \Phi_1 U_0) \right. \right.
\nonumber \\	    
	  & & + \; \left. \left. 5 (\Phi_0 U_{-1}+ \Phi_{-1} U_0) \right]
           +   9 C_-^2 \left[\Phi_0(U_1+U_{-1}) +
           (\Phi_{-1} + \Phi_1)U_0 \right] \right\},
\label{nucrust-annih-Q2}
\end{eqnarray}
where
\begin{equation}
   Q_c= { G_{\rm F}^2 \over \hbar} \,
        \left( m_e c \over \hbar \right)^9
      = 1.023 \times 10^{23} \;\;{\rm erg \; cm^{-3} \; s^{-1}}
\label{nucrust-annih-Qc}
\end{equation}
is a convenient combination of the fundamental constants
(``the electron Compton neutrino emissivity"),
while the dimensionless functions $U_k$ and $\Phi_k$ ($k$= $-1$, 0, 1, 2)
are defined by the one-dimensional integrals
\begin{equation}
   U_k= { 1 \over \pi^2} \, \int_0^\infty
   { p_e^2 \, {\rm d}p_e \over (m_e c)^3 } \, 
   \left( \epsilon_e \over m_e c^2 \right)^k \, f_e , \quad
   \Phi_k= { 1 \over \pi^2} \, \int_0^\infty
   { p^{\prime2}_e \, {\rm d}p'_e \over (m_e c)^3 } \, 
   \left( \epsilon'_e \over m_e c^2 \right)^k \, f'_e .
\label{nucrust-annih-U}
\end{equation}
They are evidently the thermodynamic functions of
the electron and positron gases, respectively.
For instance, $U_0$ and $\Phi_0$ determine
the number densities of electrons and positrons, respectively:
\begin{equation}
  n_e= \left( m_e c \over \hbar \right)^3 U_0, \quad
  n^+_e=\left( m_e c \over \hbar \right)^3 \Phi_0.
\label{nucrust-annih-ne}
\end{equation}

Therefore, the neutrino emissivity
produced by the pair annihilation
is expressed quite generally through the {\it thermodynamic functions}.
This remarkable fact is a consequence of simplicity
of the pair annihilation process.
Equation (\ref{nucrust-annih-Q2}) is valid
for the electron-positron plasma of any degree of relativism
and degeneracy. It was obtained in a somewhat different notation
in a classical paper by Beaudet et al.\ (1967).
%
%
Their result can be reproduced from Eq.\ (\ref{nucrust-annih-Q2})
by setting $C_+^2=2$ and $C_-^2=0$ in accord with the old
Feynman -- Gell-Mann theory of electroweak interactions.
They expressed the thermodynamic functions through
Fermi-Dirac integrals (which we do not do here).
Beaudet et al.\ (1967) considered also different limiting behavior of
$Q_{\rm pair}$. Subsequently, the process was
studied by a number
of authors (e.g., Dicus 1972, Munakata et al.\ 1985,
%
%
Schinder et al.\ 1987,
%
%
Blinnikov and Rudzskii 1989,
%
%
Itoh et al.\ 1989, 1996;
%
%
Kaminker and Yakovlev 1994).
We discuss some of the results below.\\

{\bf (b) Degenerate electron gas}

As noted in Sect.\ \ref{sect-nucrust-introduc},
we consider
the neutrino emission from a strongly degenerate electron
gas. In this case the electron distribution
function $f_e$ can be replaced by the step
function, and the electron thermodynamic functions
$U_k$ are evaluated analytically:
\begin{eqnarray}
  U_{-1} & = & {1 \over 2 \pi^2} \,
        \left[ y_r x_r - \ln(x_r+y_r) \right], \quad
        U_0 = { x_r^3 \over 3 \pi^2 },
\nonumber \\
   U_1 & = & {1 \over 8 \pi^2} \,
         \left[ y_r x_r (x_r^2+y_r^2) - \ln (x_r + y_r ) \right],
\quad
   U_2  =  {5 x_r^3 + 3 x_r^5 \over 15 \pi^2},
\label{nucrust-annih-U1}
\end{eqnarray}
where $x_r= p_{{\rm F}e} /(m_ec)$ 
is the familiar relativistic
parameter of the degenerate electron gas
determined by Eq.\ (\ref{pF}) and 
$y_r = \mu_e/(m_e c^2) = \sqrt{1 + x_r^2}$ is the dimensionless
chemical potential.

For the positrons we have
$f'_e \approx \exp(-(\mu_e + \epsilon'_e)/T) \ll 1$.
This means
that the positrons constitute a very dilute nondegenerate
gas. Substituting the expression for  
$f'_e$ in Eq.\ (\ref{nucrust-annih-U})
and taking the factor $\exp (-\mu_e/T)$
out of the integration, we find that the remaining integral depends
only on one parameter
$t_r=k_{\rm B}T/(m_e c^2)=0.1686 \, T_9$, 
where $T_9=T/10^9$ K, and can be expressed
through a McDonald function 
(e.g., Abramowitz and Stegun 1964). 
%
%
In the limiting cases 
$t_r \ll 1$ (nonrelativistic positrons) and
$t_r \gg 1$ (ultrarelativistic positrons) the integral
is done analytically. Kaminker and Yakovlev (1994)
%
%
propose highly accurate (error $\leq 0.08\%$)
analytic fits of $\Phi_k$ valid for any $t_r$
at which the positrons are nondegenerate:
\begin{eqnarray}
    \Phi_{-1} & = & {t_r \over 2 \pi^2} \,
    \left( 2 \pi t_r + 7.662 \, t_r^2 + 1.92 \, t_r^3
    \over 1 + 0.48 \, t_r \right)^{1/2} \,
    \exp \left(- { 1 + y_r \over t_r} \right),
\nonumber \\
    \Phi_k & = & {t_r \over 2 \pi^2} \,
    \left( 2 \pi t_r + \sum_{i=1}^{3+2 k} \,
    p_i \, t_r^{i +1} \right)^{1/2} \,
    \exp \left( - { 1 + y_r \over t_r} \right), \quad k \geq 0,
\label{nucrust-annih-fit}
\end{eqnarray}
with $p_1=23.61$, $p_2=32.11$, $p_3=16$ for $k=0$;
$p_1=42.44$, $p_2=140.8$, $p_3=265.2$, $p_4=287.9$, $p_5=144$ for $k=1$;
and $p_1=61.33$, $p_2=321.9$, $p_3=1153$, $p_4=2624$,
$p_5=4468$, $p_6=4600$, $p_7=2304$ for $k=2$.

Equations (\ref{nucrust-annih-Q2}), (\ref{nucrust-annih-U1})
and (\ref{nucrust-annih-fit}) enable one to evaluate
the neutrino emissivity $Q_{\rm pair}$ for any $T$ and $\rho$
in the case of the strongly degenerate electron gas
with any degree of relativism. All functions $\Phi_k$
contain the factor $\exp(-(1+y_r)/t_r)$ which decays
exponentially with decreasing temperature and with
increasing density in the limit of relativistic
electrons, $x_r \gg 1$. Thus the positron fraction and the
neutrino emissivity are exponentially suppressed.
Therefore,
the emissivity is highest at lower $\rho$ and
higher $T$.

Although we do not
study the case of non-relativistic degenerate electrons,
in which $x_r \ll 1$ and $t_r \ll x_r^2/2$,
we
note that in this case $U_k=x_r^3/(3 \pi^2)$ and
$\Phi_k=t_r \,(2 \pi^2)^{-1} \, (2 \pi t_r)^{1/2}$ $\exp(-2/t_r)$
for any $k$, and 
\begin{equation}
     Q_{\rm pair}= { Q_c \, x_r^3 \, t_r^{3/2} \over 6 \pi^5}
     \, \sqrt{2 \pi} \, (C_+^2 + C_-^2) \,
     \exp \left( - {2 \over t_r} \right).
\label{nucrust-annih-nonrel}
\end{equation}
This means that the neutrino emission of non-relativistic electrons
(and positrons)
goes through the vector currents ($Q_{\rm pair} \propto C_V^2$).

In the case of ultrarelativistic electrons ($x_r \gg 1$)
at $t_r \lesssim x_r$
the main contribution to the emissivity
$Q_{\rm pair}$, Eq.\ (\ref{nucrust-annih-Q2}),
comes from the two terms containing 
$U_2 \approx x_r^5/(5 \pi^2)$. Then
the emissivity is given by a remarkably simple fit 
\begin{equation}
    Q_{\rm pair} = {Q_c  \over 90 \, \pi^3} \, x_r^5 \, C_+^2 \,
           (4 \Phi_1 - \Phi_{-1}),
\label{nucrust-annih-relat}
\end{equation}
which is sufficient for studying the thermal evolution of
cooling neutron stars.
In this case the vector and axial-vector currents produce
similar contributions.

It should be stressed that there are other fit expressions for
$Q_{\rm pair}$ in the literature.
Dicus (1972), Munakata et al.\ (1985), Schinder et al.\ (1987), and
Itoh et al.\ (1989, 1996)
%
%
derived the fits which do not differ
strongly from one another and reproduce 
$Q_{\rm pair}$ sufficiently accurately at those 
values of $T$ and $\rho$ where the pair
annihilation mechanism is most efficient. However, their fit
formulae are chosen in such a way that they do not reproduce
the asymptotes of $Q_{\rm pair}$. Consequently, they give large errors
when $Q_{\rm pair}$ is small.
A very accurate analytic description of $Q_{\rm pair}$ was
obtained by Blinnikov and Rudzskii (1989). It is based
on the exact expression for $Q_{\rm pair}$, equivalent
to Eq.\ (\ref{nucrust-annih-Q2}), and on the analytic approximation
of the thermodynamic functions. Their fit expression is valid
for any degree of electron relativism and degeneracy, but it is
more complicated than the more restricted fit
presented above.

Note in passing that for a hot,
relativistic ($t_r \gg 1$) and nondegenerate plasma Eqs.\
(\ref{nucrust-annih-Q2}) and (\ref{nucrust-annih-U})
yield the asymptotic expression
\begin{equation}
    Q_{\rm pair} = {7 \zeta(5) \, Q_c  \over 12 \, \pi} \, C_+^2 \, t_r^9,
\label{nucrust-annih-hot}
\end{equation}
where $\zeta(5)=1.037$ is a value of the Riemann zeta function.
This emissivity is very large since there are plenty of positrons
in a hot plasma ($n_e^+ \approx n_e \propto T^3$). It is
independent of density and is a strong function
of temperature. An accurate fit for $Q_{\rm pair}$
in a nondegenerate plasma of any
degree of relativism was presented by Kaminker et al.\ (1994).
%
%
An interpolation procedure to calculate $Q_{\rm pair}$
for any electron degeneracy and relativism
was suggested by Kaminker and Yakovlev (1994).
It is alternative to that proposed by Blinnikov and Rudzskii (1989).

The efficiency of the pair annihilation process is
demonstrated in Figs.\ \ref{nucrust-fig93} and \ref{nucrust-fig91}
for a neutron star crust with
the three values of the
magnetic field, $B=10^{12}$, $10^{13}$ and $10^{14}$ G
(we discuss the dependence
of the emissivity on the magnetic field 
below). 
For the highest temperature,
$T=3 \times 10^9$ K, the emissivity is actually independent
of the magnetic field for $B \lesssim 10^{14}$ G.
In the domain of strong electron
degeneracy we use fit expressions
(\ref{nucrust-annih-Q2}), (\ref{nucrust-annih-U1}), and
(\ref{nucrust-annih-fit}), and at lower densities
the interpolation procedure of Kaminker et al.\ (1994).
At $T=10^9$ K, the emissivity depends noticeably
on the magnetic field as described by the
fit expressions obtained by Kaminker
et al.\ (1994) for the magnetized plasma (see below).
Comparison of Figs.\ \ref{nucrust-fig93} and \ref{nucrust-fig91} reveals
an expected sharp reduction of the pair emissivity with
decreasing temperature. This is why the `pair' curves disappear
from Figs.\ \ref{nucrust-fig83} and \ref{nucrust-fig81}.\\

{\bf (c) Pair annihilation in a magnetic field}

In principle, the pair annihilation can be affected
by strong magnetic fields. The complicated general expression for
the emissivity $Q_{\rm pair}$ in a plasma of any degree of degeneracy
and relativism in an arbitrary magnetic field $B$
was derived by Kaminker et al.\ (1992a).
%
%
The same authors obtained also practical expressions for
$Q_{\rm pair}$ in a nonrelativistic (degenerate and
nondegenerate) plasma at $B \ll 4 \times 10^{13}$ G.
Kaminker et al.\ (1992b, 1994) derived the practical expressions for
a hot, nondegenerate plasma with an arbitrary magnetic field.
Kaminker and Yakovlev (1994) considered the case of a
magnetized, strongly degenerate electron gas.
In addition, they
described the interpolation procedure for calculating $Q_{\rm pair}$
in a plasma of any relativism and
degeneracy for any value of the magnetic field.

The results of these studies are as follows.
In a hot, nondegenerate plasma ($T \gtrsim 10^{10}$ K)
one needs extremely large magnetic fields,
$B \gg 10^{15}$ G, to affect noticeably the neutrino emissivity.
Such fields amplify $Q_{\rm pair}$
by increasing the number densities of electrons and positrons
via very strong quantization of their motion.
Somewhat lower fields may also influence $Q_{\rm pair}$
but less significantly. In a strongly degenerate electron
gas, the typical electron energies are determined by the mass density of matter
while the positron energies depend on the temperature.
For instance, one needs superstrong magnetic fields
$B \gtrsim  7 \times 10^{15}$ G
to quantize the motion of degenerate
electrons at the lowest densities
$\rho \sim 10^{10}$ g cm$^{-3}$ of interest. Therefore, one can safely
ignore the effects of magnetic fields on the
electron plasma component 
in practical calculations of $Q_{\rm pair}$.
If, for example, $T \gtrsim 3 \times 10^9$ K
then the magnetic fields $B \lesssim 10^{14}$ G do not quantize
motion of positrons and do not affect
the neutrino emissivity (Fig.\ \ref{nucrust-fig93}).
However, even a magnetic field $B \sim 10^{14}$ G
may quantize the motion of positrons at $T \lesssim 10^9$ K
and increase their number density. In this way
strong magnetic fields greatly enhance $Q_{\rm pair}$
in a not too hot plasma. However,
this enhancement usually takes place where
the pair annihilation emissivity is much
lower than emissivity of other
neutrino reactions (as 
seen in Fig.\ \ref{nucrust-fig91}).

\subsection{Plasmon decay}
\label{sect-nucrust-plasmon}

{\bf (a) General equation}

In this section we consider another neutrino emission
mechanism, {\it plasmon decay} into a neutrino pair.
This mechanism is extremely efficient at high
temperatures and not too high densities in the neutron
star crusts.

It is evident that a free electron cannot emit a neutrino
pair; it is forbidden by the energy-momentum conservation. However,
an electron interacting with the surrounding medium can.
Plasmon decay is an example of this statement.
Strictly, the process can be written as $e \to e + \nu + \bar{\nu}$,
where $e$ denotes an ensemble of ``dressed" electrons
interacting with the plasma microfields. We neglect the
contribution of positrons to this process, although it can be
substantial in a very hot plasma.
While treating proper collective modes in
terms of the plasmons, the process can be written
as
\begin{equation}
    \gamma \to \nu + \bar{\nu},
\label{nucrust-plasmon-decay}
\end{equation}
where $\gamma$ stands for a plasmon. As in the pair annihilation,
the plasmons can emit neutrinos of any flavor. Also, there exist several
types of plasmons,
so that the total neutrino emissivity
$Q_{\rm pl}$ is a sum over the different types.

The formal derivation of the emissivity 
is fairly simple, although accurate calculations are complicated.
We start with the same interaction Hamiltonian [Eq.\
(\ref{nucrust-annih-H})], in which $l^\alpha$ is the 
neutrino current [Eq.\ (\ref{nucrust-neutrino-current})],
and $J^\alpha$ is given by Eq.\ (\ref{nucrust-electron-current}),
where $\psi_e$ and $\psi'_e$ are the wave functions of the
``dressed" electrons in the initial and final states, respectively.
 
The next step is to treat $J^\alpha$ as the second-quantized
4-vector of the collective electron current density, and to introduce
$j^\alpha=e \, J^\alpha$, the associated electric current
density in the presence of plasmons. 
This current density can be expanded in
the normal plasma modes,
\begin{equation}
    j^\alpha = \sum_s \left( \hat{a}_s \, j^\alpha_s \,
                    {\rm e}^{i kx} + \hat{a}_s^\dagger \,
                    \tilde{\jmath}^\alpha_s \, {\rm e}^{-i kx} \right),
\label{nucrust-plasmon-field}
\end{equation}
where $s=({\bf k},\lambda)$ identifies plasmon modes,
${\bf k}$ is the wave vector,
$\lambda$ is the polarization index,
$k=(\omega,{\bf k})$ is the 4-vector of the plasmon momentum;
$\hat{a}_s$ and $\hat{a}_s^\dagger$ are the plasmon annihilation and creation
operators, respectively; $j_s$ and $\tilde{\jmath}_s$ are
the associated 4-vector current amplitudes. Only the terms containing
$\hat{a}_s$ operate in the process of plasmon annihilation. 
The Fermi Golden Rule yields the neutrino emissivity
\begin{equation}
   Q_{\rm pl} = (2 \pi)^4 \, { G_{\rm F}^2 \over 2 e^2 } \,
      \sum_{\lambda \nu}  \int {{\rm d}k \over (2 \pi)^3} \,
     {{\rm d}p_\nu \over (2 \pi)^3} \,
     {{\rm d}p'_\nu \over (2 \pi)^3} \,
     n(\omega) \, \delta^{(4)}(k-p_\nu-p'_\nu) \, (\epsilon_\nu+\epsilon'_\nu) \,
     j_{s\alpha}^\ast j_{s\beta} \, {\cal L}^{\alpha \beta},
\label{nucrust-plasmon-Q}
\end{equation}
where $n(\omega)=({\rm e}^{\omega/T} -1)^{-1}$
is the Bose-Einstein distribution of plasmons,
and ${\cal L}^{\alpha \beta}$ is given by
Eq.\ (\ref{nucrust-ll}). The integration over ${\bf p}_\nu$
and ${\bf p}'_\nu$ is done using the Lenard identity
(\ref{nucrust-Lenard}) resulting in
\begin{equation}
   Q_{\rm pl} = { G_{\rm F}^2 \over 48 e^2 \, \pi^4} \,
           \sum_{\lambda \nu}  \int {\rm d}{\bf k} \, n(\omega) \,
           \omega \, \left[ \left|(j_sk)^2 \right|
           -(j_s j_s^\ast) k^2 \right].
\label{nucrust-plasmon-Q1}
\end{equation}
This is the general equation for the neutrino emissivity.
The energy-momentum conservation
implies $k=p_\nu+p'_\nu$, and
$k^2=\omega^2-{\bf k}\cdot{\bf k}\geq 0$. This inequality
restricts the domain of integration. If $k^2<0$, i.e., the
plasmon phase velocity is smaller than the speed of light
($\omega/k<c$), the plasmon decay process is
forbidden. This stringent constraint excludes
many plasmon modes, especially in a magnetized plasma,
from producing 
plasmon neutrinos. For instance, we can rule out all
acoustic plasma modes with the dispersion
relation $\omega = v_s k$ and the sound speed $v_s < c$.
One can expect that the plasmon
process, if allowed, is efficient in the presence of
the well-defined (weakly attenuated) plasma modes
with typical frequencies $\omega \gtrsim T$. If $T \ll \omega$,
the emissivity must be strongly suppressed due to the very small
number of available plasmons ($n(\omega)\ll 1$).\\

{\bf (b) Non-magnetized degenerate relativistic plasma}

Let us consider a non-magnetized, uniform and isotropic
plasma. There all plasma oscillations are known to
split into the longitudinal and transverse modes.
We will focus on the high-frequency electron
modes, which are the best candidates for plasmon decay.
They are of two types: the
longitudinal potential Langmuir plasma oscillations ($\lambda=l$)
associated with the electric field oscillations, and the
two transverse degenerate plasma modes
(with two different orthogonal polarizations, $\lambda=t_1$ and $t_2$)
associated with the oscillations
of the electric and magnetic fields transverse to ${\bf k}$.
The dispersion relations $\omega(k)$ follow
from the equations
\begin{equation}
     \varepsilon_l(\omega,k)=0, \quad
     \omega^2 \varepsilon_t(\omega,k)=k^2,
\label{nucrust-plasmon-dispers}
\end{equation}
where $\varepsilon_l(\omega,k)$ and $\varepsilon_t(\omega,k)$
are the longitudinal and transverse dielectric plasma functions,
respectively. These functions can be taken from the plasma
theory. In particular, for a strongly
degenerate electron gas
they were calculated in a classical paper by Jancovici (1962).
%
%
The major parameter which determines the plasmon propagation
is the electron plasma frequency $\omega_{pe}$. For strongly degenerate
electrons, it is
\begin{equation}
    \omega_{pe} = \sqrt{ 4 \pi e^2 n_e / m_e^\ast},
\label{nucrust-plasmon-omegape}
\end{equation}
where $m_e^\ast=\mu_e/c^2$ and $\mu_e$ is the electron chemical
potential. The frequencies of both plasmon modes are generally
higher than $\omega_{pe}$. For instance, in the limit
of ultrarelativistic degenerate electrons
at long wavelengths
($k \ll \omega_{pe}/c$)
Eqs.\ (\ref{nucrust-plasmon-dispers}) yield:
\begin{equation}
  \omega_l^2=\omega_{pe}^2 + {3 \over 5} \, k^2 c^2, \quad
  \omega_t^2=\omega_{pe}^2 + {6 \over 5} \, k^2 c^2.
\label{nucrust-plasmon-omega-lt}
\end{equation}
In the opposite case ($k \gg \omega_{pe}/c$), the longitudinal
plasmons experience strong Landau damping (and,
therefore, cannot exist as the well-defined quasiparticle excitations)
while the transverse plasmons
transform into the familiar electromagnetic waves almost
unaffected by the medium ($\omega_t \to ck$).

Next, we consider $j_s$, the amplitude
of the electric current  which is induced by an annihilating plasmon
and which enters Eq.\ (\ref{nucrust-plasmon-field}).
Generally, $j$ contains two terms, 
$j^\alpha=C_V \, j^\alpha_V+C_A \, j^\alpha_A$, which describe the
contributions of the vector and axial-vector currents, respectively
[see Eq.\ (\ref{nucrust-electron-current})].
The axial-vector contribution in a strongly degenerate
relativistic electron plasma
was studied by several authors
(e.g., Kohyama et al.\ 1986, 1994, and
Braaten and Segel 1993).
%
%
%
%
Although the results of these studies differ in details,
the main conclusion is the same: the axial-vector contribution
to $Q_{\rm pl}$
is negligibly small, typically, 2--4 orders of magnitude lower
than the vector one (summed over the
longitudinal and transverse modes).
Thus we 
concentrate on the vector contribution alone:
$j^\alpha_s = C_V \, j^\alpha_{sV}$. It is clear
that $j_V$ represents the familiar
{\it electron conduction current density} and can be taken
from the well known equations of 
plasma physics: $j_s^\alpha=\Pi^{\alpha \beta}_s A_{s \beta}$,
where $\Pi^{\alpha \beta}_s \equiv 
\Pi^{\alpha \beta}_\lambda({\bf k},\omega)$
is the polarization tensor, and $A_{s \beta} \equiv A_{\lambda \beta}
({\bf k},\omega)$ is the plasmon vector potential
defined by the quantization of plasma oscillations. 
The final 
expression for $Q_{\rm pl}$ 
in standard physical units is
\begin{equation}
  Q_{\rm pl}  =  { Q_c \over 96 \pi^4 \alpha_{\rm f}} \, I_{\rm pl} \,
                   \sum_\nu C_V^2 ,
                   \quad I_{\rm pl}= I_l+ I_t,
\label{nucrust-plasmon-Q2}
\end{equation}
where $Q_c$ is defined by Eq.\ (\ref{nucrust-annih-Qc}),
$\alpha_{\rm f}=e^2/(\hbar c)=1/137$, $\sum_\nu C_V^2=0.9248$.
The dimensionless 
functions $I_l$ and $I_t$ describe the contributions of the
longitudinal and transverse plasmons, respectively. They
are given by the following integrals ($\hbar=c=k_{\rm B}=1$):
\begin{eqnarray}
   I_l & = & {2  \over m_e^9} \, \int_0^\infty {\rm d}k \, k^2 \,
         \left( \partial \varepsilon_l \over \partial \omega \right)^{-1} \,
         (\omega_l^2-k^2)^2 \, \omega_l \, n(\omega_l),
\nonumber \\
   I_t & = & {4  \over m_e^9} \, \int_0^\infty {\rm d}k \, k^2 \,
         \left( \partial \omega^2 \varepsilon_t
         \over \partial \omega \right)^{-1} \,
         (\omega_t^2-k^2)^3 \, \omega_t \, n(\omega_t).
\label{nucrust-plasmon-Ilt}
\end{eqnarray}
The derivatives in the integrands come from the expressions
for $A_{\lambda \beta}({\bf k},\omega)$.
The effective upper limit of integration in $I_l$ 
is actually determined by the strong damping
of longitudinal plasmons at large $k$.

Equations (\ref{nucrust-plasmon-Q2})
and (\ref{nucrust-plasmon-Ilt}) provide the practical
expressions for calculating $Q_{\rm pl}$.
They were obtained first by Adams et al.\ (1963)
%
%
who made, however, two omissions improved by
Tsytovich (1963)
%
%
and Zaidi (1965).
%
%
The corrected equations were used
for extensive calculations in
a classical paper by Beaudet et al.\ (1967)
%
%
and several others (e.g., Munakata et al., 1985).
%
%
Nevertheless, as pointed out by Braaten (1991),
%
%
the plasmon dispersion relations, used originally
by Beaudet et al.\ (1967) and subsequently by others,
were not very accurate for ultrarelativistic
electrons. The improved calculations
were performed by Itoh et al.\ (1992), and Braaten and Segel (1993).
%
%
According to Braaten and Segel
(1993) the asymptotes of the functions $I_l$ and
$I_t$ in the high-temperature limit ($T \gg \omega_{pe}$) for
the strongly
degenerate, ultrarelativistic electrons
are
\begin{equation}
   I_t = 4 \zeta(3) \, \left( 3 \over 2 \right)^3 \,
   {T^3 \omega_{pe}^6 \over m_e^9},
   \quad
   I_l = 0.349 \, {T \omega_{pe}^8 \over m_e^9} ,
\label{nucrust-plasmon-highT}
\end{equation}
where $\zeta(3)=1.202$.
In this case, the transverse plasmons give the major
contribution to the neutrino emissivity, because they operate
in a much larger part of momentum space. In the opposite,
low-temperature limit ($T \ll \omega_{pe}$)
\begin{equation}
    I_l= \sqrt{2} \, I_t = \sqrt{\pi \over 2 } \,
    \left( 5 \over 3 \right)^{3/2}
    {\omega_{pe}^{15/2} T^{3/2} \over m_e^9 } \,
    \exp \left(- {\omega_{pe} \over T} \right),
\label{nucrust-plasmon-lowT}
\end{equation}
i.e., the emissivity is exponentially suppressed due to
the  small number of plasmons.
The latter asymptote for $I_l$ was obtained correctly
by Beaudet et al.\ (1967).

Now let us return to practical units and introduce
the dimensionless relativistic temperature
$t_r = k_{\rm B}T/(m_e c^2)$ and the dimensionless
plasma parameter $f_p= \hbar \omega_{pe}/(k_{\rm B} T )
$ = $ [4 \alpha_{\rm f} \,
x_r^3 /(3 \pi \sqrt{1 + x_r^2}) ]^{1/2}/t_r$.
Using these variables we write down the asymptotes as
$I_{\rm pl}= 16.23 \, t_r^9 \, f_p^6$ for $f_p \ll 1$, and
$I_{\rm pl}= 4.604 \, t_r^9 \, f_p^{15/2} \exp(-f_p)$ for $f_p \gg 1$.
We have verified that these asymptotes describe quite
accurately the appropriate numerical values of $Q_{\rm pl}$
obtained by Itoh et al.\ (1992). Moreover, we have verified,
that the simple interpolation formula,
\begin{equation}
      I_{\rm pl} = t_r^9 \, (16.23 \, f_p^6 + 4.604 \, f_p^{15/2}) \,
                  \exp(-f_p),
\label{nucrust-plasmon-fit}
\end{equation}
reproduces {\it all numerical values} presented in Table 1
of Itoh et al.\ (1992) for $\rho \gtrsim 10^{8}$ g cm$^{-3}$
within $\lesssim 10\%$.
Thus, Eqs.\ (\ref{nucrust-plasmon-Q2}) and (\ref{nucrust-plasmon-fit})
represent a reliable analytic fit to the neutrino emissivity
$Q_{\rm pl}$, much simpler than that proposed by
Itoh et al.\ (1992). Also, it correctly reproduces the
asymptotes, while the fit by Itoh et al.\ (1992) does not.

To summarize, plasmon decay in a relativistic,
degenerate electron gas is a selective process.
It operates most efficiently in the high-temperature
plasma as long as $T \gtrsim \omega_{pe}$
and it is suppressed exponentially at low temperatures
(see Figs.\ \ref{nucrust-fig93}--\ref{nucrust-fig81}).
Since 
$\omega_{pe} \propto \rho^{1/3}$, the emissivity depends strongly 
on density at fixed $T$. 
The emissivity grows
as $Q_{\rm pl} \propto \rho^2$ as long as $\omega_{pe} \lesssim T$, i.e.,
in the 
high-temperature 
(low-density) domain. It
reaches maximum at 
$\omega_{pe} \sim T$ and decays exponentially at
higher densities (Figs.\ \ref{nucrust-fig93}--\ref{nucrust-fig83}).
The electron plasma energy in
neutron stars is high; for instance,
$\hbar \omega_{pe} \approx 1.5$ MeV for the ground-state matter
at the neutron drip point ($\rho = 4.3 \times 10^{11}$ g cm$^{-3}$,
$A=118$ and $Z=36$).
Therefore,
plasmon decay is efficient
only at high temperatures in the early stages of neutron star cooling.
For instance, at $T= 3 \times 10^9$ K the plasmon decay
is the dominant neutrino process in the entire density
range displayed in Fig.\ \ref{nucrust-fig93}, with the maximum
emissivity at $\rho \lesssim 10^{13}$ g cm$^{-3}$.
For 
$T= 10^9$ K
(Fig.\ \ref{nucrust-fig91}), the maximum shifts to
$\rho \sim 3 \times 10^{10}$ K, and the region of dominance
narrows to $\rho \lesssim 10^{11}$ g cm$^{-3}$.
In this case, the process becomes exponentially
suppressed at densities higher than the neutron drip
density. Its emissivity becomes a very sensitive function
of $n_e$ and works as a `magnifying glass' 
reproducing the jumps of $n_e$
associated with variations of the nuclear composition 
(Sect.\ \ref{sect-nucrust-introduc}).
For $T=3 \times 10^8$ K (Fig.\ \ref{nucrust-fig83}),
the maximum is at $\rho \sim 10^9$ g cm$^{-3}$,
and the mechanism becomes unimportant for cooling
neutron stars.\\

{\bf (c) Plasmon decay in a magnetic field}

The neutrino emission via plasmon decay may be affected
by a strong magnetic field. It is well known that the
magnetic field influences plasma dispersion properties.
Generally, plasmon modes in a magnetized plasma
cannot be separated into
the longitudinal and transverse ones and the plasmon
dispersion relations may by strongly distorted.
New plasma modes may appear.
The contribution of the axial-vector currents
may become important. The study of these effects
has only started recently (Kennett and Melrose 1998).
%
%

It is clear that the magnetic field
affects strongly the electron plasma dispersion
only if $\omega_B^\ast \gtrsim \omega_{pe}$, where
$\omega_B^\ast$ is the relativistic electron gyrofrequency.
For instance, at $\rho \sim 10^{10}$ g cm$^{-3}$
one needs a very strong field, $B \gtrsim 7 \times 10^{14}$ G, to affect 
plasmon decay, and the required strength of the magnetic field
grows as $\rho^{2/3}$.
Therefore, the effect of the magnetic field
may be not very significant
in practice.\\

{\bf (d) Phonon decay into a neutrino pair}

In analogy with plasmon decay, the decay of other
elementary excitations may lead to the emission of neutrino pairs.
In particular, one could expect phonon decay 
in Coulomb crystals of atomic nuclei
(Flowers 1973).
%
%
Actually, however, phonon modes in a non-magnetized plasma
are unable to produce neutrinos {\it kinematically} (see Sect.\
\ref{sect-nucrust-annih})
because the phase velocities
of these phonons are smaller than the speed of light.
It has already been mentioned in Sect.\ \ref{sect-overview-struct} that
even the phonon mode which tends to behave as optical
($\omega(k) \approx \omega_{pi}$,
the ion plasma frequency) at rather small $k$, transforms
into the acoustic mode in the limit of very small $k$,
with the sound velocity smaller than the speed of light
(e.g., Pollock and Hansen 1973). Nevertheless, a strong magnetic field
%
%
affects the phonon modes (for $ \omega_{Bi} \gtrsim \omega_{pi}$,
$\omega_{Bi}$ being the cyclotron ion frequency), and 
the phonons distorted by the
magnetic field may produce neutrino pairs. 
This may happen at low enough densities of matter.

\subsection{Neutrino synchrotron emission by degenerate electrons}
\label{sect-nucrust-syn}

{\bf (a) Quantum formalism}

As discussed in the previous section in connection with
plasmon decay, an interacting electron
can emit a neutrino pair.
In this section we consider another example,
the emission of neutrino pairs
by relativistic electrons in a strong magnetic field ${\bf B}$.
The magnetic field is assumed to be constant and uniform
on microscopic scales, and directed along the $z$ axis.
It forces the electrons to rotate
around the magnetic field lines. The electron momentum is not conserved,
which opens the process
\begin{equation}
    e \stackrel{B}{\to} e + \nu + \bar{\nu},
\label{nucrust-syn}
\end{equation}
similar to the ordinary synchrotron emission
of photons but much weaker.
The symbol $B$ indicates that the process operates
only in the presence of the magnetic field.

The calculation of $Q_{\rm syn}$ is similar to that
of the pair annihilation process
(Sect.\ \ref{sect-nucrust-annih}). The four-tail Feynman diagram
of the synchrotron emission is complementary to that
of the pair annihilation. All neutrino flavors
may be emitted. The main difference is
that now $\psi'_e$ in Eq.\ (\ref{nucrust-electron-current})
is the wave function of the electron in the final state,
and both electron functions, $\psi_e$ and $\psi'_e$,
describe the Landau states of the electron in a quantizing
magnetic field. Using the formalism of relativistic electrons
in a quantizing magnetic field
Kaminker et al.\ (1992a) obtained the general
expression for the neutrino emissivity
%
%
\begin{eqnarray}
     Q_{\rm syn} &  = &  \frac{G_{\rm F}^2 b \, m_e^2 }{3 (2\pi)^5}
     \sum_{n=1,n'=0}^\infty
     \int_{-\infty}^{+\infty} {\rm d} p_z
     \int {\rm d} k_z \int k_\perp {\rm d} k_\perp \,
     A \omega \, f(1-f').
\label{nucrust-syn-Quantum}
\end{eqnarray}
where $b=B/B_c$ is
the dimensionless magnetic field; $B_c \equiv
m_e^2c^3/(\hbar e) \approx 4.414 \times 10^{13}$ G.
Here, $n$ and $p_z$ are, respectively, the
Landau level and the momentum along the magnetic field
of an initial electron.
The energy of this electron is
$\epsilon=(m_e^2 +2nb\, m_e^2 +p_z^2)^{1/2}$.
The primed quantities $n'$ and $p'_z$ refer to the electron
after the emission; its energy is
$\epsilon'=(m_e^2 +2n'b \, m_e^2 +p_z^{\prime 2})^{1/2}$;
$f= f(\epsilon)$ and
$f'=f(\epsilon')$ 
are the Fermi-Dirac distributions of the initial
and final electrons.
The energy and momentum
carried away by the neutrino pair are 
$\omega = \epsilon - \epsilon'$ and {\bf k}, respectively.
The component of vector {\bf k} along the magnetic field
is $k_z=p_z -p'_z$,
while the perpendicular component 
is $k_\perp$.
The summation and integration in Eq.\ (\ref{nucrust-syn-Quantum})
are over all allowed electron transitions.
The integration has to be done over the
{\it kinematically} allowed domain
$k_z^2 + k_\perp^2 \leq \omega^2$.
The differential transition rate $A$
is given by Eq.\ (17) of Kaminker et al.\ (1992a).
Taking into account that some terms
are odd functions of $p_z$ and 
therefore vanish after the integration
with the equilibrium distribution functions $f$ and $f'$,
Bezchastnov et al.\ (1997) found
%
%
%
\begin{eqnarray}
       A & = & \frac{C_{+}^2}{2 \epsilon \epsilon'}
              \left\{
              \left[
              \left( \omega^2 - k_z^2 - k_{\perp}^2   \right)^{}
              \left( p_{\perp}^2 + p_{\perp}^{\prime 2} + 2 m_e^2  \right)
              + k_{\perp}^2 \, m_e^2  \right]
              (\Psi - \Phi)  \right.
\nonumber    \\
             & - &
              \left.
             \left( \omega^2 - k_z^2 - k_{\perp}^2  \right)^2
                \Psi + m_e^2  \left( \omega^2 - k_z^2 - k_{\perp}^2
             \right) \Phi   \right\}
\nonumber   \\
             & - &  \frac{C_{-}^2\, m_e^2 }{2 \epsilon \epsilon'}
             \left[
             \left( 2 \omega^2 - 2 k_z^2 - k_{\perp}^2  \right)
             (\Psi - \Phi)  +
            3 \left( \omega^2 - k_z^2 -k_{\perp}^2  \right) \Phi
         \right].
\label{nucrust-syn-A_general}
\end{eqnarray}
Here, $p_\perp = m_e \sqrt{2nb} \,$ 
and $p'_\perp = m_e \sqrt{2n'b} \,$ are
the transverse momenta of the initial and final electrons,
respectively;
\begin{eqnarray}
    \Psi & = & F_{n'-1,n}^2(u) + F_{n',n-1}^2(u),
\nonumber  \\
    \Phi & = & F_{n'-1,n-1}^2(u) + F_{n',n}^2(u),
\label{nucrust-syn-PsiPhi}
\end{eqnarray}
$u = k_\perp^2 /(2bm_e^2)$,
$F_{n'n}(u)=(-1)^{n'-n}F_{nn'}(u)=
u^{(n-n')/2} \, {\rm e}^{-u/2} \,( n'!/n! )^{1/2} \, L_{n'}^{n-n'}(u)$,
and $L_n^s(u)$ is an associated Laguerre polynomial.

Since the process of synchrotron
emission is complementary to the process of $e^-e^+$
pair annihilation, the emissivity $Q_{\rm syn}$
can be obtained from the emissivity $Q_{\rm pair}$
in the magnetic field (Kaminker et al.\ 1992a)
%
%
by replacing $\epsilon' \to -\epsilon'$ and $p'_z \to -p'_z$.
Nevertheless despite the internal similarity of the two processes, 
their emissivities are quite different functions of the plasma parameters. 
In particular, the kinematic restrictions, absent in
the pair annihilation, forbid  the synchrotron
process as $B \to 0$.\\

{\bf (b) Quasiclassical treatment}

Let us describe the quasiclassical treatment
of the neutrino synchrotron emission
of a degenerate, ultrarelativistic electron gas
in a strong magnetic field, $B= 10^{11}$--10$^{14}$ G.
We follow the papers by Kaminker et al.\ (1991)
and Bezchastnov et al.\ (1997)
and focus on the most realistic case in which the electrons populate
many Landau levels. In this case the electron chemical potential
is nearly the same as without the magnetic field,
$\mu_e \approx p_{{\rm F}e} \approx (3 \pi^2 n_e)^{1/3}$,
where $p_{{\rm F}e}$ is the field--free Fermi momentum.

If many Landau levels are populated,
the summation over $n$ in Eq.\ (\ref{nucrust-syn-Quantum})
can be replaced by the integration over $p_\perp$.
The remaining summation over $n'$ can be
replaced by the summation over the discrete cyclotron harmonics
$s=n-n'$=1, 2, 3, \ldots
The synchrotron emission for $n' \geq n$ is kinematically forbidden.
The electron in the initial state
can be described by its quasiclassical momentum
$p$ and by the pitch-angle $\theta$ ($p_z = p \cos \theta$,
$p_\perp = p \sin \theta$).
One can set $\epsilon = \mu_e$ and
$p=p_{{\rm F}e}$ in all smooth functions in the integrals
for the strongly degenerate electrons, permitting the analytic
integration over $p$.
This reduces the rigorous quantum formalism
to the quasiclassical approximation
used by Kaminker et al.\ (1991).
%
%

According to Kaminker et al.\ (1991),
the quasiclassical neutrino
synchrotron emission of ultrarelativistic degenerate
electrons differs in the three temperature
domains ${\cal A}$, ${\cal B}$, and ${\cal C}$
separated by the two 
characteristic temperatures $T_P$
and $T_B$:
\begin{eqnarray}
     T_P & = & {3 \hbar \omega_B^\ast x_r^3 \over 2 k_{\rm B}}
               ={3 \over 2} \, T_B x_r^3
               \approx 2.02 \times 10^9 \, B_{13} \, x_r^2~~{\rm K},
\nonumber \\
     T_B & = & {\hbar \omega_B^\ast \over k_{\rm B}}
         \approx 1.34 \times 10^9 \,
         { B_{13} \over \sqrt{1+x_r^2}}~~{\rm K}.
\label{nucrust-syn-T}
\end{eqnarray}
Here, $x_r=p_{{\rm F}e}/(m_e c)$ is 
our usual relativistic parameter
($x_r \gg 1$, in the given case),
$\omega_B^\ast = \omega_B /\sqrt{1+x_r^2}=  eBc/\mu_e$ is
the electron gyrofrequency at the Fermi surface,
and $\omega_B$ is the electron cyclotron frequency.

The {\it high-temperature} domain ${\cal A}$ is defined as
$T_P \ll T \ll T_{\rm F}$, where
$T_{\rm F}$ is the electron degeneracy temperature given by
Eq.\ (\ref{TF}).
This domain exists for not too
high densities and magnetic fields, where $T_P \ll T_{\rm F}$
(see Fig.\ 1 in Bezchastnov et al.\ 1997).
For instance, at $B=10^{12}$ G it extends
to $\rho \lesssim 10^{11}$ g cm$^{-3}$.
In this domain, the degenerate electrons emit neutrinos
through many cyclotron harmonics; a typical number of 
harmonics is $ s \sim x_r^3 \gg 1$.
The corresponding neutrino energies
$\omega \sim \omega_B^\ast x_r^3 \ll T$ are not restricted
by the Pauli exclusion principle. The quasiclassical approach of
Kaminker et al.\ (1991) then yields
\begin{eqnarray}
   Q_{\rm syn}^{\rm A} & = & {2 \over 189 \pi^5} \;{ G_{\rm F}^2
               k_{\rm B} T m_e^2 \omega_B^6 x_r^8 \over c^5 \hbar^4} \,
               (25C_+^2-21C_-^2).
\label{nucrust-syn-QA}
\end{eqnarray}

The {\it moderate-temperature} domain ${\cal B}$ is defined as
$T_B \lesssim T \ll T_P$ and $T \ll T_{\rm F}$.
It covers a wide range of temperatures and densities, which is
most important for the applications.
In this domain, neutrinos are emitted
through many cyclotron harmonics,
$s \sim k_{\rm B} T/ \hbar \omega_B^\ast \gg 1$,
but their spectrum is restricted by the Pauli
principle, and the typical neutrino energies are
$\omega \sim k_{\rm B} T$.
As shown by Kaminker et al.\ (1991),
in this case the neutrino emissivity
is remarkably independent of the electron number density:
\begin{eqnarray}
   Q_{\rm syn}^{\rm B} & = & {2 \zeta(5) \over 9 \pi^5} \;
               {e^2 \, G_{\rm F}^2 \,
                B^2  \over c^7 \hbar^8 } \,
                C_+^2 \, (k_{\rm B} T)^5
\nonumber \\
     & \approx & 9.04 \times 10^{14} \,
       B_{13}^2 \, T_9^5~~{\rm erg~cm^{-3}~s^{-1}},
\label{nucrust-syn-QB}
\end{eqnarray}
where $\zeta(5) \approx 1.037$ is the value of the Riemann
zeta function.
Moreover,  $Q_{\rm syn}^{\rm B}$
is independent of the electron mass. This implies that
all plasma particles would (in principle) produce the same
neutrino emission provided they meet the appropriate conditions.
However, the only such particles 
in neutron star crusts are the electrons.

The third, {\it low-temperature} domain ${\cal C}$ corresponds
to temperatures $T \lesssim T_B$ at which the main contribution
to the synchrotron emission comes from a few lower cyclotron
harmonics $s$=1, 2, \ldots
At $T \ll T_B$ all harmonics become exponentially suppressed.
The first harmonics, $s=1$, is the least reduced 
but still weak (Bezchastnov et al.\ 1997).

The emissivity $Q_{\rm syn}^{\rm AB}$ in the combined domain
${\cal A}+{\cal B}$, including a smooth transition from
${\cal A}$ to ${\cal B}$ at $T \sim T_P$, was calculated
accurately by Kaminker et al.\ (1991).
The results which are valid at $T_B \lesssim T \ll T_{\rm F}$
are written as
\begin{eqnarray}
    Q_{\rm syn}^{\rm AB} & = & Q_{\rm syn}^{\rm B} S_{\rm AB},
\label{nucrust-syn-QAB} \\
    S_{\rm AB}  & = & {27 \xi^4 \over \pi^2 \, 2^9 \, \zeta(5)}
       \left[F_+(\xi) - {C_-^2 \over C_+^2} F_-(\xi) \right],
\label{nucrust-syn-SAB}\\
    \xi & \equiv & {T_P \over T} = {3 \over 2} z x_r^3;  \quad
     z = {T_B \over T},
\nonumber
\end{eqnarray}
where $F_\pm(\xi)$
can be expressed through the McDonald functions $K_{1/3}(x)$ and
$K_{2/3}(x)$ and an integral of $K_{1/3}(x)$. Note that
$S^{\rm AB}_{\rm syn} \to 1$ as $\xi \gg 1$.
Kaminker et al.\ (1991) obtained the convenient fit expressions
\begin{eqnarray}
  F_+(\xi) & = & D_1 \, { (1+c_1 y_1)^2 \over
        (1+ a_1 y_1 + b_1 y_1^2)^4},
\nonumber  \\
  F_-(\xi) & = & D_2 \, { 1 + c_2 y_2 + d_2 y_2^2 + e_2 y_2^3
        \over (1+ a_2 y_2 + b_2 y_2^2)^5 }.
\label{nucrust-syn-F}
\end{eqnarray}
Here
$y_{1,2} = [(1+ \alpha_{1,2} \xi^{2/3})^{2/3}-1]^{3/2}$,
$a_1=2.036 \times 10^{-4}$,
$b_1=7.405 \times 10^{-8}$,
$c_1=3.675 \times 10^{-4}$,
$a_2=3.356 \times 10^{-3}$,
$b_2=1.536 \times 10^{-5}$,
$c_2=1.436 \times 10^{-2}$,
$d_2=1.024 \times 10^{-5}$,
$e_2=7.647 \times 10^{-8}$,
$D_1=44.01$, $D_2=36.97$,
$\alpha_1 = 3172$, $\alpha_2 = 172.2$.

The neutrino emissivity
in the combined domain ${\cal B}+{\cal C}$
in the quasiclassical approximation can be
written as
\begin{equation}
    Q_{\rm syn}^{\rm BC} = Q_{\rm syn}^{\rm B} S_{\rm BC},
\label{nucrust-syn-QBC}
\end{equation}
where the function $S_{\rm BC}$ depends on the
only argument $z$. Bezchastnov et al.\ (1997) derived
the complicated analytic expression for $S_{\rm BC}$
[their Eq.\ (12)] and analyzed it in detail.
In domain ${\cal B}$
($ x_r^{-3} \ll z \ll 1$) they obtained the asymptote
$ S_{\rm BC} = 1 - 0.4535 \, z^{2/3}$.
In the opposite limit ($z \gg 1$;
domain ${\cal C}$), they found a slowly converging asymptote
\begin{equation}
           S_{\rm BC} =  {3 \over 2 \zeta(5)} \,
              \exp \left( -{z \over 2} \right) \,
              \left( 1 + {28 \over z} \right).
\label{nucrust-syn-SCas}
\end{equation}
In addition to the asymptotes, they
calculated
$S_{\rm BC}$ numerically
at intermediate values of $z$ and fitted the results as
\begin{equation}
    S_{\rm BC}= \exp(-z/2)\,{\cal D}_1(z)/{\cal D}_2(z),
\label{nucrust-syn-Sfit}
\end{equation}
where ${\cal D}_1(z)=1+0.4228 \,z +0.1014 \, z^2+ 0.006240 \, z^3$ and
${\cal D}_2(z)=1+0.4535\, z^{2/3}+0.03008 \, z - 0.05043 \,
z^2 + 0.004314 \, z^3$.
The fit reproduces both the low-$z$ and the high-$z$ asymptotes.
The rms fit error at $z \leq 70$ is about 1.6\%, and
the maximum error is 5\% at $z \approx 18$.

Afterwards, it was straightforward for Bezchastnov et al.\ (1997)
to combine Eqs.\ (\ref{nucrust-syn-QAB}) and
(\ref{nucrust-syn-QBC}) and obtain a general fit expression
for the neutrino synchrotron emissivity
which is valid 
in all domains ${\cal A}$, ${\cal B}$, ${\cal C}$ ($T \ll T_{\rm F}$),
where the electrons are degenerate,
relativistic and populate many Landau levels:
\begin{equation}
   Q_{\rm syn}^{\rm ABC} = Q_{\rm syn}^{\rm B} \,
                           S_{\rm AB} \, S_{\rm BC}.
\label{nucrust-syn-ABCfit}
\end{equation}
Here, $Q_{\rm syn}^{\rm B}$ is given by Eq.\ (\ref{nucrust-syn-QB}),
while $S_{\rm AB}$ and $S_{\rm BC}$
are defined by Eqs.\ (\ref{nucrust-syn-SAB}),
(\ref{nucrust-syn-F}) and (\ref{nucrust-syn-Sfit}).
These equations are sufficient for practical
use.

The role of the synchrotron process among other
neutrino processes in the neutron star crust is seen
on Figs.\ \ref{nucrust-fig93}--\ref{nucrust-fig81}.
In the case of zero magnetic field,
the bremsstrahlung process (Sect.\ \ref{sect-nucrust-ebrems}) dominates
in denser layers of the crust
at not very high temperatures,
$T \lesssim 10^9$~K.
Plasmon decay,
photoneutrino process, and pair annihilation
are significant at high temperatures,
$T \gtrsim 10^9$~K, but they become unimportant
as the temperature falls.

The synchrotron emission is, to some extent, similar
to the bremsstrahlung, for it persists over the wide range of temperatures
and densities.
In the presence of the strong magnetic field $B \gtrsim 10^{13}$~G,
the synchrotron emission can be
important and even dominant at any temperature in Figs.\
\ref{nucrust-fig93}--\ref{nucrust-fig81}. 
In a hot plasma (Fig.\ \ref{nucrust-fig93}),
this emission is significant at comparatively
low densities, $\rho = 10^8$--$10^9$~g~cm$^{-3}$.
With decreasing $T$ it
becomes more important at higher densities.
At $T=10^8$~K, only the bremsstrahlung and
synchrotron emissions still operate (Fig.\ \ref{nucrust-fig81}).
For $B=10^{14}$~G,
the synchrotron emission
dominates the bremsstrahlung over  a wide
density range, $\rho \gtrsim 10^9$~g~cm$^{-3}$.

The neutrino synchrotron emission of electrons
was first studied by Landstreet (1967) who gave
the estimates of the emissivity.
%
%
Yakovlev and Tschaepe (1981)
%
%
were the first who made an attempt to derive accurate expressions
for $Q_{\rm syn}$ in the quasiclassical approximation. However, they
made a mistake corrected later by Kaminker et al.\ (1991),
%
%
who derived $Q_{\rm syn}$
for the strongly degenerate relativistic electron gas
(in domains ${\cal A}$+${\cal B}$). 
The latter authors also considered in passing domain ${\cal C}$ but
missed a factor of 4 in the expression
for $Q_{\rm syn}$. The overall correct treatment
valid in all domains ${\cal A}$+${\cal B}$+${\cal C}$ and described above
was given by Bezchastnov et al.\ (1997). The same problem
was also approached by Vidaurre et al.\ (1995) but
not very accurately (as criticized by Bezchastnov et al.\ 1997).\\
%
%

{\bf (c) Other synchrotron emission regimes}

The studies of the neutrino synchrotron emission have not been
restricted to the quasiclassical consideration of
strongly degenerate electrons. 
The case of relativistic
degenerate electrons in a strongly quantizing magnetic
field which forces all electrons to occupy the
ground Landau level was analyzed by Bezchastnov et al.\ (1997).
In this case the neutrino emission is very strongly suppressed
by the kinematic effects. Several papers were devoted to the synchrotron
(cyclotron) emission of nonrelativistic electrons.
The cases of nondegenerate or degenerate electrons
occupying many Landau levels
were correctly considered by Yakovlev and Tschaepe (1981).
For instance, the emissivity
of a nondegenerate gas in a nonquantizing magnetic field
($T \gg T_B$) is
\begin{equation}
   Q_{\rm syn} = {G^2_{\rm F} \omega_B^6 n_e \over
        60 \pi^3 \hbar c^6}\, \sum_\nu C_A^2 .
\label{nucrust-syn-nonrel}
\end{equation}
The emission goes mainly through the first cyclotron
harmonics, $s=1$.
The emissivity is temperature-independent and 
proportional to $B^6$. In the case
of strong degeneracy of nonrelativistic electrons $Q_{\rm syn}$ is given
by the same expression multiplied by a
small factor $3T/(2 T_{\rm F})$.
A unified treatment of synchrotron emission of nonrelativistic electrons
at any degeneracy was given by Kaminker et al.\ (1992a).

The neutrino synchrotron emission from a
hot, relativistic gas was studied by
Kaminker and Yakovlev (1993).
%
%
In this case, the emission of positrons is important.
In a wide range of temperatures, densities,
and magnetic fields the emissivity of the hot plasma
is approximately given by Eq.\ (\ref{nucrust-syn-QB}),
$Q_{\rm syn} \propto B^2 T^5$, with the additional
factor which depends logarithmically on $T$ and $B$.
This remarkably simple neutrino emissivity
is valid almost in all cases in which the electrons
(positrons) are relativistic and occupy many Landau levels.

Combining Eq.\ (\ref{nucrust-syn-ABCfit}) with Eqs.\ (40) -- (44)
in Kaminker and Yakovlev (1993) one can calculate
$Q_{\rm syn}$ for any plasma
parameters and any values of the magnetic field.

\subsection{Other electron-photon neutrino processes}
\label{sect-nucrust-egamma}

This is the last section to discuss neutrino emission
processes which involve electrons (positrons) and photons (plasmons)
and which emissivity depends only on temperature and
electron number density.
The processes discussed in the preceding
sections are important in cooling neutron stars.
There are other similar processes which
are of no practical significance in neutron
stars but can be important in other stars (e.g., in 
presupernovae). 
We outline them for completeness. In a
hot matter, in addition to the processes mentioned below,
one must take into account their analogs involving positrons.\\

{\bf (a) Photoneutrino emission}

This process can be schematically written as
\begin{equation}
    \gamma + e \to e + \nu + \bar{\nu},
\label{nucrust-egamma-photo}
\end{equation}
where $\nu$ is a neutrino of any flavor and $\gamma$ is a photon.
It resembles plasmon decay
(Sect.\ \ref{sect-nucrust-plasmon}) but is complicated by the
presence of an additional electron.
When the thermal energy $k_{\rm B}T$ becomes lower than
the electron plasma energy $\hbar \omega_{pe}$,
the process is greatly affected by the plasma effects 
(Beaudet et al.\ 1967). The
neutrino emissivity can be expressed through
a five-dimensional integral 
which needs to be evaluated numerically.

Photoneutrino emission was proposed by Ritus (1961)
and Chiu and Stabler (1961).
%
%
%
The emissivity was calculated by
Beaudet et al.\ (1967),
%
%
Dicus (1972),
%
%
Munakata et al.\ (1985),
%
%
Schinder et al.\ (1987),
%
%
and Itoh et al.\ (1989, 1996).
%
%
%
In particular, Itoh et al.\ (1989) presented extensive
tables of the photoneutrino emissivity for $\rho \lesssim 10^{10}$
g cm$^{-3}$. Many authors proposed analytic fits; the most recent ones
were given by Itoh et al.\ (1989, 1996).

The presence
of an additional electron makes photoneutrino
emission more efficient than plasmon decay in a hot, low-density
plasma. On the contrary, the process becomes much less
efficient in a cold, high-density plasma. In the
latter case, photoneutrino process represents 
a higher order correction to plasmon decay and can be
neglected. At $T= 3 \times 10^9$ K
plasmon decay dominates photoneutrino emission
for the densities $\rho \gtrsim 10^8$ g cm$^{-3}$; 
for lower temperatures this happens
at lower densities (Figs.\ \ref{nucrust-fig93} and \ref{nucrust-fig91}).
In principle, photoneutrino emission may dominate over other neutrino
processes at $T \lesssim 10^9$ K and rather low densities
$\rho \lesssim 10^7 $ g cm$^{-3}$
(e.g., Itoh et al.\ 1996).
It is reasonable
to neglect photoneutrino emission
in the models of cooling neutron stars.\\

{\bf (b) Neutrino bremsstrahlung in $ee$-collisions}

This process can be written as
\begin{equation}
   e + e \to e + e + \nu + \bar{\nu}
\label{nucore-egamma-eebrems}
\end{equation}
for any neutrino flavor. It will be
studied in Sect.\ \ref{sect-nucore-other} as the neutrino production
mechanism which can dominate in highly superfluid neutron star
cores. It operates also in neutron star crusts but seems
to be unimportant there. For the typical conditions in the crusts, 
its emissivity is
four -- seven orders of magnitude lower than the emissivity
of the bremsstrahlung process associated with the electron-nucleus
collisions (Sect.\ \ref{sect-nucrust-ebrems}).\\

{\bf (c) Other photon processes}

In addition, there are other processes of neutrino generation considered
in the literature, particularly, the processes
proposed by Chiu and Morrison (1960)
%
%
and discussed, for instance, by Fowler and Hoyle (1964),
%
%
and Beaudet et al.\ (1967):
\begin{eqnarray}
  \gamma + \gamma  & \to & \nu + \bar{\nu},
\nonumber \\
  \gamma + \gamma  & \to &  \gamma + \nu + \bar{\nu}.
\label{nucrust-egamma-others}
\end{eqnarray}
They are unimportant in neutron star crusts.

\subsection{Neutrino bremsstrahlung in collisions of electrons with
         atomic nuclei}
\label{sect-nucrust-ebrems}

{\bf (a) Introductory remarks}

Neutrino-pair bremsstrahlung of electrons in a Coulomb liquid or a
crystal of atomic nuclei is one
of the major  energy-loss mechanisms in the neutron star crust.
Here, by bremsstrahlung we imply the
neutrino emission due to the electromagnetic
interaction of electrons with atomic nuclei.
The process can be written schematically as
\begin{equation}
    e + (Z,A) \to e + (Z,A) + \nu + \bar{\nu}.
\label{nucrust-ebrems-brems}
\end{equation}
It proceeds via neutral and charged electroweak currents and
leads to the emission of neutrinos of all flavors.
Contrary to purely electronic neutrino processes,
it depends generally on the charge distribution within atomic
nuclei and on the correlation between the nuclei (ions).

Under the conditions of interest (Sect.\ \ref{sect-nucrust-introduc}),
the electrons are strongly degenerate and ultra-relativistic,
and the nuclei form either
a Coulomb liquid or a Coulomb
crystal. For densities higher than $10^{12}$ -- $10^{13}$
g cm$^{-3}$, the melting temperature of the crystal
is so high that the
case of a Coulomb liquid is of no practical importance.
In the density range from about $10^{14}$ g cm$^{-3}$
to $1.5 \times 10^{14}$ g cm$^{-3}$, the nuclei may
resemble rods or plates, rather than spheres
(Sect.\ \ref{sect-overview-struct}).

The neutrino bremsstrahlung process,
Eq.\ (\ref{nucrust-ebrems-brems}), in the crystal is formally different
from that in the liquid. In the liquid state,
the neutrinos are generated via Coulomb
scattering of the electrons by 
the nuclei, which can be described by the
two Feynman diagrams, each containing
two  vertices: 
an electromagnetic vertex and a weak four-tail vertex.
In the solid state,
there are two contributions to the process,
the electron--phonon
scattering (electron scattering by the nuclear charge
fluctuations due to lattice vibrations, referred to as the {\it
phonon contribution}),
and the Bragg diffraction of electrons, which is commonly
called the {\it static-lattice contribution}.
The phonon contribution is described by two Feynman diagrams,
like in the liquid. The static lattice contribution is
formally described by a one-vertex diagram and represents
the neutrino emission due to the direct interband transition of
the electron (whose energy spectrum possesses the 
band structure due to the presence of the lattice).
However, while calculating the matrix element,
one should use the electron wave function distorted
by the band structure effects. This distortion
reduces the matrix element, so that the static lattice
and the phonon contributions may be of the same order
of magnitude.

Neutrino bremsstrahlung was proposed by Pontecorvo (1959)
%
%
and also by Gandel'man and Pinaev (1959).
%
%
It has been
analyzed in numerous papers
(see, e.g., Itoh et al.\ 1989, 1996;
Pethick and Thorsson 1997, and references therein).
%
%
%
%
The case of the Coulomb liquid has been thoroughly
studied by Festa and Ruderman (1969),
%
%
Dicus et al.\ (1976),
%
%
Soyeur and Brown (1979),
%
%
Itoh and Kohyama (1983),
%
%
Haensel et al.\ (1996),
%
%
and by Kaminker et al.\ (1999a).
%
%
The phonon contribution in the crystalline lattice
has been analyzed by Flowers (1973),
%
%
Itoh et al.\ (1984b, 1989),
%
%
and also by
Yakovlev and Kaminker (1996)
%
%
using the one-phonon approximation.
Multiphonon processes have been
included by Kaminker et al.\ (1999a).

The static-lattice contribution
has been considered by many authors (e.g., Flowers 1973,
Itoh et al.\ 1984a)
neglecting the finite widths of the electron
energy gaps produced by the band-structure effects.
The proper treatment of the energy gaps
has been proposed by Pethick and Thorsson (1994, 1997).
%
%
In particular, Pethick and Thorsson (1997)
derived a general expression for the static-lattice
contribution in the presence of the realistic band
structure. Extensive calculations on the basis
of this expression were done by Kaminker et al.\ (1999a).
We follow closely their
work.

According to Haensel et al.\ (1996),
the general expression for the neutrino
emissivity $Q_{\rm br}$ due to the
neutrino-pair bremsstrahlung 
of the relativistic
degenerate electrons in a plasma of spherical nuclei
can be written (in standard physical units) as
\begin{eqnarray}
        Q_{\rm br} &=&
           {8 \pi G_{\rm F}^2 Z^2 e^4 C_+^2 \over 567 \hbar^9 c^8}
           (k_{\rm B} T)^6 n_i L
\nonumber\\
        &\approx&
           3.23 \times 10^{17} \, \rho_{12} \,
           Z Y_e  \,
           T_9^6 \, L~~{\rm erg~s^{-1}~cm^{-3}},
\label{nucrust-ebrems-Q}
\end{eqnarray}
where $n_i$ is the number density of nuclei (ions), 
$Y_e=n_e/n_b$ is the number of electrons per baryon,
$\rho_{12}$ is the density in units of $10^{12}$ g cm$^{-3}$,
and $L$ is a dimensionless function
to be determined.
The numerical expression for $Q_{\rm br}$ 
is obtained using $C_+^2 \approx 1.678$,
appropriate for the emission of
three neutrino flavors
($\nu_e$, $\nu_{\mu}$, and $\nu_{\tau}$).

Let $L=L_{\rm liq}$ in the liquid of atomic nuclei.
In the Coulomb solid, $L$ consists of two parts
describing the phonon and static-lattice contributions,
\begin{equation}
   L= L_{\rm sol} = L_{\rm ph} + L_{\rm sl}.
\label{nucrust-ebrems-Lsol}
\end{equation}

Since the vibrational properties of crystals of
nonspherical nuclei at the bottom of the neutron star crust
are largely unknown, we will be able to analyze only 
$L_{\rm sl}$ for these crystals.\\

{\bf (b) Liquid phase}

The most general expression for $L_{\rm liq}$ is
(Haensel et al.\ 1996):
\begin{eqnarray}
         L_{\rm liq}& =& { 1 \over T}
         \int_0^{2 p_{{\rm F}e}} \, {\rm d} q_t \,q_t^3
         \, \int_0^\infty {\rm d} q_r \,\, {S(q) |F(q)|^2 \over
         q^4 |\epsilon(q)|^2}\, R_T(q_t,q_r) \,
         R_{\rm NB}(q_t),
\label{nucrust-ebrems-Lliq}
\end{eqnarray}
where 
${\bf q}= {\bf q}_t + {\bf q}_r$ is the momentum
transferred from the electron to the nucleus in 
a collision event;
${\bf q}_t$ corresponds to the purely elastic Coulomb
collision ($q_t < 2 p_{{\rm F}e}$), while ${\bf q}_r$
takes into account weak inelasticity due to the neutrino emission;
$q^4$ in the denominator comes from the
squared Fourier transform of the Coulomb electron-ion interaction;
$R_{\rm NB}(q_t)$ is the non-Born 
correction factor. Other functions describe
effective screening of the Coulomb interaction.
The static longitudinal dielectric function 
of the electron gas $\epsilon(q)$ (Jancovici 1962)
accounts for the electron screening at $q \lesssim k_{\rm TF}$
($k_{\rm TF} \sim 0.1 \, p_{{\rm F}e}$ being the Thomas-Fermi
electron screening momentum).
The ion structure factor $S(q)$ (e.g., Young et al.\ 1991)
describes the ion screening due
to the ion-ion correlations at $q \lesssim 1/a$
[$a$ being the ion-sphere radius, see Eq.\ (\ref{Gamma})]. 
The nuclear form factor $F(q)$ 
is responsible for the screening
due to the proton charge distribution within the nucleus at
$q \gtrsim 1/R_p$ ($R_p$ is the proton core radius).
Finally, $R_T(q_t,q_r)$ 
describes the so called thermal screening 
at  $q_r \gtrsim T$
(Haensel et al.\ 1996) associated with
inelastic effects
of shifting the momenta of the initial and final electrons
away from the Fermi surface (with account for
shift restrictions due to electron degeneracy).

In our case, the electron screening is very weak,
and it is the ion screening which is the most important.
In the presence of electron degeneracy
the thermal screening is noticeable at rather high
temperatures, $T \gtrsim 1/a$.
It  gives
more freedom to the electron momentum transfer 
and may enhance $L_{\rm liq}$
by 20--30\%.
At lower temperatures, $T \lesssim 1/a$, the thermal
effect is negligible. In the latter case,
the factor $R_T(q_t,q_r)$ becomes a sharp
function of $q_r$ which allows one to set $q_r=0$ in the remaining
functions under the integral (\ref{nucrust-ebrems-Lliq})
and to integrate over $q_r$ (Haensel et al.\ 1996):
\begin{equation}
         L_{\rm liq}=
         \int_0^{1} \, {\rm d} y
          {S(q) |F(q)|^2 \over
         y |\epsilon(q)|^2} \,\left( 1 + {2 y^2 \over 1-y^2 }
         \, \ln y  \right)
         R_{\rm NB}(q),
\label{nucrust-ebrems-LliqLowT}
\end{equation}
where $y=q/(2 p_{{\rm F}e})$.
If $R_{\rm NB}(q)=1$, this expression gives
the emissivity $Q_{\rm br}$ obtained, for instance,
by Itoh and Kohyama (1983). 
If we removed artificially all screening [$\epsilon(q)=F(q)=S(q)=1$]
but introduced an artificial miminimum momentum cutoff
$q_{\rm min} \ll p_{{\rm F}e}$ we would get $L_{\rm liq} \simeq
{\rm ln}(2 p_{{\rm F}e}/q_{\rm min})$. Thus,
$L_{\rm liq}$ has basically the same meaning as the Coulomb
logarithm in the electron transport coefficients, which is 
a slowly varying function
of plasma parameters  convenient for theoretical studies. 

The non-Born factor $R_{\rm NB}(q)$
is the ratio of the
electron-nucleus scattering cross sections 
calculated exactly and in the Born approximation.
Kaminker et al.\ (1999a)
introduced the mean factor
$
        \bar{R}_{\rm NB}= L_{\rm liq}^{\rm NB}/L_{\rm liq}^{\rm Born},
$
where $L_{\rm liq}^{\rm NB}$ and $L_{\rm liq}^{\rm Born}$
are calculated from Eq.\ (\ref{nucrust-ebrems-Lliq})
with some model factor $R_{\rm NB}$
(Haensel et al.\ 1996) and
with $R_{\rm NB}=1$, respectively.
For the wide range of  parameters
typical for neutron star crusts
at $Z \lesssim 60$ they obtained the fit expression 
%
\begin{equation}
    \bar{R}_{\rm NB}=1+0.00554 \, Z+0.0000737 \,Z^2,
\label{nucrust-ebrems-NonBorn}
\end{equation}
which enabled them to calculate $L_{\rm liq}$ 
in the Born
approximation, and introduce the non-Born
correction afterwards.
For small $Z$ the Born approximation is basically accurate,
$\bar{R}_{\rm NB} \approx 1$. For $Z \gtrsim 40$
the non-Born correction increases $L_{\rm liq}$ by more
than 20\%.

The evaluation of the non-Born corrections
in crystalline matter is difficult.
However, we will see that neutrino bremsstrahlung
in the crystal is similar to that
in the Coulomb liquid. Thus it is reasonable to adopt the same
factor (\ref{nucrust-ebrems-NonBorn}) in Coulomb crystals.\\

{\bf (c) Phonon contribution}

Under astrophysical conditions at not too
low temperatures, the main contribution
to the electron--phonon scattering comes from
the {\it umklapp} processes in which the
electron momentum transfer ${\bf q}$ jumps 
outside the first Brillouin zone. Then the
phonon (quasi)momentum is determined by the reduction of
${\bf q}$ to the first Brillouin zone.
The umklapp processes require $q \gtrsim q_0$,
contrary to the {\it normal} processes in which
${\bf q}$ remains in
the first Brillouin zone  and $q \lesssim q_0$
[$q_0 = (6 \pi^2 n_i)^{1/3}$
is the radius of the Brillouin zone
approximated by a sphere]. The parameter
$y_0 \equiv q_0/(2 \, p_{{\rm F}e}) = (4Z)^{-1/3}$
is typically small, allowing umklapp
processes to operate in much larger part of momentum space
(e.g., Raikh and Yakovlev 1982) and therefore dominate over
the normal processes.
%
%

The phonon contribution  has been commonly studied
in the one-phonon approximation.
In particular, Flowers (1973) derived the integral expression
for the neutrino emissivity containing the dynamical ion
structure factor in the one-phonon approximation.
To allow for the background lattice vibrations
the one-phonon reaction rate
has usually been multiplied by
${\rm e}^{-2W}$, where
$W=W(q)$ is the Debye--Waller factor (e.g., Baiko and Yakovlev 1996).

The multi-phonon treatment, important near the melting point,
has been developed by Kaminker et al.\
(1999a). The authors have introduced the multi-phonon dynamical
ion structure factor $S(q, \Omega)$ of a Coulomb harmonic
crystal (Baiko et al.\ 2000)
into the formalism of Flowers (1973).
%
%
The electron-phonon scattering rate is determined by
the inelastic part $S_{\rm inel}(q,\Omega)$ of $S(q,\Omega)$.
Using the same simplified semianalytical method
for calculating $L_{\rm ph}$ as
in Yakovlev and Kaminker (1996),
Kaminker et al.\ (1999a) obtained
\begin{equation}
         L_{\rm ph}=
         \int_{y_0}^{1} \, {\rm d} y
          {S_{\rm eff}(q) |F(q)|^2 \over
         y |\epsilon(q)|^2} \,\left( 1 + {2 y^2 \over 1-y^2 }
         \, \ln y   \right).
\label{nucrust-ebrems-Lph}
\end{equation}
The lower integration limit $y_0$
excludes the low-momentum transfers in which
the umklapp processes are forbidden;
\begin{eqnarray}
        S_{\rm eff}(q)&=& {63 \over 16 \pi^7 \, T^6} \,
        \int_0^\infty {\rm d}\omega \, \omega^4 \,
        \int_{-\infty}^{+\infty} \, {\rm d}\Omega \,
        { \Omega + \omega \over {\rm e}^{(\Omega + \omega)
        /T }-1} \, S_{\rm inel}(q,\Omega),
\label{nucrust-ebrems-Sv}
\end{eqnarray}
where the integration variable
$\omega$ is the neutrino-pair energy. 
Comparing Eqs.\ (\ref{nucrust-ebrems-Lph}) and
(\ref{nucrust-ebrems-LliqLowT}), we see
that $S_{\rm eff}(q)$ plays a role of the effective static
structure factor that defines the phonon contribution
to the neutrino bremsstrahlung.

The factor $S_{\rm eff}(q)$ was calculated and fitted by a simple
expression (Kaminker et al.\ 1999a);
it is almost independent
of the lattice type.
The main difference from the one-phonon approximation
occurs in the high-temperature solid, $T \gtrsim T_p$
[$T_p$ being the ion plasma temperature, Eq.\ (\ref{Tp})]. In this case
$S_{\rm eff}(q)$ tends to the asymptotic expression
\begin{equation}
    S_{\rm eff}(q) = 1 - {\rm e}^{-2 W(q)}.
\label{nucrust-ebrems-ShighT}
\end{equation}
For not too small $q$ it is noticeably larger 
than in the one-phonon
approximation, as a result of multi-phonon processes.
If $T \ll T_p$, the factor $S_{\rm eff}(q)$ reduces to
the one-phonon case.\\

{\bf (d) Static-lattice contribution}

It has been widely assumed for a long time
that the static lattice 
contribution could be studied
neglecting the finite widths of the electron energy gaps
at the boundaries of the Brillouin zones.
The importance of the gap widths
was pointed out by Pethick and Thorsson (1994).
The same authors (Pethick and Thorsson 1997) 
developed the general formalism to describe the gaps' effect.
Their Eq.\ (28) is valid for both
spherical and nonspherical nuclei and can be written as
\begin{eqnarray}
        Q_{\rm sl} &=&
        {2 \pi G_{\rm F}^2 k_{\rm F} C_+^2 \over 567 \hbar^9 c^8}
        (k_{\rm B} T)^8 J \,
\nonumber\\&\approx&
        1.254 \times 10^9 \; (\rho_{12} Y_e)^{1/3}
        T_8^8 \, J~~{\rm erg~s^{-1}~cm^{-3}}.
\label{nucrust-ebrems-Qsl}
\end{eqnarray}
The dimensionless function
\begin{equation}
     J =  \sum_{{\bf K}\neq0} \;
     {y^2 \over t_V^2} \; I(y,t_V)
\label{nucrust-ebrems-S}
\end{equation}
is given by the sum over all
reciprocal lattice vectors ${\bf K} \neq 0$
for which ${\bf K}/2$ lies within the electron Fermi sphere;
$y=K/(2 p_{{\rm F}e})$ (with $y<1$).
Each term describes neutrino emission due to the Bragg diffraction
of electrons produced by a
corresponding reciprocal lattice vector ${\bf K}$.
The number of diffracting vectors
(harmonics) is generally large ($\sim 4 Z$).
Function $I(y,t_V)$ is given by a three dimensional integral,
Eq.\ (29) in Pethick and Thorsson (1997),
whose arguments are $y$ and 
$t_V=k_{\rm B}T/(y_\perp |V_{\rm K}|)$, where
$y_\perp=\sqrt{1-y^2}$ and
\begin{equation}
        V_{\bf q} = {4 \pi e \rho_Z \, F({\bf q}) \over q^2 \,
         \epsilon(q) } \, {\rm e}^{-W(q)}
\label{nucrust-ebrems-Vq}
\end{equation}
is the Fourier transform of the
effective electron-ion interaction.
Here, $\rho_Z$ is the ion charge per unit volume.
For a crystal of spherical nuclei, one has $\rho_Z=Z e n_i  \,$, but
Eq.\ (\ref{nucrust-ebrems-Vq}) is valid also for non-spherical nuclei.
The argument $t_V$ is
the ratio of the thermal energy to the gap width
in the electron spectrum near the intersection of
the Brillouin zone boundary and the electron Fermi surface.
The gap width depends on diffracting harmonics ${\bf K}$
decreasing strongly with the increase of $K$.
For the lattice of spherical nuclei, one can use
Eq.\ (\ref{nucrust-ebrems-Q}) with
\begin{eqnarray}
     L_{\rm sl}&=&{\pi Z^2 (k_{\rm F} a) \over 3 \Gamma^2} \,
     J = {1 \over 12 Z} \sum_{{\bf K}\neq0} \;
     {y_\perp^2 \over y^2} \;
     {|F(K)|^2 \over |\epsilon(K)|^2} \;
      I(y,t_V) \; {\rm e}^{-2W(K)}.
\label{nucrust-ebrems-Lsl}
\end{eqnarray}
The Debye--Waller
factor suppresses the
electron--lattice interaction at large
reciprocal lattice vectors ${\bf K}$
and weakens the neutrino emission.

{\it The high-temperature limit}, $t_V \gg 1$,
corresponds to the negligibly small electron energy gaps.
In this case
(Pethick and Thorsson 1994, 1997) 
\begin{equation}
       I = {1 \over y_\perp^2 y} \left(
       1 + {2 \, y^2 \over y_\perp^2} \ln y  \right).
\label{nucrust-ebrems-IhighT}
\end{equation}
Inserting this asymptote into Eqs.\ (\ref{nucrust-ebrems-Lsl})
and (\ref{nucrust-ebrems-Q})
we immediately reproduce the well-known result
of Flowers (1973) for the zero electron gap.
Replacing the sum over {\bf K} by an integral
over $q$, we arrive at the expression
\begin{equation}
         L_{\rm sl}^{(0)}=
         \int_{y_0}^{1} \, {\rm d} y\,
          { |F(q)|^2 \, {\rm e}^{-2W(q)} \over
         y |\epsilon(q)|^2} \,\left( 1 + {2 y^2 \over 1-y^2 }
         \, \ln y   \right),
\label{nucrust-ebrems-Lsl1}
\end{equation}
similar to Eq.\ (\ref{nucrust-ebrems-LliqLowT})
in the liquid
and to Eq.\ (\ref{nucrust-ebrems-Lph})
in the solid (phonons).
The Debye--Waller exponent ${\rm e}^{-2W(q)}$
is seen to play the role of the
diffraction (elastic) part of the ``effective static structure factor''
that defines the static-lattice contribution; 
this effective structure factor is smoothed
over the familiar diffraction peaks by replacing the summation
with  the integration. Thus, the sum
$L_{\rm ph}+L_{\rm sl}^{(0)}$ in the crystal can be written in the same
form (\ref{nucrust-ebrems-LliqLowT})
as $L_{\rm liq}$, with an effective total
structure factor $S_{\rm sol}(q)={\rm e}^{-2W(q)}+ S_{\rm eff}(q)$.
One can easily verify that
$S_{\rm sol}(q)$ resembles the structure
factor $S(q)$ in the strongly coupled liquid  
if one smears out the diffraction peaks in $S(q)$;
the integral contributions of both factors,
$S_{\rm liq}(q)$ and $S_{\rm sol}(q)$, are nearly the same.
As a result, the neutrino-pair
bremsstrahlung in the high-temperature solid is very similar
to that in the liquid (Kaminker et al.\ 1999a).

When the temperature decreases, the thermal energy becomes smaller
than the gap width. This reduces the integral $I$ 
and, therefore, the static
lattice contribution, compared to the zero-gap
case. In the {\it low-temperature limit}, $t_V \ll 1$,
the reduction is exponential:
$I \propto \exp[-2/(t_V (1+y_\perp))]$
(Pethick and Thorsson 1994, 1997).
It is important that the reduction of larger harmonics $K$ occurs at
lower $T$ ($t_V \propto K^2 T$), 
and the overall reduction of $L_{\rm sl}$
appears to be nearly power-law as long as the most distant
harmonics involved are not greatly reduced. The overall reduction becomes
exponential only when  $t_V \ll 1$
for the most distant harmonics.

For the lattice of spherical nuclei, one can use
Eq.\ (\ref{nucrust-ebrems-Lsl}) with the Debye--Waller
factor and the nuclear form factor. 
In the case of non-spherical
nuclei, one can use more general
Eqs.\ (\ref{nucrust-ebrems-Qsl}) and (\ref{nucrust-ebrems-S})
including the form factor but
setting $W=0$,
since the Debye--Waller factor
is unknown.
The reciprocal lattice vectors in Eq.\ (\ref{nucrust-ebrems-S})
have to be taken for appropriate two- or one-dimensional lattices.\\

{\bf (e) Results of calculations}

Let us outline the main properties of neutrino-pair
bremsstrahlung by electrons
in neutron star crusts 
(Kaminker et al.\ 1999a).

\begin{figure}[th!]
\begin{center}
\leavevmode
\epsfysize=8.5cm
\epsfbox{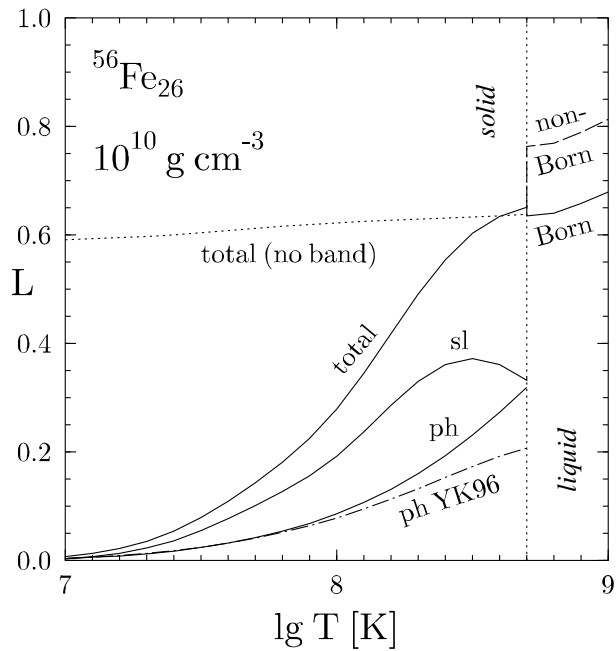}
\end{center}
\caption[ ]{
Temperature dependence of the
normalized neutrino emissivity $L$ for
iron matter
at $\rho=10^{10}$ g cm$^{-3}$ (Kaminker et al.\ 1999a). Solid lines:
Born results for the liquid phase;
phonon and static-lattice contributions, as well as
the total function (\protect{\ref{nucrust-ebrems-Lsol}}) for the
crystalline phase.
Dotted line: the total function for crystalline phase
but without band-structure effects.
Dashed line: non-Born result in the liquid phase.
Dot-and-dashed line: one-phonon approximation for the phonon
contribution (Yakovlev and Kaminker 1996).
All curves but one in the liquid phase are obtained
in the Born approximation.
}
\label{nucrust-ebrems-figfe}
\end{figure}

Figure \ref{nucrust-ebrems-figfe} shows the temperature
dependence of
the normalized neutrino emissivity $L$ for the iron matter at
$\rho = 10^{10}$ g cm$^{-3}$.
The vertical dotted line separates
liquid and solid phases. The upper (dashed) line in the liquid phase
is obtained from Eq.\ (\ref{nucrust-ebrems-Lliq}) including the
non-Born corrections, while
the lower (solid) line is obtained in the
Born approximation.
Solid lines in the crystalline phase
show $L_{\rm ph}$, $L_{\rm sl}$, 
and $L_{\rm sol}=L_{\rm ph} + L_{\rm sl}$.
Also, the dotted line gives $L_{\rm sol}$
but neglecting the band structure effects
in the static-lattice contribution.
Finally, the dot-and-dashed line
is $L_{\rm ph}$ in the
one-phonon approximation.

The phonon contribution is generally
smaller than the static-lattice one.
Strong reduction of $L_{\rm sl}$
with decreasing $T$
by the band-structure effects is seen to be most significant
being non-exponential
but rather power-law as discussed above.

The Born curve $L_{\rm liq}$ in the liquid 
matches $L_{\rm sol}$ in the solid
and makes $L$ an almost continuous
function of temperature at the melting point.
We believe that the actual non-Born curve in the solid phase,
which is difficult to calculate exactly, would match equally well
the non-Born curve in the liquid phase.
Thus the state of a Coulomb system
(liquid or solid) has little effect on
neutrino bremsstrahlung.
The one-phonon approximation is seen to be
generally quite accurate at low temperatures
but underestimates the phonon contribution near the
melting point. It is the proper inclusion of multi-phonon processes
that makes the phonon contribution larger and almost
removes the jump of the total neutrino emissivity
at the melting point which would be noticeable in the
one-phonon approximation.

\begin{figure}[th!]
\begin{center}
\leavevmode
\epsfysize=8.5cm
\epsfbox{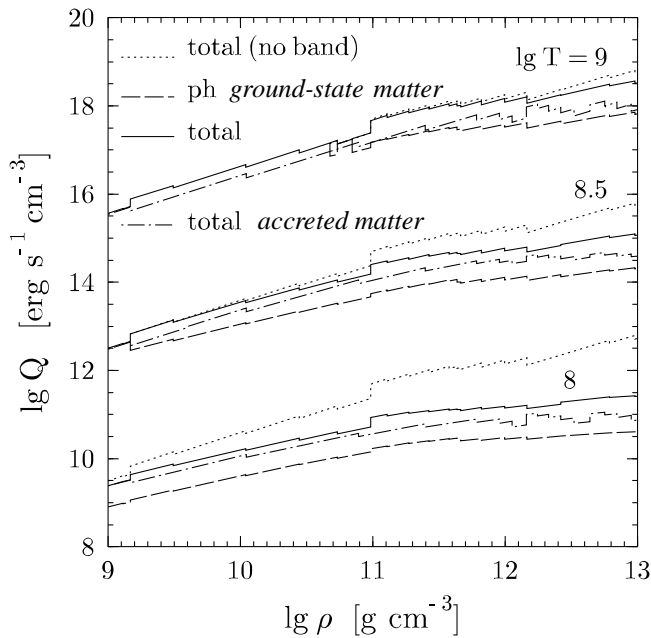}
\end{center}
\caption[ ]{
Density dependence of the neutrino bremsstrahlung emissivity
(Kaminker et al.\ 1999a)
at $T=10^8$, $10^{8.5}$ and $10^9$ K for ground state and
accreted matter.
Solid and dashed lines: the total and
phonon emissivities, respectively, for ground state matter;
dots: the total emissivity obtained neglecting
the band-structure effects. Dot-and-dashed lines: 
the total emissivity for accreted matter.
}
\label{nucrust-ebrems-figgro}
\end{figure}
%

Figure \ref{nucrust-ebrems-figgro} shows the density dependence
($10^9$ g cm$^{-3} \leq  \rho \leq  10^{13}$ g cm$^{-3}$)
of the neutrino emissivity
at three values of $T$ for the ground-state matter
(Sects.\ \ref{sect-nucrust-introduc} and \ref{sect-nucrust-beta}).
Here and below
the emissivities are calculated in the Born approximation
and multiplied by the non-Born
correction factor, Eq.\ (\ref{nucrust-ebrems-NonBorn}),
as discussed above.
For comparison, we also present the emissivity
for accreted matter which consists of lighter nuclei with lower $Z$
(Haensel and Zdunik 1990, 
Sect.\ \ref{sect-nucrust-beta}). 
Note that self-consistent models of accreted matter
(e.g., Miralda-Escud$\acute{\rm e}$ et al.\ 1990)
correspond to $T \sim 10^8$ K. We use the accreted model
for higher $T$ to illustrate how variations of nuclear
composition affect the neutrino emission.

In the liquid state we present the total neutrino 
emissivity $Q_{\rm br}$, 
while in the solid state we present the total and phonon 
emissivities for the ground state matter and the total
emissivity for the accreted matter.
We show also the total emissivity for
the ground state matter neglecting the band-structure effects.
In the displayed density range, the matter is
entirely solid for $T=10^8$ K;
there is one melting point for $T=10^{8.5}$ K
($\lg (\rho_m~[{\rm g~cm}^{-3}])=9.17$, for the ground state matter)
which separates liquid (at $\rho < \rho_m$)
and solid (at $\rho > \rho_m$);
and there are a few melting points at $T=10^9$ K
due to the non-monotonic behaviour of the melting curves
$T_m=T_m(\rho)$ associated with
strong variations of the nuclear composition.
The jumps of $Q_{\rm br}$ at melting points are small, as we remarked
earlier.
The stronger jumps are associated with variations of the
nuclear composition. The jumps of
both types may be ignored in practical applications.  The
reduction of the emissivity by the band-structure
effects becomes stronger with decreasing temperature
and reaches one order of magnitude
for $T \sim 10^8$ K and $\rho \gtrsim 10^{11}$ g cm$^{-3}$.
The band-structure reduction is power-law
(non-exponential) for the parameters
displayed.
The ratio of the phonon contribution to the static-lattice one
remains nearly constant for a wide range of temperatures much below
the melting temperature, and the static-lattice contribution
is several times larger than the phonon one.
The emissivity in the accreted matter is
lower than in the ground state
matter due to the lower $Z$, but the
difference is not large.

At $\rho \lesssim 10^{12}$ g cm$^{-3}$
one can neglect the finite
size of atomic nuclei $R_p$ and consider
the nuclei as pointlike. This is because
the form factor is $F(q) \approx 1$, for typical momentum
transfers involved, $q R_p \lesssim p_{{\rm F}e} R_p \ll 1$.
At higher densities, the finite size of the proton
distribution within the nucleus cannot be neglected.
The effect of the form factor reduces the neutrino emissivity,
and the reduction increases with growing density
reaching a factor 1.5 -- 2 at $\rho \sim 10^{13}$ g cm$^{-3}$,
and reaching 1 -- 1.5 orders of magnitude at $\rho \sim
10^{14}$ g cm$^{-3}$.
At $\rho \lesssim 10^{13}$ g cm$^{-3}$
one can get accurate results using the simplest form factor
appropriate to uniform (step-like) proton core
(with $R_p \approx 1.83 \, Z^{1/3}$ fm in the neutron drip regime
as mentioned by Itoh and Kohyama 1983).
For $\rho \gtrsim 10^{13}$ g cm$^{-3}$,
one should use more realistic
form factor based on the smoother proton charge
distribution, Eq.\ (\ref{nucrust-Oya}).
This reduces the neutrino emissivity, as compared to that
calculated for the uniform proton core
(by a factor of about 1.5 for spherical nuclei at
$\rho \sim 10^{14}$ g cm$^{-3}$).

\begin{figure}[th!]
\begin{center}
\leavevmode
\epsfysize=8.5cm
\epsfbox{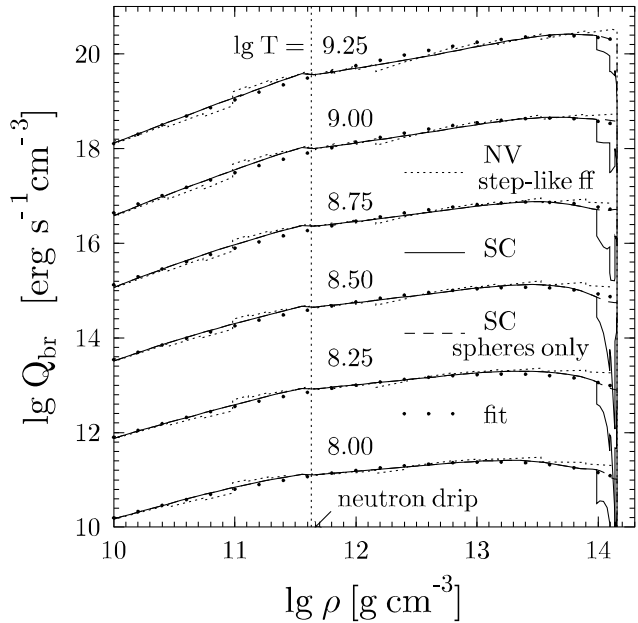}
\end{center}
\caption[ ]{
Density dependence of the bremsstrahlung  emissivity
(Kaminker et al.\ 1999a)
for the ground state matter of the neutron-star crust
at six temperatures $T$ in the model of Negele and
Vautherin (NV, dots) (1973) with the form factor (ff)
appropriate to the step-like, uniform proton core,
and in the smooth-composition
(SC) model with the realistic form factor either
including non-spherical phases (solid lines)
or assuming the nuclei be spherical to the crust bottom (dashes).
Filled circles show the fits
(\protect{\ref{nucrust-ebrems-Qfit}}) to the dashed lines.
}
\label{nucrust-ebrems-figtot}
\end{figure}
%

Figure \ref{nucrust-ebrems-figtot} 
shows the density dependence of neutrino
emissivity $Q_{\rm br}$ for the ground-state matter
at six temperatures, from
$10^8$ K to $1.8 \times 10^9$ K, in the density range
$10^{10}$ g cm $^{-3}  \leq  \rho  \leq 1.4 \times 10^{14}$ g cm$^{-3}$.
The $\rho-T$ domain displayed is the most
important one for application to
neutron star cooling.
The dotted curves are calculated using the
Negele--Vautherin model of matter and the form factors
of the uniform proton cores of atomic nuclei.
The dashed lines are obtained for the
smooth-composition (SC) model of the ground state matter
(Sect.\ \ref{sect-nucrust-introduc})
with the realistic form factor, and assuming the nuclei
to be spherical to the crust bottom.
Small jumps of the emissivity are smeared out in this model
but all the main features of the emissivity are reproduced.
The emissivities decrease abruptly
at the crust--core interface ($\rho = 1.43 \times 10^{14}$
g cm$^{-3}$, in the given model). 
The solid lines in Fig.\ \ref{nucrust-ebrems-figtot}
are also derived using the smooth composition model with the
realistic form factor but with allowance for
the phases of nonspherical nuclei (Sect.\ \ref{sect-nucrust-introduc}).
In the nonspherical phases, the Debye--Waller factor
and the phonon contribution are
presently unknown and thus neglected.
This circumstance is partly
responsible for the jumps in the emissivities
at $\rho \approx 10^{14}$ g cm$^{-3}$, the
interface between the phases with spherical and cylindrical nuclei.

Outside the neutron star crust, in the uniform matter
of the core, the bremsstrahlung process
of study transforms actually into the neutrino bremsstrahlung
of electrons which scatter off protons (remnants of atomic
nuclei). The latter process, studied in Sect.\ \ref{sect-nucore-other},
is much less efficient
due to strong proton degeneracy in the uniform matter.

The neutrino emission
at $\rho \sim 10^{14}$ g cm$^{-3}$
is very sensitive to the proton charge distribution.
The effects of possible nonspherical phases
are also rather important. Non-sphericity of the nuclei
mainly lowers the neutrino emission by reducing
the dimension of the sums over reciprocal lattice vectors
in Eq.\ (\ref{nucrust-ebrems-S}).
The reduction can exceed one order
of magnitude. More work
is required to calculate the Debye--Waller factor and
the phonon contribution, and determine accurately the
bremsstrahlung emission for nonspherical nuclei.

The calculations for the spherical nuclei with the realistic form factor
in the wide density and temperature ranges,
$10^9$ g cm$^{-3} \leq \rho \leq 1.4 \times 10^{14}$ g cm$^{-3}$
and $5 \times 10^7$ K $\leq T \leq 2 \times 10^9$ K,
were fitted by the expression
(Kaminker et al.\ 1999a)
\begin{eqnarray}
 &&    \lg Q_{\rm br} \, [{\rm erg \, cm^{-3} \, s^{-1}}]  =
         11.204+7.304 \, \tau+0.2976 \, r
\nonumber\\&&
     - 0.370 \, \tau^2  +  0.188 \, \tau r- 0.103 \,r^2 + 0.0547 \, \tau^2 r
   \nonumber\\&&
-6.77 \, \lg \left(1+ 0.228 \, \rho / \rho_0 \right),
\label{nucrust-ebrems-Qfit}
\end{eqnarray}
where $\tau = \lg T_8$, $r = \lg \rho_{12}$,
and $\rho_0 = 2.8 \times 10^{14}$ g cm$^{-3}$.
The relative error of this fit formula generally does not exceed 1\%
(in $\lg Q$) over the indicated $\rho-T$ domain.
The fitting formula reproduces the main features of the 
bremsstrahlung emissivity 
at $\rho \lesssim 10^{14}$ g cm$^{-3}$,
where the atomic nuclei are expected to be spherical,
and it probably gives a realistic upper limit of the emissivity
for higher densities, where the nuclei can be nonspherical.\\

{\bf (f) Very low temperatures, charged impurities}

The case of low temperatures is of no great
importance in practice. We outline for completeness.

If a lattice is perfect, then the static lattice
contribution becomes exponentially suppressed at
very low temperatures by the band structure effects, as discussed above.
The phonon contribution is also strongly reduced due to
freezing out of umklapp processes. 
The freezing temperature $T_u$ can be estimated as
(Raikh and Yakovlev 1982)
$T_u \sim T_p \, Z^{1/3}e^2/ (\hbar v_{\rm F})$ 
($v_{\rm F} \approx c$ being the electron Fermi velocity) and
is rather low (e.g., Baiko and Yakovlev 1996). 
At $T \ll T_u$
only a very small residual phonon contribution
survives produced by normal phonon-scattering process.
According to Flowers (1973)
the phonon contribution to $Q_{\rm br}$ behaves in this
regime as $T^{11}$.

If lattice is imperfect, the emissivity may be much larger
and its calculation is
complicated. The problem could be solved taking into account
distortion of the phonon spectrum and the electron band
structure by lattice imperfections. The result is
very sensitive to the type of imperfections and their
distribution over the crystal. However, in one
case the result is evident (e.g., Pethick and Thorsson 1994).
Let the imperfections be produced by charged
impurities, the ions of charge $Z_{\rm imp}\neq Z$ and
number density $n_{\rm imp}\ll n_i$ embedded in lattice sites.
Assume further that the impurities are distributed
randomly over the lattice. If so, the plasma electrons will
regard the impurities as Coulomb centers with charge number
$(Z_{\rm imp}-Z)$, and the neutrino emission will be produced
by nearly elastic Coulomb scattering of electrons off these
centers (like scattering off ions in Coulomb liquid).
The emissivity is then given by Eq.\ (\ref{nucrust-ebrems-Q})
with $n_i \to n_{\rm imp}$ and
$Z \to (Z_{\rm imp}-Z)$, $L \to L_{\rm imp}$, where
$L_{\rm imp} \sim 1$ is an appropriate Coulomb logarithm.
Its order-of-magnitude estimate is
$L_{\rm imp} \sim \ln (2 p_{{\rm F}e} R_c)$.
Here,
$R_c$ is a correlation length which may be determined
by impurity distribution ($R_c \sim n_{\rm imp}^{-1/3}$
for random distribution), electron screening,
etc. However
the neutrino emissivity may be strongly reduced
if the impurity ions are arranged in (quasi)ordering structures:
the plasma electrons will regard them as elements of (quasi)perfect lattice
rather than individual Coulomb centers. Since the types
of lattice imperfections in neutron star crusts are
largely unknown, the properties of neutrino bremsstrahlung
at low temperatures are generally unknown as well.
Luckily, they seem to be of purely academic interest.\\

{\bf (g) Concluding remarks}

To summarize, neutrino bremsstrahlung
in neutron star crusts is described quite
reliably at not too low temperatures
for the case of spherical atomic
nuclei. In this case the emissivity is rather
insensitive to a specific model of matter
(ground-state or accreted matter). The main principal problem
to be solved is how the phonon spectrum is affected by the
presence of free neutrons in the neutron drip regime.

As for the case of non-spherical nuclei
at the bottom of the neutron-star crust,
much work is required to evaluate
the phonon contribution and the Debye--Waller factor,
and to determine the total emissivity.

It is also important to discuss the effect of strong magnetic fields
on the neutrino bremsstrahlung process.
We expect that the neutrino emissivity
is quite insensitive to the magnetic field as long
as the field does not quantize electron motion
(does not force all the electrons to occupy the
ground Landau level) and does not change
correlation properties of Coulomb system of ions.
For instance, consider the ground state matter
at the neutron drip point ($A=118$, $Z=36$, $\rho=4.3 \times 10^{11}$
g cm$^{-3}$). One needs an enormous field $B \sim 7 \times 10^{16}$ G
to push all the electrons into the ground level
and about the same field to affect the ion correlations
(which require
$\omega_{Bi} \gtrsim \omega_{pi}$).
We do not consider these fields in neutron star
crusts and, accordingly, assume that the
neutrino bremsstrahlung emissivity
is not affected noticeably by the magnetic fields.

Generally, neutrino bremsstrahlung
is a powerful and `robust'
mechanism. The emissivity is not a strong
function of density, and nearly a power-law function
of temperature [see Eq.\ (\ref{nucrust-ebrems-Qfit}),
Fig.\ \ref{nucrust-ebrems-figtot}].
It is rather insensitive to the models of stellar matter
and to the presence of magnetic fields, and it operates
with nearly the same efficiency in all the corners
of neutron star crusts. As seen from Figs.\
\ref{nucrust-fig93}--\ref{nucrust-fig81},
it is one of the leading neutrino mechanisms
in a neutron star crust
at $T \lesssim 10^9$ K. This statement
will be additionally illustrated in Sects.\ \ref{sect-nucrust-nn}
and \ref{sect-nucrust-overlook}.\\

{\bf (h) Related neutrino processes}

Let us mention two other related mechanisms.

The first is the {\it recombination neutrino process}
which consist in neutrino pair emission
due to capture of a free electron into the K-shell
of a fully ionized atom. The process
was introduced by Pinaev (1963) and studied
by Beaudet et al.\ (1967), Kohyama et al.\ (1993)
%
%
and Itoh et al.\ (1996).
It operates at the densities
$\rho \lesssim 10^6$ g cm$^{-3}$ at which
the K-shell is not smashed by the pressure ionization.
While calculating the neutrino emissivity it is important
to use exact wave functions of electrons (in continuum)
in the Coulomb field of the ion (Kohyama et al.\ 1993).
Notice that
the present calculations do not take into account important
effects of medium on the continuum and bound
(initial and final) electron states.

The second process is the {\it photoemission of neutrino
pair} in the Coulomb field of the atomic nucleus
(see, e.g., Bisnovatyi-Kogan 1989, for references):
\begin{equation}
    \gamma + (A,Z) \to (A,Z) + \nu + \bar{\nu}.
\label{nucrust-ebrems-photo}
\end{equation}

Both mechanisms are unimportant under the conditions we
are interested in.

\subsection{Beta processes}
\label{sect-nucrust-beta}

{\bf (a) General remarks}

The beta processes are the reactions of
electron or positron captures and decays by atomic nuclei $(A,Z)$:
\begin{eqnarray}
 &&  (A,Z) + e \to (A,Z-1) + \nu_e,
     \quad (A,Z-1) \to (A,Z) + e + \bar{\nu}_e;
\nonumber \\
  &&  (A,Z) \to (A,Z-1) + e^+ + \nu_e,
     \quad (A,Z-1) + e^+ \to (A,Z) + \bar{\nu}_e.
\label{nucrust-beta-Urca}
\end{eqnarray}
The nuclei may be in the ground and excited states.
Particularly, $(A,Z)$ may stand for a nucleon (neutron or proton).
The beta processes are much slower than
interparticle collisions in stellar matter. The
latter collisions very quickly establish a quasiequilibrium
state with some temperature
$T$, and chemical potentials of electrons ($\mu_e$), positrons
($-\mu_e$), neutrons ($\mu_n$), protons ($\mu_p$),
and nuclei ($\mu(A,Z)$).
A quasiequilibrium state does not necessarily mean {\it full
thermodynamic equilibrium}.

A pair of two subsequent (direct and inverse)
reactions in each line of
Eq.\ (\ref{nucrust-beta-Urca}) is called an {\it Urca}
process. The outcome of these reactions
is a neutrino pair which escapes the star and
carries away energy without changing nuclear composition.
The composition is unchanged if
the rates of the direct and inverse reactions are equal.
This happens in the so called {\it thermodynamic beta-equilibrium} state.
In other words, the thermodynamic beta-equilibrium means
$\mu(A,Z)+\mu_e=\mu(A,Z-1)$.
In the absence of this equilibrium
the rates of direct and inverse reactions are different,
and the reactions do affect the
nuclear composition driving matter towards the equilibrium.
Urca processes involving electrons
were put forward
by Gamow and Shoenberg (1941) while those involving
positrons were introduced by Pinaev (1963).
In Sects.\ \ref{sect-nucore-Durca} and \ref{sect-nucore-Murca}
we discuss analogous Urca processes involving nucleons and electrons
which produce an efficient neutrino
cooling of neutron star cores.
The origin of the name {\it Urca} is discussed
in Sect.\ \ref{sect-nucore-Durca}.

A bath of highly energetic electrons and positrons
in stellar matter
affects stability of the nuclei.
For instance, the nuclei unstable
with respect to beta decay in
laboratory may be produced in large amounts
by beta captures of high-energy electrons;
their decay could be suppressed by Pauli blocking
provided by the Fermi sea of degenerate
electrons.
On the other hand, the nuclei stable in laboratory
may capture electrons and disappear.
In this way the beta processes
affect nuclear composition of dense matter.
The beta processes
are naturally accompanied by neutrino energy loss.
Moreover, nonequilibrium beta processes may heat matter.
Generally, all these effects
must be considered selfconsistently.

Each reaction  
given by Eq.\ (\ref{nucrust-beta-Urca}) can be 
characterized by an appropriate
reaction rate and
and neutrino energy generation rate.
These rates are
calculated in the standard manner (e.g., Bisnovatyi-Kogan 1989)
by integrating and summing the differential
reaction rates over the states of all reacting
particles. The differential rates are mainly determined
by measured, calculated or extrapolated free decay times
of parental or daughter nuclei.
Generally, calculation of the integrated reaction rates
takes into account the Pauli
principle (Fermi-Dirac statistics, blocking of occupied states)
for electrons, positrons, neutrons and protons.
For the neutron
star matter of subnuclear density at $T \lesssim 10^{10}$ K
one should also take into account the effects
of possible superfluidity of nucleons (in analogy
with the theory presented in Chapt.\ \ref{chapt-nusup}).

Once the reaction rates are known, one can write
a system of kinetic equations to describe evolution
of nuclear composition in time. These equations have to
be supplemented by the equations of charge neutrality.
Notice that the beta reactions do not change $A$,
the mass number of atomic nuclei, but only redistribute
the nuclei with given $A$ over different charge numbers $Z$.
In this way they affect the ratio of the total numbers of neutrons
and protons in matter (both free and bound). 
However, in addition to the beta processes, there are
non-leptonic {\it nuclear reactions} like neutron
emission and capture, alpha-particle emission and capture, 
nuclear fusion and fission.
They change $A$ but do not change ratio of the total numbers
of neutrons and protons.
Thus, the nuclear reactions and beta processes affect nuclear
composition in different ways.
One should, in principle, supplement
the kinetics of beta processes by the kinetics of nuclear reactions.
Matter may contain many
nuclides, neutrons, protons, alpha particles, electrons and positrons
with numerous reaction channels. Thus,
the problem of evolution of nuclear composition
is very complicated. If the problem is solved,
the neutrino emission rate due to the beta processes $Q_{\rm beta}$
can be found by summing contributions
of all reactions.\\

{\bf (b) Kinetic beta-equilibrium of hot matter}

The solution can be simplified for high temperatures
$T \gtrsim 4 \times 10^9$ K, at which
thermonuclear burning is so fast that its typical time scales
become shorter than all other macroscopic
time scales. This opens many nuclear reaction channels
and drives the matter into the state of {\it thermodynamic nuclear
equilibrium} with respect to dissociation of any nucleus into
nucleons,
          $(A,Z) \rightleftharpoons Zp + (A-Z)n$.
If so, 
the number density $n_j$ of the nuclei of type $j \equiv (A,Z)$ is determined by the
condition $\mu(Z,A)=Z \mu_p+ (A-Z) \mu_n$.
For instance, for the Boltzmann gases of all constituents
one has (e.g., Bisnovatyi-Kogan 1989)
\begin{equation}
    n_j = \left( 2 \pi \hbar^2 \over m_u k_{\rm B} T \right)^{3(A-1)/2} \,
        {A^{3/2} \over 2^A } \, (2 I +1) \, n_p^Z n_n^{A-Z} \,
        \exp \left( B \over k_{\rm B} T \right) ,
\label{nucrust-beta-nuceq}
\end{equation}
where $I$ is the spin of the given nucleus $j$, $B$ is its binding
energy, and $m_u$
is the atomic mass unit.
Then $n_j$ is expressed through the number densities of
free neutrons and protons, $n_n$ and $n_p$. The total number densities
$N_n$ and $N_p$ of neutrons and protons (free and bound)
can be written as $N_n= n_n+ \sum_j (A_j-Z_j) \, n_j$ and
$N_p = n_p + \sum_j Z_j \, n_j$. They 
are regulated by much slower beta processes:
\begin{equation}
    \dot{N}_n = - \dot{N}_p = W^+ - W^-,
\label{nucrust-beta-equil}
\end{equation}
where $W^+$ and $W^-$ are the rates of neutron production
and decay summed over all reaction channels.
Under these conditions, the beta reactions drive matter into the
state of {\it kinetic beta equilibrium} in which
$\dot{N}_n=\dot{N}_p=0$. Assuming the nuclear equilibrium and
the kinetic beta equilibrium one can determine the nuclear
composition of matter and the neutrino emissivity $Q_{\rm beta}$.
The latter is usually called the
emissivity due to Urca processes.

The kinetic beta equilibrium in an open system with escaping
neutrinos does not
generally mean thermodynamic
beta equilibrium. The rates of individual direct and inverse
reactions in Eq.\ (\ref{nucrust-beta-Urca}) may be different
(no detailed balancing), and the stationary state is then supported
by many reaction channels. The beta equilibrium would become
thermodynamic if the matter were opaque for neutrinos,
as happens at temperatures $T \gtrsim
10^{10}$ K which we do not consider.
Note that the kinetic beta equilibrium
becomes thermodynamic in purely $npee^+$ matter
(no different
reaction channels, detailed balancing has to be satisfied in
a stationary state). 

At $T \sim 4 \times 10^9$ K
and $\rho \lesssim 10^{11}$ g cm$^{-3}$ the
amount of free nucleons in matter under
kinetic beta equilibrium is small,
and the matter consists mainly of nuclei.
With increasing $T$, the nuclei
dissociate, and the free nucleons appear. Hot
matter under kinetic equilibrium consists of the nucleons, mainly
the neutrons.
The denser the matter, the higher is the temperature of
dissociation of the nuclei.

Kinetic beta equilibrium of free nucleons was considered
by Imshennik et al.\ (1966, 1967).
%
%
%
The same problem for a mixture of nucleons and Fe nuclei
was solved by Ivanova et al.\ (1969).
%
%
Chechetkin (1969)
%
%
and Nadyozhin and Chechetkin (1969)
%
%
extended this consideration taking into account alpha particles and
about 50 isotopes of Ti, V, Cr, Mn, Fe, Co, and Ni
in the ground and excited states.
In particular, Nadyozhin and Chechetkin (1969)
calculated the neutrino emissivity of the Urca processes
at the temperatures from
$4 \times 10^9$ K to $1.6 \times 10^{10}$ K and at the densities
from $10^4$ g cm$^{-3}$ to $10^{12}$ g cm$^{-3}$.

The rates of numerous beta reactions (many nuclides
in the ground and excited, initial and final states)
were calculated in a series of papers by Fuller et al. (1980, 1982a,
1982b, 1985)
%
%
%
%
%
and Aufderheide et al.\ (1994).
%
%
These results are mainly obtained for hot matter at
$\rho \lesssim 10^{12}$ g cm$^{-3}$ neglecting nucleon degeneracy.
%
%
They are most important for
studying evolution of presupernovae,
stellar collapse and supernova explosions.

Now we can turn to
Urca processes in hot matter
(nuclear equilibrium, kinetic beta equilibrium)
of neutron star crusts.
According to Nadyozhin and Chechetkin (1969),
the Urca processes produce highly efficient neutrino emission
in a hot plasma
of the temperature $T \gtrsim 10^{10}$ K in the density
range from about $10^5$ to $10^{12}$ g cm$^{-3}$.
For lower $T \lesssim 10^{10}$ K, we are interested in,
the density dependence of the neutrino emissivity
has maximum at $\rho \sim 10^9$ g cm$^{-3}$
produced by beta decays of the $^{55}$Cr and
$^{53}$V nuclei with high energy release ($\sim2.5$ MeV per
beta-decay).
The decrease of the emissivity at higher densities
is mainly explained by transformation of $^{55}$Cr and $^{53}$V
into less active $^{56}$Cr and neutrons. In addition,
the electron reaction rates are reduced by the Pauli principle
and the positron reaction rates
are reduced by the decrease of the positron number density
because of growing electron degeneracy.
For $T \sim 10^{10}$ K, the maximum emissivity
is about $10^{25}$ erg cm$^{-3}$ s$^{-1}$,
but the maximum decreases strongly
with the fall of $T$, reaching about $10^{20}$ erg cm$^{-3}$ s$^{-1}$
for $T = 5 \times 10^9$ K.
The emissivity decreases with the growth of
the density above $10^9$ g cm$^{-3}$.
The lower $T$ the sharper the emissivity drop.
For instance, at $T= 5 \times 10^9$ K and
$\rho = 10^{10}$ g cm$^{-3}$ the emissivity
becomes as low as $10^{16}$ erg cm$^{-3}$ s$^{-1}$
dropping by four orders of magnitude while $\rho$
increases by one order of magnitude.
At the densities $\rho \gtrsim 10^{10}$ g cm$^{-3}$,
we are interested in, the neutrino emissivity produced by the
Urca processes seems to be lower than the emissivities of other
major neutrino reactions (e.g., neutrino bremsstrahlung
in electron-nucleus collisions, Sect.\ \ref{sect-nucrust-ebrems}).
If so, the crustal Urca processes
are unimportant for
thermal evolution of hot neutron stars. However, 
as mentioned by Nadyozhin and Chechetkin (1969),
one should be careful in dealing 
with the Urca processes in high-density matter
since the presence of nuclei with large neutron excess and energy
release might rise the Urca emissivity at
$\rho \gtrsim 10^{10}$ g cm$^{-3}$.\\

{\bf (c) Beta reactions in cold matter}

The problem of nuclear composition and neutrino emission due to beta 
processes in dense layers of a neutron star crust at 
$T \lesssim 10^9$~K is 
most complicated. Thermonuclear reactions between nuclei, which require 
Coulomb barrier penetration, become extremely slow.
However, at sufficiently high densities some
pycnonuclear reactions occur due to zero-point motion of the nuclei.
The reactions of neutron emission and absorption 
are also possible, because they 
encounter no Coulomb barriers.
Nuclear composition ceases to depend on 
density and temperature only,  
but depends on previous history of the crust.

A model of {\it ground state matter} (equivalent term is {\it cold 
catalyzed matter}) corresponds to a
complete thermodynamic equilibrium. Clearly, this 
equilibrium holds at the neutron star birth, 
when temperature exceeds $10^{10}~$K. We can `prepare' cold
ground state matter by assuming that
the neutron star cools, remaining very close to 
complete thermodynamic equilibrium and reaching finally a state, in 
which thermal contributions to the energy and pressure are negligible. 
The properties of the cold equilibrium 
state can be calculated
in the $T=0$ approximation.  
At given nucleon density, the composition 
of such a state corresponds to the minimum energy
per nucleon. One commonly assumes the presence of one species of 
atomic nuclei at a given density (or pressure). Accordingly, the nuclear 
composition varies in a discontinuous manner at certain threshold values 
of pressure, the change of nuclides being accompanied by a density 
discontinuity of a few percent. 
The properties of ground state matter at densities
lower than the neutron drip density $\rho_{\rm d}$ 
can be studied using experimental masses of nuclei, 
and their extrapolation via semi-empirical mass formulae for
those  values of $A$ and $Z$ 
for which experimental data are not available. At $\rho < \rho_{\rm d}$
the matter consists of nuclei 
immersed in a homogeneous electron gas. This matter forms
the {\it outer crust} of the star.  
The ground state composition of the outer 
crust, based on 
present experimental  
knowledge of nuclear masses, was 
determined by Haensel and Pichon (1994).
The neutron fraction in the nuclei increases with increasing density. 
This {\it neutronization} of the nuclei results from
minimization of the energy of matter. 
At $\rho > \rho_{\rm d}\simeq 
4.3 \times 10^{11}~{\rm g~cm^{-3}}$,
in the {\it inner crust}, some of neutrons form a gas outside 
nuclei. 
The composition of the inner crust 
is studied basing on theoretical models 
of dense matter (Negele and Vautherin 1973, Lorenz et al. 1993, 
Douchin et al.  2000,
for 
a review see Pethick and Ravenhall 1995).
Notice that if thermal 
effects were taken into account, abrupt changes of nuclear species 
would be washed out, and Urca processes  could operate near the 
interfaces which separate layers of different nuclear compositions.
This neutrino emission should be very weak.

The ground state model of the crust corresponds to one  
idealized scenario of the neutron star crust formation. A very 
different scenario corresponds to an {\it accreted crust}, formed 
of a relatively cold ($T \lesssim 10^9$ K) 
matter which sinks gradually within the neutron 
star under the weight of a newly accreted material. Let us 
assume that light elements burn   
into  $^{56}$Fe in the outer layers of 
the accreting star, at $\rho\simeq 10^8~{\rm g~cm^{-3}}$.
In view of relatively low temperature further nuclear evolution
of a sinking (compressed) matter is governed by
non-equilibrium beta 
captures, neutron emissions and absorptions, and -- at high
densities -- by pycnonuclear fusion.
A basic approximation to follow this evolution 
(Bisnovatyi-Kogan and  Chechetkin 1974, 1979, Vartanyan and 
Ovakimova 1976, Sato 1979, Haensel and Zdunik 1990)
consists in 
neglecting thermal effects. In this approximation, the reactions  
have abrupt (threshold) character.
This leads to discontinuous 
variations  of nuclear composition with density,
in analogy with the ground-state matter. 
The composition  
depends evidently
on the assumed theoretical model of neutron rich nuclei, 
especially at $\rho > \rho_{\rm d}$. 
Let us discuss the results obtained by Haensel and Zdunik (1990)
using a specific 
model of nuclei. 
 
At $\rho=5.8 \times 10^8$~g~cm$^{-3}$ transformation of  
$^{56}$Fe
into  $^{56}$Cr becomes energetically possible. However, direct
transition $^{56}$Fe$\rightarrow ^{56}$Cr would require an
extremely slow double electron capture.
Therefore, the reaction proceeds in two steps. The first step is
\begin{equation}
    ^{56}{\rm Fe}+e^{-}\rightarrow ^{56}{\rm Mn}+\nu_e
\label{nucrust-Fe56}
\end{equation}
At $T\sim 10^8$K the initial $^{56}$Fe nucleus is
in its $J^{\pi}=0^+$ ground state.
The ground state and the first excited state
of the $^{56}$Mn
nucleus are $J^{\pi}=3^+$
and $J^{\pi}=2^+$, respectively. 
Thus, the electron capture should
lead to the $^{56}$Mn nucleus in the excited $J^{\pi}=1^+$ state
(e.g., Dzhelepov and Peker 1961). 
The threshold density for this reaction is 
$1.5\times 10^9~{\rm g~cm^{-3}}$. 
Because $^{56}$Mn is the odd-odd nucleus, its binding energy is 
significantly lower than that of the even-even $^{56}$Cr nucleus.
Accordingly, after de-excitation of $^{56}$Mn by the gamma emission,
the first electron capture, Eq.\ (\ref{nucrust-Fe56}), is
followed by the second one,
\begin{equation}
       ^{56}{\rm Mn}+e^-\rightarrow ^{56}{\rm Cr}+\nu_e.
\label{nucrust-Mn56}
\end{equation}
With increasing density, two-step electron captures occur every
time the threshold for a single electron capture is reached, according to
a general scheme
\begin{equation}
        (A,Z)+e^-\rightarrow (A,Z-1)+\nu_e~,
\label{nucrust-ecap1}
\end{equation}
\begin{equation}
       (A,Z-1)+e^-\rightarrow (A,Z-2)+\nu_e~.
\label{nucrust-ecap2}
\end{equation}
Usually, the first step, Eq.\ (\ref{nucrust-ecap1}), 
takes place very (infinitesimally)
close to the threshold and is accompanied by an
infinitesimally small energy release
(quasi-equilibrium process). An exception
from this rule is a process which, due to the selection rules, 
proceeds into an excited
state of the daughter nucleus [e.g.,
reaction (\ref{nucrust-Fe56})]. Notice, that because of 
the low temperature,
any nucleus undergoing an electron capture should  be 
in its ground state. If the daughter nucleus is produced in an
excited state of energy $\epsilon_{\rm exc}$, then it 
de-excites by gamma emission before the next electron
capture leading to additional 
heat release $\epsilon_{\rm exc}$ per nucleus. The second
electron capture, Eq.\ (\ref{nucrust-ecap2}), is always
non-equilibrium,
because electron capture by the odd-odd $(A,Z-1)$ nucleus
is energetically favourable. It 
is accompanied by an energy release 
$\epsilon_{\rm r}$ per nucleus. Mechanical
equilibrium requires this process to take place at constant pressure.
On average, neutrino emitted in a non-equilibrium reaction
(\ref{nucrust-ecap2}) takes away  
$\epsilon_{\nu}\simeq{5\over 6} \epsilon_{\rm r}$, 
while ${1\over 6} \epsilon_{\rm r} + \epsilon_{\rm exc}$ 
heats matter. For non-equilibrium electron capture (\ref{nucrust-Mn56}),
the neutrino energy is $\epsilon_\nu=1.9~$MeV. 

At $\rho\simeq 6\times 10^{11}~{\rm g~cm^{-3}}$ electron captures by nuclei, 
which are then $^{56}$Ar, induce neutron emission: this is  
the neutron drip 
density for the accreted crust.
At $\rho > \rho_{\rm d}$ 
general scheme of nuclear 
transformations consists of two electron captures accompanied by emission 
of several neutrons. Energies of the emitted neutrons exceed the 
Fermi energy of neutron gas outside the nuclei. 
Finally, at densities exceeding 
$10^{12}~{\rm g~cm^{-3}}$, electron capture, which decreases the  
nucleus charge, may  
trigger pycnonuclear fusion with a
neighbour nucleus, accompanied by neutron emission. Pycnonuclear
fusion reactions 
have therefore threshold character;
they are much more efficient heating sources
than nonequilibrium beta reactions in the outer crust (Haensel and Zdunik 
1990).  
   
The values of $Z$ and $A$ of nuclei in dense accreted 
matter are 
significantly lower than in ground state 
matter. Continuity of pressure at threshold densities 
implies, due to dominating contribution of electrons into the pressure, 
a noticeable density jump  
associated with the change of $Z$ and $A$. A typical density
jump  in the outer crust
is around ten per cent. It can
exceed ten percent at  $\rho \sim 10^{12}~{\rm g~cm^{-3}}$ but 
decreases then to a few per cent at  
$\rho \gtrsim 10^{13}~{\rm g~cm^{-3}}$, 
where the main contribution to the pressure comes from free neutrons. 

The scenario described above is based on several simplifying 
assumptions (one nuclear species present at once, neglect
of thermal effects). At $\rho \gtrsim \rho_{\rm d}$
it depends also on the assumed 
model of neutron rich nuclei immersed in neutron gas. 

Sinking of accreted matter is accompanied by non-equilibrium 
neutrino emission from very thin shells, where 
the electron captures occur.
The associated neutrino luminosity is proportional to the accretion rate 
and is independent of temperature, 
\begin{equation}
  L^{\rm noneq}_\nu
  = 6.03 \times 10^{33}\, \xi~{\rm erg~s^{-1}},
   \quad  \xi= \dot{M}_{-10}
   \sum_{\rm i}q_{\rm i},
\label{nucrust-Lnu.neq}
\end{equation}
where  $\dot{M}_{-10}$ is the accretion rate in
units of $10^{-10}\,M_\odot$ yr$^{-1}$ and
$q_{\rm i}$ is neutrino energy per nucleon,
expressed in MeV, from
a non-equilibrium electron capture in an i-th shell. The effect of
non-equilibrium
processes on the thermal structure of {\it steadily accreting neutron star}   
was studied by Miralda-Escud{\'e} et al.\ (1990).
The strongest heating occurs in the inner crust at $\rho\sim
10^{12}~{\rm g~cm^{-3}}$ due to the pycnonuclear fusion
triggered by electron captures. Heat deposition, about one MeV
per accreted nucleon, flows into the neutron star core. This
heat flow was shown 
to be important
in {\it transiently
accreting neutron stars}, maintaining  core temperature between 
accretion episodes (Brown et al.\ 1998).

Other scenarios of formation of a crust in an
accreting neutron star, e.g., compression of initially ground 
state crust by accreted matter, were studied by Sato (1979). 

A different model of a non-equilibrium crust was proposed
by Bisnovatyi-Kogan and Chechetkin (1979). They
start from initially very hot ($T>4\times 10^9~$K) matter
in the state of kinetic beta equilibrium (see above), 
and then cool it down,
as the star cools. The hot matter contains higher
neutron fraction, than the cold ground state matter.
After cooling below $4\times 10^9~$K,
nuclear reactions, which require Coulomb barrier penetration,
freeze out and the nuclear statistical equilibrium is violated. 
According to Bisnovatyi-Kogan and Chechetkin (1979), after cooling to
$\sim 10^9~$K the nuclear composition becomes temperature independent
and resembles the composition of the ground state
of matter, except in the density range  from about 
$10^{11}~{\rm g~cm^{-3}}$ 
to 
$2\times 10^{12}~{\rm g~cm^{-3}}$.
In the latter density range around
the neutron drip density, the excess of neutrons compared to
the ground state composition is quite significant. This could
provide storage of some nuclear energy. 
However, this conclusion 
was based on certain model assumptions
on the properties of nuclei with high neutron excess.
It would be interesting to reconsider the problem
using the recent data on the neutron-rich nuclei.

Beta processes,
which accompany cooling of matter and non-equilibrium nuclear
reactions, produce neutrino emission.
The appropriate neutrino
luminosity depends on cooling rate and
should be especially strong at $T \sim$(2--4) $\times 10^9$ K
when the main fraction of free neutrons
is captured by nuclei. However, there are many other
efficient neutrino reactions open at such temperatures,
which make the neutrino emission due to beta processes
insignificant.
On the other hand, heating produced by
non-equilibrium nuclear reactions may be more important than
non-equilibrium neutrino cooling.  

\subsection{Neutrino emission connected with 
strong interaction of free neutrons}
\label{sect-nucrust-nn}

{\bf (a) Neutrino emission and illustrative neutron superfluid models}

In this section, we describe three specific mechanisms
of neutrino emission resulting from strong interactions of
free neutrons in inner neutron star crusts:
neutrino brems\-strah\-lung in
neutron-neutron collisions,
neutrino emission due to Cooper pairing of neutrons, and
bremsstrahlung in neutron-nucleus collisions.
Two first processes are well known for uniform matter
of neutron star cores (Sects.\
\ref{sect-nucore-Brems} and \ref{sect-nusup-CP}) but they have
rarely been discussed in the literature
for the inner crusts. Nevertheless, the
properties of free neutrons in the inner crust
resemble the properties of neutrons in uniform matter.
Hence the similarity of neutrino
processes in cores and crusts.

The processes of study
depend on singlet-state superfluidity of free neutrons
in the crust (Sect.\ \ref{sect-overview-struct}).
Let us remind that the
superfluid critical temperature
$T_{cn}$ is very model dependent.
It is sufficiently low near the
neutron drip density,
increases with density,
reaches maximum at some subnuclear density and decreases
to zero in the vicinity of the core-crust interface.
While analyzing the superfluid effects we will
consider two cases, corresponding to strong (s) and weak (w)
neutron superfluids (SFs). The strong superfluid model
is based on the rather large superfluid gaps
calculated by Elgar{\o}y et al.\ (1996), 
%
%
with the maximum gap of about 2.5 MeV. The weak superfluid model
makes use of the small superfluid gaps
derived by Wambach et al.\ (1993), 
%
%
with the maximum gap of about 1 MeV.
The latter model seems to be more realistic
because it includes the effects of induced interactions.
The same superfluid models will be used in
Chapter \ref{chapt-cool} to illustrate neurton star cooling.
The density dependence of
$T_{cn}$ is plotted in Fig.\ \ref{fig-cool-tc}
in that chapter.

\begin{figure}[t!]
\begin{center}
\leavevmode
\epsfysize=8.5cm
\epsfbox[0 0 348 348]{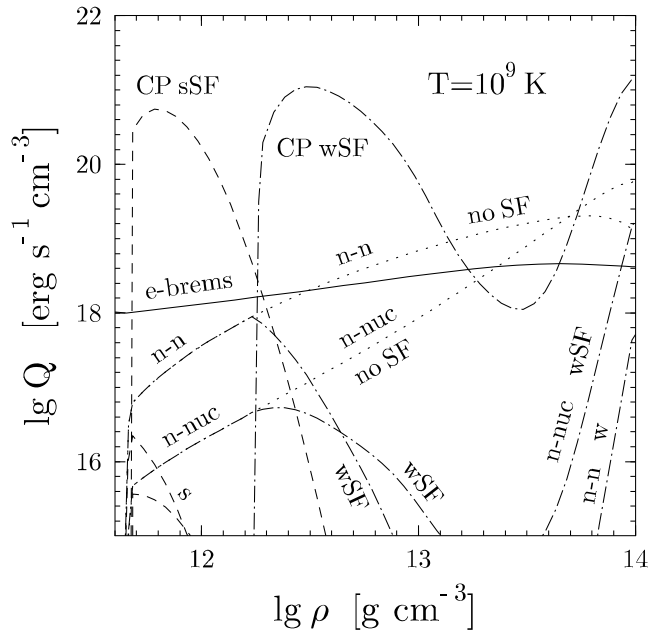}
\end{center}
\caption[ ]{
Density dependence of neutrino emissivities
in a neutron-star crust 
at $T= 10^9$ K due to
electron bremsstrahlung (e-brems, solid curve),
Cooper pairing of nucleons (CP),
neutron-neutron ($n-n$) and neutron-nucleus ($n-nuc$)
bremsstrahlung
for smooth composition -- SC --
model of ground-state matter.
The three latter mechanisms are affected
by neutron superfluidity as shown for the models of weak (w, dash-and-dots)
and strong (s, dashes) superfluidity (SF) of neutrons.
The $n-n$ and $n-nuc$ bremsstrahlung emissivities
in nonsuperfluid matter are plotted by dots (see text for
details)
}
\label{nucrust-nn-figcr9}
\end{figure}

The results of the present section will be illustrated
in Fig.\ \ref{nucrust-nn-figcr9} (as well as
in Figs.\ \ref{nucrust-overlook-figt9} and \ref{nucrust-overlook-figt8}
in the next section). In particular, Fig.\ \ref{nucrust-nn-figcr9}
shows the density dependence of the emissivity of
some neutrino processes for $T=10^9$ K in the inner
crust, from the neutron drip density
to $10^{14}$ g cm$^{-3}$. Smooth composition model
of ground state matter is employed and the atomic nuclei are
treated as spherical (Sect.\ \ref{sect-nucrust-introduc}).
The emissivities of three specific neutrino processes
discussed below are compared with the emissivity of electron
bremsstrahlung, as a reference curve.
\\

{\bf (b) Neutrino bremsstrahlung in $nn$ collisions}

The mechanism can be schematically written as
\begin{equation}
     n + n \to n + n + \nu + \bar{\nu},
\label{nucrust-nn-brems}
\end{equation}
where $\nu$ means neutrino of any flavor.
We discuss it
in Sect.\ \ref{sect-nucore-Brems} for
non-superfluid neutrons, and in Sect.\ \ref{sect-nusup-Brems}
for superfluid neutrons in neutron star cores.
The same process in the
crust is generally very similar. Thus we 
omit technical details and present the results.
The neutrino emissivity can be written as
\begin{equation}
    Q^{(nn)} = Q^{(nn)}_0 \, R^{(nn)},
\label{nucrust-nn-Q}
\end{equation}
where $Q^{(nn)}_0$ is the emissivity in non-superfluid
matter, and $R^{(nn)}$ is the factor which describes
reduction of the process by the neutron
superfluidity. One should formally set $R^{(nn)}=1$
for normal neutrons.

In Sect.\ \ref{sect-nucore-Brems}
the neutrino emissivity
of non-superfluid neutrons in the stellar core
is analyzed basing
on the matrix element of the process calculated by
Friman and Maxwell (1979) in the framework of the one-pion exchange
model.
The result is given by Eq.\ (\ref{nucore-Qnn0}).
Let us apply the same formalism for free neutrons in the crust.
A careful examination of Eq.\ (\ref{nucore-Qnn0})
shows that, for crustal matter, it has to be modified
in two respects. First,
the free neutrons in the crust occupy only some fraction of space.
Second, the factor $\alpha_{nn}$ in Eq.\ (\ref{nucore-Qnn0}),
which describes density dependence of the squared
matrix element and is nearly constant
in the stellar core, has to be
calculated from Eq.\ (52) in Friman and Maxwell (1979)
taking into account that the neutron Fermi momentum
$p_{{\rm F}n}$ in the crust may be sufficiently small.
In this way we obtain (in standard physical units)
\begin{eqnarray}
   Q^{\,(nn)} & = & {41 \over 14175} \,
             {G_{\rm F}^2 g_A^2 m_n^{\ast4} \over
             2 \pi \hbar^{10} c^8}
             \left( {f^\pi \over m_\pi} \right)^4
             p_{{\rm F}n} \alpha_{nn} \beta_{nn}
             (k_{\rm B} T)^8 {\cal N}_{\nu} \, f_v
\nonumber \\
         & \approx & 7.5 \times 10^{19}
             \left( {m_n^\ast \over m_n} \right)^4
             \left({ n_n \over n_0} \right)^{1/3}
             \alpha_{nn} \beta_{nn} {\cal N}_{\nu} T_9^8 \, f_v
             \; \;  {\rm erg\;cm^{-3}\;s^{-1}}.
\label{nucrust-Qnn0}
\end{eqnarray}
All notations are the same as in Eq.\ (\ref{nucore-Qnn0}).
In particular, ${\cal N}_\nu=3$ is the number of neutrino
flavors, $g_A \approx 1.26$ is the axial-vector
normalization constant,
$m_n^\ast$ is the neutron effective mass,
$n_n$ is the local number density of free neutrons
in space between the nuclei, and
$n_0=0.16$ fm$^{-3}$.
In addition, we introduce the filling factor $f_v$,
the fraction of space occupied by the free neutrons,
and the factor $\alpha_{nn}$ is now given by
Eq.\ (45) in Friman and Maxwell (1979):
\begin{equation}
    \alpha_{nn}= 1 - {3 \over 2} \, u \, {\rm arctan} \,
    \left( {1 \over u } \right)
     + { u^2 \over 2 (1 +  u^2 ) },
\label{nucrust-nn-alpha}
\end{equation}
with $u= m_\pi c /(2 p_{{\rm F}n})$.
One has $\alpha_{nn} \sim 1$ for the densities
at the base of the neutron star crust.
With lowering density, the factor
$u$ increases and $\alpha_{nn}$ becomes lower
reducing strongly and monotonously
the neutrino emissivity $Q^{(nn)}_0$ as the density
approaches the neutron drip point.

The neutron pairing reduces the neutrino emission rate
as described in Sect.\ \ref{sect-nusup-Brems}.
The appropriate reduction factor
$R^{(nn)}=R^{(nn)}_{n \rm A}$ is given by Eqs.\ 
(\ref{nusup-Rnpfit}) --- (\ref{nusup-New1}).
Superfluid reduction is extremely sensitive to
the critical temperature $T_{cn}$ of the neutron superfluidity.
If $T$ is constant over the
crust, the density profile of the reduction factor
$R^{(nn)}$ is strongly nonmonotonous. The highest
reduction takes place at subnuclear
density, where $T_{cn}$ has maximum, and the lowest
reduction takes place near the
neutron drip point and the core-crust interface,
where $T_{cn}$ has minima.

The two reduction sources, the low-density reduction of the matrix
element and the superfluid reduction, make the
$nn$ bremsstrahlung in the crust not very efficient
(Fig.\ \ref{nucrust-nn-figcr9}).
In the absence of superfluidity,
the emissivity $Q^{(nn)}_0$ would be
comparable to that due to the electron bremsstrahlung,
especially at subnuclear densities,
but the superfluidity significantly reduces $Q^{(nn)}$.
In the model of strong superfluidity,
$Q^{(nn)}$ becomes so small that it is
almost invisible in Fig.\ \ref{nucrust-nn-figcr9}.
In the weak-superfluid model, it is higher but nevertheless
negligible. Notice that the neutrons remain non-superfluid ($T_{cn}<T$)
in the density layer with $\rho \lesssim 1.6 \times 10^{12}$
g cm$^{-3}$ at $T=10^9$ K, for the weak-superfluid model.
Nevertheless, the emissivity $Q^{(nn)}$ is small at these
densities suppressed by
the reduction of the matrix element.
The only place where the $nn$ bremsstrahlung may be competitive is
the layer of non-spherical atomic
nuclei at the base of the inner crust where the
superfluid suppression is relatively weak. Let us stress
that we have used the simplified matrix element of the process
although it is unlikely that the improved matrix element
will change our main conclusions.\\

{\bf (c) Cooper pairing of neutrons}

The next mechanism
consists in producing neutrino pairs
(all flavors) due to Cooper pairing of free neutrons.
This mechanism
is studied in more detail in Sect.\ \ref{sect-nusup-CP}
for uniform matter in neutron star cores.
The neutrino emissivity $Q^{\rm (CP)}$
in the crust is readily given by Eq.\ (\ref{nusup-CP-Qrec})
multiplied additionally 
by $f_v$, the fraction of
space available to  free neutrons. 
In that equation
one must set $a_{n \rm A}=1$, for the singlet-state neutron
pairing,
and use the function $F_{\rm A}$ given by
Eq.\ (\ref{nusup-CP-RecFit}). The theoretical
formula for $Q^{\rm (CP)}$ is quite reliable
but numerical values of the emissivity depend strongly
on specific values of $T_{cn}$.

The temperature dependence of $Q^{\rm (CP)}$ is
demonstrated in Fig.\ \ref{fig-nusup-Qrec}
(for uniform matter).
The process starts to operate when
the temperature falls down below $T_{cn}$.
With further decrease of $T$, the emissivity $Q^{\rm (CP)}$
grows up, reaches maximum at
$T \sim 0.8 \, T_{cn}$, and then decreases exponentially.
The maximum emissivity is very large. It can exceed the
emissivity produced by electron bremsstrahlung
by 2 or 3 orders of magnitude.
However, the most intense emission
is concentrated in certain density layers.
If we assume that
$T$ is constant throughout the crust and decreasing in time
(imitating stellar cooling) we obtain that
Cooper-pairing neutrino emission starts to operate
at subnuclear densities, where $T_{cn}$ has maximum 
as a function of $\rho$. There will be a peak of
neutrino emission in a layer at these densities.
With further decrease of $T$,
the emission in the layer will weaken,
but the Cooper pairing process will
become open at larger and lower $\rho$, where $T \leq T_{cn}$.
In this way the neutrino emitting layer will split
into two layers which will shift to the upper and lower
boundaries of the inner crust. Finally, each layer will
reach the appropriate boundary and it will fade away there after $T$
becomes much lower than $T_{cn}$ near the
boundary. For different models of neutron superfluidity,
such wanderings of the neutrino emission layers
take place at different cooling stages.
For instance, in Fig.\ \ref{nucrust-nn-figcr9}
we observe two layers of intense Cooper-pairing emission
for the weak-superfluid model, centered at $\rho \sim 3 \times 10^{12}$
g cm$^{-3}$ and $\rho \sim 10^{14}$ g cm$^{-3}$. The emission
decreases abruptly at $\rho \lesssim 1.6 \times 10^{12}$ g cm$^{-3}$
where the neutrons remain non-superfluid ($T_{cn}< T$).
In the case of the strong-superfluid model,
one can see one pronounced Cooper-pairing emission peak
near the neutron drip point. There is also the second,
very narrow peak near the core-crust interface
at the densities higher
than those displayed in Fig.\ \ref{nucrust-nn-figcr9}.
In spite of very large emissivities in the layers
of intensified Cooper-pairing emission,
the layers themselves may be not too wide. Accordingly,
the process
may be not strong enough to determine
the integral neutrino luminosity of the neutron star
crust, but we 
recommend to include
it into calculations of neutron
star cooling (see Sect.\ \ref{sect-cool-super}).\\

{\bf (d) Bremsstrahlung in neutron-nucleus collisions}

This process was considered in the only paper by Flowers
and Sutherland (1977).
%
%
It can be written as
\begin{equation}
    n + (A,Z) \to n + (A,Z) + \nu  + \bar{\nu}
\label{nucrust-nn-Nbrems}
\end{equation}
(any neutrino flavors). The authors
analyzed neutron-nucleus interaction for an ensemble
of non-correlated nuclei in non-superfluid matter.
Using the approach of Chapt.\ \ref{chapt-nusup}
we can easily incorporate the effects of superfluidity
of free neutrons.
Then the neutrino emissivity can be written as
\begin{equation}
     Q^{(ni)}=Q^{(ni)}_0 \, R^{(ni)},
\label{nucrust-nn-QQ}
\end{equation}
where $Q^{(ni)}_0$ is the emissivity in the non-superfluid matter
and $R^{(ni)}$ is the superfluid reduction factor.

In our notations, the neutrino emissivity obtained
by Flowers and Sutherland (1977) can be written as
\begin{eqnarray}
  Q^{(ni)}_{0}& = & { G_{\rm F}^2 \over 63^2 \hbar^9 c^6} \,
        \left(1 + {11 \over 5} \, g_A^2 \right) \,
        {\cal N}_\nu \, n_i \left( p_{{\rm F}n} \over m_n^\ast c \right)^2
         (p_{{\rm F}n}^2 \sigma_{\rm tr})(k_{\rm B}T)^6
\nonumber \\
        & \approx & 5.5 \times 10^{18} \, {\cal N}_\nu \,
        \rho_{14} \left({200 \over A} \right)
        \left( { m_n \over m_n^\ast } \right)^2
        \left( { n_n \over n_0 } \right)^{2/3}
        \left( { p_{{\rm F}n}^2 \, \sigma_{\rm tr} \over \hbar^2 } \right) \,
        T_9^6 \quad {\rm erg \; cm^{-3} \; s^{-1}},
\label{nucrust-nn-Q0}
\end{eqnarray}
where $A$ is the number of baryons in a Wigner-Seitz cell
(confined in a nucleus plus free neutrons),
$\rho_{14}$ is the density
in units of $10^{14}$ g cm$^{-3}$,
and $\sigma_{\rm tr} = \int \, (1 - \cos \vartheta) \,
{\rm d}\sigma_{ni}$ is the transport cross section. The latter
is obtained by integration of the differential
cross section ${\rm d}\sigma_{ni}$ of neutron-nucleus scattering
(for the neutrons with the Fermi energy)
over all scattering angles $\vartheta$.
Flowers and Sutherland (1977)
used the simplest model cross section for which
$p_{{\rm F}n}^2 \sigma_{\rm tr} /\hbar^2=4 \pi \eta p_{{\rm F}n} R_i
/\hbar$, where
$R_i$ is the nucleus radius (taken as $1.4 \times A^{1/3}$ fm),
and $\eta$ is the model ``grayness" parameter
which varies from 1/4 to 1 (from elastic scattering to
full absorption of neutrons by the nuclei).

The superfluid reduction factor $R^{(ni)}$ is easily
obtained in the same manner as described in Chapt.\ \ref{chapt-nusup}
for neutrino reactions in uniform matter
(e.g., for nucleon-nucleon
bremsstrahlung,
Sect.\ \ref{sect-nusup-Brems}).
It is sufficient to consider the singlet-state neutron pairing.
Using the angular-energy decomposition in the expression for $Q^{ni}_0$
[Eq.\ (10) in Flowers and Sutherland 1977], in analogy
with Eq.\ (\ref{nusup-RNN}) we obtain
\begin{equation}
    R^{(ni)}={63 \over 4 \pi^6}
                \int_0^{+\infty}         {\rm d} x_{\nu}\, x_{\nu}^4 \,
                \int_{-\infty}^{+\infty} {\rm d} x_1 \,  f(z_1) \,
                \int_{-\infty}^{+\infty} {\rm d} x_2 \,  f(z_2) \,
                \delta (x_{\nu} - z_1 - z_2),
\label{nucrust-nn-R}
\end{equation}
where $x_1$, $x_2$ and $x_\nu$ are the dimensionless momenta
of the initial neutron, the final neutron and the neutrino-pair,
respectively, while $z_1$ and $z_2$ are the dimensionless energies
of the neutrons in the presence of superfluidity
[Eq.\ (\ref{sf-DimLessVar})].
The reduction factor $R^{(ni)}$ depends on the only
argument $v= \delta(T)/T$, the dimensionless gap parameter
defined by Eq.\ (\ref{sf-DefGap}). In the absence of
superfluidity, we naturally have $R^{(ni)}=1$. We have
calculated $R^{(ni)}$ for the superfluid matter and proposed the
analytic fit, similar to those given in Chapt.\ \ref{chapt-nusup}:
\begin{eqnarray}
   R^{(ni)} & = & {c \over 2} \,
                  \exp \left(0.6397- \sqrt{(0.6397)^2+v^2} \right)
\nonumber \\
            &  & + {d^5 \over 2} \,
                  \exp \left(1.5278 - \sqrt{(1.5278)^2+4v^2} \right),
\label{nucrust-nn-Rfit} \\
     c & = & 0.6937 + \sqrt{ (0.3063)^2 + (0.1685 \,v)^2 }, \quad \quad
     d = 0.4587 + \sqrt{ (0.5413)^2 + (0.612 \, v)^2}.
\nonumber
\end{eqnarray}

The neutrino emissivity is depicted in Fig.\ \ref{nucrust-nn-figcr9}.
We have used the expression for $\sigma_{\rm tr}$
presented above with $R_i=R_n$
[the neutron core radius in Eq.\ (\ref{nucrust-Oya})], and $\eta=0.5$.
In the absence of superfluidity, the process
could be important near the inner base of the crust,
but any superfluidity, weak or strong, reduces the emissivity
making it negligible for practical applications.

The consideration of Flowers and Sutherland (1977)
is too simplified. In principle, it would be
desirable to include
strong correlations between the
nuclei, in the same manner as in
Sect.\ \ref{sect-nucrust-ebrems} for electron-nucleus
bremsstrahlung, and to use a more advanced
model of the neutron-nucleus interaction.\\ 

{\bf (e) Second gap}

In the
inner crust, free neutrons move in a periodic
potential created by lattice of atomic nuclei.
This induces the band structure in the neutron
energy spectrum (Flowers and Itoh 1976) which can affect
kinetic and neutrino emission phenomena involving the free
neutrons.
The band structure contains energy gaps similar to those in
the energy
spectrum of the electrons (Pethick and Thorsson 1994, 1997,
Sect.\ \ref{sect-nucrust-ebrems}).
Moreover, the lattice gaps are superimposed
with the superfluid gaps.
The lattice gaps should reduce the neutrino reactions
of the bremsstrahlung type and initiate an additional
neutrino emission due to direct interband transitions
of the neutrons, in analogy with Cooper pairing of neutrons.
To our knowledge, these effects
are unexplored.\\

{\bf (f) Similar mechanisms}

In addition to the neutrino reactions discussed
above there may be several other reactions
connected with strong interactions
in the inner crust.

First of all, we mention the presence of free protons
in the possible presence of 
two last phases of matter at the bottom of the inner crust
(tubes and bubbles of neutron gas,
Sect.\ \ref{sect-nucrust-introduc}).
They initiate neutrino emission due to $pp$ and $np$ collisions,
due to nucleon collisions with nuclei,
and due to Cooper pairing of protons. The emissivities can
be obtained in the same manner as described above for the
processes involving free neutrons.

Second, some neutrino emission may come from
nucleon-nucleon interactions within the atomic nuclei.
There are plenty of nucleons confined in the nuclei;
the density profiles of neutrons and protons within
the nuclei are sufficiently smooth, especially at
higher $\rho$.
The nucleons within the nuclei can
be in a superfluid state. They can participate in the
neutrino processes which resemble nucleon-nucleon collisions
and Cooper pairing of free nucleons. However,
theoretical studies of these processes are complicated.
The simplest is the quasiclassical Thomas-Fermi model
used, for instance, by Yakovlev et al.\ (1998)
to analyze the neutrino emission due to
Cooper pairing of neutrons within atomic nuclei.
%
%
However, this model seems to be too crude
to be realistic, and
more elaborated 
many-body methods are required. 

Third, we mention one specific mechanism of
neutrino-pair emission via deexitation of excited
states of atomic nuclei through weak neutral currents.
The mechanism was
proposed by Bahcall et al.\ (1974)
%
%
as a possible source of neutrino cooling of white dwarfs.
In order to apply it for inner neutron star crusts
one should calculate the spectrum of excited energy levels
of highly unusual, neutron rich nuclei.


\subsection{Neutrino luminosity of a neutron star crust}
\label{sect-nucrust-overlook}

Let us combine the results of the present chapter
and discuss what the neutrino
emission from the crust of a cooling neutron star
would look like (Kaminker et al.\ 1999a).

To be practical, consider
the layers of the density $\rho \gtrsim 10^{10}$ g cm$^{-3}$
which produce sufficiently powerful neutrino emission
to affect stellar cooling.
As in Sect.\ \ref{sect-nucrust-nn}, we choose the
upper limit of the density range to be $10^{14}$ g cm$^{-3}$.
We assume that the atomic nuclei are spherical and use the
same weak and strong models of singlet-state pairing
of free neutrons in the inner crust.
Let us analyze the
emissivities of the most important neutrino reactions
in the absence of the magnetic field.

At high temperatures, $T= 3 \times 10^9$ K,
the most efficient neutrino emission is
provided by the plasmon decay. This is clearly seen
from Fig.\ \ref{nucrust-fig93}.

\begin{figure}[th!]
\begin{center}
\leavevmode
\epsfysize=8.5cm
\epsfbox{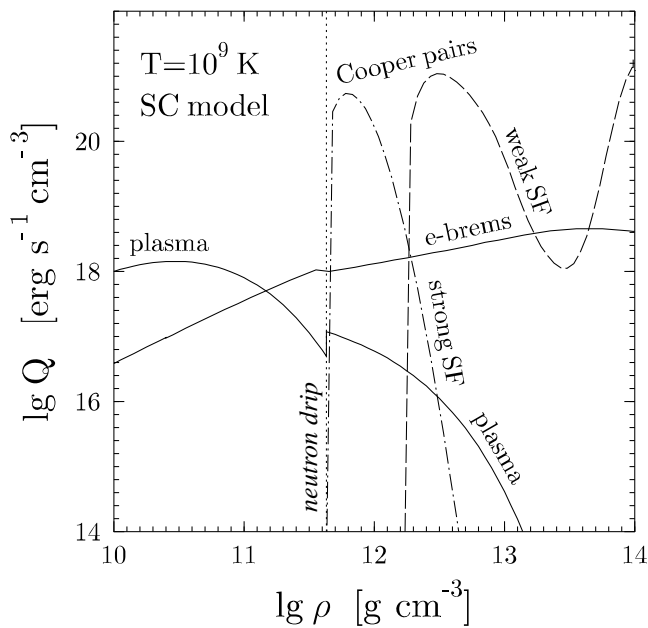}
\end{center}
\caption[ ]{
Density dependence of the neutrino emissivities
(Kaminker et al.\ 1999a) produced
in a neutron star crust (for smooth composition -- SC --
model of ground state matter)
at $T= 10^9$ K by
electron bremsstrahlung (e-brems),
plasmon decay (plasma) and by
Cooper pairing of free neutrons (Cooper pairs) in the models
of strong and weak neutron superfluidity (SF)
(see text for details)
}
\label{nucrust-overlook-figt9}
\end{figure}

\begin{figure}[th!]
\begin{center}
\leavevmode
\epsfysize=8.5cm
\epsfbox{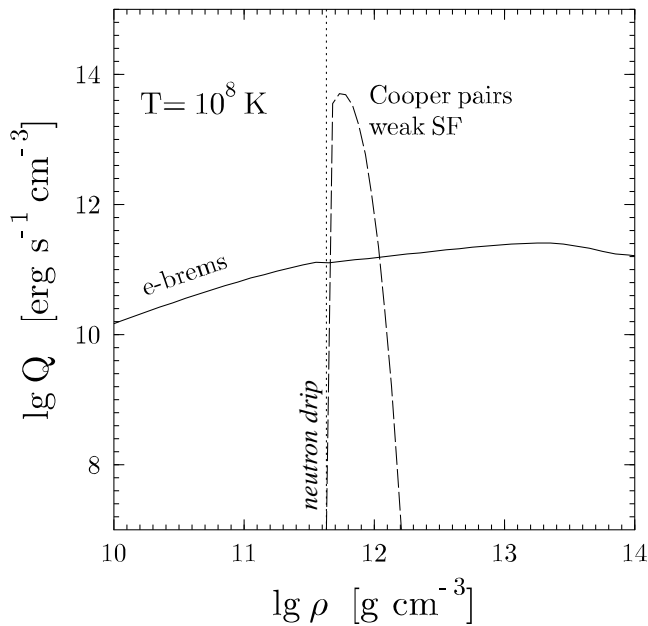}
\end{center}
\caption[ ]{
Same as in Fig.~\protect{\ref{nucrust-overlook-figt9} } but for $T=10^8$ K
(Kaminker et al.\ 1999a).
Plasmon decay and Cooper-pairing emission
in strongly superfluid model become negligible
\label{nucrust-overlook-figt8}
}
\end{figure}

Figures \ref{nucrust-overlook-figt9} and \ref{nucrust-overlook-figt8}
exhibit density dependence of the neutrino emissivity
for two lower temperatures, $T=10^9$ K and $10^8$ K,
respectively. The neutrino processes considered are:
neutrino bremsstrahlung due to electron-nucleus scattering
(Sect.\ \ref{sect-nucrust-ebrems}),
neutrino emission due to plasmon decay (Sect.\ \ref{sect-nucrust-plasmon})
and due to neutron Cooper-pair formation
(Sect.\ \ref{sect-nucrust-nn}).
The smooth composition (SC) model of ground state matter is used
which smears out the jumps of the neutrino emissivity
associated with step-like variations of nuclear composition
(cf.\ plasmon decay curves in Figs.\
\ref{nucrust-fig91} and \ref{nucrust-overlook-figt9}).
%
%
%
%
%

For $T = 10^9$ K
at $\rho \lesssim  10^{11}$ g cm$^{-3}$ (Fig.~\ref{nucrust-overlook-figt9})
the process most competitive with the electron bremsstrahlung
is again the plasmon decay. However its rate
falls exponentially with decreasing $T$,
and the process almost dies out at $T=10^8$ K
(Fig.\ \ref{nucrust-overlook-figt8}).
The neutrino emission due to Cooper pairing
in the inner crust
is also exponentially suppressed when the temperature is
much lower than the critical temperature of
neutron superfluid. Accordingly, the temperature and
density dependence of the Cooper-pair neutrino emissivity is very strong.
If $T=10^9$ K and the superfluidity is strong,
we have two peaks of Cooper-pair neutrinos: one near the
neutron drip point (at $\rho \sim 10^{12}$ g cm$^{-3}$),
and a narrow peak near the core-crust interface,
at the densities $\rho \approx 1.4 \times 10^{14}$ g cm$^{-3}$
invisible in Fig.\ \ref{nucrust-overlook-figt9}.
Both peaks are pronounced in the density ranges where  
the neutron critical temperature
is sufficiently small (only slightly exceeds $T$)
even for the strong superfluidity
(as discussed in Sect.\ \ref{sect-nucrust-nn}).
Between these density ranges the neutron critical temperature
is too high and the emission of the Cooper neutrinos
is exponentially suppressed. When the temperature
decreases the emissivity becomes smaller, and the
process dies out at $T=10^8$ K (cf.\ Figs.\ \ref{nucrust-overlook-figt9}
and \ref{nucrust-overlook-figt8}).

If $T=10^9$ K and the superfluidity is weak, the Cooper
pairing appears to be 
the dominant process
in a large fraction of the neutron star crust
since the neutron critical temperature is not much higher
than $T$. However, the process is switched
off at low densities $\rho \lesssim  1.6 \times 10^{12}$ 
${\rm g~ cm^{-3}}$,  because
a weak neutron superfluid has not yet occurred
at these densities ($T_{cn} < T=10^9$ K). When the temperature drops
to $10^8$ K, the neutrino emission due to Cooper
pairing is suppressed. Nevertheless, two high
peaks of the emissivity survive (similar to those
for the strong superfluid at $T=10^9$ K). The first one
corresponds to $\rho \sim 10^{12}$ g cm$^{-3}$, where
$T_{cn}$ is not much higher than $10^8$ K, and the second,
invisible in Fig.\ \ref{nucrust-overlook-figt8},
corresponds to $\rho \sim 1.4 \times 10^{14}$ g cm$^{-3}$, where
$T_{cn}$ is low due
to the transition from a singlet to a triplet neutron superfluid.

We conclude that the main contribution to
neutrino emission from deep layers of the crust
of the cooling neutron star
comes from {\it three} processes, the {\it plasmon decay}
at very high temperatures, and
the {\it neutrino-pair bremsstrahlung}
and {\it Cooper pairing} of neutrons at $T\lesssim 10^9$ K.

For the density-temperature range of study
($\rho \gtrsim 10^{10}$ g cm$^{-3}$, $T \lesssim 10^{10}$ K),
the neutrino emission due to electron-positron pair annihilation
(Sect.\ \ref{sect-nucrust-annih}) and the photoneutrino emission
(Sect.\ \ref{sect-nucrust-egamma})
may be comparable to
other mechanisms (particularly, to the plasmon decay)
only at lowest $\rho$ and highest $T$.
The contribution of beta-processes (Sect.\ \ref{sect-nucrust-beta})
is generally thought to be small although
they would definitely be most important for $T \gtrsim 10^{10}$ K.
The neutrino bremsstrahlung in neutron-neutron collisions
and due to neutron-nucleus collisions
(Sect.\ \ref{sect-nucrust-nn}) might make some contribution
at the crust base ($\rho \gtrsim 10^{14}$ g cm$^{-3}$)
and sufficiently high temperatures $T$.
Finally, the neutrino synchrotron
radiation by degenerate electrons
(Sect.\ \ref{sect-nucrust-syn}) may compete with
other mechanisms at not too high temperatures and densities
for the magnetic fields $B \gtrsim 10^{14}$ G.

Among two most efficient mechanisms,
the neutrino bremsstrahlung in electron-nucleus collisions
and Cooper pairing of neutrons,
the former operates in wider ranges of densities and
temperatures, and the density dependence of its emissivity
is generally smooth. The latter
mechanism is extremely sensitive
to the model adopted for calculating
the superfluid gaps in the neutron spectra.
This mechanism is more important for lower gaps
(weaker superfluid); its emissivity is a sharp
function of density and temperature.

\newpage

\section{Neutrino emission in non-superfluid cores}
\label{chapt-nucore}

\subsection{Wealth of neutrino reactions}
\label{sect-nucore-list}

In this chapter we consider the neutrino reactions
in non-superfluid, non-magnetized neutron star cores.
We focus mainly on the
non-exotic neutron star cores which consist of neutrons ($n$), protons ($p$)
and electrons ($e$).  At densities close to (1--2)$\rho_0$
muons ($\mu$) appear,
and at still higher densities hyperons are created, first of
all $\Sigma^-$ and $\Lambda$ hyperons. 
Matter composed of neutrons, protons, electrons, muons, and
$\Sigma^-$ and $\Lambda$ hyperons will be referred to as
the $npe\mu\Lambda\Sigma^-$ matter.
A list of neutrino
reactions which may be important
in the $npe\mu\Lambda\Sigma^-$ matter
is given in Table \ref{tab-nucore-list}.
These reactions can be subdivided into five groups:
(I) $4 \times 2=8$ direct Urca processes of 
the electron or muon production and capture
by baryons ({\it baryon direct Urca} processes,
Sect. \ref{sect-nucore-Durca}), (II)
$4 \times 4 \times 2=32$ more complicated
modified Urca processes, also associated with 
the electron or muon production and capture
by baryons ({\it baryon modified Urca} processes, 
Sect.\ \ref{sect-nucore-Murca}),
(III) 12 processes of neutrino-pair emission in strong
baryon-baryon collisions ({\it baryon bremsstrahlung},
Sect.\ \ref{sect-nucore-Brems}),
(IV) 4 modified Urca processes associated with
muon decay and production 
by electrons ({\it lepton modified Urca} process,
Sect.\ \ref{sect-nucore-other}),
and (V) $2 \times 2 + 3= 7$
processes of neutrino-pair emission in Coulombic
collisions ({\it Coulomb bremsstrahlung}, Sect.\ \ref{sect-nucore-other}).
Some equations of state allow also the creation of $\Delta^-$
resonances (non-strange particles of spin 3/2) in addition
to $\Sigma^-$ and $\Lambda$ hyperons. If available,
$\Delta^-$ resonances may participate in neutrino processes
as substitutes of $\Sigma^-$ hyperons. Then the number of
the possible neutrino reactions becomes even larger.

The reactions listed in Table \ref{tab-nucore-list} are very different.
Some of them switch on when the density exceeds
certain thresholds
(typically, several $\rho_0$) but some can operate
at any density in the neutron star core.
Some processes may change the composition of matter, while
others may not. We mainly consider the neutrino
emission assuming beta-equilibrium, but also study
non-equilibrium reactions in Sect.\ \ref{sect-nucore-noneq}.
The neutrino emissivities of
many reactions from Table \ref{tab-nucore-list} are
calculated quite reliably, being not very dependent
on a particular microscopic model of strong interactions.
However, other reactions do depend on the microscopic
model, and some are still almost unexplored.

\begin{table}[t]
\caption{Main neutrino processes in $npe\mu \Lambda \Sigma^-$
         matter$^{\ast)}$}
\begin{center}
  \begin{tabular}{||l||}
  \hline \hline
  \begin{tabular}{p{6cm}l}
  (I)  Baryon direct Urca & $\quad Q \sim (10^{23}$--$10^{27})\, T_9^6$
              erg cm$^{-3}$ s$^{-1}$ \\
  \end{tabular}\\
  \hline
  \begin{tabular}{llll}
   $(1)\; n \to p l \bar{\nu}_l \quad$ &
    $ p l \to n {\nu}_l  \quad \quad \quad \quad $ &
   $(2)\; \Lambda \to p l \bar{\nu}_l \quad $ &
    $ p l \to \Lambda {\nu}_l$ \\
   $(3)\; \Sigma \to n l \bar{\nu}_l \quad $ &
    $   n l \to \Sigma {\nu}_l  \quad \quad $ &
   $(4)\; \Sigma \to \Lambda l \bar{\nu}_l \quad $ &
    $   \Lambda l \to \Sigma {\nu}_l$\\
  \end{tabular}\\
  \hline \hline
  \begin{tabular}{p{6cm}l}
  (II)  Baryon modified Urca &$\quad Q \sim (10^{18}$--$10^{21})\, T_9^8$
              erg cm$^{-3}$ s$^{-1}$ \\
  \end{tabular}\\
  \hline
  \begin{tabular}{llll}
   $(1)\; n B \to p B l \bar{\nu}_l \; \;$ &
    $ p B l \to n B {\nu}_l  \quad  $ &
   $(2)\; \Lambda B \to p B l \bar{\nu}_l \;\; $ &
    $ p B l \to \Lambda B {\nu}_l$ \\
   $(3)\; \Sigma B \to n B l \bar{\nu}_l  $ &
    $   n B l \to \Sigma B {\nu}_l  \quad  $ &
   $(4)\; \Sigma B \to \Lambda B l \bar{\nu}_l $ &
    $   \Lambda B l \to \Sigma B {\nu}_l$\\
  \end{tabular}\\
  \hline \hline
  \begin{tabular}{p{6cm}l}
  (III)  Baryon bremsstrahlung
              & $\quad Q \sim (10^{16}$--$10^{20})\, T_9^8$
              erg cm$^{-3}$ s$^{-1}$ \\
  \end{tabular}\\
  \hline
  \begin{tabular}{lll}
   $\;\;(1)\; nn \to nn \nu\bar{\nu} \quad \quad$  &
   $\;\;(2)\; np \to np \nu\bar{\nu} \quad \quad$  &
   $\;\;(3)\; pp \to pp \nu\bar{\nu} \quad \quad $ \\
   $\;\;(4)\; \Sigma\Sigma \to \Sigma\Sigma \nu\bar{\nu}$ &
   $\;\;(5)\; \Sigma n \to \Sigma n \nu\bar{\nu}$ &
   $\;\;(6)\; \Sigma p \to \Sigma p \nu\bar{\nu}$  \\
   $\;\;(7)\; \Lambda \Lambda \to \Lambda \Lambda \nu\bar{\nu}$ &
   $\;\;(8)\; \Lambda n \to \Lambda n \nu\bar{\nu}$ &
   $\;\;(9)\; \Lambda p \to \Lambda p \nu\bar{\nu}$ \\
   $(10)\; \Sigma \Lambda \to \Sigma \Lambda \nu\bar{\nu}$ &
   $(11)\; \Lambda n \to \Sigma p \nu\bar{\nu} \quad \quad $  &
   $(12)\; \Sigma p \to \Lambda n \nu\bar{\nu} $ \\
  \end{tabular}\\
  \hline \hline
  \begin{tabular}{p{6cm}l}
   (IV)  Lepton modified Urca &
     $\quad Q \sim (10^{13}$--$10^{15})\, T_9^8$
     erg cm$^{-3}$ s$^{-1}$ \\
  \end{tabular}\\
  \hline
  \begin{tabular}{llll}
   $(1)\; \mu p \to e p  \bar{\nu}_e \nu_\mu \quad $&$
   e p \to \mu p \bar{\nu}_\mu \nu_e  \quad\quad $ &
   $(2)\; \mu \Sigma \to e \Sigma  \bar{\nu}_e \nu_\mu \quad $&$
   e \Sigma \to \mu \Sigma \bar{\nu}_\mu \nu_e$ \\
   $(3)\; \mu e \to e e  \bar{\nu}_e \nu_\mu \quad $&$
   e e \to \mu e \bar{\nu}_\mu \nu_e  \quad\quad $ &
   $(4)\; \mu \mu \to e \mu  \bar{\nu}_e \nu_\mu \quad $&$
   e \mu \to \mu \mu \bar{\nu}_\mu \nu_e$\\
  \end{tabular}\\
  \hline \hline
  \begin{tabular}{p{6cm}l}
   (V) Coulomb bremsstrahlung & $\quad Q \sim (10^{13}$--$10^{15})\, T_9^8$
              erg cm$^{-3}$ s$^{-1}$ \\
  \end{tabular}\\
  \hline
  \begin{tabular}{lll}
    $(1)\; l p \to l p  \nu \bar{\nu} \quad \quad \quad \quad $ &
    $(2)\; l \Sigma \to l \Sigma \nu \bar{\nu} \quad\quad \quad \quad $ &
    $(3)\; ll  \to ll \nu \bar{\nu}$\\
  \end{tabular}\\
  \hline \hline
\end{tabular}
\begin{tabular}{l}
  $^{\ast)}$  $\Sigma$ means $\Sigma^{-}$; $l$ stands for $e$ or $\mu$;
  $B$ stands for $n$, $p$, $\Sigma$ or $\Lambda$.
\end{tabular}
\label{tab-nucore-list}
\end{center}
\end{table}

The strongest are the baryon direct Urca processes,
but they are the threshold reactions
open for some equations of state
at sufficiently high densities
(Sect. \ref{sect-nucore-Durca}). If allowed, they
produce a {\it rapid} ({\it enhanced}) cooling of
neutron stars (Chapt.\ \ref{chapt-cool}). 
If they are forbidden, the main reactions
are those of the baryon modified Urca and baryon
bremsstrahlung processes which produce
a {\it slow} ({\it standard}) cooling.
These reactions are abundant.
The number of open reactions of this type
grows quickly 
with density.
However, one should not fear
this wealth of reactions: as the density grows
there are more chances that a direct Urca process
becomes open; it will determine the neutrino
luminosity since the modified Urca and
bremsstrahlung processes are negligible compared to it.
Finally, the lepton modified Urca and Coulomb
bremsstrahlung reactions are weaker than the
baryon bremsstrahlung processes. Nevertheless,
they may be important in the highly superfluid
neutron star cores since baryon superfluidity can
suppress all baryonic reactions
(Chapt.\ \ref{chapt-nusup}).

\begin{table}[t]
\caption{Leading neutrino processes in three models of exotic
         matter$^{\ast)}$}
\begin{center}
  \begin{tabular}{||lll||}
  \hline \hline
  Model   & Process   &  $Q$, erg cm$^{-3}$ s$^{-1}$ \\
  \hline
  Pion condensate &
  $\tilde{n} \to \tilde{p} l \bar{\nu}_l \quad
  \tilde{p} l \to \tilde{n} {\nu}_l  $  &
  $\quad (10^{22}$--$10^{26})\, T_9^6$  \\
   Kaon condensate &
   $\tilde{n} \to \tilde{p} l \bar{\nu}_l \quad
    \tilde{p} l \to \tilde{n} {\nu}_l  $ &
   $\quad (10^{22}$--$10^{24})\, T_9^6$ \\
   Quark matter &
   $d \to u e \bar{\nu}_e \quad  u e \to d {\nu}_e  $   &
   $\quad (10^{25}$--$10^{26})\, T_9^6$ \\
   \hline \hline
\end{tabular}
\begin{tabular}{l}
  $^{\ast)}$  $l$ stands for $e$ or $\mu$;
  $\tilde{n}$ and $\tilde{p}$ are quasinucleons \\
  (mixed $n$ and $p$ states); $u$ and $d$ are quarks.
\end{tabular}
\label{tab-nucore-exotica}
\end{center}
\end{table}

In addition to the standard models of dense matter,
in Sect.\ \ref{sect-nucore-exotica} we consider
the main neutrino reactions in a few
hypothetical exotic models (Sect.\ \ref{sect-overview-struct}).
The leading reactions for the three exotic models
(pion condensate, kaon condensate, and quark matter)
are listed in Table \ref{tab-nucore-exotica}.
They are of the direct Urca type. Their
emissivities are somewhat reduced in comparison with
the nucleon direct Urca process
(I.1) but are still much higher
than the emissivity of the modified Urca processes in the standard
neutron star matter.

Notice one important feature
of the reactions listed in Tables \ref{tab-nucore-list}
and \ref{tab-nucore-exotica}.
In non-superfluid matter the emissivity $Q$ of any reaction
can be factorized as
\begin{equation}
         Q(\rho,T)=Q_0(\rho) \, T^k,
\label{nucore-list-Trho}
\end{equation}
where $Q_0(\rho)$ describes the density dependence,
while the temperature dependence
is a power-law: $k=6$ for direct Urca
processes, and $k=8$ for all others.
As mentioned above, the density dependence may have
a threshold but otherwise it is usually not too strong.
The order-of-magnitude estimates of the emissivities
are given in Table \ref{tab-nucore-list}.

\begin{figure}[!t]
\begin{center}
\leavevmode
 \epsfysize=8.5cm
 \epsfbox[70 55 395 410]{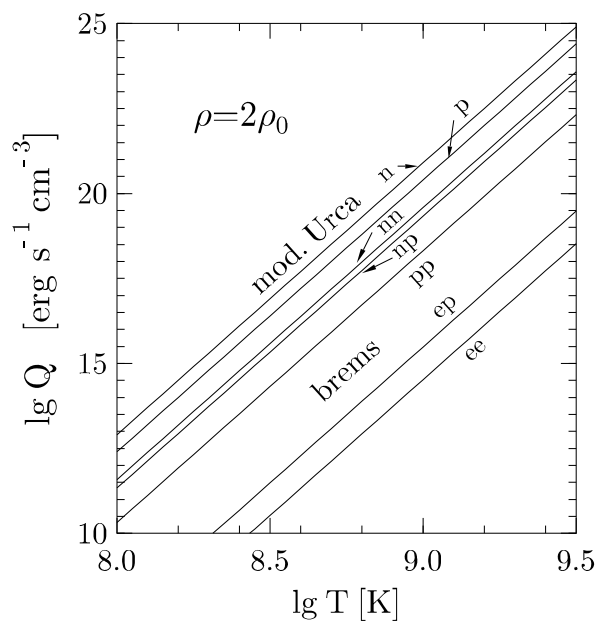}
\end{center}
\caption[]{\footnotesize
        Temperature dependence of the neutrino emissivities
        of various standard reactions
        (neutron and proton branches of the modified Urca
        process, neutrino emission due to $nn$, $np$, $pp$,
        $ep$ and $ee$ bremsstrahlung) in $npe$ matter
        for $\rho = 5.6 \times 10^{14}$ g cm$^{-2}$.
        The direct Urca process is forbidden.
        Equation of state is the same as in 
        Sect.\ \protect{\ref{sect-cool-code}} (moderate model).
}
\label{fig-nucore-Q(T)}
\end{figure}

\begin{figure}[!t]
\begin{center}
\leavevmode
\epsfysize=8.5cm
\epsfbox[70 65 395 410]{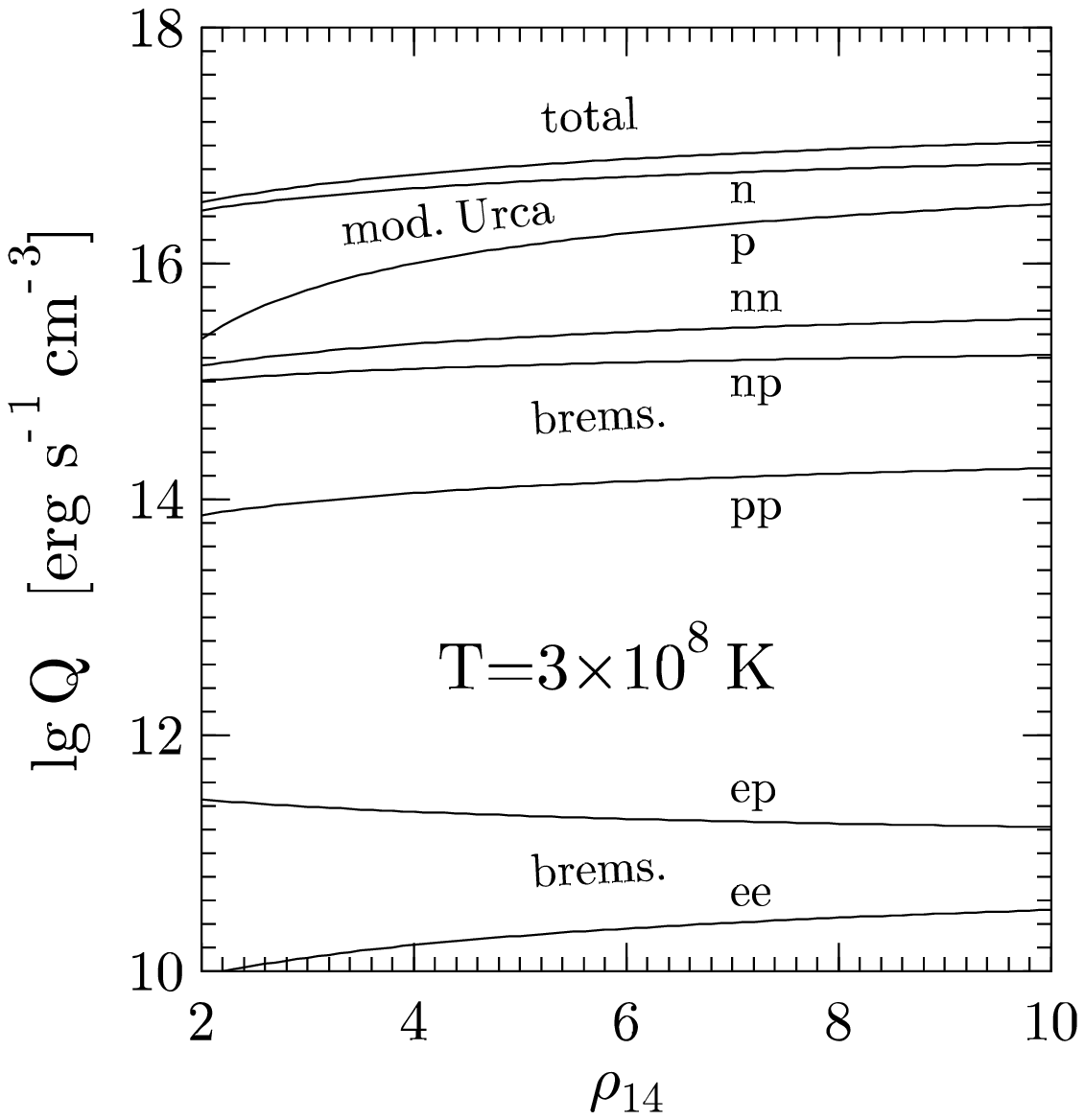}
\end{center}
\caption[]{\footnotesize
        Density dependence of the neutrino emissivities
        of the same reactions as in Fig.\ \protect{\ref{fig-nucore-Q(T)}}
        for $T = 3 \times 10^8$ K; $\rho_{14}$ is the density
        in units of $10^{14}$ g cm$^{-3}$.
}
\label{fig-nucore-Q(rho)}
\end{figure}

As an example, in Figs.\ \ref{fig-nucore-Q(T)}
and \ref{fig-nucore-Q(rho)} we show the emissivities
of various standard (slow) neutrino processes in the $npe$
matter (without muons and hyperons).  
Individual curves will be explained in the subsequent
sections. We use the same moderately stiff equation of state
of matter that will be adopted for illustrating the neutron
star cooling in Chapt.\ \ref{chapt-cool}. 
The powerful direct Urca
process operates at $\rho> \rho_{\rm crit}=1.3 \times 10^{15}$
g cm$^{-3}$, and it is forbidden for the conditions
displayed in the figures. Figure \ref{fig-nucore-Q(T)}
shows the temperature dependence of the emissivities at
$\rho=2 \, \rho_0$. As explained above, the dependence is strong
but simple: in the logarithmic variables $\lg T$--$\lg Q$
we have a sequence of parallel lines.
In Chapt.\ \ref{chapt-nusup} we show that this
strict and plain order is drastically violated by the
superfluidity of nucleons. Figure \ref{fig-nucore-Q(rho)} shows
the density dependence of the same emissivities for
a temperature $T= 3 \times 10^8$ typical for
cooling neutron stars. The density dependence
is seen to be rather weak.  This is again changed either by
superfluidity or by switching on the direct Urca process.

Notice that the weak interaction
parameters entering the expressions for $Q$
may be renormalized by the in-medium
effects. The classical example is the renormalization
(quenching) of the nucleon axial-vector constant $g_A$
in nuclear matter as known from experiments
on beta decay of atomic nuclei
(e.g., Wilkinson, 1973).
%
%
The renormalization
problem in the neutron star matter
is complicated; we will mainly adopt standard
(in-vacuum) parameters in our numerical estimates.
We will take into account explicitly only the renormalization of baryon
masses by replacing the
bare masses with the effective ones
to be taken from microscopic theories.
Baryons will be treated as {\it nonrelativistic} particles.
We will separate the results sensitive
to a model of strong interactions from the
model-independent features.

As in Chapt.\  \ref{chapt-nucrust}
we will mainly use the units in which $\hbar = c = k_{\rm B} =1$,
although we will return to the standard physical
units in the final expressions. 
The effects
of superfluidity of matter and the effects of strong magnetic
fields on the neutrino emissivity will
be studied in Chapt.\ \ref{chapt-nusup}.

\subsection{Neutron beta decay}
\label{sect-nucore-beta}

The neutrino reactions in neutron star cores can be
understood by taking the ordinary beta decay
[the first reaction of process (I.1) in Table \ref{tab-nucore-list}]
as an example:
\begin{equation}
   n \to p + e + \bar{\nu}_e.
\label{nucore-beta-decay}
\end{equation}
Here we outline briefly the derivation of the beta-decay rate.
The process is described by one 4-tail Feynman diagram.
The interaction Hamiltonian is
\begin{equation}
    \hat{H}= {G \over \sqrt{2}}\, J_\alpha l^\alpha,
\label{nucore-H}
\end{equation}
where $G=G_{\rm F} \cos \theta_{\rm C}$,
$G_{\rm F}$
is the Fermi
weak interaction constant, and $\theta_{\rm C}$ is the
Cabibbo angle ($\sin \theta_{\rm C}=0.231$).
Also,
\begin{equation}
     l^\alpha = \bar{\psi}_e \gamma^\alpha (1 + \gamma^5) \psi_\nu
\label{nucore-lepton}
\end{equation}
is the lepton weak 4-current,
$\psi_\nu$ is the neutrino wave function, and
$\psi_e$ is the electron wave function.
Other notations are the same as in Sect.\ \ref{sect-nucrust-annih}.
The wave functions are of the form
\begin{equation}
    \psi_e = {u_e \over \sqrt{2 \epsilon_e}} \; {\rm e}^{-i p_e x},
    \quad
    \psi_\nu = {u_\nu \over \sqrt{2 \epsilon_\nu}} \;
               {\rm e}^{-i p_\nu x},
\label{nucore-psi-lepton}
\end{equation}
where $p_e=(\epsilon_e,{\bf p}_e)$ is the 4-momentum of the electron,
$p_\nu=(\epsilon_\nu,{\bf p}_\nu)$ the 4-momentum of the antineutrino,
while $u_e$ and $u_\nu$ are the bispinors,
$\bar{u}_e u_e = 2 m_e$ and $\bar{u}_\nu u_\nu =0$.
Finally, $J=(J^0,{\bf J})$ in Eq.\ (\ref{nucore-H})
is the weak hadron 4-current. For non-relativistic
nucleons $J^0$  is determined
by the vector interaction, while ${\bf J}$ is determined
by the axial-vector interaction:
\begin{equation}
    J^0= f_V \, \Psi_p^\dagger \Psi_n, \quad
    {\bf J}= - g_A\, \Psi_p^\dagger {\bf \sigma} \Psi_n,
\label{nucore-hadron}
\end{equation}
where $f_V=1$ is the vector coupling constant,
$g_A=1.26$ is the Gamow-Teller axial-vector coupling constant,
${\bf \sigma}$ is the spin Pauli matrix.  Furthermore,
\begin{equation}
    \Psi_n = \chi_s \,{\rm e}^{-i p_n x}, \quad
    \Psi_p = \chi_{s'} \,{\rm e}^{-i p_p x},
\label{nucore-psi-hadron}
\end{equation}
are the non-relativistic (spinor) wave functions
of the neutron and the proton, respectively;
$p_n=(\epsilon_n,{\bf p}_n)$ and $p_p=(\epsilon_p,{\bf p}_p)$
are the nucleon 4-momenta;
$\chi_s$ and $\chi_{s'}$ are the non-relativistic unit basic spinors
($ \chi_s \chi_{s'} = \delta_{ss'}$), where
$s=\pm 1$ and $s'=\pm 1$ specify the signs of the nucleon spin projections
onto the quantization axis.

The transition rate from an initial state
$i$ to a close group of final states $f$
summed over particle spins
is given by Fermi Golden Rule
\begin{equation}
     {\rm d}W_{i \to f} = 2\pi \,
     \delta(\epsilon_n - \epsilon_p - \epsilon_e - \epsilon_\nu) \,
     \sum_{\rm spins} | H_{fi} |^2 \,
     { {\rm d}{\bf p}_p \over (2\pi)^3} \,
     { {\rm d}{\bf p}_e \over (2\pi)^3} \,
     { {\rm d}{\bf p}_\nu \over (2\pi)^3} .
\label{nucore-Golden-Rule}
\end{equation}
The expression contains the
squared matrix element of the interaction Hamiltonian (\ref{nucore-H}):
\begin{eqnarray}
     \sum_{\rm spins} | H_{fi}|^2 & = & (2 \pi)^3 \, {G^2 \over 2} \,
     \delta({\bf p}_n - {\bf p}_p - {\bf p}_e - {\bf p}_\nu) \,
     {\cal J}_{\alpha \beta} {\cal L}^{\alpha \beta},
\label{nucore-HH} \\
     {\cal J}^{\alpha \beta} & = &
     \sum_{\rm spins} (J_{fi})^{\alpha \ast}\,(J_{fi})^{\beta},\quad
     {\cal L}^{\alpha \beta}  =
     \sum_{\rm spins}  (l_{fi})^{\alpha \ast}\,(l_{fi})^{\beta},
\label{nucore-Jab}
\end{eqnarray}
where ${\cal J}^{\alpha \beta}$ and ${\cal L}^{\alpha \beta}$
contain the sums over the nucleon and electron spin states, respectively.

Using Eqs.\  (\ref{nucore-lepton}) and (\ref{nucore-psi-lepton}),
we have
\begin{equation}
    {\cal L}^{\alpha \beta} =  \sum_{\rm spins}
    { 1 \over 4 \epsilon_e \epsilon_\nu} \,
    [\bar{u}_e \gamma^\alpha (1+\gamma^5) u_\nu]^\ast
    [\bar{u}_e \gamma^\beta (1+\gamma^5) u_\nu].
\label{nucore-ll}
\end{equation}
Replacing the bilinear combination of the electron bispinors
by the electron polarization density matrix, $[(\gamma p_e)+m_e]/2$,
and the bilinear combination of the neutrino bispinors by
the neutrino polarization density matrix, $(\gamma p_\nu)$,
we reduce the
summation over the electron spin states to
calculating the trace
\begin{eqnarray}
  {\cal L}^{\alpha \beta} & = &
     { 1 \over 4 \epsilon_e \epsilon_\nu} \,
     {\rm Tr}\left\{ (\gamma p_\nu)
     \gamma^\alpha (1+\gamma^5) [(\gamma p_e)+m_e]
     \gamma^\beta (1+\gamma^5) \right\}
\nonumber \\
& = & {2 \over \epsilon_e \epsilon_\nu} \,
     \left[p_e^\alpha p_\nu^\beta + p_e^\beta p_\nu^\alpha
     - (p_e p_\nu) g^{\alpha \beta}
     + i \,e^{\alpha \beta \rho \delta} \, p_{e\rho} \,p_{\nu\delta}
     \right],
\label{nucore-ll1}
\end{eqnarray}
cf with Eq.\ (\ref{nucrust-ll}).

The matrix elements of the hadron current
in Eq.\ (\ref{nucore-HH}) are
$(J_{fi})^0= \chi_{s'}^\dagger \chi_s=\delta_{ss'}$,
and ${\bf J}_{fi} =- g_A \, \chi_{s'}^\dagger {\bf \sigma} \chi_s$.
Let us substitute them into the expression for ${\cal J}^{\alpha \beta}$
given by Eq.\ (\ref{nucore-Jab})
and sum over the neutron and proton spin states.
A straightforward evaluation shows that tensor
${\cal J}^{\alpha \beta}$ is diagonal:
\begin{equation}
  {\cal J}^{00}=2, \quad {\cal J}^{11}={\cal J}^{22}={\cal J}^{33}
  =2 \, g_A^2, \quad {\cal J}^{\alpha \beta}=0 \quad {\rm for}
  \; \alpha \neq \beta.
\label{nucore-Jab1}
\end{equation}
Combining Eqs.\ (\ref{nucore-Golden-Rule}),
(\ref{nucore-HH}), (\ref{nucore-ll1}), and (\ref{nucore-Jab1}),
we obtain the differential transition
rate [s$^{-1}$], summed over the spins of $n$, $p$ and $e$
(in standard physical units):
\begin{eqnarray}
   {\rm d}W_{i \to f} & = & (2 \pi)^4 \,
    \delta(\epsilon_n - \epsilon_p - \epsilon_e - \epsilon_\nu) \,
    \delta({\bf p}_n - {\bf p}_p - {\bf p}_e - {\bf p}_\nu)
\nonumber \\
    & \times & {2 G^2 \over \hbar^7 \epsilon_e \epsilon_\nu} \,
       \left[ \epsilon_e \epsilon_\nu + c^2 \,{\bf p}_e \cdot {\bf p}_\nu
       + g_A^2 \,(3 \epsilon_e \epsilon_\nu
       - c^2 \, {\bf p}_e \cdot {\bf p}_\nu) \right]\,
     { {\rm d}{\bf p}_p \over (2\pi)^3} \,
     { {\rm d}{\bf p}_e \over (2\pi)^3} \,
     { {\rm d}{\bf p}_\nu \over (2\pi)^3} .
\label{nucore-Golden-Rule1}
\end{eqnarray}
This is the basic expression for analyzing the properties
of beta decay, except for parity non-conservation
associated with spin polarization which
can be studied starting
from the more general expression (\ref{nucore-Golden-Rule}).

For instance, consider beta decay of a neutron at rest.
Let us integrate Eq.\ (\ref{nucore-Golden-Rule1})
 over ${\rm d}{\bf p}_p$ (to remove
the momentum conserving delta function),
over the neutrino energy ${\rm d}\epsilon_\nu$
(to remove the energy conserving delta function),
over the orientations of the neutrino and electron momenta,
and over the electron energy,
and average over the
neutron spin states (introducing the factor 1/2).
In the energy integration
it is sufficient to neglect the proton (recoil) energy:
it is negligible
due to the large proton mass.
Then we obtain the beta-decay rate 
\begin{eqnarray}
    W_\beta & = & {G^2 \,(1+3 \, g_A^2) \over 2 \hbar^7 c^5 \pi^3} \,
      \int_{m_ec^2}^\Delta {\rm d}\epsilon_e \;
         \epsilon_e \, p_e \,(\Delta -\epsilon_e)^2
\nonumber \\
      &= & { G^2 \,(1 + 3 \, g_A^2) m_e^5 c^4 \over 2 \pi^3 \hbar^7 }
      \, w_\beta,
\label{nucore-neutron-tau}\\
   w_\beta & = & \int_1^a {\rm d}x \; x \sqrt{x^2 -1} \, (a-x)^2
\nonumber \\
     & = & {a \over 4} \, \ln \left(a + \sqrt{a^2-1} \right) +
         \left( {a^4 \over 30} - {3a^2 \over 20} - {2 \over 15} \right) \,
         \sqrt{a^2 -1}=1.63,
\label{nucore-w}
\end{eqnarray}
where $\Delta=(m_n-m_p)c^2=1.29$ MeV is the neutron-proton mass
deficit (the maximum neutrino and electron energy), and
$a=\Delta/(m_ec^2)=2.53$. From Eq.\ (\ref{nucore-w}) we
obtain the $e$-folding beta decay time $1/W_\beta=966$ s. It
differs from the well-known experimental value $\approx 925$ s
because we have neglected the Coulomb interaction between
the proton and the electron.
Such effects are insignificant in neutron star cores.

The above derivation shows that the squared matrix element
summed over the spins [Eq. (\ref{nucore-HH})]
is independent of the directions of momenta of the reacting
particles if we average it over the orientations of the neutrino
or electron momentum. This enables us to treat the squared
matrix element as constant
in the subsequent calculations.

\subsection{Direct Urca processes}
\label{sect-nucore-Durca}

{\bf (a) Nucleon direct Urca}

The nucleon direct Urca process is the simplest and
most powerful neutrino process [reaction (I.1)
in Table \ref{tab-nucore-list}].
It consists of two successive
reactions, beta decay and capture:
\begin{equation}
   n \to p + e + \bar{\nu}_e, \quad    p + e  \to n + {\nu}_e.
\label{nucore-Durca}
\end{equation}
This is the basic process in the neutron star core; it
brings nucleons into the state of beta-equilibrium, in which
the chemical potentials satisfy the equality
$\mu_n= \mu_p + \mu_e$. If the equilibrium state is
not reached, one of the two reactions becomes more intense
and changes the fractions of protons and neutrons towards
the equilibrium values, in agreement with the Le 
Ch\^atelier principle
(Sect.\ \ref{sect-nucore-noneq}).
In equilibrium
both reactions have the same rate, and the direct Urca
process does not change the nucleon composition of matter.
Analogous processes for the less dense stellar matter
have been discussed in Sect.\ \ref{sect-nucrust-beta}.

Let us outline the derivation of the neutrino emissivity
$Q^{\rm (D)}$ of the direct Urca process
 (labeled by superscript ${\rm D}$) under the
condition of beta equilibrium. It is sufficient
to calculate the emissivity of the beta decay reaction
and double the result:
\begin{equation}
   Q^{\rm (D)} = 2 \int \, {{\rm d}{\bf p}_n \over (2 \pi)^3}\,
   {\rm d}W_{i \to f}\, \epsilon_\nu \, f_n \, (1-f_p) \, (1-f_e),
\label{nucore-QdDef}
\end{equation}
where ${\rm d}W_{i \to f}$ is the beta decay
differential probability, Eq.\ (\ref{nucore-Golden-Rule1}),
$f_j$ is the Fermi-Dirac function of particle species
$j$ ($j$=1, 2, and 3 refer to the neutron, proton and
electron, respectively). Formally, we have a 12-fold integral
in which 4 integrations can be removed via energy and momentum
conservation.

Luckily, the integration can be greatly simplified
using the so called {\it phase-space decomposition},
a very powerful tool in calculating the reaction rates
of strongly degenerate particles
(e.g., Shapiro and Teukolsky 1983). We will see
that the decomposition enables one to establish
the similarity relation between different neutrino reactions.
In our case, the decomposition procedure is like this. Owing
to the strong degeneracy of
nucleons and electrons,
the main contribution to the integral
(\ref{nucore-QdDef}) comes from the narrow regions of momentum space
near the corresponding Fermi surfaces. Thus, we can set
$p = p_{\rm F}$ in all smooth functions of energy and momentum
under the integral.
The energy exchange in the direct Urca reaction goes
naturally on the temperature scale $\sim T$. Accordingly,
the neutrino energy is $\epsilon_\nu \sim T$, and the neutrino momentum
$p_\nu \sim T$ is much smaller than the momenta of other particles.
We can neglect the neutrino momentum in the
momentum conserving delta-function and integrate easily 
over orientations of the neutrino momentum. Afterwards,
the integrand of Eq.\ (\ref{nucore-QdDef}) will contain
\begin{eqnarray}
   {\rm d}W_{i \to f} \; {{\rm d}{\bf p}_n \over (2 \pi)^3} 
    & = & {(2 \pi)^4 \over (2 \pi)^{12}} \;
    \delta(\epsilon_n - \epsilon_p - \epsilon_e - \epsilon_\nu) \,
    \delta({\bf p}_n - {\bf p}_p - {\bf p}_e)
\nonumber \\
    & & \times \,  \;
        |M_{fi}|^2 \; 4 \pi \, \epsilon_\nu^2 \, {\rm d}\epsilon_\nu
        \prod_{j=1}^3 p_{{\rm F}j} m^\ast_j 
        \,{\rm d}\epsilon_j \, {\rm d}\Omega_j ,
\label{nucore-Golden-Rule2}
\end{eqnarray}
where
$ {\rm d} \Omega_j $ is the solid angle 
element in the direction of ${\bf p}_j$,
$m_j^\ast=p_{{\rm F}j}/v_{{\rm F}j}$ is the effective particle mass,
$v_{{\rm F}j}=(\partial \epsilon_j/\partial p)_{p=p_{{\rm F}j}}$ 
is the Fermi velocity,
$m_e^\ast = \mu_e/c^2$. Finally,
\begin{equation}
     | M_{fi} |^2 = 2 \, G^2 \,(f_V^2+ 3 g_A^2)
              = 2 \, G_{\rm F}^2 \, \cos^2 \theta_{\rm C} \, (1+ 3 g_A^2)
\label{nucore-squared-Mdur}
\end{equation}
is the squared matrix element summed over particle spins
and averaged over orientations of the neutrino momentum.
It appears to be constant and can be taken out of the integration.
The remaining 
integrations over the directions and magnitudes of the 
particle momenta become decomposed. Then
the neutrino emissivity can be rewritten as
\begin{eqnarray}
&&    Q^{\rm (D)}  =  {2 \over (2 \pi)^{8}} \; T^6
            AI \,| M_{fi} |^2 \, \prod_{j=1}^3 p_{{\rm F}j} m_j^\ast ,
\label{nucore-decomp-Dur} \\
&&   A = 4 \pi \,  \int {\rm d} \Omega_1 \;{\rm d} \Omega_2 \; 
                {\rm d} \Omega_3 \;
                \delta({\bf p}_n - {\bf p}_p - {\bf p}_e),
\label{nucore-Adur} \\
&&   I = \int_0^\infty {\rm d} x_\nu \; x_\nu^3
       \left[ \prod_{j=1}^3 \int_{-\infty}^{+\infty}
       {\rm d} x_j \; f_j \right]
       \delta (x_1 + x_2 + x_3 -x_\nu ).
\label{nucore-Idur} 
\end{eqnarray}
The quantity $A$ contains the integrals over the orientations of
the particle momenta;
all vector lengths ${\bf p}_j$ in the delta-function
must be set equal to the corresponding Fermi momenta.
The quantity $I$, given by Eq.\ (\ref{nucore-Idur}),
includes the integration over the dimensionless energies of neutrino
$x_\nu = \epsilon_\nu /T$
and other particles $x_j = (\epsilon_j-\mu_j)/T
\simeq  v_{{\rm F}j}(p- p_{{\rm F}j})/T$.
For particles $j$=2 and 3, we have transformed
$\left[1-f(x_j) \right] \to f(x_j)$ by replacing $x_j \to -x_j$.

The integrals
$A$ and $I$ are standard (e.g.,
Shapiro and Teukolsky 1983):
\begin{equation}
    A  =  {32 \pi^3 \, \over p_{{\rm F}n}\, p_{{\rm F}p}\, p_{{\rm F}e} }
    \; \Theta_{npe},
    \quad  I  =  \int_0^\infty {\rm d} x_\nu \; x_\nu^3 \; J(x_\nu) =
     { 457 \pi^6 \over 5040},
\label{nucore-AIdur}
\end{equation}
where
\begin{equation}
  J(x)  =   \left[ \prod_{j=1}^3 \int_{-\infty}^{+\infty}
       {\rm d} x_j \; f_j \right]
       \delta (x_1+x_2+x_3 -x)
      =  { \pi^2 + x^2 \over 2 \, ({\rm e}^x + 1)},
\label{nucore-J(x)}
\end{equation}
and $\Theta_{npe}$
is the
step function: $\Theta_{npe}=1$ if the Fermi momenta
$p_{{\rm F}n}$, $p_{{\rm F}p}$ and $p_{{\rm F}e}$
satisfy the triangle condition
and $\Theta_{npe}=0$ otherwise (see below).

Thus, the neutrino emissivity of the direct Urca process
in standard physical units is
(Lattimer et al.\ 1991)
\begin{eqnarray}
 Q^{\rm(D)} &  = & {457\, \pi \over 10080} \,
                G^2_{\rm F} \, \cos^2 \theta_{\rm C} \, (1+3 g_A^2) \;
                {m_n^\ast \, m_p^\ast \, m_e^\ast \over \hbar^{10} c^3} \;
                (k_{\rm B} T)^6 \, \Theta_{npe}
\nonumber \\
        & \approx & 4.00 \times 10^{27}
                \left( {n_e \over n_0} \right)^{1/3} \;
                 {m_n^\ast \, m_p^\ast \over m_n^2} \,
                 T^6_9 \, \Theta_{npe}
                 \;\;\;\ {\rm erg\;cm^{-3}\;s^{-1}}.
\label{nucore-Qdur0}
\end{eqnarray}
Here, as before, $n_0=0.16$~fm$^{-3}$.

Taking $n_n \sim 5 \, n_0$ 
and bearing in mind that the reaction rate is $\sim Q^{\rm (D)}/k_{\rm B} T$,
we can
estimate the typical time required for one neutron to participate
in the direct Urca process,
$\tau_n \sim n_n k_{\rm B} T/Q^{\rm(D)}
\sim 3 \times 10^4 \; T_9^{-5}$ s.
Thus, the reaction is extremely slow according to the microphysical
standards, a result of the slow pace of weak
interaction processes and strong degeneracy of neutron star matter.

%
%

It is easy to explain
the strong temperature dependence of the
emissivity, $Q^{\rm(D)} \propto T^6$, from momentum space
consideration.  The reaction involves three
strongly degenerate particles which give $T^3$ ,
since the momentum space of each degenerate fermion
is restricted by a thin thermal shell around the Fermi surface.
Furthermore, there is
one non-degenerate neutrino of energy $\sim T$,
which gives an additional factor $T^2$
(after the energy conservation reduction); another factor of $T$
is provided by the neutrino energy $\epsilon_\nu$ under
the integral in Eq.\ (\ref{nucore-QdDef}),
since we consider the neutrino emissivity.
Thus, the degeneracy of matter drastically reduces 
the neutrino emissivity.\\

{\bf (b) Nucleon direct Urca threshold}

The most important feature of the direct Urca process
is its threshold, described by
the step function $\Theta_{npe}$ in Eq.\ (\ref{nucore-Qdur0}).
The step function opens the direct
Urca in the sufficiently dense matter.
Since the process is
{\it several orders of magnitude more efficient}
than other neutrino processes, it is very important
to know exactly where the threshold is placed.
As mentioned above, the reaction is allowed if
the Fermi momenta $p_{{\rm F}n}$, $p_{{\rm F}p}$ and $p_{{\rm F}e}$
satisfy the {\it triangle condition}
(can be sides of one triangle). In other words, the value of each
Fermi momentum should be smaller than the sum
of two others. In neutron star matter
$p_{{\rm F}n}$ is larger than $p_{{\rm F}p}$ and $p_{{\rm F}e}$,
and the triangle condition reads:
$p_{{\rm F}n}<p_{{\rm F}p}+p_{{\rm F}e}$.
For $\rho \sim \rho_0$ one typically has
$p_{{\rm F}n} \sim 340$ MeV/$c$, $p_{{\rm F}e} \sim p_{{\rm F}p}
\sim$ (60--100) MeV/$c$ and the condition is invalid,
i.e., the direct Urca is forbidden.
However, 
$p_{{\rm F}p}$ and $p_{{\rm F}e}$ 
may  grow
with
density faster than $p_{{\rm F}n}$, and the
process can be open at higher densities, a few times $\rho_0$.

Formally, the direct Urca is allowed
if the fraction of protons
among all baryons, $x_p = n_p/n_b$, exceeds the critical value
$x_p=x_{{\rm c}p}$. In the
$npe$ matter ($p_{{\rm F}p}=p_{{\rm F}e}$), this
corresponds to $x_{{\rm c}p} = 1/9= 0.1111$.

In the simplest model of dense matter as a gas of
non-interacting Fermi particles
(e.g., Shapiro and Teukolsky 1983)
the proton fraction is not high enough:
$x_p < x_{{\rm c}p}$ at any density. However, this may not be so
for the realistic equation of state. This was first
mentioned by Boguta (1981),
but his paper remained unnoticed for a long time.

It was a paper by Lattimer et al.\ (1991)
that opened the wide discussion of the direct Urca process.
It showed that for some realistic models of 
neutron star matter,
$x_p$ exceeded
$x_{{\rm c}p}$ at densities several times the
standard nuclear matter density $\rho_0$.
The most favorable for the direct Urca are the equations of state
with the {\it large symmetry energy}; they give the higher proton
fractions.
The nucleon
direct Urca process can then be allowed in the inner cores of
neutron stars more massive than
(1.4--1.6) $M_\odot$, see Sect.\ \ref{sect-cool-code}.

According to Eq.\ (\ref{nucore-Qdur0}), the function
$\Theta_{npe}$ switches on the direct Urca in a step-like
manner: the emissivity jumps up from zero to its finite value
as soon as the density reaches the threshold.
This is certainly an approximation associated with
the phase space decomposition in Eq.\ (\ref{nucore-decomp-Dur}).
In reality, at densities below the threshold
the direct Urca process is not
strictly forbidden but is exponentially reduced due to the strong degeneracy.
The
emissivity is suppressed approximately as $\exp (-\chi)$, where
$\chi=v_{{\rm F}p}
(p_{{\rm F}n}-p_{{\rm F}p}-p_{{\rm F}e})/T$.
This effect may be referred to as the
{\it thermal broadening of the direct Urca threshold}.
In order to account for this effect
qualitatively,
it is sufficient to replace the step function
$\Theta_{npe}$ by the approximate function of the form $({\rm e}^\chi+1)^{-1}$.
However, the thermal broadening seems to be weak and
unimportant for many applications.

\begin{figure}[!t]
\begin{center}
\leavevmode
\epsfysize=8.5cm
\epsfbox[70 65 395 410]{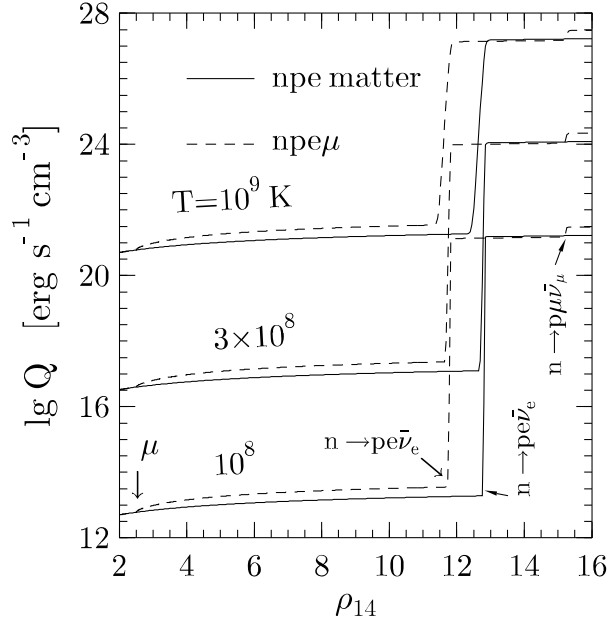}
\end{center}
\caption[]{\footnotesize
        Density dependence of the total neutrino emissivity
        in $npe$ matter (solid lines) and $npe\mu$ matter
        (dashed lines) at $T=10^8$, $3 \times 10^8$ and
        $10^9$ K. Arrows show the appearance of muons ($\mu$) in
        $npe\mu$ matter as well as the thresholds of the direct Urca process
        involving electrons ($n\to pe\bar{\nu}_e$) in
        $npe$ and $npe\mu$ matter, and involving muons
        ($n\to p\mu\bar{\nu}_\mu$) in $npe\mu$ matter.
        Equations of state are described in the text.
}
\label{fig-nucore-Q1(rho)}
\end{figure}

The threshold nature of the direct Urca process is
illustrated in Fig.\ \ref{fig-nucore-Q1(rho)},
which shows the density dependence of the total neutrino emissivity
in the $npe$ matter (solid lines) for $T=10^8$, $3 \times 10^8$
and $10^9$ K. In this figure we use the same equation of state of matter
as in Fig.\ \ref{fig-nucore-Q(rho)}, and the solid curve
at $T= 3 \times 10^8$ K is a continuation of 
the `total' curve in Fig.\ \ref{fig-nucore-Q(rho)}
to higher densities. The direct Urca reaction is switched
on at $\rho_{\rm crit}=1.298 \times 10^{15}$ g cm$^{-3}$.
The thermal broadening of the threshold
is taken into account as described above.
At $\rho < \rho_{\rm crit}$ the neutrino emissivity
is determined by the slow reactions (Fig.\ \ref{fig-nucore-Q(rho)}),
mainly by the modified Urca process (Sect.\ \ref{sect-nucore-Murca}).
At $\rho > \rho_{\rm crit}$
the direct Urca process amplifies the emissivity
by 6 -- 8 orders of magnitude.

If muons are present
for the same number density of baryons $n_b$, the proton fraction
becomes slightly higher than in the $npe$ matter, and
the electron fraction slightly lower. In this case,
the threshold proton fraction $x_{{\rm c}p}$
for the direct Urca process (\ref{nucore-Durca})
is higher and reaches
$1/[1+(1+2^{-1/3})^3]=0.1477$ for 
ultrarelativistic muons
(Lattimer et al.\ 1991).

The effect of muons is illustrated in
the same Fig.\ \ref{fig-nucore-Q1(rho)}.
The dashed lines display the total neutrino emissivities
with the same temperature but for an equation of state
which allows for the appearance of muons. This equation
of state is built on the basis of the same nuclear energy
as has been
used in the equation of state of the $npe$ matter.
The muons appear at
$\rho=2.5 \times 10^{14}$ g cm$^{-3}$. At lower densities,
both equations of state coincide and the emissivities
are naturally the same. At higher densities
the muons lower noticeably
the electron number density.  As the density grows,
the muon number density $n_\mu$ approaches $n_e$.
The presence of muons lowers the electron direct Urca
threshold to $\rho_{\rm crit}=1.18 \times 10^{15}$ g cm$^{-3}$.
The difference of the emissivities in the density range
from $2.5 \times 10^{14}$ g cm$^{-3}$ to $1.18 \times 10^{15}$
g cm$^{-3}$ is mainly attributed to switching on
the modified Urca process involving muons (see Sect.\ \ref{sect-nucore-Murca}).
The jumps associated with the
onset of the electron direct Urca process are about the same as
in the $npe$ matter, only at lower density.\\

{\bf (c) Nucleon direct Urca process with muons}

Now we turn to the direct Urca processes which involve
particles other than $n$, $p$ and $e$.
If muons are present, then the direct Urca process
involving muons may be possible,
\begin{equation}
   n \to p + \mu + \bar{\nu}_\mu, \quad    p + \mu  \to n + {\nu}_\mu.
\label{nucore-Durca-mu}
\end{equation}
along with the process involving
electrons. Its emissivity is given by the same Eq.\
(\ref{nucore-Qdur0}) as for the basic direct Urca process
since the condition of beta-equilibrium implies the equality
of chemical potentials $\mu_\mu=\mu_e$, i.e.,
$m_\mu^\ast \equiv \mu_\mu/c^2=m_e^\ast$.
The only difference is that
$\Theta_{npe}$ must be replaced by $\Theta_{np\mu}$.
Therefore, the muon process (\ref{nucore-Durca-mu})
differs from the electron process (\ref{nucore-Durca})
only by the threshold.  It
opens at a somewhat higher density
than the electron process.
Its emissivity equals exactly the emissivity
of the electron process, and the total emissivity
just doubles as demonstrated in
Fig.\ \ref{fig-nucore-Q1(rho)}.
For our particular equation of state of the $npe\mu$ matter,
the threshold density of the muon reaction (\ref{nucore-Durca-mu})
is $1.53 \times 10^{15}$ g cm$^{-3}$.
\\

{\bf (e) Hyperon direct Urca process}

If the equation of state
in the neutron star core allows for the appearance of hyperons,
direct Urca processes involving hyperons
[processes (I.2)--(I.4) in Table \ref{tab-nucore-list}] may be open
(Prakash et al.\ 1992)
\begin{equation}
   B_1 \to B_2 + l + \bar{\nu}_l, \quad    B_2 + l  \to B_1 + \nu_l,
\label{nucore-Durca-hyper}
\end{equation}
where $B_1$ and $B_2$ stand for baryons
(nucleons or $\Lambda$-, $\Sigma^-$ hyperons), while
$l$ is a lepton (electron or muon).

The neutrino emissivity of any reaction (\ref{nucore-Durca-hyper}) is
calculated exactly in the same manner as of the nucleon
direct 
Urca process, 
with somewhat different
constants $G_1$, $f_{V1}$ and $g_{A1}$
of weak hadronic currents in the weak interaction
Hamiltonian (\ref{nucore-H}).
The result can be obtained promptly using
the phase space decomposition of the emissivity,
Eq.\ (\ref{nucore-decomp-Dur}).
Indeed, the energy integral $I$, Eq.\ (\ref{nucore-AIdur}),
is evidently the same for
all direct Urca processes. The angular integrals $A$
are also similar, and the squared matrix element is given
by Eq.\ (\ref{nucore-squared-Mdur}). Then the phase space
decomposition yields {\it the rescaling rule}
which allows one to determine the
emissivity $Q^{\rm (D)}_{12l}$ of any direct Urca process
(\ref{nucore-Durca-hyper}) from the emissivity $Q^{\rm(D)}_{npe}$
of the basic reaction (\ref{nucore-Durca}):
\begin{equation}
     { Q^{\rm(D)}_{12l} \over Q^{\rm(D)}_{npe}}
     = { m_1^\ast m_2^\ast \, \Theta_{12l} \over
         m_n^\ast m_p^\ast \, \Theta_{npe}} \, r_{12},\quad
      r_{12} \equiv {G_1^2 \,(f_{V1}^2 + 3 g_{A1}^2)
               \over G_{\rm F}^2 \, \cos^2 \theta_{\rm C} \,
               (1 + 3 g_A^2)}.
\label{nucore-rescale-Dur}
\end{equation}
The result is
(Prakash et al.\ 1992)
\begin{eqnarray}
 Q^{\rm(D)}_{12l} &  = & {457\, \pi \over 10080} \,
                G_1^2 \, (f_{V1}^2 +3 g_{A1}^2) \;
                {m_1^\ast \, m_2^\ast \, m_l^\ast \over \hbar^{10} c^3} \;
                (k_{\rm B} T)^6 \, \Theta_{12l}
\nonumber \\
        & \approx & 4.00 \times 10^{27}
                \left( {n_e \over n_0} \right)^{1/3} \;
                 {m_1^\ast \, m_2^\ast \over m_n^2} \,
                 T^6_9 \, \Theta_{12l} \, r_{12}
                 \;\;\;\ {\rm erg\;cm^{-3}\;s^{-1}}.
\label{nucore-Qdur0-hyper}
\end{eqnarray}
The values of the constants for all direct Urca reactions
in the $npe\mu\Lambda\Sigma^-$ matter are listed
in Table \ref{tab-nucore-Dur} (assuming SU(3) symmetry;
after Prakash et al.\ 1992).  Those authors presented
also the constants for the reactions involving $\Sigma^0$
and $\Xi^0$, $\Xi^\pm$ hyperons, which may appear
in neutron star cores at very high densities.

Thus, the neutrino emissivity
of the direct Urca processes is determined quite accurately.
The only complication which we ignore is that
the values of $f_{V1}$ and $g_{A1}$ may be renormalized
by the in-medium effects.

\begin{table}[t]
\caption{Parameters $G_1$, $f_{V1}$ and $g_{A1}$
    and the factor $r_{12}$ in Eq.\
    (\protect{\ref{nucore-Qdur0-hyper}})
}
\begin{center}
  \begin{tabular}{||ll|llll||}
  \hline \hline
          Process&  & $G_1/G_{\rm F}$     &   $f_{V1} $
                    & $g_{A1}$           &   $r_{12} $ \\
  \hline
  $n\to pl\bar{\nu}_l \quad $                &
  $pl\to n \nu_l \quad $                &
  $          \cos \theta_{\rm C} $      &
   \protect{$\;$} 1                     &
     1.233                              &
    1                                  \\
  $\Lambda \to pl\bar{\nu}_l  $         &
  $ pl \to \Lambda \nu_l  $             &
  $         \sin \theta_{\rm C} $       &
   $ -1.225 $                           &
   $  0.893 $                         &
   $  0.039 $                        \\
  $\Sigma^- \to n l \bar{\nu}_l  $      &
  $nl \to \Sigma^- \nu_l  $             &
  $         \sin \theta_{\rm C} $       &
   $ -1 $                               &
   $ 0.279$                           &
  0.012                                \\
  $\Sigma^- \to \Lambda l\bar{\nu}_l $  &
  $\Lambda l \to \Sigma^- \nu_l $       &
  $         \cos \theta_{\rm C} $       &
  \protect{$\;$} 0                      &
    0.617                              &
  0.206                               \\
  \hline \hline
\end{tabular}
\label{tab-nucore-Dur}
\end{center}
\end{table}

Like the nucleon 
direct Urca process,
the processes with hyperons
have threshold nature and open at rather
high densities.  Naturally, these threshold densities
are higher than the densities at which
hyperons appear in the matter.  Also,
the processes involving muons have higher thresholds than
the corresponding electron reactions, 
but have the same emissivities.

Any reaction of processes
(I.1) and 
(I.4),
for which $G \propto \cos \theta_{\rm C}$,
conserves strangeness. Thus, the emissivity 
of the process (I.4) is
only slightly lower than that of the
nucleon process (I.1).  Any reaction
of processes (I.2) and (I.3), with
$G \propto \sin \theta_{\rm C}$, changes 
strangeness. Accordingly, the emissivities
of the processes (I.2) and (I.3) 
are about an order of
magnitude weaker (Table \ref{tab-nucore-Dur}).
Nevertheless,
any direct Urca process
is several orders of magnitude more efficient than
other neutrino processes
(processes II--V in Table \ref{tab-nucore-list}).

Like the nucleon 
direct Urca process,
all hyperon direct Urca processes are forbidden
in the simplest model of matter as a gas of free
particles but open in many realistic models.
One can easily see (Prakash et al.\ 1992)
that the $\Sigma^-\to nl\bar{\nu}_l$
process can only be allowed at such densities that
the nucleon Urca process is already switched on.
Thus, its emissivity is 
just a small addition
to the large emissivity
of the nucleon direct Urca.
As for the processes $\Lambda \to pl\bar{\nu}_l$ and
$\Sigma^- \to \Lambda  l\bar{\nu}_l$,
they can, in principle, be open under the conditions
that the nucleon direct Urca is forbidden,
although this would require a contrived equation of state.
The analysis of momentum conservation shows
(Prakash et al.\ 1992) that the threshold fraction
of $\Lambda$ hyperons is low, $x_{{\rm c}\Lambda}=n_{{\rm c}\Lambda}/n_b
\lesssim 0.001-0.003$, depending on the equation of state.
As a rule, the number density of $\Lambda$ hyperons
increases rapidly with density.
Thus,
the threshold of the hyperon direct Urca process
can nearly coincide with the threshold of the $\Lambda$ hyperon
creation in the matter. Prakash et al.\ (1992) verified this
using a number of equations of state. In particular, the effect is strong
in the relativistic mean-field models of dense matter,
in which hyperons appear at $\rho \sim 2 \rho_0$.

In addition to the direct Urca processes
discussed above, one may contemplate the
processes involving $\Delta^-$ isobars, the nonstrange
baryons of spin 3/2 and mass about 1232 MeV.
The $\Delta^-$ isobars may appear at
densities of several $\rho_0$ initiating the powerful direct
Urca processes of the type $\Delta^- \to n l \bar{\nu}$
and $\Delta^- \to \Lambda l \bar{\nu}$. The threshold
density of the former process is always higher than for the
nucleon direct Urca, while the latter process may, in
principle, open up when the nucleon direct Urca is forbidden.

To summarize, different 
powerful direct Urca processes can operate
in the dense neutron star matter for many realistic
equations of state.
However, as we discuss in Chapt.\ \ref{chapt-nusup},
their neutrino
emissivity can be strongly
reduced by the superfluid effects.
\\

{\bf (f) Name}

Finally, the reader should not be confused by the funny
name of the process: it was one of the jokes of George Gamow.
Indeed, a chain of reactions like
(\ref{nucore-Durca})
(beta decay and capture of atomic nuclei)
was introduced by Gamow and Schoenberg (1941)
%
%
%
to describe the neutrino emission
in evolved massive stars --- presupernovae (Sect.\ref{sect-nucrust-beta}).
They called these reactions {\it urca processes}
after the name of a casino in Rio de Janeiro (closed
by Brazilian government in 1955
along with all gambling business in Brazil).
Gamow (1970) narrated that ``We called it the Urca Process,
partially to commemorate the casino in which we first met,
and partially because the Urca Process results in a rapid
disappearance of thermal energy from the interior of a star,
similar to the rapid disappearance of money from the pockets
of the gamblers on the Casino da Urca". If {\it Physical
Review} asked to explain the name,
the authors had the solution --- the abbreviation of
``{\bf u}n{\bf r}ecordable {\bf c}ooling {\bf a}gent" ---
but they were not asked.
This instance,
however, justifies the spelling {\it URCA}
used by many authors. The authors of this review
cannot resist to remind the reader that George Gamow ---
born in Odessa, one of the most colorful Russian (now Ukrainian)
cities, full of jokes, humor and irony ---
had an excellent knowledge of the Russian language
and used to practice it in his jokes.
He could not be unaware of the fact that the word
``urca" was popular in Russia in the 1930s and 1940s
and meant {\it thief} who could borrow
money from your pockets until you said Jack Robinson.
The striking similarity of gambling and stealing
allows us to suspect that Gamow, introducing
the new word into physical language, 
related hot Rio
to the cold Russian North. 

Returning to the neutron star physics, we can
remark that one has to distinguish between the powerful
direct Urca process and the much weaker modified Urca
process.  Following Gamow's tradition
to use beautiful (Russian) names, our colleague,
Kseniya Levenfish, proposed to call the direct Urca
process as {\it Durca}, and the modified Urca process as {\it Murca}.
The latter process is the subject of the next section.

\subsection{Modified Urca process}
\label{sect-nucore-Murca}

{\bf (a) Emissivity}

As shown in the preceding section, the direct Urca process in
the $npe$ matter is allowed for the equations of state
with the large symmetry energy
at densities several times the nuclear density.
In other cases, the main neutrino reaction is the
modified Urca process. It is similar to the
direct Urca, but involves an
additional nucleon spectator.  In the
$npe$ matter, the reaction can go through two channels
\begin{eqnarray}
    n + n &\to& p + n + e + \bar{\nu}_e \, , \quad
    p + n + e \to n + n  + \nu_e \, ;
\label{nucore-Murcan} \\
    n + p &\to& p + p + e + \bar{\nu}_e  \, , \quad
    p + p + e \to n + p  + \nu_e \, ,
\label{nucore-Murcap}
\end{eqnarray}
which we define as the {\it neutron} and {\it proton}
branches of the modified Urca process, respectively
[reactions (II.1) with $B=n$ and $p$ in Table \ref{tab-nucore-list}].
The additional nucleon is required to conserve
momentum of the reacting particles; it will do the job even if
the direct Urca is forbidden.
The extra particle relaxes the momentum conservation condition
but slows the reaction rate.

The modified Urca processes
were introduced in the context of neutron star cooling
by Chiu and Salpeter (1964). They
were 
studied by
Bahcall and Wolf (1965a, b),
Flowers et al.\ (1975),
Friman and Maxwell (1979),
as well as by Maxwell (1987).
The most detailed paper
seems to be that
by Friman and Maxwell,
which however neglected the proton branch.
The latter branch was analyzed by
Yakovlev and Levenfish (1995) using the same
formalism.
Below we will mainly follow the consideration of
Friman and Maxwell (1979) and Yakovlev and Levenfish (1995).

The neutron and proton 
processes (\ref{nucore-Murcan}) and (\ref{nucore-Murcap})
are described by a set of Feynman diagrams. The diagrams for
the neutron reaction are given, for instance, by
Friman and Maxwell (1979), and the diagrams for the proton
reaction are similar.
Each diagram contains two three-tail vertices tied by
the strong-interaction line and one four-tail vertex associated
with weak interaction. In other words, we have a nucleon-nucleon
collision accompanied by beta decay or beta capture.

The modified Urca process will be labeled by upperscripts
(${\rm M}\!N$),
where $N=n$ indicates the neutron branch
(\ref{nucore-Murcan}) of the process, and
$N=p$ indicates the proton branch (\ref{nucore-Murcap}).
Both branches consist of the direct and inverse reactions.
In beta-equilibrium the rates of the two reactions
are equal; thus, it is sufficient to
calculate the rate of any reaction and double the result.
The general expression for the emissivity can be written
in the form
\begin{eqnarray}
    Q^{({\rm M}\!N)} & = & 2  \int
              \left[ \prod_{j=1}^4 { {\rm d} {\bf p}_j
              \over (2 \pi)^3} \right]
              { {\rm d}{\bf p}_e \over (2 \pi)^3} \;
              { {\rm d}{\bf p}_\nu \over (2 \pi)^3} \;
              \epsilon_\nu \,
              (2 \pi)^4 \, \delta(E_f-E_i) 
\nonumber \\
           & & \times \; \delta({\bf P}_f - {\bf P}_i ) \,
               f_1 f_2 (1-f_3)(1-f_4)(1-f_e) \, {1 \over 2} \, | M_{fi} |^2 ,
  \label{nucore-Q_mur}
\end{eqnarray}
where ${\bf p}_j$ is the nucleon momentum
($j=1$,~2,~3,~4).
The delta functions $\delta(E_f-E_i)$
and $\delta( {\bf P }_f - {\bf P }_i )$ describe energy and
momentum conservation;
subscripts $i$ and $f$ refer to the initial and final particle
states, respectively;
$| M_{fi} |^2$ is the squared matrix element summed over 
spins states. The symmetry factor
1/2 before the matrix element is introduced
to avoid double counting of the same collisions of identical particles.

The emissivity (\ref{nucore-Q_mur}) can be written in the form
similar to the direct Urca emissivity
[see Eq.\ (\ref{nucore-decomp-Dur}) and explanations afterwards]:
\begin{eqnarray}
&&    Q^{({\rm M}\!N)}  =  {1 \over (2 \pi)^{14} }\; T^8
            A I  \, \langle |M_{fi}|^2 \rangle
              \, \prod_{j=1}^5 p_{{\rm F}j} m_j^\ast ,
\label{nucore-DecompMur} \\
&&   A = 4 \pi \;  \left[ \prod_{j=1}^5 \int {\rm d} \Omega_j \; \right]
            \delta \left( {\bf P}_f - {\bf P}_i  \right) ,
\label{nucore-A} \\
&&   \langle |M_{fi}|^2 \rangle
            = { 4 \pi \over A}  \;  
           \left[ \prod_{j=1}^5 \int {\rm d} \Omega_j \; \right]
           \delta \left( {\bf P}_f - {\bf P}_i  \right) \, |M_{fi}|^2,
\label{nucore-AMfi} \\
&&   I = \int_0^\infty {\rm d} x_\nu \; x_\nu^3
       \left[ \prod_{j=1}^5 \int_{-\infty}^{+\infty}
       {\rm d} x_j \; f_j \right]
       \delta \left( \sum_{j=1}^5 x_j-x_\nu \right).
\label{nucore-I}
\end{eqnarray}
The quantities $A$ and $\langle |M_{fi}|^2 \rangle$
contain 
the integrals over the orientations of the particle momenta
($j$=5 corresponds to the electron);
all lengths of the momenta ${\bf p}_j$ of the nucleons and the electron
in the delta function are set equal to the appropriate Fermi momenta.
In the momentum conservation condition the neutrino momentum is neglected
and the integration over its orientation
yields $4 \pi$. 
As in the direct Urca process,
the squared matrix element averaged over orientations of ${\bf p}_\nu$
can be used, but now it generally depends on orientations
of other momenta and is left under the integral. Thus it is relevant
to introduce
$\langle |M_{fi}|^2 \rangle$, the squared matrix element
averaged over orientations of the nucleon momenta.
The quantity $I$, given by Eq.\ (\ref{nucore-I}),
includes the integrals over the dimensionless energies of the neutrino
($x_\nu$)
and other particles ($x_j$), where
the blocking factors $(1-f_j)$
are converted into $f_j$ in the same manner as in Eq.\
(\ref{nucore-Idur}).

The integration in Eq.\ (\ref{nucore-I}) is similar to
that in Eq.\ (\ref{nucore-AIdur}), yielding
\begin{equation}
     I  =  \int_0^\infty {\rm d} x_\nu \; x_\nu^3 \; J(x_\nu) =
       {11513 \, \pi^8 \over 120960} ,
\label{nucore-AImur}
\end{equation}
where
\begin{equation}
  J(x)  = { 9 \, \pi^4 + 10 \pi^2 \, x^2 + x^4
          \over 24 \, ({\rm e}^x + 1)}.
\label{nucore-J(x)-mur}
\end{equation}

Subsequent analysis is different for
the neutron and proton branches of the modified Urca
process, and these branches are considered separately.\\

{\bf (b) Neutron branch}

To be specific, 
consider the first reaction of the process 
(\ref{nucore-Murcan}). Let 1 and 2 label the initial neutrons,
3 be the final neutron and 4 the final proton.
The angular factor $A$ is well known
(e.g., Shapiro and
Teukolsky 1983)
\begin{equation}
    A = A_n= {2 \pi \, (4 \pi)^4 \over p_{{\rm F}n}^3}.
\label{nucore-AAn}
\end{equation}
Notice that this expression is modified 
(Yakovlev and Levenfish 1995) at the densities
higher than the direct Urca threshold but at these
densities the direct Urca process dominates and the modified Urca
processes are insignificant.

The main problem is to calculate the matrix element
$M_{fi}$ since it involves strong interaction.
We will base our consideration on the calculation performed by
Friman and Maxwell (1979).
The long-range (small momentum transfer) part of the nucleon--nucleon
interaction
was described using the one--pion--exchange (OPE) interaction model,
while the short-range (large momentum transfer) part
was described in the framework of the Landau Fermi--liquid theory
(e.g.,
Baym and Pethick 1991). 
The OPE part of the matrix element was 
obtained neglecting the electron momentum.
In the practical expressions,
the Fermi-liquid contribution was neglected.

Since the OPE result is basic for more advanced models
we have rederived the OPE matrix element, making less
number of simplifying assumptions.
Treating the nucleons as non-relativistic
particles, assuming the neutrino momentum to be much
smaller than the momenta of other particles
and averaging over orientations of the neutrino momentum
we have obtained:
\begin{eqnarray}
  |M^{({\rm M}n)}_{fi}|^2 & = & { 16 \, G^2 \over \epsilon_e^2 } \, 
             \left( f^\pi \over m_\pi \right)^4 \,
             (g_V^2 \, F_V + g_A^2 \, F_A),
\label{nucore-nMurca-Mfi} 
\end{eqnarray}
where $g_V=1$ and
$g_A = 1.26$ are the vector and axial--vector 
constants of weak hadron current,
$f^\pi \approx 1$ is the OPE
$\pi N$-interaction constant in the $p$-state, and
$m_\pi$ is the pion mass. 
Furthermore,
\begin{eqnarray}
    F_V & = & q_1^4 + q_2^4 + q_3^4 + q_4^4 + 
              q_1^2 q_3^2 + q_2^2 q_4^2 - q_1^2 q_4^2 - q_2^2 q_3^2
\nonumber \\
       &  & - 2({\bf q}_1\cdot{\bf q}_2)^2 -
        2({\bf q}_1\cdot{\bf q}_3)^2 - 2({\bf q}_3\cdot{\bf q}_4)^2
        -2({\bf q}_2\cdot{\bf q}_4)^2 + 2({\bf q}_2\cdot{\bf q}_3)^2
        + 2({\bf q}_1 \cdot{\bf q}_4)^2,
\label{nucore-S_V} \\
   F_A & = & q_1^4 + 3 q_2^4 +  q_3^4 + 3 q_4^4 
             - q_2^2 q_3^2 - q_2^2 q_4^2 - q_1^2 q_4^2 
\nonumber \\
       &  & + 2({\bf q}_2\cdot{\bf q}_3)^2 -
        ({\bf q}_1\cdot{\bf q}_3)^2 + 2({\bf q}_1\cdot{\bf q}_4)^2
        -2({\bf q}_2\cdot{\bf q}_4)^2, 
\label{nucore-S_A}
\end{eqnarray}
with
${\bf q}_i \equiv {\bf k}_i/\sqrt{k_i^2 + m_\pi^2}$,
${\bf k}_1={\bf p}_1 - {\bf p}_3$,
${\bf k}_2={\bf p}_4 - {\bf p}_2$,
${\bf k}_3={\bf p}_3 - {\bf p}_2$, and
${\bf k}_4={\bf p}_1 - {\bf p}_4$.

The Friman and Maxwell approximation
corresponds to setting ${\bf p}_e$ =0, ${\bf k}_2={\bf k}_1$ and
${\bf k}_4={\bf k}_3$ in the matrix element. 
Then $F^{\rm FM}_V=0$, i.e., the weak vector
currents do not contribute to the neutrino emissivity, while the
axial vector contribution reduces to
\begin{equation}  
    F_A^{\rm FM}= {4 \, k_1^4 \over (k_1^2+m_\pi^2)^2}
                 +{4 \, k_3^4 \over (k_3^2+m_\pi^2)^2}
                 +{({\bf k}_1 \cdot {\bf k}_3)^2 - 3 k_1^2 k_3^2
                   \over (k_1^2+m_\pi^2)(k_3^2+m_\pi^2)}.
\label{nucore-Murca-FM}
\end{equation} 
The latter result corresponds to Eq.\ (71) in Friman and Maxwell (1979).
The first term comes from the squared amplitude of the reaction
diagrams in which the nucleon 1 transforms into 3, and the
nucleon 2 transforms into 4. The second term is the squared
amplitude of the transition $1 \to 4$, $2 \to 3$, and
the third term describes interference of two amplitudes.
In the absence of the interference term, Eq.\ (\ref{nucore-Murca-FM})
reproduces the OPE part of the squared matrix element
given by Eq.\ (39) in Friman and Maxwell (1979).

At the next step Friman and Maxwell 
neglected also the proton momentum in the matrix element.
This resulted in $k_1 \approx k_2 \approx p_{{\rm F}n}$
and ${\bf k}_1 \cdot {\bf k}_3 \approx p_{{\rm F}n}^2/2$.
In this simplified model the squared matrix element is
\begin{equation}
  |M^{({\rm M}n)}_{fi}|^2  =   16 \, G^2 \, 
             \left( f^\pi \over m_\pi \right)^4 \,
              {g_A^2 \over \epsilon_e^2} \,  
              {21 \over 4} \, { p_{{\rm F}n}^4 
                  \over (p_{{\rm F}n}^2 + m_\pi^2)^2},
\label{nucore-Murca-FM1}
\end{equation}
where $21/4=8-11/4$, and $-11/4$ comes from the interference term.
This squared matrix element
is remarkably independent of 
the orientations of the particle momenta and
can be taken out of the angular integration
in Eq.\ (\ref{nucore-AMfi}) just as for the direct Urca process.
In this approximation,
$
\langle |M^{({\rm M}n)}_{fi}|^2 \rangle =   |M^{({\rm M}n)}_{fi}|^2.
$
Notice that our definition of $|M^{({\rm M}n)}_{fi}|^2$
differs from the definition of Friman and Maxwell (1979)
by a factor of $(4 \epsilon_e \epsilon_\nu)^{-1}$.

The exact OPE
neutrino emissivity (\ref{nucore-DecompMur})
contains the squared matrix element averaged
over the orientations of the particle momenta in accordance
with Eq.\ (\ref{nucore-A}). We have used the exact
OPE matrix element given by Eqs.\ 
(\ref{nucore-nMurca-Mfi})--(\ref{nucore-S_A})
and performed the angular 
averaging numerically.
We have compared these results
with the simplified results obtained 
in the approximation of constant
matrix element.
The agreement is excellent. The deviation does not exceed
several percent at $\rho \sim \rho_0$, and increases
to about 10\% at the densities $\rho \sim 3 \rho_0$ at which the OPE model
definitely becomes invalid by itself.

It is well known that the OPE 
model does not treat properly the short-range part of the nucleon-nucleon
interaction and the correlation effects. It is thought to
be qualitatively adequate at $\rho \lesssim \rho_0$  
and becomes less accurate with increasing $\rho$.
More advanced models of nucleon--nucleon interactions
and many--body theories are required at supranuclear
densities. In the absence of exact
calculations of the neutrino emissivity
it seems reasonable to use the practical expression
obtained by Friman and Maxwell (1979) on the basis
of their simplified approach, 
Eq.\ (\ref{nucore-Murca-FM1}). 
Their final result can be written as
(in standard physical units):
\begin{eqnarray}
     Q^{({\rm M}n)} & = & {11513 \over 30240} \,
               {G_{\rm F}^2 \, \cos^2 \theta_{\rm C} \, 
               g_A^2 m_n^{\ast 3} m_p^\ast \over 2 \pi}
               \left( {f^\pi \over m_\pi } \right)^4
               {p_{{\rm F}p} (k_{\rm B} T)^8 \over \hbar^{10} c^8} \,
               \alpha_n \beta_n
      \nonumber \\
      & \approx & 8.1 \times 10^{21}
                \left( {m_n^\ast \over m_n } \right)^3
                \left( {m_p^\ast \over m_p } \right)
                \left( {n_p  \over n_0 } \right)^{1/3} \!
                T_9^8 \, \alpha_n \beta_n \; \; \;
                {\rm erg \; cm^{-3} \; s^{-1}}  \, .
    \label{nucore-Qn0}
\end{eqnarray}
Here, $m_\pi$ is the $\pi^\pm$ mass.
The factor
$\alpha_n$ comes from the estimation of the squared matrix element
(\ref{nucore-Murca-FM1}) in which the interference term
has been neglected: 21/4 replaced with $4 \times 2=8$. 
The factor $\beta_n$ contains other
corrections introduced in an approximate manner.
In their final Eq.\ (65c) for $Q^{({\rm M}n)}$
Friman and Maxwell (1979)
used the value $\alpha_n = 1.13$, calculated at $\rho = \rho_0$
for some particular equation of state of dense matter,
and set $\beta_n$ = 0.68 to account for the correlation effects.
Setting $\rho=\rho_0$ in the squared matrix element
could be a better approximation than using exact OPE results
at $\rho > \rho_0$. We will adopt the latter approximation
in numerical examples 
although more advanced consideration of the modified Urca
process would be desirable.\\

{\bf (c) Proton branch}

For certainty, 
consider the second reaction of
the process (\ref{nucore-Murcap}). Let particles 1 and 2 be
the initial protons, while 3 and 4 be
the final proton and neutron. 

Calculation of the angular factor $A$ for the
proton reaction is more sophisticated
and yields  
\begin{equation}
     A_p  =  {2 (2 \pi)^5 \over p_{{\rm F}n} p_{{\rm F}p}^3 p_{{\rm F}e}} \,
       (p_{{\rm F}e}+3 p_{{\rm F}p}- p_{{\rm F}n})^2 \,  \Theta_{{\rm M}p},
\label{nucore-Anp}
\end{equation}
where $\Theta_{{\rm M}p}=1$ if the proton branch is allowed by momentum
conservation, and $\Theta_{{\rm M}p}=0$ otherwise.
Since
$p_{{\rm F}e}$ and $p_{{\rm F}p}$ are smaller than
$p_{{\rm F}n}$ in the outer neutron star core 
(Sects.\ \ref{sect-overview-struct} and \ref{sect-nucore-Durca}),
we have $\Theta_{{\rm M}p}= 1$ for
$p_{{\rm F}n} <$ $3 p_{{\rm F}p} + p_{{\rm F}e}$.
Notice that the expression for $A_p$ should be modified
at densities above the direct Urca threshold
at which the modified Urca processes become insignificant.

Let us adopt the same OPE model
to analyse the matrix element. It is easy to verify that
the exact squared OPE matrix element is given by the
same Eqs.\ (\ref{nucore-nMurca-Mfi})--(\ref{nucore-S_A}).
Using these expressions
we have calculated factor 
$\langle |M_{fi}^{({\rm M}p)}|^2 \rangle$ 
numerically from Eq.\ (\ref{nucore-AMfi}) 
and compared the numerical results with those
obtained in the approximation of constant matrix element.
As in the case of the neutron reaction branch we have found good agreement
of the exact numerical results with the approximate ones.
It turns out that in the present case 
we can take the same squared matrix element as
in Eq.\ (\ref{nucore-Murca-FM1}) replacing
$21/4 \to 6$ and $p_{{\rm F}n} \to p_{{\rm F}n}-p_{{\rm F}p}$.
The former replacement corresponds to ${\bf k}_1 \cdot {\bf k}_3 =- 
(p_{{\rm F}n}-p_{{\rm F}p})^2$
in Eq.\ (\ref{nucore-Murca-FM}) while the latter one
introduces the maximum momentum transfer 
$(p_{{\rm F}n}-p_{{\rm F}p})$ in the $np$ collisions,
an appropriate value to be substituted
into the matrix element.

Therefore, the approximation of angle-independent
matrix element holds quite well for both, the neutron
and proton reaction branches. 
The emissivities
in these branches differ by the matrix elements,
the angular factors and the densities of states
of nucleons in Eq.\ (\ref{nucore-DecompMur}). Accordingly,  
the expression for the neutrino emissivity in the
proton branch is immediately obtained from
the phase-space decomposition. It gives
the following rescaling rule [cf Eq.\ (\ref{nucore-rescale-Dur})]:
\begin{equation}
    { Q^{({\rm M}p)} \over Q^{({\rm M}n)}} =
    { \langle | M_{fi}^{({\rm M}p)}  |^2 \rangle
    \over \langle | M_{fi}^{({\rm M}n)} |^2 \rangle } \,
    \left( m_p^\ast \over m_n^\ast \right)^2 
     {(p_{{\rm F}e}+3 p_{{\rm F}p}- p_{{\rm F}n})^2 \over
      8 p_{{\rm F}e} p_{{\rm F}p}} \,
    \Theta_{{\rm M}p} \approx
    \left( m_p^\ast \over m_n^\ast \right)^2 
      {(p_{{\rm F}e}+3 p_{{\rm F}p}- p_{{\rm F}n})^2 \over
      8 p_{{\rm F}e} p_{{\rm F}p}} \,
    \Theta_{{\rm M}p}.
\label{nucore-rescale-Mur}
\end{equation}
The rule is not based on any particular
model of strong interactions.
For practical applications, it is 
reasonable to neglect
insignificant difference of the matrix elements and
calculate $Q^{({\rm M}p)}$ from $Q^{({\rm M}n)}$
setting $\langle | M_{fi}^{({\rm M}p)} |^2 \rangle
    = \langle  | M_{fi}^{({\rm M}n)} |^2 \rangle $ as indicated
in the last approximate expression in Eq.\ (\ref{nucore-rescale-Mur}).
  
Thus, the emissivities 
of the neutron  and proton
branches of the process are similar.
The main difference is in the threshold
for the proton branch; it is allowed at
$p_{{\rm F}n} < 3p_{{\rm F}p} + p_{{\rm F}e}$.
In the $npe$ matter, this inequality is
equivalent to $p_{{\rm F}n} < 4p_{{\rm F}e}$, i.e.,
to the proton fraction $x_p$ exceeding the critical
value $x_{{\rm c}p} =1/65=0.0154$.
The latter condition is satisfied almost anywhere in the
neutron star core. It can be violated only for the equations
of state with very low symmetry energy
at $\rho \lesssim \rho_0$, forbidding
the proton branch in the outermost
part of the core.
Contrary to the case of the direct Urca process,
the emissivity $Q^{({\rm M}p)}$ increases smoothly
from zero while the density exceeds the threshold value.
The proton process is especially 
efficient at higher densities, near the threshold
of the direct Urca process. For instance, in the $npe$ matter 
near this threshold
($p_{{\rm F}e} = p_{{\rm F}p}= p_{{\rm F}n}/2$), we find
$Q^{({\rm M}p)} = 0.5 \, Q^{({\rm M}n)} \, (m_p^\ast / m_n^\ast)^2$,
i.e., the proton branch is nearly as efficient
as the neutron branch. The importance of the proton branch
is also illustrated in Figs.\
\ref{fig-nucore-Q(T)} and \ref{fig-nucore-Q(rho)}.
The figures show that both branches of the
modified Urca process are the {\it leading standard (slow)
neutrino generating mechanisms} in non-superfluid neutron star
cores, provided the direct Urca processes are forbidden.

Notice, that the emissivity  $ Q^{({\rm M}\!N)}$
depends on temperature as $T^8$. An extra factor $T^2$
with respect to the direct Urca process ($Q^{(\rm D)} \propto T^6$)
appears because now two more degenerate particles are involved.

The potential efficiency of the proton branch was outlined by
Itoh and Tsuneto (1972) who, however, did not calculate $Q^{({\rm M}p)}$.
Later the neutrino emissivity $Q^{({\rm M}p)}$ was calculated by
Maxwell (1987)
who found it negligibly small compared to $Q^{({\rm M}n)}$.
That conclusion is erroneous because of several inaccuracies made by
Maxwell (1987)
and analyzed by
Yakovlev and Levenfish (1995).
In particular, Maxwell (1987) incorrectly neglected the electron momentum
in momentum conservation.
It should be added that Yakovlev and Levenfish (1995),
in their turn, incorrectly
calculated the angular integral $A_p$.
Their result is equivalent to replacing
$(p_{{\rm F}e}+3 p_{{\rm F}p}- p_{{\rm F}n})^2/
(8 p_{{\rm F}e} p_{{\rm F}p}) \to 
(4 p_{{\rm F}p}- p_{{\rm F}e})/(4 p_{{\rm F}p})$
in Eq.\ (\ref{nucore-rescale-Mur}). Accordingly, they
overestimated the efficiency of the proton
branch at the densities just above the proton Urca threshold density. 

In a series of papers initiated by Voskresensky
and Senatorov (1984, 1986)
the neutrino reactions
of the Urca type have been studied
for the models of nucleon-nucleon interaction with
highly polarized pion degrees of freedom.
Pion condensation in such a matter is very efficient and takes place at
$\rho \sim \rho_0$. According to those authors, even at lower
density before the condensation occurs,
the neutrino emissivity appears to be several orders of magnitude
higher than in the standard modified Urca
process due to the polarizability of the pion field.
Then the strong difference between the direct and modified
Urca processes disappears, which may be called the
{\it broadening of the direct Urca threshold due to pion
polarization}.  This is the second type of broadening, after
the thermal effect discussed in Sect.\ \ref{sect-nucore-Durca}.
In Sect.\ \ref{sect-nusup-durmag} we will also consider
the {\it magnetic broadening}.
The cooling of neutron stars with the neutrino emissivity
intensified by the very high pion polarization
was simulated
by Schaab et al.\ (1997b).
We will not discuss these models and refer the reader to the above
references.\\

{\bf (d) Modified Urca process with other particles}

If muons are present in the dense matter, the muon
modified Urca processes are possible, similar to
processes (\ref{nucore-Murcan}) and
(\ref{nucore-Murcap}) with the electrons
replaced by muons. It is easy to verify that,
in the formalism of Friman and Maxwell (1979),
the emissivities of the neutron and
proton branches of the muon modified Urca
are given by the same Eqs.\
(\ref{nucore-Qn0}) and (\ref{nucore-rescale-Mur}) with the following
modifications. First, one should replace
$\Theta_{{\rm M}p} \to
\Theta_{{\rm M}p\mu}$ (muon threshold function) and
$p_{{\rm F}e} \to p_{{\rm F}\mu}$.
Second, one should include an additional factor
$v_{{\rm F}\mu}/c=(n_\mu /n_e)^{1/3}$
into both expressions for the emissivities,
where $v_{{\rm F}\mu}$ is the Fermi velocity of muons.
Strictly speaking, the analogous electron factor,
$c p_{{\rm F}e}/\epsilon_{{\rm F}e}
=v_{{\rm F}e}/c$, should have been present in Eqs.\
(\ref{nucore-Qn0}) and (\ref{nucore-rescale-Mur}),
but it was omitted because $v_{{\rm F}e} \approx c$
in the neutron star cores.
The muon neutron branch of the modified Urca process is
switched on at the densities above the threshold
density $\rho_\mu$ at which muons appear.  Since
the emissivity $Q^{({\rm M}n\mu)} \propto v_{{\rm F}\mu}$,
it vanishes at the threshold $\rho_\mu$ and
grows smoothly with increasing $\rho$. The emissivity
of the muonic proton branch contains the step function
$\Theta_{{\rm M}p\mu}$ and is also naturally restricted
by $\rho > \rho_\mu$. The step function allows the
process at
density high enough that
$p_{{\rm F}n}<3 p_{{\rm F}p}+p_{{\rm F}\mu}$.
However, it seems that
for many realistic equations of state, the step function opens the
process for all densities $\rho > \rho_\mu$,
and therefore plays no role.

The above statements are illustrated in Fig.\ \ref{fig-nucore-Q1(rho)}.
For our model of the $npe\mu$ matter,
both the muon neutron and muon proton modified Urca processes are
switched on and operate at
$\rho> \rho_\mu=2.5 \times 10^{14}$ g cm$^{-3}$.
They switch on smoothly, without any jump.
They are chiefly responsible for the increase of the neutrino emissivity
of the $npe\mu$ matter relative to the $npe$ matter
at densities
from $\rho_\mu$ to about $1.18 \times 10^{15}$ g cm$^{-3}$.
Comparing the solid and dashed curves in Fig.\ \ref{fig-nucore-Q1(rho)}
we may conclude that the presence of muons makes the matter
more ``neutrino luminous".

In the presence of hyperons,
the modified Urca reactions of the type
\begin{equation}
   B_1 + B_3 \to B_2 + B_3 + l + \bar{\nu},
   \quad    B_2 + B_3 + l  \to B_1 + B_3+ \nu,
\label{nucore-Murca-hyper}
\end{equation}
can operate, where $B_1$, $B_2$, $B_3$ are baryons,
and $l$ is either electron or muon [processes
(II.1)--(II.4) in Table \ref{tab-nucore-list}].
The hyperons may act either as beta-decaying particles or as
spectators. Processes (II.1) and (II.4)
conserve strangeness
and may be nearly as efficient as the main
modified Urca reaction (\ref{nucore-Murcan}).
They were analyzed by Maxwell (1987).
Notice that one should be careful in using his results
because of some oversimplifications
made in his analysis (see above).
Processes (II.2) and (II.3) change strangeness
and are expected to be about one or two orders
of magnitude less efficient ($Q \propto \sin^2 \theta_{\rm C}$).
To the best of our knowledge, their emissivities
have not yet been calculated.

Thus, the rigorous treatment of the modified Urca processes can be
a subject of future work.  In any case, however,
the modified Urca processes are negligible
if the direct Urca processes are allowed.
The modified Urca processes with hyperons are
certainly less efficient than those
with nucleons: the hyperons appear at rather high densities
at which the much more powerful
direct Urca processes may already operate.

\subsection{Non-equilibrium Urca processes}
\label{sect-nucore-noneq}

Let us consider now the Urca processes in the absence of
beta equilibrium.
Here, by beta equilibrium we mean
 the thermodynamic beta equilibrium discussed in
Sect.\ \ref{sect-nucrust-beta}. The inter-particle collisions, 
mediated by strong and electromagnetic interactions, nearly 
instantaneously establish
a local thermodynamic quasi-equilibrium at a given temperature $T$,
which still does not imply the full equilibrium.
For the sake of simplicity, let us study
the $npe$ matter. As we will see,
the relaxation time towards
beta equilibrium, $\tau_{\rm rel}$, may be quite large.
It is much longer than the typical timescale
$\sim 10^{-3}$~s,
corresponding to
the local compression of matter during the collapse of a
neutron star into the black hole (Gourgoulhon and Haensel 1993)
or to the radial pulsations of the neutron star.
In sufficiently old pulsars,
$\tau_{\rm rel}$ can also be much longer than the timescale
of the local compression of matter, implied by the slowing-down 
of pulsar rotation (Reisenegger 1995). 
In all these cases, one has to consider the Urca processes in 
neutron star matter out of beta equilibrium (Haensel 1992, 
Reisenegger 1995). 

In the absence of beta equilibrium,
there is a finite difference of the chemical
potentials, $\delta \mu \equiv \mu_n - \mu_p - \mu_e \neq 0$ 
(our definition of $\delta\mu$ agrees with that of Haensel 1992,
and differs in sign from that of Reisenegger 1995).
To be specific, we set $\delta \mu >0$, which means an excess
of neutrons and deficit of protons and electrons
with respect to the fully equilibrium values.
Then the direct and inverse reactions of the Urca processes
have different rates.
It is clear that
the beta decay rate $\Gamma_{n \to p}$ [cm$^{-3}$ s$^{-1}$]
will exceed the beta capture rate $\Gamma_{p \to n}$
bringing the matter towards beta equilibrium, in 
accordance with Le Ch{\^a}telier's principle.
We do not consider large departures from the equilibrium,
assuming $\delta \mu$ to be much smaller than the chemical
potentials of $n$, $p$, and $e$, but we allow $\delta \mu$
to be larger than $T$.
Along with the neutrino emissivity
$Q$, we will also analyze the net rate of changing of the
number density of neutrons, $\dot{n}_n=-\Delta \Gamma$, where
$\Delta \Gamma = \Gamma_{n \to p} - \Gamma_{p \to n}$.
Clearly, this rate 
$\Delta \Gamma \propto \delta \mu$
for small $\delta \mu$. Therefore,
the results
can be written in the form
\begin{equation}
      Q_{\rm noneq}  =  Q_{\rm eq} \,F(\xi), \quad
      \Delta \Gamma  =  b \, \xi \,H(\xi)  \, Q_{\rm eq} / T,
\label{nucore-noneq-Q}
\end{equation}
where $Q_{\rm eq}$ is the emissivity of the
direct Urca process (Sect.\ \ref{sect-nucore-Durca})
or any branch of the modified
Urca process (Sect.\ \ref{sect-nucore-Murca})
in beta equilibrium,
$Q_{\rm eq} \, \xi /T$ is a typical
value of $\Delta \Gamma$ at small $\delta \mu$,
$\xi=\delta \mu / T$ is the dimensionless measure
of the departure
from beta equilibrium, $b$ is a numerical coefficient
to be determined, and the functions
$F(\xi)$ and $H(\xi)$ describe the effects of suprathermal
departure from the equilibrium, $\delta \mu \gtrsim T$.

Let us start with the direct Urca process.
The expression for the neutrino emissivity is similar to
that given by Eqs.\ (\ref{nucore-QdDef}) and (\ref{nucore-decomp-Dur}),
but contains two different terms describing the direct and
inverse reactions. Each term is given by a 12-fold integral
which can be decomposed into the phase space integrals
as in Eq.\ (\ref{nucore-decomp-Dur}). The angular
integrals remain the same, as do the products of the
densities of state, but the energy integrals differ due to
the chemical potential shift in the
energy conserving delta functions. Comparing the
emissivity obtained in this way with
the equilibrium value, Eq.\ (\ref{nucore-Qdur0}),
we arrive at Eq.\ (\ref{nucore-noneq-Q}) with the
function $F(\xi)=F_{\rm D}(\xi)$ given by
\begin{equation}
      F_{\rm D }(\xi) =  { 2520 \over 457 \pi^6} \,
           \int_0^\infty {\rm d}x_\nu \; x_\nu^3 \,
           \left[J(x_\nu - \xi) + J(x_\nu + \xi) \right],
\label{nucore-noneq-Fdur}
\end{equation}
where $J(x)$ is defined by Eq.\ (\ref{nucore-J(x)}).
The integral is taken analytically (Reisenegger 1995):
\begin{equation}
      F_{\rm D}(\xi) =  1 + {1071 \, \xi^2 \over 457 \, \pi^2}
                   + {315 \, \xi^4 \over 457 \, \pi^4 }
                   + {21 \, \xi^6 \over 457 \, \pi^6}.
\label{nucore-noneq-Fdur1}
\end{equation}

The expressions for the partial reaction rates
$\Gamma_{n \to p}$ and $\Gamma_{p \to n}$
are given by the similar integrals
with the only difference --- there should be
one less power of neutrino energy under the integral.
This leads us to Eq.\ (\ref{nucore-noneq-Q}) with
$b_{\rm D}  =  714/( 457 \, \pi^2)= 0.158$,
\begin{eqnarray}
      H_{\rm D}(\xi) &  = & { 60 \over 17 \pi^4 \xi} \,
           \int_0^\infty {\rm d}x_\nu \; x_\nu^2 \,
           \left[J(x_\nu - \xi) - J(x_\nu + \xi) \right]
\nonumber \\
          & = & 1+ {10 \, \xi^2 \over 17 \, \pi^2}
                     + { \xi^4 \over 17 \pi^4}.
\label{nucore-noneq-Hdur}
\end{eqnarray}

The analysis of the modified Urca process is quite similar
(Reisenegger 1995).
We again come to Eq.\ (\ref{nucore-noneq-Q}),
where $b_{\rm M}  = 14680 /( 11513 \, \pi^2)= 0.129$,
\begin{eqnarray}
      F_{\rm M}(\xi) & = & 1 + {22020 \, \xi^2 \over 11513 \, \pi^2}
                   + {5670 \, \xi^4 \over 11513 \, \pi^4 }
                   + {420 \, \xi^6 \over 11513 \, \pi^6}
                   + {9 \, \xi^8 \over 11513 \, \pi^8},
\nonumber \\
      H_{\rm M }(\xi) &  = &
                  1+ {189 \, \xi^2 \over 367 \, \pi^2}
                     + { 21 \, \xi^4 \over 367 \pi^4}
                     + { 3 \, \xi^6 \over 1835 \pi^6}.
\label{nucore-noneq-FHmur}
\end{eqnarray}
These expressions are equally valid for the neutron and
proton branches of the modified Urca process.

Let us stress that Eqs.\ (\ref{nucore-noneq-Q})--(\ref{nucore-noneq-FHmur})
 are based on
the phase space decomposition. Thus they
are model-independent, insensitive to the details of
a specific strong interaction model employed to calculate
the matrix element of the process.

In order to illustrate these results, consider the process of 
beta-relaxation
in the $npe$  matter.  From the thermodynamical point of view, the
chemical potential excess
is a function of three variables,
$\delta \mu = \delta \mu (n_b,x_p,T)$.
The Urca reactions
do not change the baryon number density $n_b$.
Let us also fix the temperature $T$.  Then the
Urca processes would only affect the proton fraction $x_p=n_p/n_b$, 
driving it to the equilibrium
value. We have $\delta \dot{\mu} = (\partial \delta \mu
/ \partial x_p) \, \dot{x}_p = - \chi \, \Delta \Gamma$,
where $\chi=- (\partial \delta \mu/ \partial x_p) /n_b$.
Introducing $\dot{\xi}=\delta \dot{\mu} /T$, we arrive at
the equation
\begin{equation}
    \dot{\xi}= - \Gamma_{\rm rel} \, \xi, \quad
    \Gamma_{\rm rel}=\chi b \, Q_{\rm eq} \, H(\xi)/T^2,
\label{nucore-noneq-relax}
\end{equation}
which describes the {\it relaxation to beta equilibrium}.
Accordingly, the time $\tau_{\rm rel}=1/\Gamma_{\rm rel}$
can be called the beta equilibration time. Strictly speaking, 
$\tau_{\rm rel}$ is a standard (independent of $\xi$)
relaxation time only for $\xi\ll 1$.

\begin{figure}[!t]
\begin{center}
\leavevmode
\epsfysize=8.5cm
\epsfbox[60 50 305 305]{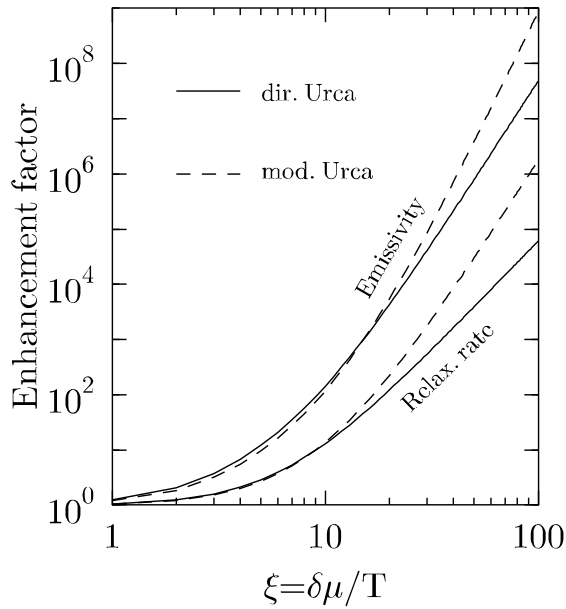}
\end{center}
\caption[]{\footnotesize
        Enhancement factors $F$ and $H$ of the neutrino
        emissivity and of the beta relaxation rate,
        respectively, for the direct
        (solid lines) and modified (dashes) Urca processes.
}
\label{fig-nucore-FH}
\end{figure}

Equations (\ref{nucore-noneq-Q}) and
(\ref{nucore-noneq-relax}) determine the neutrino emissivity
and the beta relaxation rate in the absence of beta equilibrium.
The results depend strongly on the departure from the
equilibrium, described by the parameter $\xi$
(Fig.\ \ref{fig-nucore-FH}).

For small $\xi \ll 1$, we have $F(\xi) \approx H(\xi) \approx 1$.
Accordingly, the neutrino emissivity is nearly
the same as in beta equilibrium.
Let us adopt the same moderate equation of state of the $npe$ matter
that will be used to illustrate the neutron star cooling
(Sect.\ \ref{sect-cool-code}); it opens the direct Urca process at
$\rho_{\rm crit}=1.30 \times 10^{15}$ g cm$^{-3}$.
Consider, for instance,
the density $\rho=1.38 \times 10^{15}$ g cm$^{-3}$
for which the direct Urca is allowed.
In this model the energy density of matter is
$n_b \epsilon_0(n_b,x_p)+3 \, \mu_e n_e/4$,
where $\epsilon_0(n_b,x_p)$ is the nucleon energy per baryon.
Furthermore, the nucleon energy is given by
the familiar expression $\epsilon_0(n_b,x_p)=
W(n_b)+4S_V(n_b)(x_p-1/2)^2$, where $W(n_b)$ 
is the energy per nucleon in the symmetric nuclear matter  
($x_p=1/2$) 
and $S_V(n_b)$ is the symmetry energy. Then
$\delta \mu=4 S_V(1-2 x_p)-\mu_e$,
$\chi =(8x_p S_V +\mu_e/3)/n_p$,
and the relaxation time is $\tau_{\rm rel}^{\rm (D)} \approx
20 \; T_9^{-4}$ s.
We see that in a not too hot neutron star, the
relaxation is very slow as a result of weakness of the
beta processes.  Yet, at densities below $\rho_{\rm crit}$
beta equilibrium
is reached via the modified Urca processes.
The corresponding relaxation time can be estimated as
$\tau_{\rm rel}^{\rm(M)} \sim T_9^{-6}$ months,
which is naturally much longer than the relaxation via the direct Urca
process.

A standard assumption made in the cooling simulations of
the non-rotating neutron stars is that the matter in the neutron star
core is in beta equilibrium. Using the thermal balance equation,
one can easily estimate the typical age $\tau_{\rm cool}$
of a neutron star with the given internal temperature $T$
at the neutrino cooling stage for the cases of
slow cooling via the modified Urca processes or
fast cooling via the direct Urca process.  These estimates show
that $\tau_{\rm cool}$
is of the same order of magnitude
as $\tau_{\rm rel}$
for both cases of the slow and fast cooling, with
$\tau_{\rm rel}$ being several times shorter.
Using this fact Reisenegger (1995) concluded that the beta relaxation
occurs much faster than the cooling. From our point of view,
the difference by a factor of several is not decisive
and the non-equilibrium state may
persist for some time, i.e, the value of $\xi$ may decrease slowly
as the cooling proceeds. 
The problem requires a
detailed numerical study since our consideration
of the beta relaxation is purely qualitative.
In reality, the relaxation in a cooling
neutron star is additionally affected by the temperature variation
neglected in Eq.\ (\ref{nucore-noneq-relax}), 
and the cooling is complicated by the ``chemical" heating discussed below.

Large departures from beta equilibrium,
$\xi \gg 1$, strongly enhance the neutrino emissivity
and accelerate the relaxation.  The enhancement of the emissivity
is described by the function $F(\xi)$, while the acceleration
of the relaxation is described by $H(\xi)$
(Fig.\ \ref{fig-nucore-FH}).
For instance, at $\xi=10$  the enhancement factors
in the direct and modified Urca processes are
$H_{\rm D}=13$, $F_{\rm D}=143$, $H_{\rm M}=14$, and $F_{\rm M}=117$,
while at $\xi=100$ they are
$H_{\rm D}=6.1 \times 10^4$, $F_{\rm D}=4.8 \times 10^7$,
$H_{\rm M}=1.8 \times 10^6$, and $F_{\rm M}=8.6 \times 10^8$.
It is important that at $\xi \gg 1$ the neutrino emissivity
and the relaxation time cease to depend on $T$, but instead
depend on $\delta \mu$ approximately in such a way as
if $T$ was replaced by $\mu$. This is natural since,
under the equilibrium condition, it is the temperature $T$
that determines the momentum space available to the reacting particles.
For large departures from the equilibrium, the allowed momentum
space is determined by $\delta \mu$ instead.
In these cases Urca processes become highly
asymmetric: the beta decay rate
strongly exceeds the beta capture rate for $\delta \mu>0$; the flux of the
generated antineutrinos is much larger than that of the neutrinos.
For $\delta \mu<0$, on the contrary, the beta capture rate would be
much higher, producing the much larger flux of neutrinos.
The neutrino spectra and neutrino opacities are
also modified (Haensel 1992).

It should be stressed that the non-equilibrium Urca processes, 
accompanying the relaxation towards beta equilibrium, produce also
the heating of the matter (Haensel 1992, Gourgoulhon and Haensel 1993,
Reisenegger 1995). The total rate of change of the thermal energy
per unit volume of the $npe$ matter is
    ${\dot {\cal E}_{\rm th}}=\Delta \Gamma \, \delta \mu - Q
     = Q_{\rm eq} \, [b \, \xi^2 \, H(\xi)-F(\xi)]$.
For $\xi \lesssim 4.5$, the deviation from beta equilibrium enhances
the energy loss of the matter, with the maximum enhancement factor 
of about 1.5 at
$\xi\simeq 3.5$.  However,
the chemical heating term $\Delta\Gamma \, \delta \mu$
prevails  for $\xi \gtrsim 5.5$,
leading to the net heating of the
matter. Such a situation is relevant for the neutron star collapse
(Gourgoulhon and Haensel 1993), and for the interplay of cooling and 
spin-down of older pulsars (Reisenegger 1995).  

Let us mention in passing that the beta relaxation
in hyperonic matter is more complicated.
Consider, for instance, the $npe\Lambda\Sigma^-$ matter.
Strong non-leptonic collisions
$n\Lambda \rightleftharpoons p \Sigma^-$ almost instantaneously
produce partial equilibration, in which $\mu_n+\mu_\Lambda=\mu_p+
\mu_{\Sigma}$.
In addition, there are non-leptonic collisions
of the type $p\Sigma^-\rightleftharpoons nn$.
They do not conserve strangeness and therefore involve
the weak interaction. Accordingly, they are much
slower than the strong collisions indicated above but still
several orders of magnitude faster than the beta processes
(Langer and Cameron 1969). 
%
%
They will produce the equilibration of the type
$\mu_p+\mu_\Sigma=2 \mu_n$, which still does not imply
the full equilibrium. After this equilibration is achieved,
the non-equilibrium state of the matter will be determined
by the quantity $\delta \mu=\mu_n-\mu_p-\mu_e$.
At this stage the final relaxation to the full
equilibrium will go via the beta processes. It can
be described by the formalism similar to that presented above.

\subsection{Neutrino bremsstrahlung in baryon-baryon collisions}
\label{sect-nucore-Brems}

In the absence of the direct Urca process,
the standard neutrino luminosity of the $npe$ matter
is determined not only by the modified Urca processes
but also
by the processes of
{\it neutrino bremsstrahlung radiation} in
nucleon--nucleon collisions
[processes (III.1)--(III.3) in Table \ref{tab-nucore-list}]:
\begin{equation}
   n+n   \rightarrow   n+n + \nu + \bar{\nu} \, , \quad
   n+p   \rightarrow   n+p + \nu + \bar{\nu} \, , \quad
   p+p   \rightarrow   p+p + \nu + \bar{\nu}.
\label{nucore-Brems}
\end{equation}
These reactions go via weak neutral currents and
produce neutrinos of any flavor;
neutrino pairs are emitted
in strong nucleon-nucleon collisions.
Each reaction is described by a set of Feynman diagrams
(Friman and Maxwell 1979).
In analogy with the modified Urca process, the emissivities
depend on the employed model of nucleon-nucleon interaction.
Contrary to the modified Urca, 
an elementary 
act of the nucleon-nucleon bremsstrahlung does not change
the composition of matter. The nucleon-nucleon
bremsstrahlung has evidently no thresholds associated with momentum
conservation and operates at any density in the uniform matter.
The $nn$ bremsstrahlung is also allowed for free neutrons
in the inner neutron star crust (Sect.\ \ref{sect-nucrust-nn}).

The neutrino emissivities of the bremsstrahlung processes
(\ref{nucore-Brems}) are given by the expressions similar
to Eq.\ (\ref{nucore-Q_mur}):
\begin{eqnarray}
    Q^{(N\!N)} & = &  \int
              \left[ \prod_{j=1}^4 { {\rm d} {\bf p}_j
              \over (2 \pi)^3} \right]
              { {\rm d}{\bf p}_\nu \over (2 \pi)^3} \;
              { {\rm d}{\bf p}'_\nu \over (2 \pi)^3} \;
              \omega_\nu \,
              (2 \pi)^4 \, \delta(E_f-E_i)
\nonumber \\
           & & \times \, \delta({\bf P}_f - {\bf P}_i ) \,
               f_1 f_2 (1-f_3)(1-f_4) \, s \, | M_{fi} |^2 ,
\label{nucore-Q_NN}
\end{eqnarray}
where $j$ from 1 to 4 labels the nucleons,
${\bf p}_\nu$ and ${\bf p}'_\nu$ are the momenta
of neutrino and antineutrino, $\omega_\nu=\epsilon_\nu
+ \epsilon'_\nu$ is the neutrino pair energy,
and $s$ is the symmetry factor introduced to 
avoid double counting the same collisions of
identical nucleons. One has $s=1$ for the $np$ bremsstrahlung
process, and $s=1/4$ for the $nn$ and $pp$ processes.
The squared matrix element summed over the particle
spins may be written as
%
\begin{equation}
      |M_{fi}|^2 = | \widetilde{M}_{fi}|^2/\omega_\nu^2.
\label{nucore-brems-Mfi}
\end{equation}
The neutrino-pair energy in the denominator comes
from the propagator of a virtual nucleon which enters
the matrix element (simplified in the approximation
of nonrelativistic nucleons). Therefore, it is
convenient to operate with the reduced matrix element
$\widetilde{M}_{fi}$. In analogy with the Urca
processes (Sects.\ \ref{sect-nucore-Durca} and \ref{sect-nucore-Murca})
it is sufficient to use the squared matrix element
averaged over the directions of the neutrino momenta.
In this case $| \widetilde{M}_{fi}|^2$ is totally independent
of the neutrino momenta as long as the latter momenta are
much smaller than the nucleon ones.
Then the integration over the neutrino momenta
is reduced to
\begin{equation}
   \int_0^\infty \epsilon_\nu^2 \; {\rm d}\epsilon_\nu
   \int_0^\infty \epsilon_\nu^{\prime 2} \; {\rm d}\epsilon'_\nu \, \ldots
   = {1 \over 30} \int_0^\infty \omega_\nu^5 \, {\rm d}\omega_\nu \, \ldots,
\label{nucore-brems-pair}
\end{equation}
and the angular-energy decomposition yields
\begin{eqnarray}
    Q^{(N\!N)} & = &  { (2 \pi)^4 \over (2 \pi)^{18} } 
              \, {s \over 30} \;
              A I \, \left\langle | \widetilde{M}_{fi} |^2 \right\rangle T^8
               \, \prod_{j=1}^4 m^\ast_j p_{{\rm F}j},
\label{nucore-Q_NN1}
\end{eqnarray}
where
\begin{eqnarray}
&&   A = (4 \pi)^2 \;  \int {\rm d} \Omega_1 \;
     {\rm d} \Omega_2 \; {\rm d} \Omega_3 \;
     {\rm d} \Omega_4 \;
     \delta \left( {\bf p}_1 + {\bf p}_2 - {\bf p}_3 - {\bf p}_4
     \right) \, ,
\label{nucore-ANN} \\
&&   \left\langle  | \widetilde{M}_{fi}|^2 \right\rangle
     = {(4 \pi)^2 \over A} \;  \int {\rm d} \Omega_1 \;
     {\rm d} \Omega_2 \; {\rm d} \Omega_3 \;
     {\rm d} \Omega_4 \;
     \delta \left( {\bf p}_1 + {\bf p}_2 - {\bf p}_3 - {\bf p}_4  \right) \,
      | \tilde{M}_{fi}|^2,
\label{nucore-MNNmean} \\
&&     I = \int_0^\infty {\rm d} x_\nu \; x_\nu^4
       \left[ \prod_{j=1}^4 \int_{-\infty}^{+\infty}
       {\rm d} x_j \; f_j \right]
       \delta \left( \sum_{j=1}^4 x_j-x_\nu \right) =
       {164 \, \pi^8 \over 945},
\label{nucore-bremas-I}
\end{eqnarray}
$\left\langle  | \widetilde{M}_{fi}|^2 \right\rangle$
being the reduced squared matrix element averaged
over orientations of the particle momenta.
As for the modified Urca process (Sect.\ \ref{sect-nucore-Murca}),
$x_j$ is the dimensionless
energy of the nucleon $j$, while $x_\nu=\omega_\nu/T$
is now the dimensionless
energy of the {\it neutrino pair}.
For the $nn$, $np$ and $pp$ processes the angular integral is
$A_{nn}=(4 \pi)^5/(2 p_{{\rm F}n}^3)$,
$A_{np}=(4 \pi)^5/(2 p_{{\rm F}n}^2 p_{{\rm F}p})$,
$A_{pp}=(4 \pi)^5/(2 p_{{\rm F}p}^3 )$, respectively.
Using the phase-space decomposition
we can again, as in Sects.\ \ref{sect-nucore-Durca}
and \ref{sect-nucore-Murca},
obtain the rescaling rules, independent of the strong interaction model:
\begin{equation}
    { Q^{(np)} \over Q^{(nn)}} = 4 \,
    { \langle  | \widetilde{M}_{fi}^{(np)} |^2 \rangle
    \over  \langle | \widetilde{M}_{fi}^{(nn)} |^2 \rangle } \,
    \left( m_p^\ast \over m_n^\ast \right)^2 \,
    {p_{{\rm F}p} \over p_{{\rm F}n}} , \quad
    { Q^{(pp)} \over Q^{(nn)}} =
    { \langle | \widetilde{M}_{fi}^{(pp)} |^2 \rangle
    \over  \langle | \widetilde{M}_{fi}^{(nn)} |^2 \rangle } \,
    \left( m_p^\ast \over m_n^\ast \right)^4 \,
   { p_{{\rm F}p} \over  p_{{\rm F}n}} .
\label{nucore-rescale-Brems}
\end{equation}

However, the matrix elements $| \widetilde{M}_{fi}^{(N\!N)} |^2$ do
depend on the strong interaction model.
In particular, in the OPE model
after averaging over orientations of the neutrino momenta
one has
[cf with Eqs. (\ref{nucore-nMurca-Mfi}) and (\ref{nucore-Murca-FM})]:
\begin{equation}
  | \widetilde{M}_{fi}^{(N\!N)}|^2  =   16 \, G_{\rm F}^2 \, g_A^2 \,
        \left( f^\pi \over m_\pi \right)^4 \, F_{N\!N},
\label{nucore-brems-reduceM}
\end{equation}
where
\begin{equation}
    F_{N\!N}=  {k^4 \over (k^2 + m_\pi^2)^2 }+
                {k_1^4 \over (k_1^2 + m_\pi^2)^2 }+
                {k^2 k_1^2 -3 ({\bf k} \cdot {\bf k}_1)^2
                \over (k^2 + m_\pi^2) (k_1^2 + m_\pi^2) },
\label{nucore-brems-Mnn}
\end{equation}
for the $nn$ and $pp$ processes, and
\begin{equation}
     F_{np} = {k^4 \over (k^2 + m_\pi^2)^2 }+
              {2 \, k_1^4 \over (k_1^2 + m_\pi^2)^2 } - 2 \,
                { k^2 k_1^2 - ({\bf k} \cdot {\bf k}_1)^2
                \over (k^2 + m_\pi^2) (k_1^2 + m_\pi^2) },
\label{nucore-brems-Mnp}
\end{equation}
with ${\bf k}={\bf p}_1 - {\bf p}_3$ and
      ${\bf k}_1={\bf p}_1 - {\bf p}_4$
(1 and 3 refer to the neutrons in the case of
the $np$ process).
One can easily prove that ${\bf k} \cdot {\bf k}_1=0$
for all three processes if the nucleons are strongly degenerate.
The quantity $ |M_{fi}^{(nn)}|^2$ was calculated
by Friman and Maxwell (1979) for the strongly degenerate matter,
without the $({\bf k} \cdot {\bf k}_1)^2$ term.
The latter term has been mentioned in a number
of papers (e.g., Hannestad and Raffelt 1998 and
references therein).
%
%
Notice that the third term in Eq.\
(\ref{nucore-brems-Mnn}) comes from interference of two 
different reaction amplitudes.
The quantity $ |M_{fi}^{(np)}|^2$ was also calculated
by Friman and Maxwell (1979) for the strongly degenerate matter,
without the $({\bf k} \cdot {\bf k}_1)^2$ term.
However, in Eq.\ (\ref{nucore-brems-Mnp})   [their Eq.\ (70)]
they missed the factor 2 in the third term.
This seems to be a misprint as can be deduced from their
subsequent Eq.\ (80). Their statement below Eq.\ (70)
that Eq.\ (\ref{nucore-brems-Mnp}) has to be supplemented
by additional
terms with $k$ and $k_1$ interchanged is in error.
Equation (\ref{nucore-brems-Mnp}) has also been presented
in several papers (e.g., Hannestad and Raffelt 1998)
including the $({\bf k} \cdot {\bf k}_1)^2$ term
but with the wrong sign of the third term ($+2$ instead of $-2$).

The OPE square matrix elements can be averaged analytically
over the orientations of the nucleon momenta in accordance with
Eq.\ (\ref{nucore-MNNmean}). This gives the quantity which
enters the neutrino emissivity:
\begin{equation}
  \left\langle | \widetilde{M}_{fi}^{(N\!N)}|^2 \right\rangle
         =   16 \, G_{\rm F}^2 \, g_A^2 \,
        \left( f^\pi \over m_\pi \right)^4 \, \langle F_{N\!N} \rangle,
\label{nucore-brems-averMfi}
\end{equation}
where
\begin{equation}
    \langle F_{N\!N} \rangle = 3 - {5 \over q} \;
     {\rm arctan} \, q + {1 \over 1 + q^2}
     + { 1 \over q \sqrt{2 + q^2}} \;
     {\rm arctan} \, \left( q \sqrt{2+ q^2} \right),
\label{nucore-brems-Fnn}
\end{equation}
for the $nn$ or $pp$ bremsstrahlung
(with $q=q_N=2 p_{{\rm F}\!N}/m_\pi$),
and
\begin{eqnarray}
    \langle F_{np} \rangle & = & 1 - 
     {3 \, {\rm arctan} \, q_p \over 2 q_p}
     + {1 \over 2 (1 + q_p^2) } +
     { 2 \, p_{{\rm F}n}^4 \over (p_{{\rm F}n}^2 + m_\pi^2)^2 }
\nonumber \\
     & & -  \left(1 - {{\rm arctan} \, q_p \over q_p} \right)
     { 2 \, p_{{\rm F}n}^2 \over p_{{\rm F}n}^2 + m_\pi^2 }.
\label{nucore-brems-Fnp}
\end{eqnarray}
Equation (\ref{nucore-brems-Fnn}) is exact, within the formulated
model; it was obtained by Friman and Maxwell (1979).
Equation (\ref{nucore-brems-Fnp}) is approximate, derived
assuming $p_{{\rm F}p} \ll p_{{\rm F}n}$. We have verified
that this is an excellent approximation for the densities
$\rho \lesssim 3 \rho_0$. Actually the factor $\langle F_{np} \rangle$
can be calculated exactly in a closed analytic form
but we do not present this cumbersome expression here.
Our numerical estimates show that $\langle F_{nn} \rangle$
and $\langle F_{np} \rangle$ are slow functions of the density.
We have $\langle F_{np} \rangle \approx 1$ for 
$\rho \gtrsim 2 \times 10^{14}$ g cm$^{-3}$,
and $\langle F_{nn} \rangle \approx 2$ in the range from $4 \times 10^{14}$
to about $10^{15}$ g cm$^{-3}$. This is a consequence of
the fact that $F_{nn}$ and $F_{np}$ are nearly constant 
in the indicated density ranges. Particularly, they
depend only slightly on the orientations of the
nucleon momenta. Thus, the approximation of constant
(angle independent) matrix elements may work quite satisfactorily
for the $nn$ and $np$ processes. On the other hand,
$\langle F_{pp} \rangle$ is rather small but
grows rapidly with the density at $\rho \sim \rho_0$.
The growth slows down only at 
the densities $\rho \gtrsim 6 \times 10^{14}$
g cm$^{-3}$ at which $\langle F_{pp} \rangle$
becomes $\approx 1$. The actual applicability of
the OPE model for the $nn$ and $np$ processes seems to be
restricted to $\rho \lesssim \rho_0$. However, the
$pp$ process can be described accurately by this model 
at higher $\rho$
due to smaller number density of the protons in the neutron star cores.

Now we can turn to evaluation of the neutrino bremsstrahlung
emissivities in the neutron star core. As in the case
of the modified Urca process (Sect.\ \ref{sect-nucore-Murca})
we will follow the prescriptions of Friman and Maxwell (1979).
For the $nn$ processes, they neglected the exchange
term in the squared matrix element (\ref{nucore-brems-Mnn}),
averaged the result over orientations of nucleon momenta,
and set $\rho=\rho_0$, using some specific equation of
state of dense matter. They inserted this density independent
averaged squared matrix element into the expression for
$Q^{(nn)}$ and introduced rather arbitrarily
a correction factor $\beta_{nn}$ to account for numerous
effects omitted in their analysis
(correlations, repulsive part of the 
nucleon-nucleon interaction, etc.).
They adopted the same procedure for the $np$ process,
neglecting the third (interference) term in the
squared matrix element (\ref{nucore-brems-Mnp})
but they did not consider the $pp$ process.
The emissivity of the latter process was estimated
in the same manner by Yakovlev and Levenfish (1995).
The resulting expressions for the emissivities
are
(Friman and Maxwell 1979;
Yakovlev and Levenfish 1995):
\begin{eqnarray}
   Q^{\,(nn)} & = & {41 \over 14175} \,
             {G_{\rm F}^2 g_A^2 m_n^{\ast4} \over
             2 \pi \hbar^{10} c^8}
             \left( {f^\pi \over m_\pi} \right)^4
             p_{{\rm F}n} \alpha_{nn} \beta_{nn}
             (k_{\rm B} T)^8 {\cal N}_{\nu}
\nonumber \\
         & \approx & 7.5 \times 10^{19}
             \left( {m_n^\ast \over m_n} \right)^4
             \left({ n_n \over n_0} \right)^{1/3}
             \alpha_{nn} \beta_{nn} \, {\cal N}_{\nu} T_9^8
             \; \;  {\rm erg\;cm^{-3}\;s^{-1}},
\label{nucore-Qnn0} \\
   Q^{\,(np)} & = & {82 \over 14175} \,
              { G_{\rm F}^2 g_A^2 m_n^{\ast2}
             m_p^{\ast2} \over
             2 \pi \hbar^{10} c^8} \left( {f^\pi \over m_\pi} \right)^4
             p_{{\rm F}p} \alpha_{np} \beta_{np}
             (k_{\rm B} T)^8 {\cal N}_{\nu}
             \rule{0em}{5.5ex}
\nonumber \\
         & \approx & 1.5 \times 10^{20} \left(
             {m_n^\ast \over m_n}\, {m_p^\ast \over m_p} \right)^2
             \left({ n_p \over n_0} \right)^{1/3}
             \alpha_{np} \beta_{np} \, {\cal N}_{\nu} T_9^8
             \; \; {\rm erg\;cm^{-3}\;s^{-1}},
\label{nucore-Qnp0}   \\
   Q^{\,(pp)} & = & {41 \over 14175} \,
             { G_{\rm F}^2 g_A^2 m_p^{\ast4} \over
             2 \pi \hbar^{10} c^8} \left( {f^\pi \over m_\pi} \right)^4
             p_{{\rm F}p} \alpha_{pp} \beta_{pp}
             (k_{\rm B} T)^8 {\cal N}_{\nu}
             \rule{0em}{5.5ex}
\nonumber \\
         & \approx & 7.5 \times 10^{19}
             \left( {m_p^\ast \over m_p} \right)^4
             \left({ n_p \over n_0} \right)^{1/3}
             \alpha_{pp} \beta_{pp} \, {\cal N}_{\nu} T_9^8
             \; \; {\rm erg\;cm^{-3}\;s^{-1}},
\label{nucore-Qpp0}
\end{eqnarray}
where $m_{\pi}$ is the $\pi^0$ mass and
${\cal N}_{\nu}$ is the number of neutrino flavors.
The dimensionless factors $\alpha_{N\!N}$ come from the estimates
of the square matrix elements at $\rho=\rho_0$:
$\alpha_{nn} = 0.59$, $\alpha_{np} =1.06$, $\alpha_{pp}= 0.11$.
The correction factors $\beta_{N \!N}$ were taken as
$\beta_{nn} = 0.56$, $\beta_{np} = 0.66$
and $\beta_{pp} \approx 0.7$.
Hereafter we will use ${\cal N}_{\nu}=3$
in Eqs.\ (\ref{nucore-Qnn0})--(\ref{nucore-Qpp0}), whereas
Friman and Maxwell (1979)
took into account only two neutrino flavors.
All three bremsstrahlung processes are
of comparable intensity, with $Q^{(np)} < Q^{(np)} < Q^{(nn)}$,
as seen in Figs.\ \ref{fig-nucore-Q(T)} and \ref{fig-nucore-Q(rho)}.

The overall structure of the expressions for
the emissivities
$Q^{(N \!N)}$ of the bremsstrahlung processes is similar to
the structure of the expressions
for the modified Urca processes, $Q^{({\rm M}\!N)}$.
In particular, the temperature dependence of the
emissivity is the  same, $Q^{(N \!N)} \propto T^8$,
as can be explained from the momentum space
consideration. Indeed, any bremsstrahlung process involves
four degenerate fermions instead of five
in the modified Urca (minus one $T$)
but one additional neutrino (an extra factor $T^3$);
moreover, the squared matrix element (\ref{nucore-brems-Mfi})
is inversely proportional to $\omega_\nu^2$ which removes $T^2$
keeping the same factor $T^8$ as for the modified Urca.
However, the expressions for $Q^{(N \!N)}$
contain smaller numerical coefficients.
As a result,
in a non-superfluid matter, the neutrino bremsstrahlung is about two
orders of magnitude less efficient than the modified
Urca process. This is seen in Figs.\ \ref{fig-nucore-Q(T)}
and \ref{fig-nucore-Q(rho)}.
We will show (Chapt.\ \ref{chapt-nusup}) that the bremsstrahlung can
be more important in the presence of nucleon superfluidity.

The presence of muons does not have any direct influence
on the nucleon-nucleon bremsstrahlung. In hyperonic matter
a variety of baryon-baryon bremsstrahlung reactions
is possible [reactions (III) in Table \ref{tab-nucore-list}]:
\begin{equation}
     B_1 + B_2 \to B_3 + B_4 + \nu + \bar{\nu}.
\label{nucore-baryon-brems}
\end{equation}
Here $B_1$, $B_2$, $B_3$ and $B_4$ stand for the baryons
which can experience the non-leptonic collisions.
In the majority of these reactions
the baryons do not undergo transmutation,
  $B_3=B_1$ and $B_4=B_2$
[see reactions (III.1)--(III.10) in Table \ref{tab-nucore-list}],
while reactions (III.11) and (III.12) are accompanied
by the changes of baryons species.
Notice that the two latter reactions do not control 
the baryon composition of matter, since the
equilibrium with respect to the transformation
$n \Lambda \rightleftharpoons p \Sigma^-$
is governed by the strong non-leptonic collisions on 
microscopic timescales (Sect.\ \ref{sect-nucore-noneq}). 
Reactions (III) were
studied by Maxwell (1987), but one should use his
results with caution (as discussed in Sect. \ref{sect-nucore-Murca}).
The reactions with hyperons
are somewhat less efficient than the nucleon
reactions (\ref{nucore-Brems}). The emissivities of all 
reactions are model-dependent, and it would be interesting to reconsider them
using the modern
models of strong interaction.

Notice that the neutrino bremsstrahlung can be
greatly enhanced near the threshold of appearance of
the reacting hyperons. The effect is especially important for
the processes involving hyperons of the same species,
e.g., $\Sigma^- \Sigma^- \to \Sigma^- \Sigma^- \nu \bar{\nu}$.
This is because
in the near vicinity of the threshold
the number density of the hyperons is so small that 
the particles are yet non-degenerate. This removes the stringent 
suppression of the emissivity associated with the degeneracy
(Sect.\ \ref{sect-nucore-Durca}). However,
such an enhancement
occurs only in the very thin layers of neutron star
cores and seems to have no effect on the total neutrino luminosity.

\subsection{Other neutrino reactions}
\label{sect-nucore-other}

We have considered the main neutrino reactions
in neutron star cores with the standard (non-exotic)
composition of matter. All of them involve baryons.
In Chapt.\ \ref{chapt-nusup} we will see that these reactions
can be strongly suppressed by possible superfluidity
of baryons. Therefore, it is reasonable to mention some
other, weaker neutrino reactions; they can be negligible in the
 non-superfluid matter,
but they can dominate if the main reactions are
suppressed by superfluidity.\\

{\bf (a) Neutrino-pair bremsstrahlung in Coulomb collisions}

These reactions [processes (V) in Table \ref{tab-nucore-list}]
can be schematically written as
\begin{equation}
   l + C   \to   l + C + \nu + \bar{\nu},
\label{nucore-other-brems}
\end{equation}
where $l$ is a lepton ($e$ or $\mu$), $C$ is any charged fermion
(lepton or baryon), and $\nu\bar{\nu}$
denote the neutrino pair of any flavor. These processes
are similar to the neutrino bremsstrahlung in
baryon collisions (Sect.\ \ref{sect-nucore-Brems}) but
involve Coulomb collisions instead of the strong baryon-baryon
collisions. They are ``well defined" processes in a sense that
they do not depend directly on the model of strong interaction.

First consider the case in which $C$ is a baryon.
The simplest reaction of this type,
$e+p \to e+p+\nu+\bar{\nu}$
[reaction (V.1) with $l=e$],
has been studied
by Kaminker and Haensel (1999). The emissivity is
\begin{eqnarray}
      Q^{(ep)} & = & {41 \, \pi^4 G_{\rm F}^2 e^4 C^2_{+} m_p^{\ast 2}
               \over 11340  \, \hbar^9 c^8 y_s p_{{\rm F}e}^4}  \,
               n_p \,(k_{\rm B}T)^8
\nonumber \\
           & \approx &
           {3.7 \times 10^{14} \over y_s} \,
           \left( m_p^\ast \over m_p \right)^2
           \left( n_0 \over n_p \right)^{1/3} T_9^8 \; \;
           {\rm erg\;cm^{-3}\;s^{-1}},
\label{nucore-other-ep}
\end{eqnarray}
where $C^2_{+}=\sum_\nu (C_{V}^2+C_{A}^2) \approx 1.678$ is the sum
of the squared constants of the 
vector and axial-vector lepton weak interaction, 
$y_s=q_s/(2 p_{{\rm F}e})$, and $q_s$ is the screening momentum
($\hbar / q_s$ being the screening length of the Coulomb interaction).
Generally, one has
\begin{equation}
      y_s^2 = { e^2 \over \pi \hbar c} \,
      \sum_j {m_j^\ast p_{{\rm F}j}
      \over m_e^\ast p_{{\rm F}e}},
\label{nucore-other-ys}
\end{equation}
where the sum extends over all charged fermions $j$
of spin $1/2$
in the stellar matter, including the electrons.  Each term in the sum
describes plasma screening by particle species $j$
(e.g., $j=e$ and $p$, for the $npe$ matter).
We see that the temperature dependence of the emissivity,
$Q^{(ep)} \propto T^8$, is the same as in the bremsstrahlung
due to nucleon-nucleon collisions. This is natural
since the temperature dependence arises from the phase space
restrictions determined by the number of reacting degenerate
fermions.

The above process
is similar to the neutrino
bremsstrahlung due to the electron scattering off atomic nuclei
in the neutron star crust (Sect.\ \ref{sect-nucrust-ebrems}).
In both cases,
neutrinos are emitted in the Coulomb collisions of electrons with
heavy charged particles.
A proton in the stellar core plays the same
role as an atomic nucleus at not too low temperatures
in the molten crust, 
with the two differences. First,
the recoil energy of the nucleus in a scattering event
in the crust is negligible while the recoil energy of the proton
in the stellar core is important.
Second, the nuclei in the crust form a strongly coupled Coulomb plasma while
the protons in the core constitute a degenerate Fermi liquid.
The strong link between
the two processes allowed Kaminker et al.\ (1997) and Kaminker
and Haensel (1999) to propose a simple similarity
criterion for the emissivities,
\begin{equation}
     Q^{(ep)} / Q^{(eZ)} \sim  \nu_{ep} / \nu_{eZ},
\label{nucore-other-similar}
\end{equation}
where $\nu_{ep}$ and $\nu_{eZ}$ are
the effective electron collision
frequencies in the core and the crust, respectively.
These are
the same frequencies
that determine, for instance,
the electric conductivity. This and similar
expressions enable one to estimate the emissivities of different
neutrino emission processes without complicated
calculations.  In the $npe\mu$ matter with 
ultrarelativistic 
muons
one evidently has $Q^{(\mu p)}=Q^{(ep)}$. Notice that
Kaminker et al.\ (1997) proposed the inaccurate
formulation of the similarity criterion
(using inadequate expressions for $\nu_{ep}$);
this inaccuracy was fixed
by Kaminker and Haensel (1999).

According to Eq.\ (\ref{nucore-other-ep})
and Figs.\ \ref{fig-nucore-Q(T)} and \ref{fig-nucore-Q(rho)},
the neutrino emissivity
of the $ep$ bremsstrahlung is really weak, several orders
of magnitude smaller than the emissivity of the
nucleon-nucleon bremsstrahlung processes, 
a natural consequence of
the weakness of the Coulomb collisions relative
to the nuclear ones.

In analogy with Eq.\ (\ref{nucore-other-ep}) one can easily
estimate the emissivity $Q^{(lC)}$ of any bremsstrahlung process
(V.1) or (V.2), where $C$ is a charged
baryon. Being very weak by themselves, these processes are
additionally suppressed by the superfluidity of baryons. As a result, they
seem to be of no importance for the neutron star cooling.

Now let us turn to processes (V.3), in which
both colliding particles are leptons.
Namely, we have three processes of this type,
associated with the $ee$, $e\mu$ and $\mu\mu$ collisions.
They have been analyzed in the same paper by
Kaminker and Haensel (1999). For instance, the emissivity
of the most important of them, the $ee$ bremsstrahlung process is
\begin{eqnarray}
      Q^{(ee)}& = & {41 \, \pi^4 G_{\rm F}^2 e^4 C^2_{+}
               \over 7560 \,\hbar^9 c^{10} y_s p_{{\rm F}e}^2} \,
	       n_e\,(k_{\rm B}T)^8
\nonumber \\
           & \approx &
	   {0.69 \times 10^{14} \over y_s} \,
	   \left( n_e \over n_0 \right)^{1/3} T_9^8 \; \;
	   {\rm erg\;cm^{-3}\;s^{-1}}.
\label{nucore-other-ee}
\end{eqnarray}
The emissivities of the $e\mu$ and $\mu\mu$ processes are
comparable or smaller.
Expressions for $Q^{(e\mu)}$ and
$Q^{(\mu\mu)}$ are obtained by Kaminker and Haensel (1999) for some cases,
while the other cases can be considered using the
similarity criteria either 
analogous to Eq.\ (\ref{nucore-other-similar})
or based on the phase-space decomposition.
In the limit of ultrarelativistic muons one has
$Q^{(\mu\mu)}=Q^{(ee)}= Q^{(e\mu)}/4$,
cf Eq.\ (\ref{nucore-rescale-Brems}).
These processes are even weaker than
the $ep$ bremsstrahlung (as seen from Figs.\ \ref{fig-nucore-Q(T)} and
\ref{fig-nucore-Q(rho)}),
but their great advantage is
being almost unaffected by the baryon superfluidity.
Therefore, these processes provide the
{\it ``residual'' neutrino emissivity of the highly superfluid
neutron star cores}, as we discuss in Chapt.\ \ref{chapt-nusup}.\\

{\bf (b) Lepton modified Urca}

The lepton direct Urca process in the $npe\mu$ matter,
associated
with muon decay and creation
($\mu \to e + \nu_\mu + \bar{\nu}_e$,
$e \to \mu + \bar{\nu}_\mu + \nu_e$), is clearly forbidden
by momentum conservation due to the same
arguments as in Sect.\ \ref{sect-nucore-Murca}.
However, the lepton modified Urca processes
[processes (IV) from Table \ref{tab-nucore-list}]
in the presence of an additional charged fermion $C$
are allowed:
\begin{equation}
    \mu + C  \to e + C + \nu_\mu + \bar{\nu}_e, \quad
    e + C \to \mu + C + \nu_e + \bar{\mu}_\mu.
\label{nucore-other-muMur}
\end{equation}
They are similar to the modified Urca processes discussed in
Sect.\ \ref{sect-nucore-Murca}, but involve purely lepton
beta reactions. Their emissivity
is also proportional to $T^8$.

The lepton modified Urca processes have not been
studied so far, to our knowledge,
although they have been mentioned in the literature
(e.g., Bahcall and Wolf 1965b).
The muon beta decay can be as efficient
as the baryon one. However, processes (IV.1) and (IV.2)
involving baryons
are weaker than the baryon modified Urca processes
discussed in Sect.\ \ref{sect-nucore-Murca} since
they involve
the electromagnetic 
interaction. In addition, such reactions
are suppressed by baryon superfluidity.

Therefore, the only lepton modified Urca processes that may be
important are those which involve leptons alone
[processes (IV.3) and (IV.4)].
Our order-of-magnitude estimates show that the emissivity
of these processes cannot be much larger than the emissivity
of the lepton-lepton bremsstrahlung discussed above. Like the latter,
the purely lepton
modified Urca processes are almost independent of baryon
superfluidity and contribute to the residual neutrino emissivity
of the highly superfluid neutron star cores.\\

{\bf (c) Other reactions}

One can imagine a number of other neutrino
processes in the neutron star core with the non-exotic
composition of matter. For instance, we can mention
the plasmon decay process which is efficient in the
neutron star crust (Sect.\ \ref{sect-nucrust-plasmon}).
However, the electron plasmon energy in the
core is much larger than
the thermal energy ($\hbar \omega_p \sim 10$ MeV), and
the process is exponentially suppressed.
Another process, the electron-positron pair
annihilation, which can also be efficient at lower
densities (Sect.\ \ref{sect-nucrust-annih}),
can operate in the stellar cores
but it is again negligible due to the exponentially small
number of positrons in a strongly degenerate relativistic
electron gas.

\subsection{Neutrino emission from exotic phases of dense matter}
\label{sect-nucore-exotica}

Although this review concentrates on the processes in the standard
neutron star matter,
our analysis would
be incomplete without a brief discussion of the neutrino
emission from the exotic matter.
The main neutrino reactions in the exotic matter are thought
to be similar to the direct Urca processes (Sect.\ \ref{sect-nucore-Durca}).
Our discussion will follow closely the
comprehensive review by Pethick (1992).
The main results of this section are summarized in Table
\ref{tab-nucore-exotica}.\\
%
%

{\bf (a) Pion condensed matter}

The hypothetical pion condensate represents a macroscopic
condensed pion field, whose excitation is energetically
favorable at densities above several $\rho_0$
in some models of dense matter with 
the high polarizability of 
pion degrees of freedom.
Pion condensation does not lead to the creation of free
pions, but it is accompanied by
the appearance of the coherent field excitations
(waves) with the same quantum numbers
(spin, isospin) as the pions.
First of all, these are $\pi^-$ mesons, although
a mixed $\pi^- \, \pi^0$ condensate also may appear
at densities above the threshold of $\pi^-$ condensation.
The strength of the pion field is characterized by
the so called condensate angle $\theta_\pi$
(the condensate vanishes at $\theta_\pi=0$).
Field oscillations
are specified by the oscillation frequency $\mu_\pi$ and the
wavevector ${\bf k}_\pi$.
The pion field is thought to be essentially
non-stationary and non-uniform ($\mu_\pi \neq 0$, $k_\pi \neq 0$).
The condensate parameters $\theta_\pi$,
$\mu_\pi$ and $k_\pi$ depend on the density and
are determined by the microscopic model of stellar matter.
Typically, $\theta_\pi^2 \sim 0.1$, and $\mu_\pi \sim k_\pi \sim m_\pi$.

The pion field strongly mixes the neutron and
proton states. The proper treatment of these states
requires the quasiparticle formalism. Instead
of free neutrons and protons, it is more adequate
to introduce two types of quasinucleons, $\tilde{n}$ and
$\tilde{p}$, which are the coherent
superpositions of the neutron and proton states
(plus various virtual excitations).
Because of the strong mixing, both quasiparticles may
have the (quasi)momenta close to the Fermi momenta
of free neutrons. Therefore, the direct Urca process
involving quasinucleons becomes open:
\begin{equation}
   \tilde{n} \to \tilde{p} + e + \bar{\nu}_e, \quad
   \tilde{p} +e \to \tilde{n} + \nu_e.
\label{nucore-exotica-pions}
\end{equation}

The neutrino emissivity of this process
in the simplest model of the pion condensed field and
quasiparticle states 
was calculated by Maxwell et al.\ (1977).
The calculation is similar to that for the ordinary
nucleon direct Urca process (Sect.\ \ref{sect-nucore-Durca}).
The interaction Hamiltonian has the same structure as
(\ref{nucore-H}).  Pion condensate modifies the
nucleon weak current (\ref{nucore-hadron}). In particular,
the current acquires a plane-wave factor $\exp(-i \mu_\pi t
+ i {\bf k}_\pi \cdot {\bf r})$, which introduces
an additional energy shift $\mu_\pi$
in the energy conserving delta function and
the additional momentum
${\bf k}_\pi$ in the momentum conserving delta
function in the differential transition probability,
Eq.\ (\ref{nucore-Golden-Rule1}). Moreover, pion condensate
changes the nucleon wave functions (\ref{nucore-psi-hadron}).
The neutrino emissivity obtained by
Maxwell et al.\ (1977) can be written as
\begin{equation}
   Q^{(\pi)}  =  Q^{\rm(D)} \, \left( m_n^\ast \over m_p^\ast \right) \,
          \left( \mu_\pi^2 \over k_\pi m_e^\ast \right) \,
          { \theta_\pi^2 \over 16} \;
          \left[ 1 + \left( g_A k_\pi \over \mu_\pi \right)^2 \right],
\label{nucore-exotica-Qpi}
\end{equation}
where $Q^{\rm(D)}$ is the emissivity of the ordinary
nucleon direct Urca process given formally
by Eq.\ (\ref{nucore-Qdur0}) without the step function.
The above expression is valid for $\theta_\pi \lesssim 1$.
The 
process (\ref{nucore-exotica-pions}) 
is allowed for any proton fraction
and the neutrino emissivity depends on temperature as
$Q^{(\pi)} \propto T^6$. Equation (\ref{nucore-exotica-Qpi})
may be treated as
a rescaling formula, analogous to Eq.\ (\ref{nucore-rescale-Dur}), 
provided by the space-phase decomposition.

In the presence of muons, the muon
version of process (\ref{nucore-exotica-pions}) is also
open. It doubles the neutrino emissivities $Q^{\rm(D)}$ and
$Q^{(\pi)}$, as explained in Sect.\ \ref{sect-nucore-Durca}.

For the typical pion condensate parameters, Maxwell et al.\
(1977) obtained $Q^{(\pi)} \lesssim 0.1 \, Q^{\rm (D)}$.
This result is in a very good agreement with a crude
estimate 
of the neutrino emissivity
by Bahcall and Wolf (1965b)
in a simplified model
of 
pion-condensed matter
containing free pions.

The neutrino emissivity in the $\pi^-$ condensed
matter was also considered by Tatsumi (1983)
%
%
using a more complex model of the condensate
and quasinucleon states.
He found the emissivity to be about an order
of magnitude lower than calculated by Maxwell et al.\
(1977). The main discrepancy
was in Tatsumi's use of the pion condensed model with the
smaller values of $k_\pi$ and $\mu_\pi$ which reduced the allowed
momentum space of the reacting particles and
the reaction rate. Moreover, Tatsumi (1983) used the value of the
axial-vector constant $\tilde{g}_A$ strongly reduced
by the baryon-baryon short-range correlations
which also lowered the numerical value of $Q^{(\pi)}$.
In subsequent papers
Tatsumi (1987)
%
%
and Muto and Tatsumi (1988)
%
%
analyzed the neutrino emission under the combined effect
of $\pi^-\pi^0$ condensates.  The inclusion of the
$\pi^0$ condensate further reduces the emissivity
by an order of magnitude.
Let us stress that these
reductions of $Q^{(\pi)}$
are model-dependent and, therefore, uncertain.
Nevertheless, even after these corrections
the presence of pion condensates
enhances the neutrino emission over that given by
the modified Urca processes.\\

{\bf (b) Kaon condensed matter}

Like the pion condensate, the kaon condensate
manifests itself as a macroscopic field excited
in the matter of rather high density. The excitations
are characterized by the same quantum
numbers as $K^-$ mesons, which are strange particles,
and thus form the ``{\it strangeness condensation}".
Unlike the pion condensate, the kaon condensate is likely to be
stationary and uniform, $\mu_K=k_K=0$.
Its strength is determined by the condensate
angle $\theta_K$, analogous to $\theta_\pi$.
The condensate strongly affects the nucleon states
transforming them into quasiparticle states, which are
the coherent superpositions of the states of nucleons
and the hyperon-like excitations.
These quasiparticles have the (quasi)momenta about
$p_{{\rm F}n}$. The main neutrino reaction
is again the direct Urca process of the type
(\ref{nucore-exotica-pions}). Its formal calculation
is even simpler than for the pion condensed matter
because of the stationarity and uniformity of the kaon field.
For $\theta_K \lesssim 1$, Brown et al.\ (1988) find
%
%
\begin{equation}
   Q^{(K)}  =  Q^{\rm(D)} \, \left( m_n^\ast \over m_p^\ast \right) \,
               { \theta_K^2 \over 16} \; \tan^2 \theta_{\rm C},
\label{nucore-exotica-QK}
\end{equation}
where $Q^{\rm(D)}$ refers to the emissivity of the
ordinary nucleon direct Urca process. Like
in the pion condensed matter, this process
is open for any proton fraction, and the presence
of muons doubles $Q^{(K)}$. The emissivity
$Q^{(K)}$ is proportional to the small
factor $\sin^2 \theta_{\rm C}$
since each elementary reaction of the process changes strangeness.
For $\theta_K^2 \sim 0.1$, the emissivity $Q^{(K)}$
is about three orders of magnitude lower than
$Q^{\rm(D)}$ but still remains much higher than that of the modified
Urca process. Thus, kaon condensation increases
the neutrino emission over its standard level but
not as strongly as the nucleon direct Urca process.\\

{\bf (c) Quark matter}

The  quark
matter consists of deconfined degenerate $u$, $d$, $s$ quarks
with a small admixture of relativistic degenerate electrons
(e.g., Weber 1999).
Typical Fermi energies of quarks are expected to be
$\sim 300 $ MeV and higher. Accordingly, the $u$ and $d$
quarks can be treated as massless particles, while
the $s$ quark is moderately relativistic and, generally,
its mass cannot be neglected. The beta equilibrium conditions
imply
\begin{equation}
        \mu_d=\mu_s=\mu_u+\mu_e.
\label{nucore-quarks-beta}
\end{equation}

Neutrino emission from the quark matter was considered
by Iwamoto (1980, 1982).
%
%
%
%
The most important process is the direct Urca involving
$u$ and $d$ quarks:
\begin{equation}
    d \to u + e + \bar{\nu}_e, \quad u + e \to d + \nu_e.
\label{nucore-quarks-d}
\end{equation}
It is essentially the ordinary direct Urca 
process, Eq.\ (\ref{nucore-Durca}), 
considered on the quark level.
The derivation of its 
emissivity, $Q^{({\rm D}d)}$, is analogous to
the ordinary nucleon direct Urca (Sects.\ \ref{sect-nucore-beta} and
\ref{sect-nucore-Durca}). The major difference comes from
the fact that the $u$ and $d$ quarks are relativistic particles.
If we treated them as free, non-interacting particles,
Eq.\ (\ref{nucore-quarks-beta}) would immediately yield
$p_{{\rm F}d}= p_{{\rm F}u}+p_{{\rm F}e}$, which is
at the edge of the triangle condition imposed by momentum
conservation. This would imply that the reaction is almost
forbidden: beta decay of the $d$ quark would
be allowed only if $u$ and $e$ were emitted
in the forward direction. This would greatly
reduce the kinematically allowed momentum space and
the neutrino emissivity.

In reality, the deconfined $u$ and $d$ quarks are not entirely free
but weakly coupled. In this regime, their chemical potentials
are given by
$
    \mu_u / (c p_{{\rm F}u})= \mu_d / (c p_{{\rm F}d}) =
    1 + 8 \alpha_c /( 3 \pi ), 
$
where $\alpha_c=g^2/(16 \pi) \sim 0.1$ is the QCD coupling
constant ($g$ being the quark-gluon coupling constant),
a parameter of the microscopic model of strange quark matter.
The coupling of quarks allows the
$d$ quark Urca process to operate in a small but
finite momentum space determined by the specific value of $\alpha_c$.
The calculation of the emissivity is straightforward
and yields (Iwamoto 1982)
\begin{equation}
   Q^{({\rm D}d)} = {914 \, G_{\rm F}^2 \cos^2 \theta_{\rm C}
             \over 315\, \hbar^{10} c^6 }\;
             \alpha_c \, p_{{\rm F}d} \, p_{{\rm F}u} \, p_{{\rm F}e} \,
             (k_{\rm B} T)^6.
\label{nucore-quarks-Q}
\end{equation}
Comparing this with the emissivity of the nucleon direct
Urca process, Eq.\ (\ref{nucore-Qdur0}), we see that
the numerical coefficient is somewhat larger (phase space integration
is different) but, generally, the emissivity is
reduced by a factor of several by
the presence of the small constant $\alpha_c$ and the
factors $c^2 p_{{\rm F}d}\, p_{{\rm F}u}$ which are smaller
than the analogous quantities $m_n^\ast m_p^\ast$ in
Eq.\ (\ref{nucore-Qdur0}).  Still, despite this reduction
the emissivity $Q^{({\rm D}d)}$ is much larger than that
of the modified Urca processes in the $npe$ matter.

In addition to the $d$ quark direct Urca (\ref{nucore-quarks-d}),
another direct Urca process is also allowed,
\begin{equation}
    s \to u + e + \bar{\nu}_e, \quad u + e \to s + \nu_e.
\label{nucore-quarks-s}
\end{equation}
associated with the strange $s$ quark. However,
the beta process involving the $s$ quark
changes strangeness, i.e., the emissivity
is proportional to the small factor $\sin^2 \theta_{\rm C}$.
The calculations are more complicated because
the $s$ quark cannot be considered as massless [see
Eq.\ (4.11) in Iwamoto 1982], and give the smaller emissivity 
than in the main quark process (\ref{nucore-quarks-d}).

Aside from the direct Urca processes, there are also the modified
Urca and neutrino
bremsstrahlung processes due to the quark-quark collisions
(Iwamoto 1982), which are naturally
much weaker than the direct Urca process. Notice that in some
models of quark matter the number density of electrons
is so small that the direct Urca processes are forbidden
by momentum conservation. In these cases, the modified quark
Urca processes play the leading role.
Also, Iwamoto (1982) estimated the neutrino emissivity
in the quark plasmon decay.  This
process is exponentially suppressed due to the very high
quark plasma frequency.\\

{\bf (e) Other exotic models}

Another example of the exotic matter in the neutron star
core is the model with localized protons.
Such matter consists of the neutron liquid with a small
admixture of strongly degenerate relativistic electrons
and the same amount of protons localized in the self-consistent
potential wells (of polaron type) produced by their
interaction with the neutrons.
The neutrino emission from this matter was analyzed by
Baiko and Haensel (1999), assuming the localized protons
are distributed in a disordered manner.
%
%
The direct Urca process in such matter is forbidden by the
proton localization. Some neutrino emission is produced
by neutrino bremsstrahlung due to the $nn$ collisions, which
is the same as in the ordinary matter
(Sect.\ \ref{sect-nucore-Brems}). However, the main process
is the neutrino-pair emission due to the $np$ collisions
in which the protons act as the localized scattering centers.
The neutrino emissivity varies with temperature as
$T^6$, and it may be one or two orders of magnitude
higher than the emissivity of the modified Urca
process in the ordinary $npe$ matter.

\newpage

\section{Neutrino emission from superfluid and magnetized cores}
\label{chapt-nusup}

\subsection{Reduction and enhancement
         of neutrino emission by superfluidity and magnetic fields}
\label{sect-nusup-reduction}

{\bf (a) Effects of superfluidity}

As we have shown in Chapt. \ref{chapt-nucore}, the main neutrino reactions
in neutron star cores involve baryons. It is important that the baryonic
component of matter may be in a superfluid state
(Sect. \ref{sect-overview-struct}). The presence of 
the superfluid gap
in the baryon energy spectrum {\it reduces}
the reaction rates.
Generally, the neutrino emissivity $Q$ of any baryonic process
(direct or modified Urca, bremsstrahlung) can be written as
\begin{equation}
        Q = Q_0 \, R,
\label{nusup-reduct}
\end{equation}
where $Q_0$ is the emissivity in the non-superfluid matter
considered in Chapt.\ \ref{chapt-nucore}, and $R$ is {\it the superfluid
reduction factor}. In the absence of superfluidity, one
can formally set $R=1$, while in the superfluid matter $R<1$.
We analyze the reduction factors $R$ for the main
neutrino reactions in subsequent sections.
In our analysis, we will consider the neutron star matter in beta equilibrium.
Although the neutrino emissivities $Q_0$ themselves
can be very sensitive to the details of the microscopic model
of dense matter (Chapt.\ \ref{chapt-nucore}),
the reduction factors are rather insensitive to the
models. In many cases, they can be calculated more reliably
than the non-superfluid emissivities $Q_0$.
Although much work has already been done and the main features
of the superfluid suppression of neutrino processes
are understood, some cases have not been considered yet.
Following the traditional point of view (Sect.\ \ref{sect-overview-struct})
we use the Bardeen-Cooper-Schrieffer (BCS) model of baryon
superfluidity. 
Some results of this model are summarized in Sect.\ \ref{sect-sf-gaps}.
Generally, the
reduction factors depend on temperature $T$ and superfluid gaps
(or critical temperatures $T_c$) of the baryons involved
in the neutrino emission processes. Baryons of different species
may be superfluid at the same time. In these cases 
we have to deal with several
superfluidities with different $T_c$.
We will express $R$ in terms of the
simple dimensionless superfluidity parameters to
facilitate application of the results. Notice that
the superfluidity breaks the simple power-law temperature
dependence (\ref{nucore-list-Trho}) of the emissivity. As a rule,
the emissivity decreases sharply after the temperature falls
below $T_c$. Since the critical temperature depends on the density,
the emissivity becomes a complicated function of density as well.
%
%
The values of the critical temperatures are rather uncertain
(Sect.\ \ref{sect-overview-struct}), so that
we will treat them as free parameters.

The overall analysis of the reduction factors of various
neutrino reactions is done in Sect.\ \ref{sect-nusup-similar}.
We present not strict but sufficiently
accurate approximate {\it similarity criteria} which enable us to
construct new reduction factors from the known ones.

The superfluidity
not only reduces the emissivity
of the well-known neutrino reactions but also
initiates a specific neutrino emission
{\it due to Cooper pairing of baryons}, which cannot occur
in the non-superfluid matter. In this way the superfluidity
{\it enhances} the neutrino emission. This specific
neutrino mechanism is considered in
Sect.\ \ref{sect-nusup-CP}. Its emissivity is also
a strong function of temperature. 
The neutrino emission due
to Cooper pairing of free neutrons operates also in the inner
crust (Sect.\ \ref{sect-nucrust-nn}).

The overall analysis of the effects of superfluidity
on various neutrino reactions in neutron star cores
is carried out in Sect.\ \ref{sect-nusup-all}.
We will see that the superfluidity {\it drastically}
changes the neutrino emissivity. In particular,
it {\it reduces the strong difference between the standard and
enhanced neutrino emission}. Moreover, in some cases it
makes the standard emission look like the enhanced and
vice versa.  This has very important consequences
for the neutron star cooling (Yakovlev et al.\ 1999b, 
Chapt. \ref{chapt-cool}).\\

{\bf (b) Effects of magnetic fields}

Another complication into the neutrino emission from
neutron star cores is introduced by the possible presence
of the core magnetic fields. If protons or other charged
baryons are superfluid (superconducting), the magnetic
field exists most likely in the form of fluxoids  --- the quantized
magnetic flux tubes, Sect.\ \ref{sect-nusup-fluxa}. In the absence
of superconductivity the magnetic field is
uniform on microscopic scales. We will briefly
consider both cases.

%
First of all, notice that contrary to some speculations
appearing in the literature from time to time, the effects
of the magnetic field on the neutrino emission from
neutron star cores are not overwhelming.
There is a simple physical reason for this: the Fermi energies of
the particles participating in neutrino reactions
are so high
that the particles occupy many Landau levels.
For instance, in the field $B \sim 10^{16}$ G
at the density about several $\rho_0$ the electrons and
protons occupy $\sim 300$ Landau levels.
In such cases, the effects of magnetic quantization
on the neutrino emissivities are usually weak and the
emissivities are about the same as in the
non-magnetized matter. The effects
would be significant in the superstrong magnetic fields $B \gtrsim 10^{18}$ G,
in which the particles would occupy one or several Landau
levels. However, one can hardly expect the presence of such
fields in the cores of cooling neutron stars, and we will
not consider them in detail. The effects of the weaker, realistic fields
are very selective and can be pronounced in those rare
occasions where the reaction rates are
strongly modified even if the Landau states are treated
in the quasiclassical approximation.
To our knowledge, these effects are of two types,
and we will consider each of them.

The effects of the {\it first type} are associated with opening
those neutrino reactions that
are suppressed by momentum conservation
at $B=0$. Strong, non-quantizing magnetic fields
can relax the suppression conditions.
As an example, in Sect.\ \ref{sect-nusup-durmag}
we consider the direct Urca process in the strongly
magnetized non-superconducting matter. We show
that the magnetic field can, indeed, broaden
noticeably the direct Urca threshold.

The effects of the {\it second type} are associated with 
the neutrino emission of charged particles moving in the magnetic
field. Clearly, such emission does not require
the strong quantization of particle orbits.  One example, the neutrino
synchrotron emission of electrons in the uniform magnetic field,
has been considered in Sect.\ \ref{sect-nucrust-syn}
for the conditions
in the neutron star crust. In Sect.\
\ref{sect-nusup-fluxa} we extend these results
to the neutron star core
and also analyze a variation of the same process,
the neutrino emission due to the scattering of electrons off
fluxoids in the matter with the highly superfluid
(and therefore highly superconducting) protons.

The results of these studies show that the presence of the
magnetic field $B \lesssim 10^{11}$ G
in the neutron star core has practically
no effect on the neutrino emission.
Stronger fields induce a `synchrotron-fluxoid' neutrino
emission of electrons which can dominate in the strongly
superfluid and sufficiently cold neutron star cores.
Still stronger fields, $B \gtrsim 3 \times 10^{14}$ G,
can produce a noticeable `magnetic broadening' of the
direct Urca threshold and the thresholds of other
reactions.

As in Chapters  \ref{chapt-nucrust} and \ref{chapt-nucore}
we will mainly use the units in which $\hbar = c = k_{\rm B} =1$,
returning to the standard physical
units in the final expressions.

\subsection{Bogoliubov transformation,
energy gaps and critical temperatures}
\label{sect-sf-gaps}

{\bf (a) Three superfluid models}

Before studying the superfluid reduction of various reaction
rates, let us outline the main properties of superfluidity
in neutron star cores.
As discussed in Sect.\ \ref{sect-overview-struct},
the cases of $^1{\rm S}_0$ or $^3{\rm P}_2$ pairing
are of special interest.
The $^3{\rm P}_2$ pairing in the $npe$ matter
occurs mainly in the system of neutrons.
While studying the pairing of this type, one should take into account
the states with different projections $m_J$ of
the total angular momentum of 
a neutron pair on the quantization axis:
$|m_J|=0$, 1, 2. The actual (energetically favorable)
state may be a superposition of states with different
$m_J$ (see, e.g.,
Amundsen and {\O}stgaard 1985;
Baldo et al., 1992). Owing to the uncertainties of
microscopic theories this state is still unknown;
it may vary with the density and temperature.
In simulations of neutron star cooling, one usually assumes
the triplet-state pairing with $m_J=0$
(except for the recent paper by
Schaab et al., 1998b). Below we will analyze the
$^3{\rm P}_2$ superfluids
with $m_J=0$ and $|m_J|=2$, since their effects on 
the neutrino emissivities
are qualitatively different.

\begin{table}[t!]
%
%
\caption{Three types of superfluidity}
\begin{center}
  \begin{tabular}{||c|cccc||}
  \hline \hline
         & Superfluidity type  & $\lambda$ &   $F(\vartheta)$
         & $k_{\rm B} T_c/\Delta(0)$  \\
  \hline
  {\rm A} & $^1{\rm S}_0$  &     1     &        1
         &   0.5669   \\
  {\rm B} & $^3{\rm P}_2\ (m_J =0)$ & $1/2$
          & $(1+3\cos^2\vartheta)$  & 0.8416  \\
  {\rm C} & $^3{\rm P}_2\ (|m_J| =2)$ & $3/2$
          & $\sin^2 \vartheta$      & 0.4926  \\
  \hline \hline
\end{tabular}
\label{tab-sf-ABC}
\end{center}
\end{table}

To be specific, we will study the BCS superfluidity
for an ensemble of almost free (qua\-si-)par\-tic\-les.
The superfluidity types
$^1{\rm S}_0$,
$^3{\rm P}_2$ ($m_J=0$) and  $^3{\rm P}_2$ ($m_J=2$)
will be denoted as
{\rm A}, {\rm B} and {\rm C}, respectively (Table \ref{tab-sf-ABC}).

The Cooper pairing appears as a result of the attraction of particles
with the anti-parallel momenta. Its effect is most pronounced near
the Fermi surface ($|p- p_{\rm F}| \ll p_{\rm F}$).
We consider the most important case of
the superfluid energy gap $\delta$ much smaller
than the chemical potential $\mu$ of the superfluid particles. In this case
superfluidity has little effect on the bulk properties
of matter because they are
determined by the entire Fermi sea
of the particles.
However, superfluidity may strongly influence the processes
associated with the particles near the
Fermi surface, such as the heat capacity, transport and neutrino
emission.

In a superfluid, the wave functions of quasiparticles
represent the coherent superpositions
of the wave functions
of the pairs of particles with the anti-parallel momenta.
The onset of superfluidity leads to the rearrangement of
the quantum states, from the single-particle states to their superpositions,
and to the appearance of
the energy gap $\delta$ in the particle energy
spectrum. These effects are studied using the {\it Bogoliubov
transformation} discussed in many textbooks
(e.g., Lifshitz and Pitaevskii 1980) for the
singlet-state pairing. In the case of triplet-state pairing it
is called the {\it generalized Bogoliubov transformation}.
For the particles in neutron stars, the generalized transformation
was studied in a classical paper by Tamagaki (1970).
The formalism is based on the second quantization
of the particle field.

To avoid possible confusion,
we note that the choice of the quasiparticle
(or the associated hole)
states is not unique, as will be discussed briefly
at the end of this section. It is a matter of taste and
convenience to choose from the different descriptions of
the quasiparticle states, all of which 
lead to the same physical conclusions.
In any case, the BCS theory predicts that
the energy of either quasiparticles or the holes
near the Fermi surface is
\begin{equation}
      \epsilon - \mu = \pm  \tilde{\epsilon},\quad
      \tilde{\epsilon} = \sqrt{\delta^2 + v_{\rm F}^2(p-p_{\rm F})^2},
\label{sf-spectrum}
\end{equation}
where $v_{\rm F}$ is the particle Fermi velocity.
Thus, we have two energy branches, above and below
the Fermi level, which never intersect
for the finite value of the gap $\delta$. The minimum
separation is the doubled gap, $2 \delta$, at $p=p_{\rm F}$.\\

{\bf (b) Superfluid gaps}

The gap $\delta$ is the fundamental quantity in the BSC theory.
In principle, it is a slowly varying function of $\epsilon$, but one
usually ignores this dependence by calculating the gap at $\epsilon=\mu$
since the superfluid effects are most pronounced
near the Fermi surface.
We will follow this convention.  Then, the gap depends
on the temperature and the position of the quasiparticle momentum
at the Fermi surface, as well as on the type of superfluidity.
For the cases of interest, $\delta^2 = \Delta^2 (T) F(\vartheta)$, where
$\Delta(T)$ is the amplitude which determines the
temperature dependence of the gap,
and the function $F(\vartheta)$ describes the dependence
on the angle $\vartheta$ between
the particle momentum ${\bf p}$ and
the quantization axis (axis $z$).
In our cases, the gap possesses azimuthal
symmetry with respect to the equator of the Fermi sphere.
Both functions $\Delta(T)$ and $F(\vartheta)$
depend on the type of superfluidity
(Table \ref{tab-sf-ABC}).
In case {\rm A} the gap is isotropic,
$\delta=\Delta(T)$, while in cases {\rm B} and {\rm C}
it is anisotropic.
In case {\rm B} the gap decreases from the poles of the Fermi sphere
towards the equator, becoming twice as small at the equator.
In case {\rm C} the gap
vanishes at the poles 
and increases towards the equator;
thus the superfluidity does not
affect the particles moving along the quantization axis.
Notice that for the triplet-state superfluidity with $|m_J|=1$ or
for a triplet-state superfluidity described by
the superposition of states with different $m_J$, the anisotropic
gap $\delta$ depends not only on $\vartheta$, but also on
the azimuthal angle $\phi$ of the particle momentum ${\bf p}$.
The effects of such superfluidity on the properties of
dense matter have not been studied yet.

In the cases of study,
the gap amplitude $\Delta(T)$ is determined by the BCS equation
(see, e.g.,
Lifshitz and Pitaevskii 1980,
Tamagaki 1970) which can be written as
\begin{equation}
     \ln \left[ \frac{\Delta_0}{\Delta(T)} \right] =
     2 \lambda \int {{\rm d} \Omega \over 4 \pi}
     \int^\infty_0 \, \frac{{\rm d} x}{z} \, f(z) \, F(\vartheta),
\label{sf-BCS}
\end{equation}
where $\Delta_0=\Delta(0)$,
${\rm d} \Omega$ is the solid angle element in the direction of ${\bf p}$,
$f(z)= (1 + {\rm e}^z)^{-1}$
is the Fermi--Dirac distribution,
$\lambda$ is numerical coefficient (Table \ref{tab-sf-ABC}), and
\begin{equation}
    z=\frac{\tilde{\epsilon}}{k_{\rm B} T}= \sqrt{x^2 + y^2},
    \quad \quad
    x=\frac{v_{\rm F}(p-p_{\rm F})}{k_{\rm B} T},
    \quad \quad  y=\frac{\delta}{k_{\rm B} T}.
\label{sf-DimLessVar}
\end{equation}
Using Eq.\ (\ref{sf-BCS}) one can easily obtain the values of
$k_{\rm B} T_c / \Delta_0$ presented in Table \ref{tab-sf-ABC}.
It is convenient to introduce the variables
\begin{equation}
   v = \frac{\Delta(T)}{k_{\rm B} T} , \quad \quad
   \tau = \frac{T}{T_c}.
\label{sf-DefGap}
\end{equation}
The dimensionless gap amplitude $v$ describes the temperature
dependence of the gap. It is determined by the superfluidity type
and the dimensionless temperature
$\tau$. In case {\rm A} the amplitude $v$ corresponds to the
isotropic gap, in case {\rm B} it corresponds to the minimum
and in case {\rm C} to the maximum gap at the Fermi surface.
In these notations, the dimensionless gap $y$ has the form:
\begin{equation}
  y_{\rm A} = v_{\rm A}, \quad \quad
  y_{\rm B} = v_{\rm B} \, \sqrt{1+3 \cos^2 \vartheta},  \quad \quad
  y_{\rm C} = v_{\rm C} \, \sin \vartheta.
\label{sf-y_v}
\end{equation}

Using Eq.~(\ref{sf-BCS}) one can obtain the asymptotes of the
gap amplitude near the critical temperature and in the
limit of the so called strong superfluidity ($T \ll T_c$, $v \gg 1$).
For instance, at $T \to T_c$ (with $T < T_c$)  
one has:
$\,v=\beta \sqrt{1-\tau}$, where
$\beta_{\rm A}= 3.063$,
$\beta_{\rm B}= 1.977$,
$\beta_{\rm C}= 3.425$
(see, e.g.,
Lifshitz and Pitaevskii 1980,
Levenfish and Yakovlev 1994a).
For $T \ll T_c$, one has
$v=\Delta_0/(k_{\rm B} T)=\Delta_0/(k_{\rm B} T_c \, \tau)$.
Levenfish and Yakovlev (1993, 1994a)
calculated $v=v(\tau)$ for the intermediate values of $\tau$
and obtained the analytical fits of the numerical results:
\begin{eqnarray}
  v_{\rm A} & = & \sqrt{1-\tau}
          \left( 1.456 - \frac{0.157}{\sqrt{\tau}} + \frac{1.764}{\tau}
          \right),
\nonumber \\
  v_{\rm B} & = & \sqrt{1-\tau} \left( 0.7893 + \frac{1.188}{\tau} \right),
\nonumber \\
  v_{\rm C} & = & \frac{\sqrt{1-\tau^4}}{\tau}
          \left(2.030 - 0.4903 \tau^4 + 0.1727 \tau^8 \right).
\label{sf-FitGaps}
\end{eqnarray}
These fits reproduce also the above asymptotes; they 
will be useful for evaluating 
the neutrino luminosity in superfluid matter
(Sect.\ \ref{sect-nucrust-nn}, subsequent sections
of this chapter).

The analytic fits presented here and below in this chapter
reproduce numerical data with the mean error about
1--2\%, while the maximum error does not exceed 5\%.
This fit accuracy is more than sufficient for
many applications.\\

{\bf (c) Bogoliubov transformation}

Now let us describe briefly the possible choice of quasiparticle
states. One often assumes that quasiparticles
have the energy
\begin{equation}
     \epsilon = \mu + \tilde{\epsilon}
\label{sf-choice1}
\end{equation}
for any $p$ (below and above $p_{\rm F})$, and
the energy of the corresponding quasiholes is
$\epsilon = \mu - \tilde{\epsilon}$.
In this language
the second-quantized particle wave function can be written as
(Lifshitz and Pitaevskii 1980, Tamagaki 1970)
\begin{equation}
  \hat{\Psi} = \sum_{ {\bf p} \sigma \eta} \, \chi_\sigma \left[
        {\rm e}^{ -i
        \tilde{\epsilon} t+ i {\bf p} \cdot {\bf r} } \,
        U_{\sigma \eta}({\bf p}) \, \hat{\alpha}_{{\bf p} \eta}+
        {\rm e}^{ i
        \tilde{\epsilon} t- i {\bf p} \cdot {\bf r} } \,
        V_{\sigma \eta}(-{\bf p}) \, \hat{\alpha}_{{\bf p} \eta}^\dagger
        \right].
\label{sf-CP-Psi}
\end{equation}
A basic spinor $\chi_\sigma$
describes the particle state with the fixed spin projection
($\sigma\! =\! \pm 1$)
onto the quantization axis;
$\eta= \pm 1$ enumerates the quasi-particle spin states;
$\hat{\alpha}^\dagger_{{\bf p}\eta}$
and $\hat{\alpha}_{{\bf p}\eta}$
are the creation and annihilation operators
of the quasiparticle in the
$({\bf p} \eta)$ state, respectively.
Also, $\hat{U}({\bf p})$ and
$\hat{V}({\bf p})$ are the operators of the
generalized Bogoliubov transformation.
For $|p-p_{\rm F}| \ll p_{\rm F}$,
their matrix elements obey the relationships
\begin{equation}
    U_{\sigma \eta}({\bf p}) = {\rm u} \, \delta_{\sigma \eta}, \quad
    \sum_{\sigma \eta} |V_{\sigma \eta}({\bf{p}})|^2 =2 \,{\rm v}^2, \quad
    \sum_{\sigma \eta}  |U_{\sigma \eta}({\bf{p}})|^2 +
    |V_{\sigma \eta}({\bf{p}})|^2  =2,
\label{sf-CP-Bogoliubov}
\end{equation}
where
\begin{equation}
  {\rm u} = \left[ {1 \over 2}
           \left( 1 + { v_{\rm F}  (p- p_{\rm F}) \over
            \tilde{\epsilon}} \right) \right]^{1/2},\quad
  {\rm v} = \left[ {1 \over 2}
           \left( 1 - { v_{\rm F}  (p- p_{\rm F}) \over
           \tilde{\epsilon}} \right) \right]^{1/2}
\label{sf-CP-uv}
\end{equation}
are the coefficients of the Bogoliubov transformation.
Thus, the matrix $U_{\sigma \eta}({\bf p})$ is
diagonal and identical for all three superfluidity types {\rm A}, {\rm B},
and {\rm C}.  The matrix $V_{\alpha \beta}$
is more complicated. In case {\rm A} it has the form
\begin{equation}
   \hat{V}({\bf p})=
   \left(
     \begin{tabular}{rl}
       0   & v  \\
       $-$v  & 0
     \end{tabular}
   \right)
\label{sf-V}
\end{equation}
and possesses the following symmetry properties:
$V_{\sigma \eta}(-{\bf p})=V_{\sigma \eta}({\bf p})$,
$V_{\sigma \eta}({\bf p})=- V_{ \eta \sigma}({\bf p})$.
For the triplet pairing, according to
Tamagaki (1970),
$V_{\sigma \eta}(-{\bf p})=-V_{ \sigma \eta}({\bf p})$,
$V_{\sigma \eta}({\bf p})=V_{ \eta \sigma}({\bf p})
= {\rm v} \, \Gamma_{\sigma \eta}({\bf p})$,
where $\Gamma_{\sigma \eta}({\bf p})$ is a unitary $(2 \times 2)$ matrix.
In case {\rm C} matrices $V_{\sigma \eta}({\bf p})$
and $\Gamma_{\sigma \eta}({\bf p})$ are diagonal.

An alternative choice of the quasiparticle states is to put
%
\begin{equation}
   \epsilon = \mu - \tilde{\epsilon} \quad
                     \mbox{at} \quad p<p_{\rm F}; \quad \quad
   \epsilon = \mu + \tilde{\epsilon} \quad
                     \mbox{at} \quad p \geq p_{\rm F} .
\label{sf-Dispers}
\end{equation}
In this language, one must introduce quasiholes
with the energies $\epsilon = \mu + \tilde{\epsilon}$ for
$p < p_{\rm F}$ and $\epsilon = \mu - \tilde{\epsilon}$ for
$p \geq p_{\rm F}$.

\subsection{Superfluid reduction of direct Urca process}
\label{sect-nusup-Durca}

In this section we
study the superfluid reduction of the direct Urca
processes [processes (I) in Table \ref{tab-nucore-list}].
For simplicity, we discuss the nucleon direct Urca process,
Eq.\ (\ref{nucore-Durca}),
involving $n$, $p$ and $e$, but the same approach
is valid for any direct Urca process.\\

{\bf (a) Matrix elements}

In order to understand the effects of superfluidity on the
matrix elements,
let us rederive the neutrino emissivity of 
the direct Urca process (Sects.\ \ref{sect-nucore-beta}
and \ref{sect-nucore-Durca})
in the superfluid matter. For simplicity,
let the neutrons be
superfluid but the protons are not.
Since we assume beta equilibrium, the rates of the
beta decay and beta capture reactions of the
direct Urca process are equal. To be specific,
we can focus on beta decay.

Looking through the derivation of the beta decay
transition rate in Sect.\ \ref{sect-nucore-beta}, one
sees that superfluidity affects only
the matrix elements of the weak hadron current $J^\alpha_{fi}$.
The hadron current itself is defined by
Eq.\ (\ref{nucore-hadron}), regardless of the superfluid state
of the reacting particles.
The most convenient way to calculate $J^\alpha_{fi}$
in the superfluid matter
is to use the formalism of second quantization of the
proton and neutron states.
Instead of the wave function
(\ref{nucore-psi-hadron}) of a free neutron, let us
employ the second-quantized Bogoliubov operator-function
given by Eq.\ (\ref{sf-CP-Psi}).
This function describes the neutron field in terms
of the annihilation and creation operators of quasiparticles
with the momentum ${\bf p}_n$, spin states $\eta=\pm 1$, and
the energy spectrum (\ref{sf-choice1}) containing the superfluid gap $\delta$.

Consider, for instance, beta decay (annihilation) of a quasineutron
with the reduced energy $\tilde{\epsilon}$ (with respect
to the Fermi level) higher than $\delta$.
We are naturally interested in the case of
${\bf p}_n$ near the Fermi surface.
If the direction of ${\bf p}_n$ is fixed and the matter
is non-superfluid, we have only one value of the
momentum $p_n>p_{{\rm F}n}$ for a given $\tilde{\epsilon}>0$.
In the superfluid matter with the energy spectrum (\ref{sf-choice1}),
we have {\it two} possible quasineutron momenta,
$p_n>p_{{\rm F}n}$ and $p_n<p_{{\rm F}n}$,
inside and outside the Fermi sphere
at the same distance $|p_n-p_{{\rm F}n}|$ from it.
Accordingly, beta decay of the quasineutron with given $\tilde{\epsilon}$
can go through {\it two channels}.
Let us denote them by ``$+$'' and ``$-$''.
The total reaction rate
includes the contributions from both channels.

The direct calculation of $J^\alpha_{fi}$
is easily performed with the aid of familiar
relationships for the annihilation
operators. The beta decay probability
contains the bilinear combinations of $J^\alpha_{fi}$ summed over
the spin states [similar to those given by Eq.\ (\ref{nucore-Jab})]
and additionally over the two reaction channels. Let us
denote these bilinear combinations
by ${\cal J}^{\alpha \beta}$, like for beta decay
in the non-superfluid matter (Sect.\ \ref{sect-nucore-beta}).
For instance, we have
\begin{equation}
     {\cal J}^{00} = \sum_{\sigma \eta \pm}
        |U_{\sigma \eta}({\bf p}^\pm_n)|^2= 2({\rm u}^2_+ +
         {\rm u}^2_-)=2,
\label{nusup-beta_J00}
\end{equation}
exactly the same as in the non-superfluid matter 
[cf Eq.\ (\ref{nucore-Jab1})].
Here, $U_{\sigma \eta}({\bf p})$ is the matrix element
of the Bogoliubov transformation 
given by Eq.\ (\ref{sf-CP-Bogoliubov}).
The sum over the two channels is ${\rm u}^2_+ + {\rm u}^2_-=1$,
which is evident from Eq.\ (\ref{sf-CP-uv}).
Strictly speaking, the expression for the neutrino emissivity
contains this sum multiplied by the
energy and momentum conserving delta functions,
and the combination
of Fermi-Dirac distributions $f_n\,(1-f_p)$.
However, all these factors are
the same for both channels ``$+$'' and ``$-$'',
since the quasineutron energy $\tilde{\epsilon}$ is the same
and since we can shift the neutron and
proton Fermi momenta to the appropriate Fermi surfaces
in the momentum conserving delta function due to
the strong degeneracy.
Accordingly, we can pull all these factors
outside the summation sign.
%

In a similar way
we can prove that all tensor components
${\cal J}^{\alpha \beta}$ are given by the same
Eqs.\ (\ref{nucore-Jab1}) as for the standard beta decay.
This means that even though superfluidity
affects the matrix element,
the squared matrix element
summed over the two reaction channels and the nucleon spin states
remains the same as in the non-superfluid matter.
This is true in the presence of proton superfluidity as well.

Therefore, the expression for
the neutrino emissivity (\ref{nucore-QdDef}) of the direct Urca process
{\it is almost the same as in the non-superfluid matter
with only one modification: one should introduce the superfluid gaps
into the neutron and proton energy spectra}.
This conclusion greatly simplifies our analysis.\\

{\bf (b) General expression for the reduction factor}

The reduction factor of the direct Urca process
can be obtained using the phase-space decomposition of the emissivity,
Eq.\ (\ref{nucore-decomp-Dur}).  The reduction factor
is model-independent, insensitive to
the details of the theory used to calculate the matrix element.
Let the protons ($j=2$) undergo Cooper pairing of type A, while
the neutrons ($j=1$) undergo pairing A,
B or C.  Superfluidity may affect only the
phase space integrals $A$ and $I$.
Since the neutron gap is generally anisotropic, the angle
and energy integrations are not really decomposed
(we refer to the integration over ${\rm d}x_j$ as energy integration,
for brevity; in principle it is the integration
over the dimensionless particle momentum).
In a non-superfluid matter, the energy
integral is $I =I_0 = 457 \pi^6 / 5040$, cf.\ Eq.\ (\ref{nucore-AIdur}).
In order to incorporate the superfluidity into the energy integral,
we adopt the energy spectrum of quasiparticles
in the form (\ref{sf-Dispers}). Then
it is sufficient to replace $x_j \to z_j$ for $j=1$ and 2
in the energy conserving delta function and in the Fermi-Dirac
functions in Eq.\ (\ref{nucore-Idur}).
Here $z \equiv (\epsilon - \mu)/T= {\rm sign}\,(x) \;
\sqrt{x^2 + y^2}$ is the dimensionless nucleon
energy [see Eqs.\ (\ref{sf-DimLessVar})].
One should
perform the angular integration first and the energy integration
afterwards. 
In the angular integral, we can temporarily fix the orientation
of the neutron momentum ${\bf p}_1$
and integrate over
the orientations of the other momenta. In the non-superfluid matter we
obtain $A_0/(4\pi)$, where $A_0$ is the angular
integral, Eq.\ (\ref{nucore-AIdur}).
Thus, the reduction factor
can be written as
\begin{equation}
    R^{\rm (D)}(v_1,v_2) =
             {A I \over A_0 I_0} \!= \!
             \int \frac{{\rm d} \Omega}{4\pi} \,J(y_1,y_2)  = \!
             \int_0^{\pi/2} \!\! {\rm d} \vartheta \, \sin \!\vartheta \,
             J(y_1,y_2),
\label{nusup-RdurcaDef}
\end{equation}
where $v_1$ and $v_2$ are the dimensionless amplitudes
of the neutron and proton superfluid gaps, respectively,
defined by Eq.\ (\ref{sf-DefGap}),
$y_2 \equiv v_2$, $y_1$ is given by Eq.\ (\ref{sf-DimLessVar})
and (\ref{sf-y_v}),
d$\Omega$ is the solid angle element in the direction of ${\bf{p}}_1$,
$\vartheta$ is the angle between ${\bf{p}}_1$ and the
quantization axis,
\begin{equation}
 J(y_1,y_2) \! = \!  { 1 \over I_0}
                \int_0^{+\infty}         \!\!\! {\rm d} x_{\nu}\, x_{\nu}^3 \,
                \int_{-\infty}^{+\infty} \!\!\! {\rm d} x_1 \,  f(z_1) \,
                \int_{-\infty}^{+\infty} \!\!\! {\rm d} x_2 \,  f(z_2) \,
                \int_{-\infty}^{+\infty} \!\!\! {\rm d} x_e \, f(x_e) \,
                 \delta (x_{\nu}\!-\! z_1\! -\! z_2\! -\!x_e).
\label{nusup-JDef}
\end{equation}
The integrals (\ref{nusup-RdurcaDef}) and (\ref{nusup-JDef})
were calculated by
Levenfish and Yakovlev (1993, 1994b)
for various combinations of the neutron and proton superfluids.
The results are discussed below.\\

{\bf (c) Superfluidity of neutrons or protons}

The two cases where either
neutrons or protons become superfluid are similar.  For example,
let the neutrons be superfluid. Then we can set $z_2=x_2$ in Eqs.\
(\ref{nusup-RdurcaDef}) and (\ref{nusup-JDef}), and
$R^{\rm (D)}$ depends only on
$v_1=v$ and the type of superfluidity.
If the dimensionless
temperature $\tau \equiv T/T_{c} \ge 1$, then, as mentioned above,
$R^{\rm (D)}=1$.  For the strong superfluidity ($\tau \ll 1$, $v \gg 1$)
the neutrino emission is greatly suppressed, $R^{\rm (D)} \ll 1$.
The appropriate asymptotes of $R^{\rm (D)}$ can be obtained
in the following
way. The energy integrals can partly be done analytically.
The quantity $z_1=(x_1^2 + y_1^2)^{1/2}$
becomes generally large, and its presence in the
exponent arguments in the remaining integral makes the
integrand very small. This strongly modifies the
momentum space which contributes to $R^{\rm (D)}$.
In cases A or B, it is sufficient to set
$z_1 = y_1 + x_1^2/(2y_1)$ in the exponents of the
integrand, $z_1=y_1$ in the pre-exponent factors, and treat
$y_1$ as a large parameter. In case A, the main contribution
comes from the momentum space of the neutron with 
$0< x_1 \lesssim \sqrt{v_1}$ (positive energy
states with $f(z_1) \approx {\rm e}^{-z_1}$),
regardless of the direction of neutron momentum on the Fermi
sphere. In case B, the main contribution also comes
from the momentum space with $0<x_1 \lesssim \sqrt{v_1}$
but in the narrow cone of angles $|\cos \vartheta|
\lesssim v_1^{-1/2}$ near the equator of the neutron Fermi
sphere, where the neutron gap has minimum
as a function of $\vartheta$. 
Finally, in case
C the main contribution comes from the values $|x_1| \lesssim 1$
within the narrow cones near the poles of the Fermi sphere,
$ \sin \vartheta \lesssim 1/v_1 \ll 1$, where the gap
has nodes (the gap effects are weaker there and do not
suppress exponentially the neutrino reaction).
More details can be found in Levenfish and Yakovlev (1994b).
As a result, the asymptotes
for the strong superfluidities A, B and C read:
\begin{eqnarray}
 R_{\rm A}^{\rm (D)}
               & = & \frac{252}{457 \,\pi^6} \, \sqrt{\frac{\pi}{2}} \,
                v^{5.5}\, \exp(-v)
                =  \frac{0.0163}{\tau^{5.5}} \;
                            \exp \left(- \frac{1.764}{\tau} \right),
\nonumber \\
 R_{\rm B}^{\rm (D)} & = & \frac{126}{457\, \pi^5 \sqrt{3}} \, v^5 \exp(-v)
               = \frac{0.00123}{\tau^5} \,
                \exp \left(- \frac{1.188}{\tau}  \right),
\nonumber \\
 R_{\rm C}^{\rm (D)} & = & \frac{6029\, \pi^2}{5484 \, v^2}
              = 2.634 \, \tau^2.
\label{nusup-RcAsy} 
\end{eqnarray}
The asymptotes of $R_{\rm A}^{\rm (D)}$
and $R_{\rm B}^{\rm (D)}$ are exponentials, while
the asymptote of $R_{\rm C}^{\rm (D)}$ is a power-law.
The exponential character in cases A and B is
due to the exponentially small probability of finding
a neutron with the energy $\sim \delta \gg T$ above the
Fermi level to undergo beta decay. This means that the strong
superfluidity drastically modifies the neutrino emission
process as a whole. In particular, the typical
neutrino
energies become $\epsilon_\nu \sim \delta \gg T$ whereas
they were $\sim T$ in the non-superfluid matter.
It is now the gap $\delta$ that mainly
determines the momentum space available for the neutrino emission.
If we fix $T$ but
increase $\delta$, the number of interacting
particles becomes very small.
On the other hand,
the momentum domain responsible for the neutrino
emission becomes larger
(in cases A and B)
than for $\delta=0$,
leading to the large pre-exponential factors in Eqs.\ (\ref{nusup-RcAsy}).
The power-law reduction of the neutrino emissivity
in the strong superfluidity C is clearly
associated with the nodes of the gap at
$\vartheta=0$ and $\pi$. In this case the typical
neutrino energies remain $\sim T$ and the superfluid reduction
of the emissivity is due to
the reduction of the `active' momentum space near the
neutron Fermi surface. It is evident that the angular distributions
of the emitted neutrinos in cases B and C become anisotropic;
the direction of anisotropy is specified by the quantization axis.
Thus, strong superfluidity modifies the spectrum and
angular distribution of neutrinos emitted in this and other
reactions.

\begin{figure}[!t]
\begin{center}
\leavevmode
\epsfysize=8.5cm
\epsfbox[135 495 410 785]{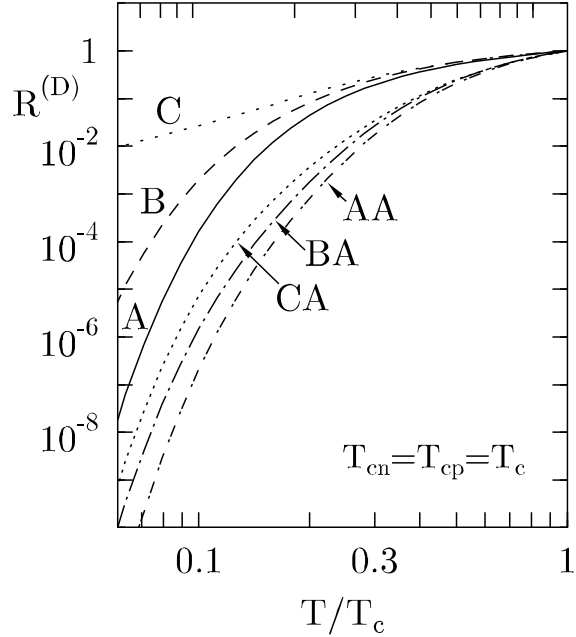}
\end{center}
\caption[]{\footnotesize
          Superfluid reduction factors of the direct Urca process
          versus $\tau=T/T_c$ (Yakovlev et al.\ 1999b).
          Letters near the curves indicate the type of superfluidity
          (see Table \protect{\ref{tab-sf-ABC}}): A, B and C for superfluidity
          of neutrons or protons, while AA, BA and CA
          for superfluidity of neutrons and protons with
          $T_{cn}=T_{cp}$.}
\label{fig-nusup-Rone}
\end{figure}

Asymptotes (\ref{nusup-RcAsy}) and numerical values of integrals
(\ref{nusup-RdurcaDef}) and (\ref{nusup-JDef})
for the intermediate values of $v$ can be fitted by:
\begin{eqnarray}
 R_{\rm A}^{\rm (D)} & = &
                   \left[ 0.2312 + \sqrt{ (0.7688)^2+(0.1438\,v)^2}
                   \right]^{5.5} \,
         \exp \! \left( 3.427 - \sqrt{ (3.427)^2+v^2 } \right),
 \nonumber \\
 R_{\rm B}^{\rm (D)} & = &
                    \left[ 0.2546 + \sqrt{ (0.7454)^2+(0.1284\,v)^2}
                    \right]^5 \,
         \exp \! \left( 2.701 - \sqrt{ (2.701)^2+v^2 } \right),
\nonumber \\
 R_{\rm C}^{\rm (D)} & = &
            \frac{ 0.5 + (0.09226\,v)^2}{1 + (0.1821\, v)^2
                             + (0.16736\, v)^4} \,
       + \, \frac{1}{2} \,\exp \! \left( 1 - \sqrt{1+(0.4129\,v)^2} \right).
\label{nusup-RcFit}
\end{eqnarray}
Equations (\ref{nusup-RcFit}), along with (\ref{sf-FitGaps}),
enable one to calculate $R^{\rm (D)}$ at any temperature.
The results are illustrated in Fig.~\ref{fig-nusup-Rone}.

In simulations of neutron star cooling prior to publication
of the above results
(Levenfish and Yakovlev 1993, 1994b)
one usually used
(e.g., Page and Applegate 1992)
the simplified reduction factors of the direct Urca process
by superfluidity  A or B,
$R^{\rm (D) \ast}=\exp(-v^\ast)$. Here,
$v_{\rm A}^\ast =\delta_{\rm A}/(k_{\rm B} T)=1.764/\tau$,
$v_{\rm B}^\ast=\delta_{\rm max}/(k_{\rm B} T)
=2.376/\tau$, $\delta_{\rm max}=
2 \Delta_{\rm B}(0)$ is the maximum energy gap
$\delta_{\rm B} (0,\vartheta)$ as a function of $\vartheta$ at $T=0$.
These factors were proposed by
Maxwell (1979)
from a simple (but inadequate) consideration.  The comparison
of the accurate and simplified factors shows that the latter
strongly overestimate the 
reduction effect.
Indeed, the correct asymptotes (\ref{nusup-RcAsy})    
in the limit $T \ll T_c$
contain large pre-exponents omitted in the simplified expressions.
In addition, factors $R^{\rm (D)}_{\rm B}$ 
and $R^{\rm (D) \ast}_{\rm B}$
differ by the exponent arguments. Instead of the correct argument
$v_{\rm B}=\delta_{\rm min}/(k_{\rm B} T)$,
the simplified factor
contains $v_{\rm B}^\ast=\delta_{\rm max}/(k_{\rm B} T)
=2 \delta_{\rm min}/(k_{\rm B} T)$,
which is twice as large. As a result, at $T=0.1 \, T_c$
the accurate factor
$R_{\rm A}^{\rm (D)}$ appears to be about four orders of magnitude
larger than the simplified one,
and the accurate factor $R_{\rm B}^{\rm (D)}$ is more
than seven orders of magnitude larger. 
\\

{\bf (d) Superfluidity of neutrons and protons}

Consider now reduction of the direct Urca process by
the combined action of superfluidities of the neutrons and protons of types
AA, BA and CA.
In these cases, the reduction factor
$R^{\rm (D)}$ depends on two arguments,
$v_1=v_n$ and $v_2=v_p$.

A direct calculation of $R^{\rm (D)}$ in the presence of
neutron and proton superfluids
is complicated; let us analyze the factor
$R^{\rm (D)}_{\rm AA}$ as an example.
According to Eqs.\  (\ref{nusup-RdurcaDef}) and (\ref{nusup-JDef})
for the singlet-state pairing,
we have $R^{\rm (D)}_{\rm AA}(v_1,v_2)$ = $ J(v_1,v_2) = J(v_2,v_1)$,
where $y_1=v_1$, $\, y_2 = v_2$. Clearly,
$R^{\rm (D)}_{\rm AA}(0,0)=1$. If both superfluidities are strong
($v_1 \gg 1$ and $v_2 \gg 1$) and
$v_2-v_1  \gg \sqrt{v_2}$,
the asymptote of the reduction factor is:
\begin{eqnarray}
 &&    R^{\rm (D)}_{\rm AA} \!\! = \!\! J(v_1,v_2) = {1 \over I_0}
              \left( \frac{\pi}{2} \, v_2 \right)^{1/2}
              \exp(-v_2) \, K,
\label{nusup-RaaAsy} \\
 &&  K = \frac{s}{120} \,
      \left( 6v_2^4+83v_2^2v_1^2 +16v_1^4 \right)
       -  \frac{1}{8}\,v_2v_1^2
          \left( 3v_1^2+4v_2^2 \right)
          \ln \left( \frac{v_2 + s}{v_1} \right),
\label{nusup-KaaAsy}
\end{eqnarray}
where $s \! = \! \sqrt{v_2^2 - v_1^2}$.
In the limit $v_1 \ll v_2$ Eq.\ (\ref{nusup-KaaAsy}) gives
$K=v_2^5/20$, which corresponds to the asymptote
(\ref{nusup-RcAsy}) of $R^{\rm (D)}_{A}$.
In another limit $ \sqrt{v_2} \ll v_2-v_1 \ll v_2$
we obtain $ K=(2/315) \, s^9/v_2^4$.
The asymptote (\ref{nusup-KaaAsy}) fails
in the range $ | v_2-v_1 | \lesssim \sqrt{v_2}$.
As shown by
Levenfish and Yakovlev (1994b)
$K \sim \sqrt{v_2}$ for $v_1=v_2$.

A general fit that reproduces the asymptote
(\ref{nusup-RaaAsy}) and the numerical values of
$R^{\rm (D)}_{\rm AA}$ calculated for a wide range of arguments,
$\sqrt{v_1^2+v_2^2\,} \lesssim 50$, is
\begin{equation}
      R^{\rm (D)}_{\rm AA}= J(v_1,v_2) = \frac{u}{u+0.9163}\, S + D,
   \label{nusup-RaaFit}
\end{equation}
where
\begin{eqnarray}
&&  S = {1 \over I_0} \;
         (K_0+K_1+0.42232\,K_2)
           \,\left( \frac{\pi}{2} \right)^{1/2}
        p_s^{1/4} \exp(- \sqrt{p_e}),
\nonumber  \\
&& K_0 = \frac{ \sqrt{p-q}}{120} \; (6p^2+83pq+16q^2) -
        \sqrt{p} \, \frac{q}{8}\,(4p+3q)\;
        \ln \left( \frac{ \sqrt{p}+ \sqrt{p-q}}{ \sqrt{q}} \right),
\nonumber  \\
&& K_1 = \frac{\pi^2 \sqrt{p-q}}{6} \, (p+2q) \; - \; \frac{\pi^2}{2} \, q
      \sqrt{p}\,
      \ln \left( \frac{ \sqrt{p}+ \sqrt{p-q}}{ \sqrt{q}} \right),
\nonumber  \\
&& K_2 = \; \frac{7 \, \pi^4}{60} \, \sqrt{p-q},
\nonumber \\
&& 2p  =  u+12.421 + \sqrt{w^2+16.350 \, u+45.171},
\nonumber \\
&& 2q  =  u+12.421 - \sqrt{w^2+16.350 \, u+45.171},
\nonumber \\
&& 2p_s  =  u + \sqrt{w^2+5524.8\,u+6.7737},
\nonumber \\
&&  2p_e  =  u + 0.43847 + \sqrt{w^2+8.3680\,u+491.32},
\nonumber \\
&&  D  =  1.52 \, \mbox{\rule{0cm}{0.8cm}} (u_1 u_2)^{3/2} \; (u_1^2+u_2^2)\,
    \exp(-u_1-u_2),
\nonumber \\
&&  u_1  =  1.8091 + \sqrt{v_1^2+(2.2476)^2},
\nonumber \\
&&  u_2 =  1.8091 + \sqrt{v_2^2+(2.2476)^2},
\label{nusup-RaaFit_}
\end{eqnarray}
with $u = v_1^2+v_2^2$ and $w = v_2^2-v_1^2$.
For $v_2=0$, the factor $R^{\rm (D)}_{\rm AA} (v_1,0)$ agrees with
the factor $R^{\rm (D)}_A (v_1)$ given by Eq.\ (\ref{nusup-RcFit}).

\begin{figure}[!t]
\begin{center}
\leavevmode
\epsfysize=17.5cm
\epsfbox[40 70 310 775]{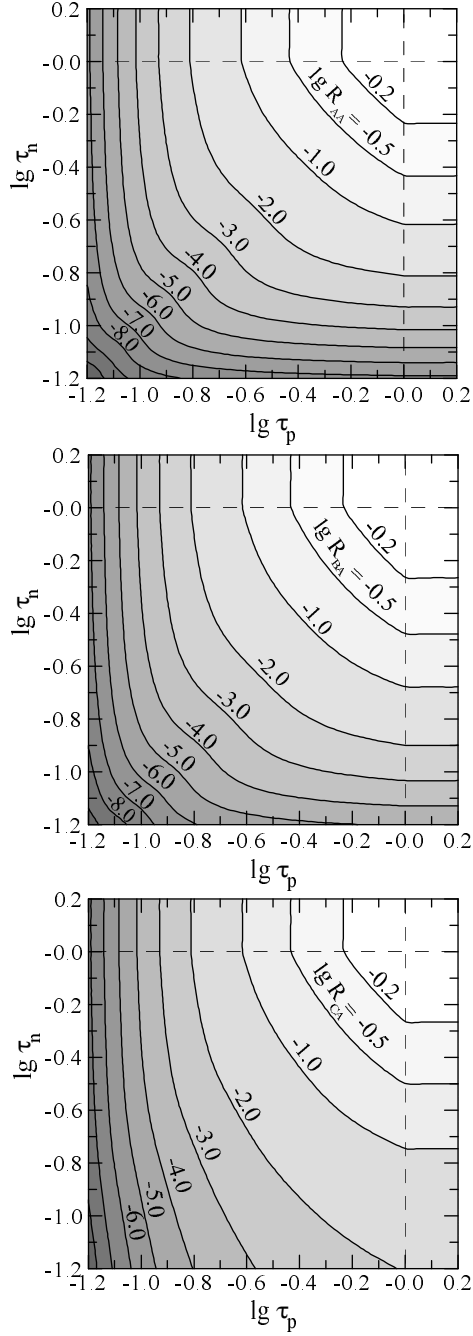}
\end{center}
\caption[]{\footnotesize
          Contours of constant reduction factors
          $R^{\rm (D)}$ of the direct Urca process
          in the presence of neutron and proton superfluidities
          AA, BA and CA (Yakovlev et al.\ 1999b). 
          The curves are labeled by the values $\lg R^{\rm (D)}$.
          In the region $\tau_n \ge 1$ and $\tau_p \ge 1$,
          neutrons and protons are normal and $R^{\rm (D)}=1$.
          In the region $\tau_n < 1$, $\tau_p \ge 1$,
          where only neutrons are superfluid, and in the region
          $\tau_n \ge 1$, $\tau_p < 1$, where only protons are
          superfluid,
          $R^{\rm (D)}$ depends on only one parameter, 
          either $\tau_n$ or $\tau_p$.
          In the region
          $\tau_n < 1,\; \tau_p < 1$, both neutrons and protons
          are superfluid.
          }
\label{fig-nusup-Rboth}
\end{figure}

Figure \ref{fig-nusup-Rboth} shows the contours of
$R^{\rm (D)}_{\rm AA}\!=\,$const versus
$\tau_1=T/T_{c1}$ and $ \tau_2=T/T_{c2}$.
The behavior of $R^{\rm (D)}_{\rm AA}$ at $\tau_1^2+ \tau_2^2 \ll 1$,
where both superfluidities are strong, is of special interest.
In this case one can obtain an approximate relationship
(Lattimer et al.\ 1991;
Levenfish and Yakovlev 1994b)
\begin{equation}
      R^{\rm (D)}_{12} \sim
           \min\left( R^{\rm (D)}_1 \! , \, R^{\rm (D)}_2 \right)\, ,
\label{nusup-Estimation}
\end{equation}
where $R^{\rm (D)}_1$ and $R^{\rm (D)}_2$ are the reduction factors
for either type of superfluidity.  Roughly speaking, this means that
$R^{\rm (D)}_{12} \sim \exp(- \Delta /T)$, where
$\Delta$ is the largest of the two gaps, $\Delta_1$ and $\Delta_2$,
since it is the largest gap that opens the process
with respect to energy conservation.
Accordingly, the factor
$R^{\rm (D)}_{12}$ is mainly determined by the stronger
superfluidity.
The presence of the second,
weaker superfluidity affects
$R^{\rm (D)}_{12}$, but not to a great extent.

Using Eq.\ (\ref{nusup-RdurcaDef}), it is not difficult to evaluate
$R^{\rm (D)}$ for cases BA and CA, in which
the protons undergo singlet-state pairing
while the neutrons undergo triplet-state pairing.
The calculation is reduced to the one-dimensional integration in Eq.\
(\ref{nusup-RdurcaDef}) of the quantity
$J(y_1,y_2)$ fitted by Eqs.\ (\ref{nusup-RaaFit})
and (\ref{nusup-RaaFit_}).
The results
%
%
for any $T$, $T_{c1}$ and $T_{c2}$
are shown in Fig.~\ref{fig-nusup-Rboth}.
The dependence of $R^{\rm (D)}_{\rm BA} $ and
$R^{\rm (D)}_{\rm CA} $ on $\tau_1$ and $\tau_2$ is similar to the
dependence of $R^{\rm (D)}_{\rm AA}$, shown in the same figure,
but now $R^{\rm (D)} (\tau_1,\tau_2) \neq R^{\rm (D)} (\tau_2,\tau_1)$.
The approximate expression (\ref{nusup-Estimation})
is also valid in these cases
(Levenfish and Yakovlev 1994b).
Since superfluidity 
C reduces the neutrino emission much weaker
than superfluidities A or B,
in the case CA with $v_1 \gg 1$
the transition from the one dominating superfluidity to the other
 takes place at
$v_2 \sim \ln v_1 $. Moreover, for
$v_1 \gtrsim v_2 \gg 1$ the factor
$R^{\rm (D)}_{\rm CA}$ appears to be much larger than $R^{\rm (D)}_{\rm AA}$
or $R^{\rm (D)}_{\rm BA}$.
Accordingly, the lines of constant $R_{\rm CA}$ are lower
than the corresponding contours of $R_{\rm AA}$ or $R_{\rm BA}$
(cf.\ Figs.\ \ref{fig-nusup-Rone} and \ref{fig-nusup-Rboth}).

For not very strong superfluidities BA,
with $\sqrt{v_1^2 + v_2^2\,} \lesssim 5 $,
the reduction factor is fitted by
%
\begin{equation}
     R^{\rm (D)}_{\rm BA} = \frac{10^4-2.839 \, v_2^4 - 5.022
                         \, v_1^4}{ 10^4 + 757.0 \, v_2^2
              + 1494 \, v_1^2 + 211.1 \, v_2^2v_1^2 + 0.4832\,v_2^4v_1^4}.
\label{nusup-RabFit}
\end{equation}
In the case of not too strong superfluidities
CA, with $\sqrt{v_1^2 + v_2^2\,} \lesssim 10$,
the fit is:
%
\begin{eqnarray}
  R^{\rm (D)}_{\rm CA} &=&
       10^4 \times \left( 10^4 + 793.9\,v_2^2 + 457.3 \, v_1^2
       + 66.07 \, v_2^2v_1^2 +2.093 \, v_1^4 + \right.
\nonumber \\
       &+& \left. 0.3112 \, v_2^6 + 1.068 \, v_2^4v_1^2 +
     0.01536 \, v_2^4v_1^4 +0.006312 \, v_2^6v_1^2 \right)^{-1} \, .
\label{nusup-RacFit}
\end{eqnarray}
If $v_1=0$ or $v_2=0$, the above fits agree with Eqs.\ (\ref{nusup-RcFit}).
The tables of $R^{\rm (D)}_{\rm BA}$
and $R^{\rm (D)}_{\rm CA}$, as well as the asymptotes of these factors
in the limit of strong superfluidity, are given in
Levenfish and Yakovlev (1994b).

Before publication of the above results
(Levenfish and Yakovlev 1993, 1994b),
as far as we know, the only simulation of the fast neutron star cooling
with the account of superfluidity of neutrons and protons
was carried out by
Van Riper and Lattimer (1993).
They used a simplified reduction factor,
$R^{\rm (D)}_{12} = R^{\rm (D)}_1 R^{\rm (D)}_2$,
which strongly overestimates the
effect of superfluidity.\\

{\bf (c) Other direct Urca processes}

We have discussed the
superfluid reduction of the nucleon direct Urca
process only as an example. The same
formulae are valid for any baryon direct Urca process
[processes (I) in Table \ref{tab-nucore-list},
Sect.\ \ref{sect-nucore-Durca}] involving electrons or muons
and the two baryon species, one of which undergoes singlet
pairing. Thus, the presented results are sufficient to describe
the superfluid reduction of all reactions with hyperons,
since hyperon pairing is expected to be of singlet type
(Sect.\ \ref{sect-overview-struct}).

The reduction of the emissivity in the case of
both particles undergoing triplet
pairing has not yet been studied,
to the best of our knowledge.
It would also be interesting to analyze
the effects of the triplet-state superfluidity which is
a coherent superposition of states with the different
projections $m_J$ of the total angular momentum 
of the pair on the quantization axis.

We expect that the above
formalism can also be used for a qualitative
description of the superfluid reduction of the leading
direct-Urca-type reactions in the hypothetic exotic
pion condensed or kaon condensed matter
(Table \ref{tab-nucore-exotica}, Sect.\ \ref{sect-nucore-exotica}).
In these cases we recommend to use the factor $R^{\rm (D)}_{\rm AA}$,
given by Eq.\ (\ref{nusup-RaaFit}) as a function
of the superfluid gap parameters, $v_1$ and $v_2$, and calculate
these parameters for the appropriate critical temperatures
of the quasiparticle species $\tilde{n}$ and $\tilde{p}$,
provided the superfluidities of these quasiparticles can be regarded
as independent. This would reflect one of the similarity
criteria formulated in Sect.\ \ref{sect-nusup-similar}.

In addition, we expect that similar formalism can be
used to study direct Urca processes in quark
matter (Sect.\ \ref{sect-nucore-exotica}) with superfluidity
of like and unlike quarks (Sect.\ \ref{sect-overview-struct}).
In the latter case the critical temperature may be so high,
$T_c \lesssim 50$ MeV, that the superfluidity reduces the direct
Urca emissivity to negligibly small level.

  \subsection{Modified Urca processes in superfluid matter}
\label{sect-nusup-Murca}

In this section we analyze superfluid reduction
of the modified Urca processes [processes (II) from
Table \ref{tab-nucore-list}, Sect.\ \ref{sect-nucore-Murca}].
As an example, we consider the nucleon modified
Urca processes involving electrons,
Eqs.\ (\ref{nucore-Murcan}) and (\ref{nucore-Murcap}), but 
our analysis will actually be more general (see below).
Like for the direct Urca process, it
is based on the phase-space decomposition and therefore
is rather insensitive to the details of the strong interaction model.
We adopt the traditional assumption
that proton superfluidity is of type A,
while neutron superfluidity is of type B.
Our results will also be valid for neutron pairing of type A
(at $\rho \lesssim \rho_0$), as will be mentioned later.
Reduction of modified Urca reactions by neutron pairing of type
C has not been considered so far.  We follow the consideration
of Yakovlev and Levenfish (1995).\\

{\bf (a) Reduction factors}

Reduction factors $R^{({\rm M}n)}$ and
$R^{({\rm M}p)}$ of the neutron and proton branches
of modified Urca process can be derived
in the same manner as for the direct Urca process
(Sect.\ \ref{sect-nusup-Durca}) from
the phase-space decomposition. In analogy with the direct
Urca process, we include the effects of superfluidity
by introducing the superfluid gaps into the nucleon
dispersion relations.
Generally, the reduction factors
are given by
\begin{equation}
   R^{({\rm M}\!N)} = J_N / (I_{N0} A_{N0}),
\label{nusup-MurcaReduc}
\end{equation}
where $I_{N0}$ and $A_{N0}$ are the energy and angular
integrals in the non-superfluid matter, given by Eqs.\
(\ref{nucore-A}), (\ref{nucore-I}) and (\ref{nucore-Anp}),
and
\begin{equation}
   J_N   =  4\pi \int \prod_{j=1}^5 {\rm d} \Omega_j \;
          \int_0^{\infty} {\rm d} x_\nu \, x_\nu^3
          \left[ \prod_{j=1}^5 \int_{-\infty}^{+\infty}
          {\rm d} x_j \, f(z_j) \right]
         \delta \left( x_\nu - \sum_{j=1}^5 z_j \right)
          \delta \left( \sum_{j=1}^5 {\bf{p}}_j \right),
\label{nusup-JN}
\end{equation}
where $z_j$ is the same as in Eq.\ (\ref{nusup-JDef}) ($j \leq 4$),
and $z_5 =x_5 = x_e$ for the electron.

Equation (\ref{nusup-JN}) enables one to calculate the reduction factors
$R^{({\rm M}n)}$ and $R^{({\rm M}p)}$ as function of $T$,
$T_{cn}$ and $T_{cp}$.
Below we present the results
for proton superfluidity of type A and
normal neutrons, as well as for
neutron superfluidity of type B and normal protons.
The behavior of $R^{({\rm M}n)}$ and $R^{({\rm M}p)}$
under the combined effect of the $n$ and $p$ superfluids
is discussed in Sect.\ \ref{sect-nusup-similar}.\\

{\bf (b) Singlet-state proton pairing}

Since the singlet-state gap is isotropic, the angular
and energy integrations in Eq.\ (\ref{nusup-JN})
are decomposed and the angular integration remains the same
as in the non-superfluid matter.

Just as in the case of direct Urca process,
the reduction factors of the neutron
and proton branches of modified Urca process,
$R_{p{\rm A}}^{({\rm M}n)}$ and $R_{p{\rm A}}^{({\rm M}p)}$,
can be expressed in terms of
the dimensionless energy gap
$v_p= v_{\rm A}=v$; hereafter
the subscripts in
$R^{({\rm M}\!N)}$ indicate the superfluid particle species and
superfluidity type.  Clearly,
$R_{p{\rm A}}^{({\rm M}n)}  =   1 $  and
$R_{p{\rm A}}^{({\rm M}p)}  =   1 $
for $T \ge T_{cp}$ ($v_{\rm A}  =   0$).
In the case of strong superfluidity
($T \ll T_{cp}$, $v_{\rm A}   \to   \infty$),
the asymptotes of the two factors
are obtained in the same manner as in Sect.\ \ref{sect-nusup-Durca}
%
\begin{eqnarray}
 R_{p{\rm A}}^{({\rm M}n)} &=&  {72 \sqrt{2 \pi} \over 11513 \pi^8} \,
                               v^{7.5} \exp (-v)
           =  {1.166  \times 10^{-4} \over \tau^{7.5}}
            \exp \left(- {1.764 \over \tau} \right) \, ,
\label{nusup-Rn1asy} \\
    R_{p{\rm A}}^{({\rm M}p)} &=& {120960 \over 11513 \pi^8} \,
       \xi \, v^7 \exp(-2v)
    = {0.00764 \over \tau^7}
    \exp \left( - {3.528 \over \tau} \right),
\label{nusup-Rp1asy}
\end{eqnarray}
where $\tau=T/T_{cp}$ and $\xi=0.130$.

Yakovlev and Levenfish (1995) calculated
$R_{p{\rm A}}^{({\rm M}n)}$ and $R_{p{\rm A}}^{({\rm M}p)}$
for intermediate $v$ and fitted the results by the analytic expressions,
which reproduced the asymptotes
in the limit of $v \to \infty$ and obeyed the condition
$R^{({\rm M}\!N)}(0)=1$:
\begin{eqnarray}
    R_{p{\rm A}}^{({\rm M}n)} \! &=& \! { a^{7.5} + b^{5.5} \over 2}
            \exp \left( 3.4370 - \sqrt{ (3.4370)^2 + v^2} \, \right),
\label{nusup-Rn1fit} \\
   a \! &=& \! 0.1477 + \sqrt{ (0.8523)^2 + (0.1175 \, v)^2}, \quad
   b = 0.1477 + \sqrt{ (0.8523)^2 + (0.1297 \, v)^2}; \hspace{0.5cm}
\nonumber \\
   R_{p{\rm A}}^{({\rm M}p)} \! &=& \!
      \left[0.2414 \! + \!
      \sqrt{ (0.7586)^2 \! + \! (0.1318 \, v)^2} \right]^7
      \exp \left( 5.339 \! - \! \sqrt{ (5.339)^2 + (2v)^2} \, \right) .
\label{nusup-Rp1fit}
\end{eqnarray}
Equations (\ref{nusup-Rn1fit}) and
(\ref{nusup-Rp1fit}), together with (\ref{sf-FitGaps}),
fully determine the dependence of
$R_{p{\rm A}}^{({\rm M}n)}$ and $R_{p{\rm A}}^{({\rm M}p)}$ on $\tau$.

The above results
are valid also for the singlet-state superfluidity of neutrons.
Evidently, in that case one should set
$v_n = v_{\rm A}$, and
\begin{equation}
     R_{n \rm A}^{({\rm M}p)}(v_{\rm A})
     =R_{p{\rm A}}^{({\rm M}n)}(v_{\rm A}), \quad
     R_{n \rm A}^{({\rm M}n)}(v_{\rm A})=R_{p{\rm A}}^{({\rm M}p)}(v_{\rm A}).
\label{NewR}
\end{equation}
%

Wolf (1966) 
and Itoh and Tsuneto (1972) were the first
to consider the reduction factor
$R_{n \rm A}^{({\rm M}n)}=R_{p{\rm A}}^{({\rm M}p)}$
of the neutron branch of modified Urca process
by the singlet-state neutron superfluidity.
Note that Itoh and Tsuneto analyzed only the asymptote
(\ref{nusup-Rp1asy}).
In both papers the same asymptote
(\ref{nusup-Rp1asy}) was obtained but with different
numerical factors $\xi$.
Wolf (1966) found $\xi =0.123$, while
Itoh and Tsuneto (1972)
obtained $\xi =\pi /15 \approx 0.209$.
Recently, $R_{n \rm A}^{({\rm M}n)}$ was independently calculated by
Pizzochero (1998)
under the artificial assumption that the superfluid gap
is temperature-independent.  His results can be described
by the factor $R_{n \rm A}^{({\rm M}n)}(v_{\rm A})$ given above
with $v_{\rm A}=1.764/\tau$.\\

{\bf (c) Triplet-state neutron pairing}

In this case the neutron gap is anisotropic.
The proton branch of modified Urca process is
analyzed easily since the only one superfluid particle is involved.
The expression for $R_{n{\rm B}}^{({\rm M}p)}$
reduces to a one-dimensional integral over the angle
$\vartheta_n$
(between neutron momentum and quantization axis) of factor
$R_{p{\rm A}}^{({\rm M}n)}(v)$ fitted by Eq.\ (\ref{nusup-Rn1fit}).
Argument $v_{\rm A}$ in the latter expression should formally be
replaced by $y_{\rm B} = v_{\rm B} \,(1 + 3 \, \cos^2 \vartheta)^{1/2}$
in accordance with Eq.\ (\ref{sf-y_v})
(Sect.\ \ref{sect-sf-gaps}).
It is evident that $ R_{n{\rm B}}^{({\rm M}p)}(v)=1$ for $v=v_{\rm B}=0$.
In the limit $v \to \infty$ one can use the asymptote
(\ref{nusup-Rn1asy}) in the integrand.
Then for $T \ll T_{cn}$ ($v \gg 1$),
according to Yakovlev and Levenfish (1995),
\begin{equation}
  R_{n{\rm B}}^{({\rm M}p)} = {72 \over 11513 \,
      \pi^7 \sqrt{3} }\, v^{7} \exp(-v) =
      {3.99 \times 10^{-6} \over \tau^7}
      \exp \left(- {1.188 \over \tau} \right) \, ,
\label{nusup-Rp2asy}
\end{equation}
where $\tau=T/T_{cn}$.
The same authors
calculated also $R_{n{\rm B}}^{({\rm M}p)}$
for intermediate values of $v$
and fitted the results by the analytic expression:
\begin{eqnarray}
   R_{n{\rm B}}^{({\rm M}p)} & = & { a^7 + b^5 \over 2} \exp \left(2.398 -
                \sqrt{ (2.398)^2 +v^2} \, \right),
\label{nusup-Rp2fit} \\
 a & = & 0.1612 + \sqrt{ (0.8388)^2 + (0.1117 \, v)^2},
   \quad b = 0.1612 + \sqrt{ (0.8388)^2 + (0.1274 \, v)^2}.
\nonumber
\end{eqnarray}
%

\begin{figure}[!t]
\begin{center}
\leavevmode
\epsfysize=8.5cm
\epsfbox[40 30 250 280]{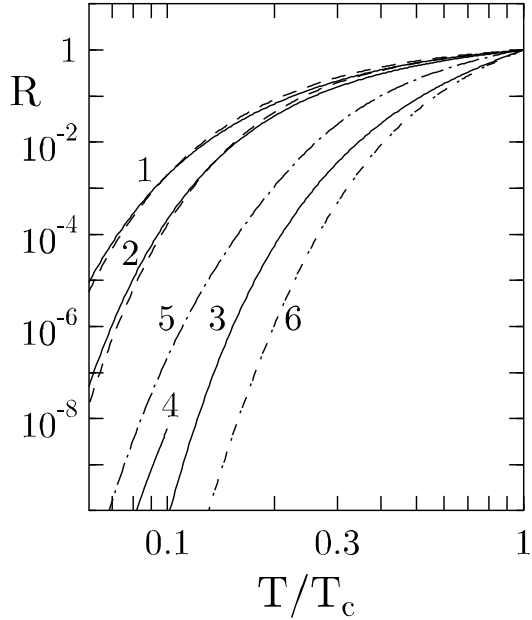}
\end{center}
\caption[]{\footnotesize
         Reduction factors of various neutrino emission
         processes by superfluidity of neutrons or protons versus $T/T_c$
         (Yakovlev et al.\ 1999b). 
         Curves 1 show reduction of the
         $p$-branch of the modified Urca process (solid line)
         and the direct Urca process (dashed line)
         by neutron superfluidity of type B.
         Curves 2 correspond to reduction of the
         $n$-branch of the modified Urca (solid line)
         and the direct Urca (dashed line) by proton
         superfluidity of type A.
         Dot-and-dash lines 5, 6 and the solid line
         3 refer to the $np$, $pp$-scattering and $p$-branch
         of the modified Urca processes, respectively,
         for the same proton superfluidity. Solid line
         4 is the asymptote of the reduction factor for
         the $n$-branch of the modified Urca process
         due to neutron superfluidity of type B.
}
\label{fig-nusup-Rstandard_tau}
\end{figure}

Exact calculation of $R_{n{\rm B}}^{({\rm M}n)}$
of the neutron branch of modified Urca process
by the triplet-state neutron superfluidity
for intermediate values of $v$ is complicated;
an approximate expression will be given in Sect.\ \ref{sect-nusup-similar}.
Here we present only the asymptote of $R_{n{\rm B}}^{({\rm M}n)}$
in the limit $\tau \ll 1$
%
\begin{equation}
  R_{n{\rm B}}^{({\rm M}n)} \! = \!
     {120960 \over 11513 \pi^8} \; {2 \over 3 \sqrt{3}} \;
      \xi  \, v^6 \exp(-2v)
    =  {1.56 \times 10^{-4} \over \tau^6}
    \exp \left( - {2.376 \over \tau} \, \right).
\label{nusup-Rn2asy}
\end{equation}
In this case, as for the proton reaction involving superfluid protons
(\ref{nusup-Rp1asy}), the effect of superfluidity is
very strong: the exponent argument in
$R_{n{\rm B}}^{({\rm M}n)}$ contains the doubled gap.
This is because the
three neutrons participating in the reaction
belong to the superfluid component of matter. Two of them have positive energies
[$z \gg 1$, $f(z) \approx {\rm e}^{-z}$] while the third one
has negative energy [$z<0$, $f(z) \approx 1$].

\begin{figure}[!t]
\begin{center}
\leavevmode
\epsfysize=8.5cm
\epsfbox[50 35 255 275]{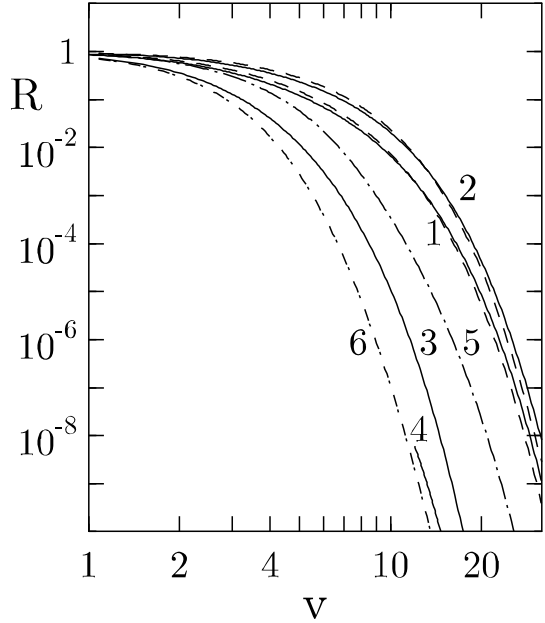}
\end{center}
\caption[]{\footnotesize
           Same as in Fig.~\protect{\ref{fig-nusup-Rstandard_tau}}, but
           versus dimensionless gap parameter
           (Yakovlev et al.\ 1999b).}  
\label{fig-nusup-Rstandard_v}
\end{figure}

The dependence of reduction factors
$R_{p{\rm A}}^{({\rm M}n)}$,
$R_{p{\rm A}}^{({\rm M}p)}$, $R_{n{\rm B}}^{({\rm M}p)}$, and
$R_{n{\rm B}}^{({\rm M}n)}$ on the dimensionless temperature
$\tau$ and the dimensionless gap parameter $v$
is plotted in Figs.\ \ref{fig-nusup-Rstandard_tau}
and \ref{fig-nusup-Rstandard_v}.
For comparison, we present also reduction factors
of direct Urca process, $R^{\rm (D)}_{\rm A}$ and $R^{\rm (D)}_{\rm B}$.
Let us mention that in the case of the strong
neutron superfluidity ($T \ll T_{cn}$) and normal protons
the proton branch of modified Urca process
becomes much more efficient than the neutron branch
(factors ${\rm e}^{-v}$ and ${\rm e}^{-2 v}$
in Eqs.\ (\ref{nusup-Rp2asy}) and (\ref{nusup-Rn2asy})).
However in this case the main neutrino emission comes from
neutrino bremsstrahlung due to $pp$-scattering,
which is not affected by the neutron superfluidity.\\

{\bf (d) Other modified Urca processes}

We have considered superfluid reduction of modified
Urca processes involving $n$, $p$, and $e$. It is clear
that the same reduction factors describe the
processes with muons [processes (II.1) with $B=n$ or $p$
from Table \ref{tab-nucore-list}]. In hyperon matter,
they are also valid for all baryon modified Urca processes
(II.1)--(II.4) which involve two types of baryons, for
instance, $\Lambda \Lambda \to p \Lambda l \bar{\nu}_l$,
$p \Lambda l \to \Lambda \Lambda \nu_l$ [process (II.2)
with $B=\Lambda$ and $l=e$ or $\mu$].

Superfluid suppression of
the processes involving three different baryons (e.g.,
$\Lambda \Sigma^- \to p \Sigma^- l \bar{\nu}_l$,
$p \Sigma^- l \to \Lambda \Sigma^- \nu_l$) has not been
considered yet. If, however, only one
beta decaying baryon belongs to the superfluid component,
the present results are valid again. For the cases
in which two (of the three) baryons are superfluid,
one can construct approximate reduction factors
using the similarity criteria formulated in Sect.\
\ref{sect-nusup-similar}. 

Finally, there are also modified Urca processes (IV) based
on muon decay. Their rate has not been determined accurately
so far, and their reduction has not been analyzed.
However, these processes produce very weak neutrino
emission. If they involve a charged baryon as a spectator,
they will be suppressed by the baryon superfluidity
and will have negligible neutrino emissivity. If they involve
an additional lepton (electron or muon), they will
not be suppressed at all although they may be slightly
affected by the baryon superfluidity through the
plasma screening length. Similar effects will be described
in the next section with regard to neutrino bremsstrahlung
in $ee$ and $ep$ collisions.

\subsection{Neutrino bremsstrahlung in superfluid matter}
\label{sect-nusup-Brems}

In this section we analyze
the effect of superfluidity on neutrino generation
in baryon brems\-strah\-lung processes [reactions
(III) from Table \ref{tab-nucore-list}, Sect.\ \ref{sect-nucore-Brems}] and
Coulomb bremsstrahlung processes
[reactions (V), Sect.\ \ref{sect-nucore-other}].\\

{\bf (a) Nucleon-nucleon bremsstrahlung}

We start with neutrino-pair bremsstrahlung
in $nn$, $np$ and $pp$-collisions [reactions (III.1)--(III.3),
Table \ref{tab-nucore-list} and Sect.\ \ref{sect-nucore-Brems}] following
the approach of Yakovlev and Levenfish (1995).
The general expression for reduction factors
$R^{(N\!N)}$ can be obtained in analogy with Urca reactions
from phase-space consideration by inserting
the superfluid gap in the nucleon dispersion relation.

Consider singlet superfluid
A of neutrons or protons. Then the modification of the
dispersion relation by superfluidity affects only
the energy integral given by Eq.\ (\ref{nucore-bremas-I}).
Accordingly the reduction factors take the form:
\begin{equation}
   R^{(N\!N)}  =  {945 \over 164 \pi^8}
       \int_0^\infty {\rm d} x_\nu \; x_\nu^4
       \left[ \prod_{j=1}^4 \int_{-\infty}^{+\infty}
       {\rm d} x_j \; f(z_j) \right]
       \delta \left( \sum_{j=1}^4 z_j - x_\nu \right).
\label{nusup-RNN}
\end{equation}
It is clear that $R^{(N\!N)} = 1$ for $\tau \ge 1$.

In the case of proton superfluid A
suppression factors of 
$np$ and $pp$ processes,
$R^{\,(np)}_{p{\rm A}}$ and $ R^{\,(pp)}_{p{\rm A}}$, are
reduced to two dimensional integrals which have been calculated
numerically.
In the limit of strong superfluidity
($\tau \ll 1$, $v=v_{\rm A} \to \infty$) the asymptotes of these factors are:
\begin{eqnarray}
     R^{\,(np)}_{p{\rm A}} & = &
       {945 \over 164 \pi^8} \, \xi_1 \, v \exp(-v)
       =
       {0.910 \over \tau} \exp \left( - { 1.764 \over \tau} \right),
\label{nusup-Rnpasy} \\
     R^{\,(pp)}_{p{\rm A}} &=& { 8505 \over 41 \pi^6} \;
       v^2 \exp(-2v)
       =
       {0.671 \over \tau^2} \exp \left( - {3.528 \over \tau} \right),
\label{nusup-Rppasy}
\end{eqnarray}
where $\xi_1 \approx 849$.
These asymptotes as well as
numerical values of
$R^{\,(np)}_{p{\rm A}}$ and  $R^{\,(pp)}_{p{\rm A}}$
calculated for intermediate
values of $v$ are described by the fits:
\begin{eqnarray}
      R^{\,(np)}_{p{\rm A}} & = & {1 \over 2.732}
      \left[ \, a \exp \left( 1.306 -
      \sqrt{ (1.306)^2 + v^2} \right) \right.
\nonumber \\
&&     \left. + \; 1.732 \, b^7
      \exp \left( 3.303 - \sqrt{(3.303)^2 + 4v^2} \right) \right],
\label{nusup-Rnpfit} \\
 a& = & 0.9982 + \sqrt{(0.0018)^2 + (0.3815 \, v)^2}, \quad
      b = 0.3949 + \sqrt{(0.6051)^2 + (0.2666 \, v)^2} \, ;
\nonumber
\end{eqnarray}
and
\begin{eqnarray}
    R^{\,(pp)}_{p{\rm A}} & = & {1 \over 2}
      \left[ \, c^2
      \exp \left( 4.228 - \sqrt{ (4.228)^2 + (2 v)^2} \right) \right.
\nonumber \\
&&    +  \left. d^{7.5} \exp
      \left( 7.762 - \sqrt{ (7.762)^2 + (3v)^2} \right) \right],
\label{nusup-Rppfit} \\
    c & = & 0.1747 + \sqrt{ (0.8253)^2 + (0.07933 \, v)^2}, \quad
      d  =  0.7333 + \sqrt{ (0.2667)^2 + (0.1678 \, v)^2}.
\nonumber
\end{eqnarray}
For the singlet-state neutron pairing, we evidently have
\begin{equation}
  R^{\,(np)}_{n \rm A}(v_{\rm A})=  R^{\,(np)}_{p \rm A} (v_{\rm A}), \quad
  R^{\,(nn)}_{n \rm A}(v_{\rm A})=  R^{\,(pp)}_{p \rm A}(v_{\rm A}).
\label{nusup-New1}
\end{equation}

Reduction factors $R^{\,(np)}_{p \rm A}$
and  $R^{\,(pp)}_{p \rm A}$ are shown in
Fig.\ \ref{fig-nusup-Rstandard_tau} versus
$T/T_c$ and in Fig.~\ref{fig-nusup-Rstandard_v} versus
the dimensionless gap parameter $v$. We will discuss these
figures and the case of the triplet-state pairing in the next section.\\

{\bf (b) Hyperon bremsstrahlung processes}

The results obtained above are independent
of the model of strong interaction and can be
used not only for nucleon-nucleon bremsstrahlung
reactions (III.1)--(III.3) from Table \ref{tab-nucore-list} 
but for all baryon
bremsstrahlung reactions (III.1)--(III.10) which involve baryons
of one or two types. Reduction of reactions
(III.11) and (III.12) involving
baryons of four different types has not been
considered yet. If, however, only one or two of the
four particles belong to superfluid component, one can
construct approximate reduction factors using
the similarity criteria formulated in Sect.\
\ref{sect-nusup-similar}.\\

{\bf (c) Coulomb bremsstrahlung processes}

For non-superfluid matter, these processes 
[processes (V), Table \ref{tab-nucore-list}]
have been analyzed in Sect.\ \ref{sect-nucore-other}.
Strictly speaking, all these reactions are affected by
superfluidity, and the effect is {\it twofold}.
The {\it first}, most significant effect is that the
reaction is suppressed by superfluidity of reacting baryons.
This effect can be described introducing
the superfluid reduction factor in accordance
with Eq.\ (\ref{nusup-reduct}). {\it Second},  superfluidity
of charged baryons affects the plasma screening momentum $q_s$,
given by Eq.\ (\ref{nucore-other-ys}), and hence
the neutrino emissivity $Q_0$ itself in Eq.\ (\ref{nusup-reduct}).
The second effect is much weaker than the first one,
but nevertheless noticeable.
One can easily show that in superfluid (superconducting)
matter, instead of Eq.\ (\ref{nucore-other-ys})
we have
\begin{equation}
      y_s^2 \equiv { q_s^2 \over 4 p_{{\rm F}e}^2}=
       { e^2 \over \pi \hbar c} \,
      \sum_j {m_j^\ast p_{{\rm F}j}
      \over m_e^\ast p_{{\rm F}e}} \; D_j,
\label{nusup-brems-ys}
\end{equation}
where summation is over all charged fermions $j$
in stellar matter, and $D_j$ describes superfluid
reduction of corresponding plasma screening.
For leptons ($e$ or $\mu$) and for charged baryons $j$ at
$T \geq T_{cj}$ this reduction is absent and $D_j=1$.
For singlet-state
pairing of charged baryons at $T<T_{cj}$ reduction factor $D_j(v_j)$
is given by 
(e.g., Gnedin and Yakovlev 1995)
\begin{equation}
     D= \left(0.9443 + \sqrt{(0.0557)^2 + (0.1886 \, v)^2} \right)^{1/2}
     \exp \left(1.753 - \sqrt{(1.753)^2 + v^2} \right),
\label{nusup-brems-D}
\end{equation}
where $v$ is the dimensionless gap. In the presence
of strong superfluidity ($v \gg 1$) the reduction
is exponential, and the charged 
superfluid (superconducting) baryons of species $j$
produce no screening decreasing thus $y_s$.
However, this decrease is not crucial since there are
always electrons (and possibly muons) which are
non-superfluid and determine the residual
screening momentum in a strongly superconducting 
matter.

Consider, for instance, neutrino bremsstrahlung
in $ep$ collisions [process (V.1) 
from Table \ref{tab-nucore-list} with $l=e$]
discussed in Sect.\ \ref{sect-nucore-other} for non-superfluid matter.
In the presence of superfluidity, its emissivity $Q^{(ep)}$
is given by Eq.\ (\ref{nusup-reduct}),
$Q^{(ep)}=Q^{(ep)}_0 \, R^{(ep)}$, where $Q^{(ep)}_0$
is formally given by the same Eq.\ (\ref{nucore-other-ep})
as in the non-superfluid matter but with the parameter $y_s$
affected by the superfluidity and determined by Eqs.\
(\ref{nusup-brems-ys}) and (\ref{nusup-brems-D}).
Reduction factor $R^{(ep)}$
for this process can be found using phase space
decomposition; it coincides (Kaminker and Haensel 1999)
with factor $R^{(np)}_{p{\rm A}}$ for $np$ bremsstrahlung,
Eq.\ (\ref{nusup-Rnpfit}).
For instance, in non-superfluid
$npe$ matter the main contribution into screening parameter $y_s$ comes
from the protons, while the contribution from the electrons is several
times smaller. When temperature $T$ decreases below $T_{cp}$
the proton superfluidity enhances
emissivity $Q^{(ep)}_0$ by a factor of several,
with respect to its purely non-superfluid value by
decreasing $y_s$, but reduces full emissivity
$Q^{(ep)}$ much stronger owing to the exponential superfluid reduction
described by factor $R^{(ep)}$. Similar equations
describe the neutrino emissivities due to Coulomb collisions
of electrons or muons with all charged baryons
[all processes (V.1)--(V.2)].

Another example is neutrino bremsstrahlung
due to collisions between leptons [processes (V.3)]
which experience no direct superfluid reduction. For instance,
consider bremsstrahlung in $ee$ collisions
[process (V.3) with $l=e$] which always operates
the neutron star cores.
Without superfluidity, it has been discussed in
Sect.\ \ref{sect-nucore-other}. Its emissivity
$Q^{(ee)}$ is given by Eq.\ (\ref{nucore-other-ee})
and contains the same screening parameter $y_s$.
If the baryon superfluidity is switched on,
the same equation remains true ($R^{(ee)} \equiv 1$)
but the parameter $y_s$ becomes reduced
in accordance with Eq.\ (\ref{nusup-brems-ys}).
This effect enhances $Q^{(ee)}$ over
its purely non-superfluid value 
by a factor of several as discussed above. Therefore,
the emissivity of lepton-lepton bremsstrahlung
processes shows no superfluid reduction. Although
these neutrino reactions are really weak
(Table \ref{tab-nucore-list}) they can
be leading neutrino processes in highly superfluid
neutron star cores (Sects.\ \ref{sect-nusup-all} and \ref{sect-nusup-fluxa}).

The latter statement can be especially true in
strange quark matter with pairing of unlike quarks
(Sect.\ \ref{sect-overview-struct}). This pairing may
be very strong to lock all the neutrino processes
involving quarks. However, strange quark matter
is known to contain a small admixture of electrons.
Their neutrino bremsstrahlung would dominate. 

\subsection{Similarity criteria}
\label{sect-nusup-similar}

As we have seen in Chapt.\ \ref{chapt-nucore}
the number of neutrino reactions is large. 
Calculation of superfluid reduction factors for all reactions
is naturally complicated.
However, if a high accuracy
is not needed, one can avoid the calculation by noticing
similarity of the reduction factors for different reactions
(Yakovlev and Levenfish 1995, Levenfish and Yakovlev 1996).

Let us demonstrate these {\it similarity criteria} for
neutrino reaction in $npe$ matter, although
analogous formulae can be proposed for hyperonic matter
containing muons.

As seen from Figs.\ \ref{fig-nusup-Rstandard_tau}
and \ref{fig-nusup-Rstandard_v},
the reduction factors of various processes
in $npe$ matter are different. The main difference comes
from the exponent arguments
in the asymptotes of the reduction factors
in the limit of strong superfluidity. An examination
of the asymptotes presented in Sects.\
\ref{sect-nusup-Durca} -- \ref{sect-nusup-Brems}
reveals that the exponent argument
contains a single gap if one or two reacting particles
belong to superfluid component of matter.
Under the {\it number of superfluid reacting particles} we mean
{\it the total number of particles} (in the initial and final states
of a bremsstrahlung process, in the initial and final states
of either direct or inverse reaction of an Urca process)
belonging to the superfluid components.
For instance, the gap in the exponent argument
is doubled if three or four superfluid particles are
involved. According to
Fig.\ \ref{fig-nusup-Rstandard_v}, factor
$R^{\,(np)}_{p{\rm A}}$ (two superfluid particles)
decreases with the growth of superfluidity
strength $v$
much more rapidly than 
$R_{p{\rm A}}^{({\rm M}n)}$
or $R_{n{\rm B}}^{({\rm M}p)}$ (one superfluid particle).
Accordingly, $R^{\,(pp)}_{p{\rm A}}$ (four superfluid particles)
decreases faster than
$R_{p{\rm A}}^{({\rm M}p)}$ (three superfluid particles).
Therefore, {\it superfluid suppression is most strongly
affected by the number of superfluid reacting particles}.
Let us stress that we discuss neutrino reactions of
Urca or bremsstrahlung type.
The formulated rule must be modified for 
neutrino emission due to Cooper pairing of nucleons
(Sect.\ \ref{sect-nusup-CP}).

At the next step it is important to emphasize
that the reduction factors
in $npe$ matter
for the processes involving
one superfluid particle
($R^{\rm (D)}_{\rm A}$,
$R_{p{\rm A}}^{({\rm M}n)}$, $R^{\rm (D)}_{\rm B}$,
and $R_{n{\rm B}}^{({\rm M}p)}$)
{\it are close to one another as functions of the dimensionless
gap parameter} $v$ (Fig.\ \ref{fig-nusup-Rstandard_v}).
This enabled
Levenfish and Yakovlev (1996)
to construct approximate
reduction factors for the proton and neutron branches
of modified Urca process in the presence of the neutron and proton
superfluids:
\begin{eqnarray}
R^{({\rm M}p)}_{\rm BA}(v_n,v_p) &\approx&
                    { R^{\rm (D)}_{\rm BA}(v_n,2v_p) \over
          R^{\rm (D)}_{\rm B}(v_n) } \, R_{n{\rm B}}^{({\rm M}p)}(v_n) \, ,
\label{nusup-RpAB} \\
R^{({\rm M}n)}_{\rm BA}(v_n,v_p) &\approx&
          { R^{\rm (D)}_{\rm BA}(2v_n,v_p) \over
         R^{\rm (D)}_{\rm A}(v_p) } \, R_{p{\rm A}}^{({\rm M}n)}(v_p).
\label{nusup-RnAB}
\end{eqnarray}
We expect that these factors, as functions of corresponding
parameters $v$
(corrected due to the number of superfluid particles)
do not differ strongly from reduction factor
$R^{\rm (D)}_{\rm BA}(v_n,v_p)$ for the direct Urca process.
If protons are normal ($v_p=0$), then the expression for
$R^{({\rm M}p)}_{\rm BA} (v_n,v_p)$ becomes exact; if neutron are normal
($v_n=0$), factor $R^{({\rm M}n)}_{\rm BA} (v_n,v_p)$ is exact.
In addition, the approximate factors satisfy the relationships
analogous to Eq.\ (\ref{nusup-Estimation}).

Therefore, new reduction factors $R^{({\rm M}p)}_{\rm BA}(v_n,v_p)$
and $R^{({\rm M}n)}_{\rm BA}(v_n,v_p)$ are expressed through
the factors obtained in Sects.\ \ref{sect-nusup-Durca}
and \ref{sect-nusup-Murca}.

One can also expect similarity
of reduction factors for other neutrino reactions.
For instance, reduction factor 
of the neutron branch of modified Urca
process, $R_{n{\rm B}}^{({\rm M}n)}$, by a moderate neutron superfluidity
($v \lesssim10$) should not deviate strongly from
reduction factor of the proton branch of modified Urca process,
$R_{p{\rm A}}^{({\rm M}p)}$, by the proton superfluidity:
\begin{equation}
    R_{n{\rm B}}^{({\rm M}n)} \approx R_{p{\rm A}}^{({\rm M}p)}(v_n).
\label{nusup-RmnnB}
\end{equation}
Approximate reduction factor
$R^{\,(nn)}_{n{\rm B}}$ of $nn$ scattering
in the presence of $n$ superfluidity
and approximate reduction factor
$R^{\,(np)}_{\rm BA}$ of 
$np$ scattering by neutron and proton superfluidities
can be written as
\begin{eqnarray}
     R^{\,(nn)}_{n{\rm B}} &\approx& R^{\,(pp)}_{p{\rm A}}(v_n)\, ,
\label{nusup-Rnn_B} \\
     R^{\,(np)}_{\rm BA} &\approx&  {R^{\rm (D)}_{\rm BA}(v_n,v_p) \over
     R^{\rm (D)}_{\rm A}(v_p) } \, R^{\,(np)}_{p{\rm A}} (v_p) \, .
\label{nusup-Rnp_BA}
\end{eqnarray}
In the absence of the neutron superfluidity, Eq.\ (\ref{nusup-Rnp_BA})
becomes exact.

We can warn the reader once more that the similarity criteria are
not exact and it would be interesting to 
calculate accurately all reduction
factors in the future. Nevertheless
it is worth to mention that one does not need very
accurate values of the neutrino emissivity for simulations
of neutron star cooling: some theoretical
uncertainties in the emissivity are easily compensated by
very small variations of the temperature due to strong temperature
dependence of the emissivity.

\subsection{Neutrino emission due to Cooper pairing of baryons}
\label{sect-nusup-CP}

In contrast to the neutrino emission processes considered
above this process is allowed only in the presence
of superfluidity:
the superfluidity distorts the baryon dispersion relation
near the Fermi surface and opens the reaction
\begin{equation}
    B  \to B + \nu + \bar{\nu}  \, ,
\label{nusup-CP-Recomb}
\end{equation}
where $B$ stands for any baryon (nucleon or hyperon)
which belongs to superfluid component of matter.
This reaction is forbidden
without the superfluid gap by energy-momentum conservation.
Thus the process consists in emission of a neutrino pair
by a baryon whose energy spectrum contains
a gap. However, in theoretical studies, it is
convenient to use the formalism of quasi-particles
(Sect.\ \ref{sect-sf-gaps})
and treat the process
(Flowers et al.\  1976)
as annihilation of two quasi-baryons
$\widetilde{B}$ with nearly antiparallel momenta 
into a neutrino pair (which may be considered as 
Cooper pair formation):
\begin{equation}
     \widetilde{B} + \widetilde{B}  \to \nu + \bar{\nu}.
\label{nusup-CP-Rec}
\end{equation}
The reaction is described by the simplest one-vortex four-tail
Feynman diagram; it goes via weak neutral currents and
produces neutrinos of all flavors.

To be specific, we discuss the reaction involving
nucleons (neutrons or protons) and mention 
hyperonic reactions later.
Following Yakovlev et al.\ (1999a)
we outline derivation of the
neutrino emissivity due to singlet-state or triplet-state
pairing of non-relativistic nucleons. The reaction is described
by the Hamiltonian 
similar to that for ordinary beta decay [Eq.\ (\ref{nucore-H}),
Sect.\ \ref{sect-nucore-beta}]
\begin{equation}
     \hat{H} = -{ G_{\rm F} \over 2 \sqrt{2} } \,
     \left( c_V \, J_0 l_0 - c_A \, {\bf{J}} {\bf{l}} \right),
\label{nusup-CP-H}
\end{equation}
where $c_V$ and $c_A$ are, respectively, the vector and axial-vector
constants of neutral hadron currents.
For the neutron current, we have
(see, e.g.,
Okun' 1984) 
$c_V=1$, $c_A=g_A=1.26$, while for the proton current
$c_V=4 \, \sin^2 \Theta_{\rm W} -1 \approx -0.08$, $c_A=-g_A$,
$\Theta_{\rm W}$ being the Weinberg angle ($\sin^2 \Theta_{\rm W} = 0.23$).
Strong difference of $c_V$ for neutrons and protons
comes from different quark structure of these particles.
Furthermore,
\begin{equation}
      J_0 \equiv \hat{\Psi}^\dagger \hat{\Psi}, \quad
      {\bf J} \equiv  -  \hat{\Psi}^\dagger {\bf \sigma } \hat{\Psi},
    \quad
    l^\alpha = \overline{\psi}_\nu \gamma^\alpha (1 + \gamma^5) \psi_\nu,
\label{nusup-CP-l}
\end{equation}
$l^\alpha$ being the 4-vector  of
the neutrino current
($\alpha$=0,1,2,3);
other notations are the same as in Sect.\ \ref{sect-nucore-beta}.
In the present discussion, as in Sect.\ \ref{sect-nusup-Durca},
we use second--quantized quasi-nucleon wave function
$\hat{\Psi}$ described in Sect.\ \ref{sect-sf-gaps},
Eq.\ (\ref{sf-CP-Psi}).

Let $p_\nu = (\epsilon_\nu, {\bf{p}}_\nu)$ and
$p'_\nu = (\epsilon'_\nu, {\bf{p}}'_\nu)$ be 4-momenta of
neutrino and antineutrino, respectively,
while
$p=(\epsilon,{\bf{p}})$ and $p'=(\epsilon', {\bf{p}}')$ be
4-momenta of annihilating quasi-nucleons.
Using Golden Rule of quantum mechanics, we can present
the neutrino emissivity due to Cooper pairing
(CP) as:
\begin{eqnarray}
   Q^{\rm (CP)} & = & \left( {G_{\rm F} \over 2 \sqrt 2} \right)^2 \,
       {1 \over 2} \, {\cal N}_{\nu} \,
       \int { {\rm d}  {\bf{p}} \over (2 \pi)^3 } \;
       {{\rm d} {\bf{p}}' \over (2 \pi)^3 } \;
       f(\epsilon)f(\epsilon ')
\nonumber \\
     &  & \times 
     \int { {\rm d} {\bf{p}}_\nu \over 2 \epsilon_\nu (2 \pi)^3 } \;
     { {\rm d} {\bf{p}}'_\nu \over 2 \epsilon'_\nu (2 \pi)^3 } \;
     \left[c_V^2 {\cal J}_{00} |l_0|^2 + c_A^2 \,
     {\cal J}_{jk} l_j l_k^\ast \right]
\nonumber \\
     &  & \times 
     (2 \pi)^4 \, \delta^{(4)} \left( p + p' - p_\nu - p'_\nu \right) \,
     (\epsilon_\nu + \epsilon'_\nu),
\label{nusup-CP-Qgen}
\end{eqnarray}
where ${\cal N}_{\nu}$=3 is the number of neutrino flavors, and
the factor $1/2$ before ${\cal N}_{\nu}$ excludes double counting
of the same quasi-nucleon collisions. The integral is taken over
the range
$(p_\nu + p'_\nu)^2 >0$, where the process is open kinematically;
$f(\epsilon)=1/[\exp(\epsilon/T)+1]$, $j,k\, = \, 1,2,3$;
\begin{equation}
   {\cal J}_{00} = \sum_{\eta\eta'} \,
   |(\hat{\Psi}^+ \hat{\Psi})_{fi} |^2,\quad
   {\cal J}_{jk} = \sum_{\eta\eta'} \,
   (\hat{\Psi}^+ \sigma_j \hat{\Psi})_{fi} \,
   (\hat{\Psi}^+ \sigma_k \hat{\Psi})_{fi}^\ast.
\label{nusup-CP-tensor}
\end{equation}
Here $i$ stands for an initial state of the quasi-particle
system in which the one-particle states
$({\bf{p}},\eta)$ and $({\bf{p}}',\eta')$ are occupied, and
$f$ stands for a final state of the system in which
the indicated one-particle states are empty.

Integral (\ref{nusup-CP-Qgen}) is simplified 
using the standard technique.
The simplifications imply that nucleons are
non-relativistic and strongly degenerate as well as 
that the process is open kinematically in a small domain of momentum space,
where the quasi-nucleon momenta
$\bf{p}$ and ${\bf{p}}'$ are almost antiparallel.
The latter circumstance allows one to take smooth functions
${\cal J}_{00} ({\bf{p}}, {\bf{p}}')$ and
${\cal J}_{jk}({\bf{p}}, {\bf{p}}')$
out of the integral over
${\rm d} {\bf{p}}'$ putting ${\bf{p}}' = - {\bf{p}}$.
After some transformations
the final expression for the neutrino emissivity
can be written
(in standard physical units)
as (Yakovlev et al.\  1998, 1999a)
\begin{eqnarray}
 Q^{\rm (CP)} & = & {4 G_{\rm F}^2
           m_N^\ast p_{\rm F} \over 15 \pi^5 \hbar^{10}
           c^6} \, (k_{\rm B} T)^7 \, {\cal N}_{\nu} \; a F(v) =
\nonumber \\
     & = & 1.170 \times 10^{21} \, \left( { m_N^\ast \over m_N } \right)
     \left( { p_{\rm F} \over m_N c } \right) \, T_9^7 \, {\cal N}_\nu \;
     aF(v) \; \; {\rm erg \;cm^{-3} \; s^{-1}},
\label{nusup-CP-Qrec}
\end{eqnarray}
where $a$ is a numerical factor (see below), and the function
$F(v)$, in our standard notations (\ref{sf-DimLessVar}), is given
by the integral
\begin{equation}
     F(v) = {1 \over 4 \pi} \, \int {\rm d}\Omega \, y^2
     \int_0^\infty \, {z^4 \, {\rm d}x \over ({\rm e}^z +1)^2},
\label{nusup-CP-F}
\end{equation}
$v$ being the gap parameter, Eq.\ (\ref{sf-DefGap}).
The singlet-state gap is isotropic; thus, integration
over all orientations of the nucleon momentum
at the Fermi surface (over ${\rm d} \Omega$)
is trivial and gives $4 \pi$.
In the triplet case,  
$F(v)$ contains averaging over positions of a quasi-nucleon on the
Fermi surface. Using Eq.\ (\ref{nusup-CP-Qrec}),
one can take into account that
$p_{\rm F}/(m_N c) \approx 0.353 \, (n_N/n_0)^{1/3}$,
where $n_N$ is the number density of nucleons $N$,
$n_0 = 0.16$~fm$^{-3}$.

The emissivity
$Q^{\rm (CP)}$ depends on superfluidity type through
factor $a$ and function $F(v)$.
For the singlet-state neutron pairing,
$a$ is determined by the only vector constant
$c_V$: $a_{n \rm A}=c_V^2=1$.
If we used similar expression for the singlet-state pairing
of protons, we would obtain a very small factor
$a_{p \rm A}=0.0064$, which mean weakness of neutrino emission.
Under these conditions, one should take into account
relativistic correction to
$a$, produced by the axial-vector proton current.
Calculating and adding this correction for the singlet-state pairing
of protons, Kaminker et al.\ (1999b) obtained
%
\begin{equation}
a_{p \rm A} = c_V^2+ c_A^2 \left( v_{{\rm F}p} \over c \right)^2
         \left[ \left(\frac{m_p^\ast}{m_p} \right)^2 +\frac{11}{42} \right]
       =
         0.0064 + 1.59 \left( v_{{\rm F}p} \over c \right)^2
   \left[ \left(\frac{m_p^\ast}{m_p} \right)^2 +\frac{11}{42} \right],
\label{nusup-CP-AP}
\end{equation}
where $v_{{\rm F}p}/c = p_{{\rm F}p}/(m_p^\ast c)$.
The relativistic correction appears to be about 10 -- 50 times
larger than the main non-relativistic term.
This enhances noticeably neutrino emission
produced by singlet-state proton pairing although
it remains much weaker than the emission due to the neutron pairing.

In the case of the triplet-state pairing,
$a$ is determined by both, the vector and the axial-vector,
constants of neutral hadron currents:
$a=a_{N \rm B}=a_{N \rm C}=c_V^2+2 c_A^2$
(Yakovlev et al.\ 1999a).
For neutron pairing, we obtain $a_{n \rm B}=a_{n \rm C}=4.17$.
Notice that in a not very realistic case of triplet pairing of protons
we would obtain $a_{p \rm B}=a_{p \rm C}=3.18$.
Under such exotic conditions, neutrino emission
due to proton pairing would be almost as efficient as
the emission due to neutron pairing.

The result for the singlet-state pairing of neutrons,
presented above, coincides with that obtained
in the pioneering paper by Flowers et al.\ (1976)
(for two neutrino flavors, ${\cal N}_\nu = 2$).
Similar expressions obtained by
Voskresensky and Senatorov (1986, 1987)
for ${\cal N}_{\nu} = 1$ contain an extra factor $(1+3g_A^2)$.
In addition, the expression for
$Q^{\rm (CP)}$
derived by
the latter authors
contains a misprint:
$\pi^2$ in the denominator instead of $\pi^5$,
although numerical estimate of
$Q^{\rm (CP)}$ is obtained with the correct factor $\pi^5$.
The cases of singlet proton pairing and
triplet neutron pairing
were analyzed, for the first time, by
Yakovlev et al.\ (1998, 1999a) and Kaminker et al.\ (1999b).

Let us remind that the values of
$c_V$, $c_A$ and $a$ can be renormalized in dense
matter due to in-medium effects. The renormalization
is a difficult task which we neglect.
However one should bear in mind that
the medium effects may be stronger than the relativistic
correction to the proton pairing and determine
actually $a_{p \rm A}$.

Function $F(v)$, given by Eq.\ (\ref{nusup-CP-F}),
depends on the only argument $v$, the gap parameter. Using Eq.\
(\ref{nusup-CP-F}) one can easily obtain the asymptote of this function
and calculate its dependence on
$\tau=T/T_c$ for superfluids
A, B and C in analogy with calculations presented in
Sects.\ \ref{sect-nusup-Durca}--\ref{sect-nusup-Brems}.
This was done by
Yakovlev et al.\ (1998, 1999a).

In a small vicinity of $T \approx T_c$,
in which $v \ll 1$ and $\tau \to 1$, these authors obtained:
\begin{eqnarray}
   F_{\rm A}(v) & = & 0.602 \, v^2 = 5.65 \, (1-\tau),
\nonumber \\
   F_{\rm B}(v) & = & 1.204 \, v^2 = 4.71 \, (1-\tau),
\nonumber \\
   F_{\rm C}(v) & = & 0.4013 \, v^2 = 4.71 \, (1-\tau).
\label{nusup-CP-low_v}
\end{eqnarray}
For low temperatures $T \ll T_c$ (i.e., for $v \gg 1$)
the asymptotes of $F(v)$ are
\begin{eqnarray}
   F_{\rm A}(v) & = & {\sqrt{\pi} \over 2} \, v^{13/2} \,
     \exp(-2v) = {35.5 \over \tau^{13/2}} \,
     \exp\left( -\frac{3.528}{\tau} \right) ,
\nonumber \\
   F_{\rm B}(v) & = & {\pi \over 4 \sqrt{3}} \, v^6 \,
     \exp(-2v) = {1.27 \over \tau^6} \,
     \exp \left( -\frac{2.376}{\tau} \right) ,
\nonumber \\
   F_{\rm C}(v) & = & {50.03 \over v^2} = 12.1 \, \tau^2.
\label{nusup-CP-high_v}
\end{eqnarray}

Neutrino emission due to nucleon pairing
differs from other neutrino reactions: first, is has a temperature
threshold,
$T = T_c$; second, its emissivity
is a nonmonotonic function of temperature.
The emissivity grows rapidly with decreasing
$T$ just after superfluidity onset, reaches maximum and decreases.
According to Eq.\ (\ref{nusup-CP-high_v}), a strong superfluidity
reduces considerably the emissivity
just as it reduces direct
Urca process and other standard reactions
(Sects.\ \ref{sect-nusup-Durca}--\ref{sect-nusup-Brems}): 
the reduction is exponential if the gap is nodeless
(cases A and B) and it is power-law
otherwise (case C).
For cases A and B the asymptotes (\ref{nusup-CP-high_v})
contain the doubled gap in the exponent. This is natural because
our process consists actually of
nucleon transition from a state with the energy above the Fermi
level into a state with the energy below the Fermi level.
The minimum separation of the energies of the
initial and final states is given by the doubled gap.

The asymptotes (\ref{nusup-CP-low_v}) and (\ref{nusup-CP-high_v}),
as well as numerical values of
$F(v)$ for intermediate $v$,
are fitted by the simple expressions
(Yakovlev et al.\  1998, 1999a)
%
\begin{eqnarray}
  F_{\rm A}(v) & = & (0.602 \, v^2 + 0.5942\, v^4 +
     0.288 \, v^6) \,
     \left( 0.5547 + \sqrt{(0.4453)^2 + 0.0113 \,v^2} \right)^{1/2}
\nonumber \\
   & & \times \,
     \exp \left(- \sqrt{4 \, v^2 + (2.245)^2 } + 2.245 \right),
\nonumber \\
  F_{\rm B}(v) & = & {1.204 \, v^2 + 3.733 \, v^4 +
     0.3191 \, v^6 \over 1 + 0.3511 \, v^2} \,
     \left( 0.7591 + \sqrt{ (0.2409)^2 + 0.3145 \, v^2 } \right)^2
\nonumber \\
    & & \times \,
    \exp \left( - \sqrt{ 4 \, v^2 + (0.4616)^2} + 0.4616 \right),
\nonumber \\
  F_{\rm C}(v) & = &
    { 0.4013 \, v^2 - 0.043 \, v^4 + 0.002172 \, v^6 \over
     1 - 0.2018 \, v^2 + 0.02601 \, v^4 - 0.001477 \, v^6
     + 0.0000434 \, v^8}.
\label{nusup-CP-RecFit}
\end{eqnarray}

\begin{figure}[!t]
\begin{center}
\leavevmode
\epsfysize=8.5cm
\epsfbox[95 60 440 390]{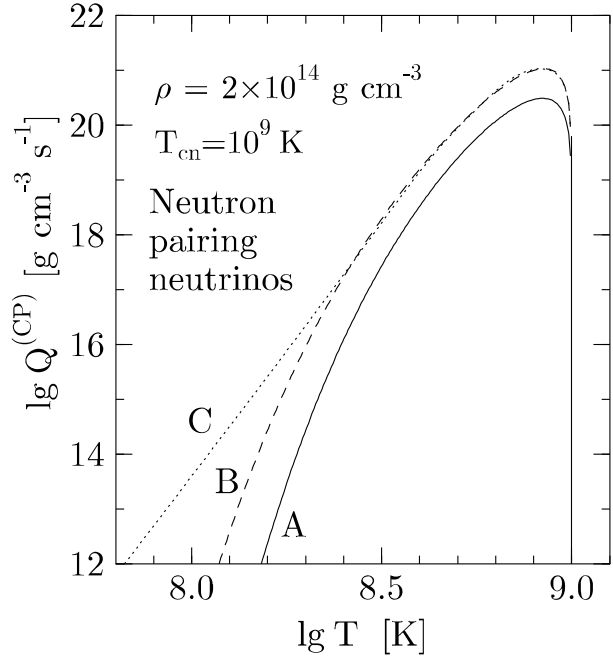} 
\end{center}
\caption[]{\footnotesize
        Temperature dependence of the neutrino emissivity
        due to Cooper pairing of neutrons for
        $\rho = 2 \times 10^{14}$ g cm$^{-3}$
        and $T_{cn} = 10^9$ K for superfluidity A (solid line),
        B (dashes) and C (dots); from Yakovlev et al.\ (1999b).
}
\label{fig-nusup-Qrec}
\end{figure}
Equations (\ref{nusup-CP-Qrec}) and (\ref{nusup-CP-RecFit}) enable one
to calculate easily neutrino emissivity
$Q^{\rm (CP)}$ due 
for superfluidity A, B and C.

Figure \ref{fig-nusup-Qrec}
shows temperature dependence of the emissivity
$Q^{\rm (CP)}$ due to neutron pairing in a neutron star core for
$\rho = 2 \times 10^{14}$ g cm$^{-3}$.
We adopted the moderate equation of state of matter
described in Sect.\ \ref{sect-cool-code}. The effective nucleon masses are
set equal to
$m_N^\ast = 0.7 \, m_N$, and the critical temperature is
$T_{cn} = 10^9$ K. The density of study is typical for
transition between the single-state pairing and the triple-state one
(Sect.\ \ref{sect-overview-struct}). Thus, different models of nucleon--nucleon
interaction may lead to different neutron superfluidity type,
and we show the curves for all three superfluidity types
considered above.

When the temperature falls down below
$T_{cn}$, the neutrino emissivity produced by the Cooper
pairing sharply increases. The main neutrino energy release
takes place in the temperature interval
$0.2 \, T_{cn} \lesssim T \lesssim 0.96 \,T_{cn}$, 
with the maximum at $T \approx  0.8\, T_{cn}$.
The emissivity may be sufficiently high, compared with
or even larger than the emissivity of modified Urca process
in non-superfluid matter (Sect.\ \ref{sect-nucore-Murca}).
The reaction may be noticeable even in the
presence of direct Urca process in the inner neutron star core
if the direct Urca process is strongly reduced by the
proton superfluidity (see below).
Neutrino emission
due to pairing of neutrons may also be significant
in inner neutron star crusts
(Sect.\ \ref{sect-nucrust-nn}).

Equations similar to (\ref{nusup-CP-Qrec}) describe
neutrino emission due to Cooper pairing of hyperons.
In this case $m_N^\ast$ should be replaced by the
effective baryon mass $m_B^\ast$, and $p_{\rm F}$
should be treated as the baryon Fermi momentum, while
$F(v)$ is given by the same
Eq.\ (\ref{nusup-CP-RecFit}); the case of singlet-state
hyperon pairing is most important for applications.
The required values of
$a$ are listed in Yakovlev et al.\ (1999a).
For instance, in $npe\mu\Lambda\Sigma^-$ matter
it would be reasonable to assume the singlet state
pairing of $\Sigma^-$, and $\Lambda$ hyperons
(e.g., Balberg and Barnea 1998).
For $\Sigma^-$, Yakovlev et al.\ (1999a) yield
$a_{\Sigma}=(2-4 \sin^2 \Theta_{\rm W})^2=1.17$.
The situation with $\Lambda$-hyperons is similar
to that with protons: one formally has
$a_{\Lambda} = 0$ in the non-relativistic approximation
including the vector neutral currents but
one obtains small finite value of $a_\Lambda$ by
including either non-relativistic correction or
in-medium effects. In any case one can expect
that neutrino emission due to pairing of
$\Lambda$ particles is very weak.
Note that similar formalism can be used to describe
neutrino emission due to pairing of quarks in
quark matter (Sect.\ \ref{sect-overview-struct}).
Pairing of unlike quarks (if allowed)
with very large superfluid gaps would 
occur just at the neutron star birth (at the
proto-neutron-star stage)  
producing a very strong neutrino outburst.  

\subsection{Leading reactions in superfluid cores}
\label{sect-nusup-all}

Now we are ready
to summarize the results of Sects.\  \ref{sect-nusup-Durca}--
\ref{sect-nusup-CP} and compare the neutrino emissivities
of different reactions in superfluid cores of neutron stars.

For illustration, we restrict ourselves to $npe$ matter,
and use the same moderate equation of state
in the neutron star core as
in the cooling simulations presented in
Chapt.\ \ref{chapt-cool} with the nucleon effective masses
$m_N^\ast = 0.7 \, m_N$.
We assume further the neutron pairing of type
B and the proton pairing of type A.
We follow consideration of Yakovlev et al.\ (1999b). 

\begin{figure}[!t]                         
\begin{center}
\leavevmode
\epsfysize=8.5cm
\epsfbox[40 15 540 270]{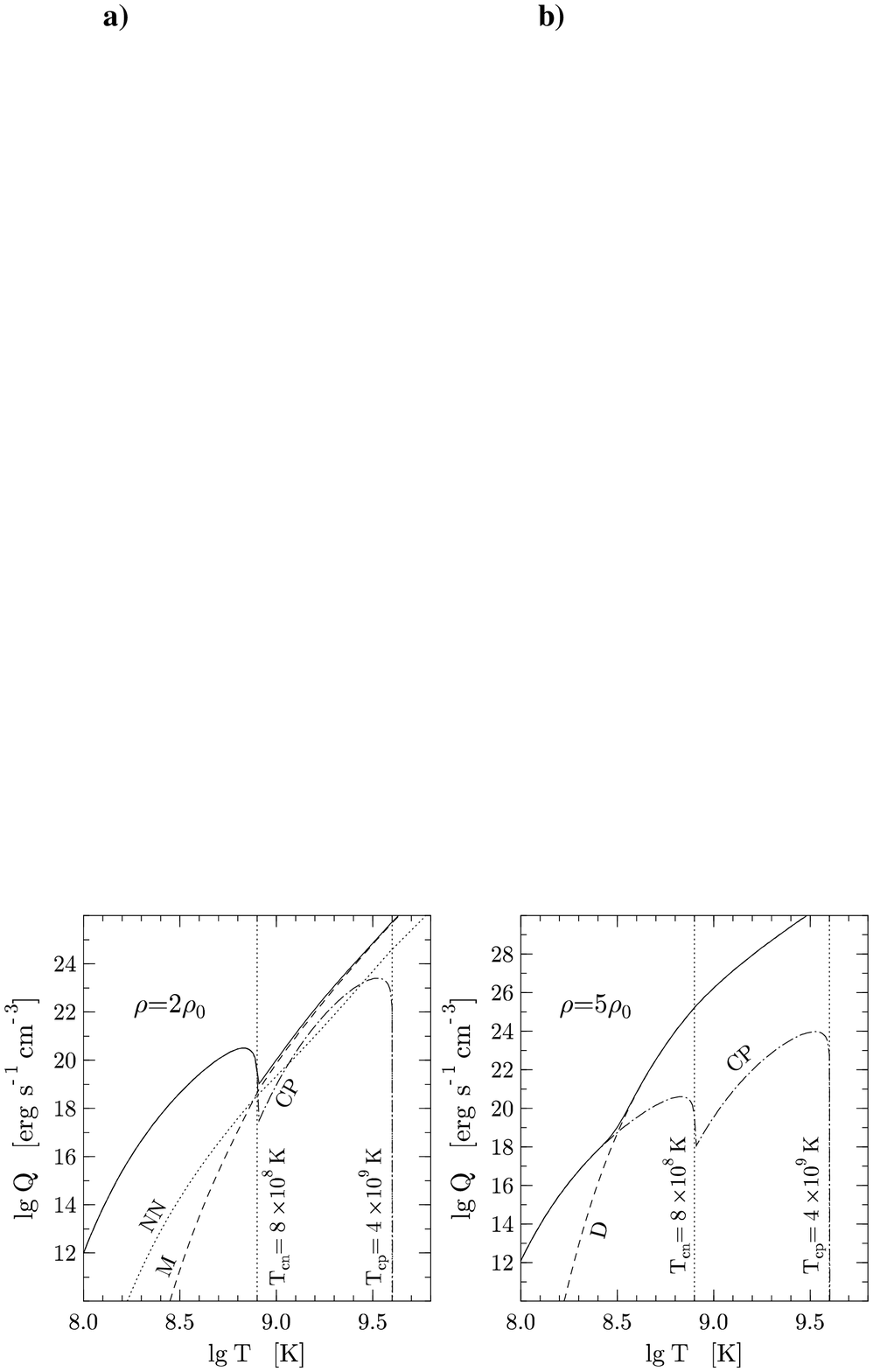}
\end{center}
\caption[]{\footnotesize
        Temperature dependence of neutrino emissivity
        in different reactions for neutron
        superfluidity  B with
        $T_{cn}= 8 \times 10^8$~K
        and proton superfluidity  A with
        $T_{cp} = 4 \times 10^9$~K
        at $\rho = 2\, \rho_0$ (standard neutrino reactions, left panel)
        and $\rho = 5\, \rho_0$ (direct Urca process is allowed, right panel).
        Dot-and-dash line shows the emissivity due to Cooper
        pairing (CP) of neutrons plus protons; solid line presents
        the total emissivity.
        On the left panel the dashed line (M) 
        gives the total emissivity of two
        branches of modified Urca process;
        the dotted line (NN) exhibits total
        bremsstrahlung emissivity in
        $nn$, $np$ and $pp$ collisions.
        The right panel: 
        dashed line (D) corresponds to direct Urca process.
         }
\label{fig-nusup-QrecMurDur}
\end{figure}

Figures \ref{fig-nusup-QrecMurDur}
show temperature dependence of 
the neutrino emissivities in different reactions for
$T_{cn} = 8 \times 10^8$~K and
$T_{cp}  =  4 \times 10^9$~K.
The left panel corresponds to
$\rho  =  2\rho_0$ and can be compared with analogous
Fig.\ \ref{fig-nucore-Q(T)} for non-superfluid matter.
The direct Urca process is forbidden at this density
being allowed at
$\rho_{\rm crit}=4.64 \, \rho_0= 1.298 \times 10^{15}$ g cm$^{-3}$,
for the given equation of state.
In the absence of the neutron superfluidity
($T> T_{cn}$) the most efficient is modified Urca process.
If, however, the temperature decreases from
$T  =  T_{cn}$ to
$T \approx  10^{8.8}$~K,
the total neutrino emissivity increases by about
two orders of magnitude due to neutrino emission produced by
Cooper pairing of neutrons.
Thus the appearance of superfluidity
may accelerate neutron star cooling instead of slowing it.

\begin{figure}[!t]                         
\begin{center}
\leavevmode
\epsfysize=17.0cm
\epsfbox[45 75 520 740]{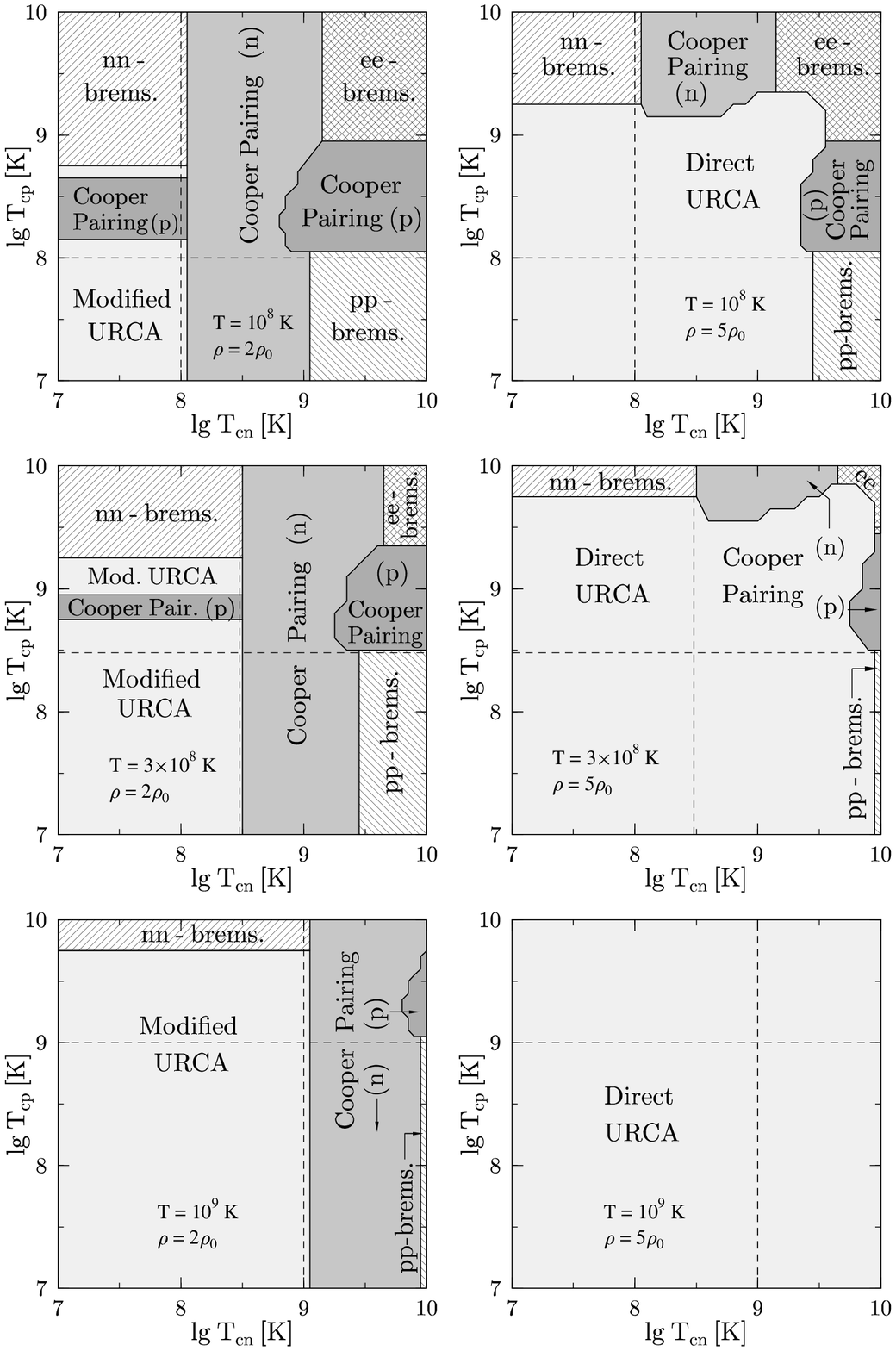}
\end{center}
\caption[]{\footnotesize
        Regions of $T_{cn}$ (superfluidity of type B)
        and $T_{cp}$ (type A), in which different neutrino
        reactions dominate at $T=10^9$,
        $3 \times 10^8$ and $10^8$ K (horizontal and vertical dashed lines)
        in matter of density
        $\rho = 2\,\rho_0$ (standard cooling)
        and $\rho = 5\, \rho_0$ (enhanced cooling);
        from Yakovlev et al.\ (1999b).
        }
\label{fig-nusup-qq_murdur}
\end{figure}

The right panel of
Fig.\ \ref{fig-nusup-QrecMurDur} corresponds to a denser matter,
$\rho = 5 \rho_0$, where the powerful direct Urca process
operates. In this case, the neutrino emissivity is
actually determined by the two processes, direct Urca and
Cooper pairing of nucleons. The direct Urca process dominates at
$T \gtrsim 3 \times 10^8$~K. With further decrease of $T$,
the Urca and nucleon-nucleon bremsstrahlung processes 
are reduced so strongly that Cooper-pairing 
neutrino emission becomes dominant.

As discussed in Chapt.\ \ref{chapt-nucore},
neutrino emission from a non-superfluid neutron star
core is mainly determined by a single, most powerful neutrino
emission mechanism: direct Urca process
for the enhanced (fast) cooling, or modified Urca
process for the standard (slow) cooling.
However, as seen from Fig.\ \ref{fig-nusup-QrecMurDur},
superfluidity violates this ``simplicity".
Different neutrino mechanisms can dominate at different
cooling stages depending on
$T$, $T_{cn}$, $T_{cp}$, and $\rho$.

Figures \ref{fig-nusup-qq_murdur} show which neutrino mechanisms
dominate for different
$T_{cn}$ and $T_{cp}$ in $npe$ matter.
In addition to neutrino processes
which involve nucleons we have included also 
neutrino bremsstrahlung in electron--electron
collisions (Sect.\ \ref{sect-nusup-Brems}).
This mechanism
is weak and, as a rule, it can be neglected
in cooling simulations.
Three left Figs.\ \ref{fig-nusup-qq_murdur} illustrate
standard cooling at
$\rho  =  2\, \rho_0$ for
the three values of the internal stellar temperatures:
$10^8$, $3 \times 10^8$ and $10^9$~K;
three right figures
correspond to enhanced cooling
at $ \rho =  5\, \rho_0$ for the same $T$.
The chosen values of $T$ cover the region
most interesting for practice. Our calculations show
that topology of the figures varies only slightly
with $\rho$ as long as $\rho$ does not cross
the threshold value $\rho=\rho_{\rm crit}$. Therefore, the
presented figures reflect adequately the efficiency
of all neutrino processes in the cores of cooling neutron stars.
One can see that, in the presence of superfluidity,
many different mechanisms can dominate in certain parameter ranges.

Notice that if the direct Urca process is open, the modified Urca
is always insignificant.
In the presence of the neutron superfluidity alone,
neutrino bremsstrahlung in $pp$ collisions
becomes the main mechanism at
$T \ll T_{cn}$ being unaffected by the neutron superfluidity.
In the presence of the proton superfluidity alone,
neutrino bremsstrahlung in $nn$ collisions dominates at
$T \ll T_{cp}$. Cooper-pairing neutrino emission of neutrons exceeds
neutrino emission in other reactions for
$T  \lesssim  10^9$~K
and for not too strong neutron superfluidity
($0.12  \lesssim T/T_{cn} \lesssim  0.96$).
This parameter range is very interesting for applications.
Neutrino emission due to Cooper pairing of neutrons
is also significant in a hotter matter
(at earlier cooling stages), when
$T \gtrsim 10^9$~K, but in a narrower temperature range near
$T  \approx  0.8 \, T_{cn}$, or in the presence of the
proton superfluidity.
Although neutrino production in proton pairing is much weaker,
it can also dominate. Neutrino emission due to
pairing of neutrons and protons can dominate
in rapid cooling as well
in the cases in which the nucleons of one species
are strongly superfluid while the other ones are moderately superfluid.
Very strong superfluidities of neutrons and protons
(upper right corners of the figures) switch
off all neutrino processes involving nucleons. As a result,
neutrino bremsstrahlung in electron--electron collisions,
which is practically unaffected by superfluidity,
becomes dominant producing residual neutrino emissivity
from highly superfluid neutron star cores. This case
is discussed further in Sect.\ \ref{sect-nusup-fluxa}.

\subsection{Direct Urca process in strong magnetic fields}
\label{sect-nusup-durmag}

Now we are turning from the effects of superfluidity to
the effects of strong magnetic fields on neutrino
emission from neutron star cores.
In this section
we consider the simplest and most powerful
direct Urca process in a non-superfluid but strongly
magnetized core. We will
discuss the basic nucleon direct Urca
process (\ref{nucore-Durca})
but the same technique can be used to analyze other direct
Urca processes involving hyperons and muons.

Beta--decay and related reactions in strong
magnetic fields have been studied since late 1960's
(e.g., Canuto and Chiu 1971, Dorofeev et al.\ 1985,
Lai and Shapiro 1991, and references therein).
%
%
%
%
However, these results have been obtained under
various simplified assumptions (constant matrix elements,
non--degenerate nucleons, etc.); they do not give explicitly the
emissivity of direct Urca process
in neutron star cores.
Works on the subject have been appearing later from time to
time with rather contradictory results.

From our point of view, the problem was solved
with considerable accuracy by Baiko and Yakovlev (1999).
%
%
We summarize their results below.
We mainly concentrate on the realistic case, in which
the magnetic field is not extremely high
(although still high, $B \lesssim 3 \times 10^{16}$ G),
and charged particles populate many Landau levels.
We will mention also the case of superstrong fields.\\

{\bf (a) Quantum formalism}

First, we outline derivation
of the neutrino emissivity of 
direct Urca process valid at any
magnetic field $B \ll 10^{20}$ G
(at higher fields protons become relativistic).
We consider neutrino emission in a pair
of reactions (\ref{nucore-Durca})
under the conventional assumption
that reacting electrons are relativistic, while
protons and neutrons are nonrelativistic.
Calculations are similar to those in Sects.\
\ref{sect-nucore-beta} and \ref{sect-nucore-Durca}.
The interaction Hamiltonian is that for the
ordinary beta decay, Eq.\ (\ref{nucore-H}).
The rate of transition from an initial
state $i$ to a final state $f$ is again
given by Fermi Golden Rule which can be
schematically written as
$W_{i\to f} = 2 \pi |H_{fi}|^2
\delta (\epsilon_n - \epsilon_p - \epsilon_e - \epsilon_\nu)$,
cf with Eq.\ (\ref{nucore-Golden-Rule}).
We will use the wave functions of electrons and protons
in a constant quantizing magnetic field ${\bf B}=(0,0,B)$
(the Landau states) choosing the Landau
gauge of the vector potential
${\bf A} = (-By,0,0)$.
To evaluate the neutrino emissivity we need
to sum $W_{i \to f}$ times the energy of the newly born
antineutrino over all initial and final states.
Calculating the squared matrix element,
performing summation over electron spin states
and integration over $y$-coordinates of the
Larmor guiding centers of electrons and protons
Baiko and Yakovlev (1999) obtained the general expression for
the emissivity
(including the inverse reaction which doubles
the emissivity)
\begin{eqnarray}
  Q^{\rm (D)} &=& {2 e B G^2 \over (2 \pi)^7} \sum_{nn's_p s_n} \int
          {\rm d}{\bf p}_n \, {\rm d}{\bf p}_\nu \,
          {\rm d}p_{pz} \, {\rm d}p_{ez} \,
            f_n \, (1-f_p) \, (1-f_e) \, \epsilon_\nu
\nonumber \\
          &  &\times \,
          \delta(\epsilon_n-\epsilon_e-\epsilon_p-\epsilon_\nu) \,
          \delta (p_{nz}-p_{ez}-p_{pz}-p_{\nu z}) \, {\cal P}.
\label{nusup-durmag-Qnu}
\end{eqnarray}
In this case, $G=G_{\rm F} \, \cos \Theta_{\rm C}$
[see Eq.\ (\ref{nucore-H})], $e=|e|$ is the elementary charge,
$\epsilon_j$ is the particle energy, $f_j$ is its Fermi-Dirac
distribution,
$p_{zj}$ is a momentum component along the magnetic field,
${\bf p}_\nu$ and ${\bf p}_n$ are, respectively,
the full neutrino and neutron momenta,
$s_n=\pm 1$ and $s_p = \pm 1$ are the
doubled $z$ projections of neutron and proton
spin, while
$n\geq 0$ and $n'\geq 0$
enumerate the electron and proton Landau levels, respectively.
The particle energies are given by the familiar expressions
\begin{eqnarray}
          \epsilon_e &=& \sqrt{m^2_e + p_{ez}^2 + 2 e B n },
\nonumber \\
          \epsilon_p &=&  {k^2_{pz} \over 2 m_p^\ast} +
          \left[n' + \frac{1}{2}
          \left(1 - g_p s_p {m_p^\ast \over m_p} \right)
          \right] {eB \over m_p^\ast},
\nonumber \\
          \epsilon_n &=& {k^2_n \over 2 m_n^\ast} -
           {g_n s_n eB \over 2 m_p}, \quad \quad
          \epsilon_\nu = p_\nu,
\label{nusup-durmag-energs}
\end{eqnarray}
with the proton and neutron gyromagnetic
factors $g_p=2.79$ and $g_n=-1.91$.
Axial-vector weak interaction constant $g_A$, 
as well as factors $g_p$ and
$g_n$ can be renormalized by medium effects which we
ignore, for simplicity.

Furthermore, the quantity ${\cal P}$ in Eq.\ (\ref{nusup-durmag-Qnu})
represents the dimensionless squared matrix element
\begin{eqnarray}
     {\cal P} &=& \frac{1}{2} \,\, \delta_{s_p s_n} \, (1+g_A^2) \,
     \left[  \left(1-{p_{ez} \over \epsilon_e} \right)
     \left(1-{p_{\nu z} \over \epsilon_\nu} \right) \, F^{\prime 2}
     \right.
\nonumber \\
     & & +   \left.
     \left(1+{p_{ez} \over \epsilon_e} \right)
     \left(1+{p_{\nu z} \over \epsilon_\nu} \right) \, F^2  \right]
     - \delta_{s_p s_n} \, g_A \, s_p \,
     \left[ \left(1-{p_{ez} \over \epsilon_e} \right)
     \left(1-{p_{\nu z} \over \epsilon_\nu} \right) \, F^{\prime 2}
     \right.
\nonumber \\
      & &-   \left.
     \left(1+{p_{ez} \over \epsilon_e} \right)
     \left(1+{p_{\nu z} \over \epsilon_\nu} \right) \, F^2  \right]
     + 2 \,\, \delta_{s_p, 1} \, \delta_{s_n, -1} \, g_A^2 \,
     \left(1+{p_{ez} \over \epsilon_e} \right)
     \left(1-{p_{\nu z} \over \epsilon_\nu} \right) \, F^2
\nonumber \\
     & &+  2 \,\, \delta_{s_p, -1} \, \delta_{s_n, 1} \, g_A^2 \,
     \left(1-{p_{ez} \over \epsilon_e} \right)
     \left(1+{p_{\nu z} \over \epsilon_\nu} \right) \, F^{\prime 2}
     +  \delta_{s_p s_n} \, (1-g_A^2) {p_{e \bot} \over
     \epsilon_e} {{\bf p_\nu}\cdot{\bf q} \over \epsilon_\nu q}
     \, F F' .
\label{nusup-durmag-M}
\end{eqnarray}
where $p_{e \bot} = \sqrt{2eBn}$ and
${\bf q} = (p_{nx}-p_{\nu x},p_{ny}-p_{\nu y},0)$. Finally,
$F=F_{n',n}(u)$ and $F'=F_{n',n-1}(u)$ are the
normalized Laguerre functions 
%
%
of argument $u=q^2 / (2eB)$ 
(e.g., Kaminker and Yakovlev 1981). If any index $n$ or $n'$ is negative,
then $F_{n,n'}(u)=0$. Notice a misprint in the
paper by Baiko and Yakovlev (1999): $g_A$ has to be replaced by $-g_A$
in all expressions. However, this misprint affects the practical
expressions only in the limit of superstrong magnetic fields. 

Since the neutrino momentum is much smaller
than the momenta of other particles
(Sect.\ \ref{sect-nucore-Durca}) we
can neglect it in $z$-momentum conserving delta
function and in the definition of the vector {\bf q}.
Then, using isotropy
of Fermi-Dirac distributions
under the integral in Eq.\ (\ref{nusup-durmag-Qnu}),
we can further simplify
the expression for ${\cal P}$:
\begin{eqnarray}
     {\cal P} &=& 2 g_A^2 \,
       \left( \delta_{s_p, 1} \, \delta_{s_n, -1} \, F^2
     + \delta_{s_p, -1} \, \delta_{s_n, 1} \, F^{\prime 2} \right)
\nonumber \\
      & & +  \frac{1}{2} \,\, \delta_{s_p s_n} \, (1+g_A^2) \,
     \left( F^{\prime 2} + F^2 \right)
     - \delta_{s_p s_n} \, g_A \, s_p \,
     \left( F^{\prime 2} -  F^2 \right),
\label{nusup-durmag-Ms2}
\end{eqnarray}
where $F$ and $F'$ depend now on
$u=
(p_{nx}^2+p_{ny}^2)/(2eB) \equiv p_{n \bot}^2/(2eB)$.\\

{\bf (b) Quasiclassical case}

First of all consider the most realistic case
of not too high magnetic fields, in which electrons and
protons populate many Landau levels.
These fields do not affect particle Fermi energies,
and one can safely use the same Fermi momenta $p_{\rm F}$
as for $B=0$.
In this case the transverse wavelengths of
electrons and protons are much smaller than their
Larmor radii. This situation may be referred to
as quasiclassical, and corresponding techniques apply.
If the main contribution
comes from large $n$ and $n'$, the difference between
$F^2$ and $F^{\prime 2}$ can be neglected.
Moreover, we can neglect the contributions
of magnetic momenta of particles to their energies.
Thus, we can pull all the functions of energy
out of the sum over $s_n$ and $s_p$, and evaluate the latter
sum explicitly:
$ \sum_{s_n s_p} {\cal P} = 2 \, (1+3 g_A^2) \, F^2$
[cf with Eq.\ (\ref{nucore-squared-Mdur})].
Inserting this expression into Eq.\ (\ref{nusup-durmag-Qnu})
and integrating over orientations of neutrino momentum,
we get
\begin{eqnarray}
  Q^{\rm (D)} &=& {16 \pi G^2 \, (1+3 g_A^2) \,
          e B \over (2 \pi)^7} \int
          {\rm d} \epsilon_\nu \,
          {\rm d}{\bf p}_n \, {\rm d}p_{pz} \, {\rm d}p_{ez}
          \, \epsilon_\nu^3
\nonumber \\
         & & \times \sum_{nn'} \, F^2_{n',n} (u)\,
          f_n \, (1-f_p) \, (1-f_e)
          \delta(\epsilon_n-\epsilon_e-
                 \epsilon_p-\epsilon_\nu) \,
          \delta (p_{nz}-p_{ez}-p_{pz}) ,
\label{nusup-durmag-Qnu2}
\end{eqnarray}
where
$\epsilon_p = (p_{pz}^2 + p_{p \bot}^2)/(2 m_p^\ast)$,
and $\epsilon_n = p_n^2/( 2 m_n^\ast)$,
$p_{p \bot} = \sqrt{2eBn'}$,
while the electron energy is still given
by Eq.\ (\ref{nusup-durmag-energs}).

If the magnetic field is not too large,
the transverse electron and proton momenta, $p_{e\bot}^2$ and
$p_{p\bot}^2$, are sampled
over a dense grid of values, corresponding to integer
indices $n$ and $n'$. Thus, the double
sum in Eq.\ (\ref{nusup-durmag-Qnu2}) can be replaced
by the double integral over $p_{e \bot}^2$
and $p_{p \bot}^2$.
Afterwards
Eq.\ (\ref{nusup-durmag-Qnu2}) can be considerably simplified.
Note, that
${\rm d}p_z \, {\rm d}p_{\bot}^2 = {\rm d}{\bf p}/ \pi =
2 m^\ast \, {\rm d}\epsilon \, p \, \sin{\theta} \, {\rm d}\theta$,
where $\theta$ is the particle pitch--angle. Then
the energy integral $\int {\rm d}\epsilon_n \, {\rm d}\epsilon_p \,
{\rm d}\epsilon_e \, {\rm d}\epsilon_\nu \ldots$ is taken
by assuming that the temperature
scale is small and provides the sharpest
variations of the integrand. If so,
we can set $p=p_{\rm F}$
in all smooth functions.
Finally, we integrate over the azimuthal angle of the neutron momentum
and over its polar angle to eliminate the momentum
conserving delta function and obtain:
\begin{eqnarray}
    Q^{\rm (D)} &=& Q^{\rm (D)}_0 \, R_B^{\rm (D)}; \quad
         Q^{\rm (D)}_0 =
         {457 \pi \,  G^2 \,(1+3 g_A^2) \over
         10080} \, m_n^\ast \, m_p^\ast \, m_e^\ast \, T^6,
\nonumber \\
        R_B^{\rm (D)} & = &
        2 \int \int^{1}_{-1} {\rm d}\cos{\theta_p} \, {\rm d}\cos{\theta_e}
        \, {p_{{\rm F}p} \, p_{{\rm F}e} \over 4eB} \,
        F^2_{Np,Ne}(u)
\nonumber \\
        & &\times \, \Theta(p_{{\rm F}n}-
        |p_{{\rm F}p} \cos{\theta_p}+p_{{\rm F}e} \cos{\theta_e}|),
\label{nusup-durmag-Qnu3}
\end{eqnarray}
where $Q^{\rm (D)}_0$ is the field--free emissivity
given by Eq.\ (\ref{nucore-Qdur0}) but with the {\it removed}
step function $\Theta_{npe}$,
factor $R_B^{\rm (D)}$ describes the effect of
the magnetic field; $N_p=p_{p\bot}^2/(2eB)$, $N_e=p_{e\bot}^2/(2eB)$;
$p_{p,e \bot} = p_{{\rm F}p,e} \sin{\theta_{p,e}}$,
and $p_{n\bot}^2=p_{{\rm F}n}^2 - (p_{{\rm F}p}
\cos{\theta_p}+p_{{\rm F}e} \cos{\theta_e})^2$;
$\Theta(x)=1$ for $x \ge 0$, $\Theta(x)=0$ for $x<0$.

Therefore, the problem is reduced to
evaluating $R_B^{\rm (D)}$.
This was done by Baiko and Yakovlev (1999)
using various asymptotic formulae for the
Laguerre functions $F_{nn'}(u)$.
Moreover, the authors calculated
emissivity $Q^{\rm (D)}$ directly
from Eq.\ (\ref{nusup-durmag-Qnu2}) and determined
$R_B^{\rm (D)}$ in this way.
The factor in question depends on two parameters,
$x$ and $y$, which characterize the reaction kinematics with respect
to the magnetic field strength:
\begin{equation}
   x  =  { p_{{\rm F}n}^2 - (p_{{\rm F}p}+p_{{\rm F}e})^2
           \over  p_{{\rm F}p}^2 \, N_{{\rm F}p}^{-2/3}}, \quad
   y = {N_{{\rm F}p}}^{2/3},
\label{nusup-durmag-xy}
\end{equation}
where $N_{{\rm F}p}= p_{{\rm F}p}^2/(2eB)$ is the
number of the Landau levels populated by protons.
The quasiclassical approach corresponds to the limit
of large $y \gg 1$ (many populated levels).
In this limit, $R^{\rm (D)}_B$ depends on the
only argument
$x \approx (4 \Delta p/ p_{{\rm F}p}) N_{{\rm F}p}^{2/3}$,   
where
$\Delta p \equiv p_{{\rm F}n} - p_{{\rm F}p} - p_{{\rm F}e}$.
We present the final results only in this limit
since they are sufficient for most of applications,
and we refer to the original paper by Baiko
and Yakovlev (1999) for a discussion of a more general case.
The result depends on
sign of $x$, which determines if the direct Urca is
open ($x<0$) or forbidden ($x>0$) at $B=0$.

If $x<0$, then
$p_{{\rm F}n}<p_{{\rm F}p}+p_{{\rm F}e}$ (i.e., $\Delta p <0$),
and direct Urca process is open for $B=0$.
Applying the quasiclassical approach,
Baiko and Yakovlev (1999)
obtained for $R_B^{\rm (D)}$ an oscillating curve.
These oscillations
are of quasiclassical nature and have nothing in common
with the quantum oscillations associated with
population of new Landau levels. 
From the practical point of view, they
have hardly
any effect on neutron star cooling.
For $x \to -\infty$, the oscillation amplitude
vanishes, and $R_B^{\rm (D)} \to 1$, reproducing the
field-free results for the open direct Urca process.
The results of numerical calculations
for $-20 \leq x \leq 0$
are accurately fitted by 
%
%
\begin{eqnarray}
       R_B^{\rm (D)} &=& 1 - { \cos \varphi \over (0.5816 + |x|)^{1.192}},
\label{nusup-durmag-fitperm} \\
       \varphi &=& { 1.211 + 0.4823 \,|x| + 0.8453 \, |x|^{2.533}
                    \over 1 + 1.438 \, |x|^{1.209}}.
\nonumber
\end{eqnarray}
At the threshold of the direct
Urca process for $B=0$ we have $x=0$
and $R_B^{\rm (D)}=1/3$.

If $x>0$, then $p_{{\rm F}n}>p_{{\rm F}p}+p_{{\rm F}e}$
(i.e., $\Delta p >0$)
and the direct Urca process is forbidden at $B=0$.
The factor $R_B^{\rm (D)}$
decreases exponentially with increasing $x$ in this
domain. This behavior is accurately described
by the fit:
\begin{equation}
       R_B = { 3 x + 6.8
             \over (x_c + 6.8) (3 + x \sqrt{12})} \,
             \exp{\left(-{x_c \over 3}\right)},
\label{nusup-durmag-fitforb}
\end{equation}
where
$x_c = x \, \sqrt{x + 0.4176} - 0.04035 \, x$.
Therefore, the magnetic field
smears out the threshold between the open and
closed direct Urca regimes producing
{\it magnetic broadening of the direct Urca threshold}.
The reaction becomes quite efficient
as long as $x \lesssim 10$, i.e., $\Delta p / p_{{\rm F}n}
\lesssim N_{{\rm F}p}^{-2/3}$. If, for instance, $B = 10^{16}$ G,
and the density of matter is near the direct
Urca threshold, one typically has $N_{{\rm F}p} \sim 300$ and
$\Delta p / p_{{\rm F}p} \lesssim 1/25$.
The broadening effect is associated with momentum-nonconservation
of reacting particles moving in a magnetic field. At large $x$
it allows some particles to participate in
the direct Urca process that would be impossible in the field
free case. The reaction rate for these particles is
exponentially reduced but even the exponential reduction
may keep the powerful direct Urca process to be
the main neutrino emission mechanism.

Notice that the field--free direct Urca process can also be allowed
beyond the domain $\Delta p < 0$ due to the
{\it thermal broadening} of the threshold,
as discussed in Sect.\ \ref{sect-nucore-Durca}.
Comparing the thermal and magnetic broadenings
we conclude
that the magnetic broadening is more important
if $N_{{\rm F}p}^{2/3} \lesssim \mu_p/T$
that can often be the case in inner cores of neutron stars.

\begin{figure}[!t]                         
\begin{center}
\leavevmode
\epsfysize=8.0cm
\epsfbox[70 65 395 410]{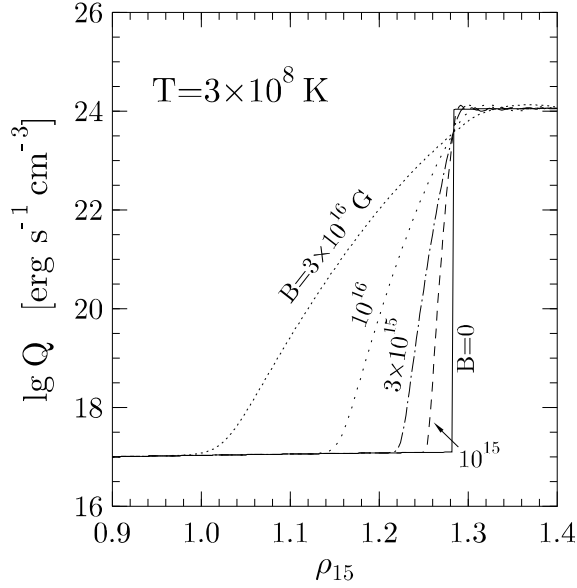}
\end{center}
\caption[]{\footnotesize
        Density profile of the total neutrino emissivity
        from non-superfluid $npe$ matter at $T= 3 \times 10^8$ K and
        $B=0$, $10^{15}$, $3 \times 10^{15}$, $10^{16}$ and
        $3 \times 10^{16}$ G. The case of $B=0$ is shown
        in more detail in Fig.\ \protect{\ref{fig-nucore-Q1(rho)}}.
        }
\label{fig-nusup-durmag}
\end{figure}

The effect is demonstrated in Fig.\ \ref{fig-nusup-durmag}
where we show density profile of the total neutrino emissivity
of $npe$ matter at $T= 3\times 10^8$ K for several
values of $B$. We have already shown the
density profile for $B=0$  in Fig.\ \ref{fig-nucore-Q1(rho)}
under the same conditions.
The magnetic fields $B \lesssim 3 \times 10^{16}$ G
only broaden the direct Urca threshold and do not affect
direct Urca as well as other neutrino processes otherwise,
for the conditions displayed in Fig.\ \ref{fig-nusup-durmag}.
The thermal broadening of the threshold is weak by itself although
stronger
than the magnetic one as long as $B \lesssim 3 \times 10^{14}$ G.
However, for higher $B$ the magnetic broadening becomes
stronger and is seen to be quite substantial.
It may affect cooling of magnetized neutron stars with central
densities which are slightly lower than the direct Urca
threshold at $B=0$ (Sect.\ \ref{sect-cool-durmag}).

Notice also that the direct quantum calculation of $Q^{\rm (D)}$ from
Eq.\ (\ref{nusup-durmag-Qnu2}) shows that along with
the quasiclassical effects described above the
magnetic field induces also traditional quantum
oscillations of the neutrino emissivity
associated with population of new Landau levels
under variations of parameters of matter.
If $T$ is larger than the energy spacing between the Landau levels,
the oscillations are washed out, and a smooth
quasiclassical curve emerges.
In the case of lower temperatures
the actual
neutrino emissivity does oscillate but the quantity
of practical significance is the emissivity
averaged over the oscillations, a smooth curve again.
Notice also that the magnetic field may affect the spectrum
of neutrino emission and make this emission anisotropic.\\

{\bf (c) The case of superstrong magnetic fields}

Using the exact quantum formalism presented in the beginning
of this section
we can outline the main features of
direct Urca process produced by electrons and protons occupying
the lowest Landau levels. 
Notice, however,
that one needs superhigh
magnetic field, $B \gtrsim 10^{18}$ G,
to force all electrons and protons to occupy  their ground
Landau levels. 
Contrary to the field-independent Fermi momenta
obtained for not too high fields in the quasiclassical approximation,
the limiting
momenta along the superhigh magnetic field are field--dependent,
$p^{(\rm F)}_{e,p} = 
2 \pi^2 n_{e,p} / eB$, $p^{(\rm F)}_e \approx \mu_e$.
The distribution
of neutrons can be characterized by two Fermi momenta $p^{(\rm F)}_n(s_n)$
for particles with spins along and against the magnetic field, 
with $p^{(\rm F)}_n(+1) < p^{(\rm F)}_n(-1)$.
Our starting point
is Eq.\ (\ref{nusup-durmag-M}), which reduces to
${\cal  P} = \, F^2 \,
          [0.5 \, (1+g_A)^2 \, \delta_{s_n,1} +
          2 g_A^2 \, \delta_{s_n,-1} ],
$
where $F=F_{00}(u)={\rm e}^{-u/2}$ and $u=p_{n\bot}^2/(2eB)$.
Then from Eq.\ (\ref{nusup-durmag-Qnu}), neglecting $p_{\nu z}$
in the momentum conserving delta function,
one obtains
\begin{eqnarray}
      Q^{\rm (D)} &=& {457 \pi \, G^2 \, (1+3 g_A^2) \over 10080}
              m_n^\ast m_p^\ast \, \mu_e T^6
\nonumber \\
            &\times&
           {eB \over p_p^{(\rm F)} p_e^{(\rm F)} \, (1+3 g_A^2)}
            \sum_{\alpha=\pm 1}
           \left[ \frac{1}{4} \, (1+g_A)^2 \,
           \Theta(u_{1 \alpha}) \, e^{-u_{1 \alpha}}+
           g_A^2 \,
           \Theta(u_{-1 \alpha}) \, e^{-u_{-1 \alpha}} \right],
\label{nusup-durmag-Q00}
\end{eqnarray}
where $2 e B \, u_{s\alpha} = [p_n^{(\rm F)}(s)]^2 -
[p^{(\rm F)}_p + \alpha p^{(\rm F)}_e]^2$, $s=s_n$,
and $\alpha = \pm 1$ corresponds to
two different reaction channels in which the electron and
proton momenta along the $z$-axis
are either parallel ($\alpha=1$) or antiparallel ($\alpha=-1$).
The step functions indicate that the channels are open
if $u_{s\alpha} \geq 0$. Channel $\alpha=-1$
is always open in $npe$ dense matter with
the superstrong magnetic field, while
channel $\alpha=1$ is open if
$p^{(\rm F)}_n(s) \geq p^{(\rm F)}_p + p^{(\rm F)}_e$. The latter condition
is opposite to the familiar condition
$p_{{\rm F}n} \leq p_{{\rm F}p}+p_{{\rm F}e}$
in the field--free case.
One has $\exp(-u_{s\alpha}) \leq 1$ and
$Q^{\rm (D)} \propto B^2$, but one cannot expect $Q^{\rm (D)}$
to be much larger than the field-free emissivity
$Q^{\rm (D)}_0$ as long as $B \lesssim 10^{19}$ G.
Similar result for the superstrong magnetic fields
was obtained 
by Leinson and P\'erez (1998).
Notice that they got $(1+g_A^2)$ instead of $(1+g_A)^2$
in the expression for $Q^{( \rm D)}$.

\subsection{Neutrino emission due to scattering of electrons
            off fluxoids}
\label{sect-nusup-fluxa}

{\bf (a) Preliminaries}

In this section we study neutrino-pair emission by electrons
in magnetized neutron star cores. The magnetic field
curves the
electron trajectories and
induces thus specific neutrino emission that is absent
in the field free case. This emission depends on superfluid
(superconducting) state of matter.

In the absence of superfluidity of protons (and other
available charged baryons) the matter is
non-superconducting, the magnetic field is
uniform on microscopic scales, and electrons
emit neutrino synchrotron radiation
considered already in Sect.\
\ref{sect-nucrust-syn} for neutron star crusts. The equations for the
emissivity presented there are equally valid
for neutron star cores. Under typical conditions
in the neutron star cores, the synchrotron emissivity
is given by Eq.\ (\ref{nucrust-syn-QB});
it scales as $T^5 B^2$ being
independent of density.

However, it is likely that the protons (and other charged
baryons) can be in the superfluid (superconducting) state.
It is commonly thought that superfluid protons form
the type II superconductor (see Sauls 1989, 
Bhattacharya and Srinivasan 1995
and references therein).
%
%
A transition to a superconducting state
in the course of stellar cooling is
accompanied by a dramatic change in the spatial structure of the
magnetic field. Initially homogeneous field
splits into an ensemble
of fluxoids which contain
a superstrong magnetic field, embedded in the field-free
superconducting medium.
Neutrino synchrotron radiation is then modified
and may be treated as neutrino pair emission
due to scattering of electrons off magnetic fluxoids $f$,
\begin{equation}
         e + f \to e + f + \nu + \bar{\nu},
\label{nusup-fluxa-ef}
\end{equation}
where $\nu \bar\nu$ mean a neutrino pair of any
flavor. This mechanism was studied by Kaminker et al.\ (1997)
for superconducting protons.
%
%
We present their results below.\\

{\bf (b) Superconducting neutron star cores}

%
%
An important parameter of the theory
is the $pp$ correlation length
$\xi$; it measures the size of a $pp$ Cooper pair.
In the BCS model, $\xi$ is related to the proton superfluid gap $\Delta_p$
and the proton Fermi velocity $v_{{\rm F}p}$ by
$\xi= \hbar v_{{\rm F}p}/ ( \pi \Delta_p)$.
The zero temperature value of $\xi$ will be denoted by $\xi_0$.
Another important parameter is the penetration depth $\lambda$
of the magnetic field in a proton superconductor.
In the case of $\lambda \gg \xi$ for the singlet-state proton
pairing, the zero
temperature penetration depth $\lambda_0=c/\omega_p$ is determined by the
proton plasma frequency, $\omega_p=(4 \pi n_p e^2/m_p^\ast)^{1/2}$
(e.g., Lifshitz and Pitaevskii 1980).
Simple estimates yield typical values of
$\xi_0$ of a few fm, and $\lambda_0$ of a few hundred fm.
If so, $\lambda_0 \gg \xi_0$,  which means that proton
superconductivity is of type II. Notice that the values of
$\lambda_0$ and $\xi_0$ are model dependent, and one cannot
exclude the case of $\lambda_0 \sim \xi_0$, in which the
proton superconductivity is of type I, although we will not
consider this case here.

For $T \to T_{cp}$, both $\xi$ and $\lambda$ diverge
as $(T_{cp}-T)^{-{1/2}}$, while for $T\ll T_{cp}$ they can be
replaced by their zero temperature values, $\xi_0$ and
$\lambda_0$. However,  we need to
know $\lambda$ for all
temperatures below $T_{cp}$. Let us 
approximate this dependence
by the Gorter--Casimir formula (e.g.\ Tilley and Tilley 1990),
%
%
%
\begin{equation}
    \lambda= \lambda_0/\alpha_T, \quad
    \alpha_T \equiv \sqrt{1 - (T/T_{cp})^4}.
\label{nusup-fluxa-lambda}
\end{equation}

The transition to the type II superconductivity
during neutron star cooling
is accompanied by formation of
quantized flux tubes (Abrikosov fluxoids), parallel to
the initial local magnetic field $\bar{\bf B}$.
Each fluxoid carries an
elementary magnetic flux $\phi_0= \pi \hbar c/e
= 2.059 \times 10^{-7}$ G cm~$^2$.
The number of fluxoids per unit area perpendicular
to the initial field is
${\cal N}_{\rm f}= \bar{B} / \phi_0$,
and the mean distance
between the fluxoids is
$d_{\rm f} = [2\phi_0 /( \sqrt{3} \bar{B})]^{1/2}$.
An estimate yields
$d_{\rm f} \approx 1500 / (\bar{B}_{13})^{1/2}~{\rm fm}$,
where $\bar{B}_{13} \equiv \bar{B} /(10^{13}~{\rm G})$. A fluxoid
has a small central core
of radius $\sim \xi$ containing normal protons.
A typical fluxoid radius is $\lambda$. Just after the
superconductivity onset ($T < T_{cp}$) this radius
is large and the fluxoids
fill all the available space. When temperature
drops to about $0.8 \, T_{cp}$, radius $\lambda$  reduces nearly
to its zero-temperature value $\lambda_0$. Typically, $\lambda_0$
is much
smaller than the inter-fluxoid distance $d_{\rm f}$ for
the magnetic fields $\bar{B} < 10^{15}$ G
to be considered in this section.
The maximum value of $B$ is reached at the
fluxoid axis, $B_{\rm max} \simeq$
$[\phi_0 /( \pi \lambda^2)]\, \log(\lambda/\xi)$.
In our case $\lambda \gg \xi$, and the
magnetic field profile at  $r\gg \xi$
is given by (e.g., Lifshitz and Pitaevskii 1980):
%
%
\begin{equation}
           B(r)={\phi_0\over 2\pi\lambda^2}
           K_0\left({r\over \lambda}\right),
\label{nusup-fluxa-Br}
\end{equation}
where $K_0(x)$ is a McDonald function.
In particular, for $r \gg \lambda$ one has
$B(r) \approx $
$\phi_0 (8\pi r\lambda^3)^{-1/2} \, {\rm e}^{-r/\lambda}$.

The superconducting state is destroyed when
$d_{\rm f} < \xi$, which corresponds to the magnetic field $\bar{B}>
B_{c2}=\phi_0/(\pi\xi^2) \sim 10^{18}$ G.
We do not consider such strong fields.

Let us mention that
the fluxoids can migrate slowly
outward the stellar core
due to the buoyancy forces
(Muslimov and Tsygan 1985, Jones 1987, Srinivasan et al.\ 1990).\\
%
%
%

{\bf (c) Formalism}

The process Eq.\ (\ref{nusup-fluxa-ef})
can be studied using the standard
perturbation theory with free electrons in nonperturbed states
in analogy to
the well known neutrino-pair bremsstrahlung
due to scattering of electrons off atomic nuclei,
Sect.\ \ref{sect-nucrust-ebrems}.
It is described by two second-order diagrams,
where one (electromagnetic) vertex is associated
with electron scattering by the fluxoid magnetic field,
while the other (four-tail) vertex is due to
the neutrino-pair emission.

The neutrino emissivity
can be written as
\begin{eqnarray}
      Q_{\rm flux} & = & {{\cal N}_{\rm f} \over (2 \pi)^{10}}
            \int \! {\rm d} {\bf p} \! \int \! {\rm d} {\bf p}'
            \! \int \! {\rm d} {\bf p}_\nu \!
            \int \! {\rm d} {\bf p}'_\nu
            \nonumber \\
            &  & \times \,
            \delta(\epsilon - \epsilon' -\omega) \,
            \delta(p_z - p'_z - k_z) \,
            \omega f(1-f') W_{i \to f},
\label{nusup-fluxa-Qgeneral}
\end{eqnarray}
where ${\cal N}_{\rm f}$
is the fluxoid surface number density,
$p = (\epsilon, {\bf p})$ and
$p'= (\epsilon',{\bf p}')$ are
the electron 4-momenta in
the initial and final states, respectively;
$p_\nu = (\epsilon_\nu, {\bf p}_\nu)$  and
$p'_\nu =(\epsilon'_\nu, {\bf p}'_\nu)$
are the 4-momenta of the neutrino and the
antineutrino, and
$k = p_\nu + p'_\nu = (\omega, {\bf k})$ is the 4-momentum
of the neutrino pair $(\omega = \epsilon_\nu + \epsilon'_\nu$,
${\bf k} = {\bf p}_\nu + {\bf p}'_\nu)$.
Furthermore, $f$
and
$f'$ are the Fermi-Dirac functions
for the initial and final electrons.
The delta functions describe
energy conservation and momentum
conservation along the fluxoid axis
(the axis $z$). Finally,
\begin{equation}
      W_{i \to f} = {G_{\rm F}^2 \over 2} \; \frac{1}
         {(2 \epsilon_\nu)(2 \epsilon'_\nu)
         (2 \epsilon)(2 \epsilon')} \;
         \sum_{\rm \sigma,\nu} | M_{fi} |^2
\label{nusup-fluxa-W}
\end{equation}
if the $i \to f$ transition rate, determined by the square
matrix element
$|M_{fi}|^2$. Summation is over the
electron spin states $\sigma$ before and after scattering
and over the neutrino flavors ($\nu_e$, $\nu_\mu$, $\nu_\tau$).
The standard approach yields (in our standard notations
introduced in Sect.\ \ref{sect-nucrust-annih})
\begin{eqnarray}
     \sum_{\sigma} |M_{fi}|^2 & = & e^2 \,
           \int {\rm d} {\bf q} \;
          \delta({\bf p} - {\bf q} - {\bf p}' - {\bf k}) \,
          A^i A^{j\ast}
\nonumber \\
     &  &\times   {\rm Tr}
          \left[ (p_\nu \gamma) \gamma^\alpha (1 + \gamma^5)
              (p'_\nu \gamma) \gamma^\beta (1 + \gamma^5)
               \right]
\nonumber \\
     &  & \times 
              {\rm Tr} \left\{ \bar{L}_{\beta j} \, [(p'\gamma)+m_e] \,
              L_{\alpha i} \,[(p\gamma)+m_e]  \right\} ,
\label{nusup-fluxa-M2} \\
        L_{\alpha i} & = & \Gamma_{\alpha} G(p-q) \gamma_i
              +   \gamma_i G(p'+q) \Gamma_{\alpha},
\label{nusup-fluxa-OL} \\
    G(P) & = & \frac{(p\gamma) + m_e}{p^2 - m_e^2},
              \quad \Gamma^{\alpha} = C_V \gamma^{\alpha} +
              C_A \gamma^\alpha \gamma^5.
\label{nusup-fluxa-GG}
\end{eqnarray}
Here, ${\bf q} = {\bf p} - {\bf p}' - {\bf k}$ is the
momentum transfer in the $(x,y)$-plane
from the electron to the fluxoid,
$q  = p - p' -k = (\Omega, {\bf q})$
is the appropriate 4-momentum transfer
(with vanishing energy transfer, $\Omega=0$),
${\bf A} \equiv {\bf A}({\bf q})$ is
the two-dimensional Fourier transform of the
magnetic-field vector-potential, which lies in the
$(x,y)$ plane.
Greek indices $\alpha$ and $\beta$ run over 0, 1, 2, and 3, while
Latin ones $i$ and $j$ refer to spatial components 1, 2, and 3;
$G(p)$ is the free-electron propagator
(Berestetskii et al.\ 1982).
%
%

Using the Lenard identity, Eq.\ (\ref{nucrust-Lenard}),
one has
\begin{eqnarray}
        Q_{\rm flux} & = &
             \frac{e^2 \, G_{\rm F}^2 \,{\cal N}_{\rm f} }{12 (2 \pi)^{9}}
             \int  {\rm d} {\bf p} \int  {\rm d} {\bf p}'
             \int  {\rm d} {\bf k} \;
             \delta(p_z - p'_z - k_z)
             \,  A^i A^{j\ast}
             \frac{\omega}{ \epsilon \epsilon'} \,
             J_{i j} \, f (1 - f') ,
\label{nusup-fluxa-Q1} \\
          J _{i j}  & = & \sum_\nu (k^\alpha k^\beta -
              k^2 g^{\alpha \beta}) \,
              {\rm Tr} \left[((p'\gamma) + m_e) L_{\alpha  i}
              ((p \gamma) + m_e)
              \bar{L}_{\beta  j}  \right].
\label{nusup-fluxa-Jgeneral}
\end{eqnarray}
The integration has to be
carried out over the kinematically allowed domain defined by
$k^2=\omega^2- {\bf k}^2 \geq 0$.

The two-dimensional Fourier transform $B(q)=B_z(q)$
of the fluxoid magnetic field (\ref{nusup-fluxa-Br})
in cylindrical coordinates is
\begin{eqnarray}
       B(q)  & = &
       \frac{\phi_0}{\lambda^2}
       \int_0^{\infty} {\rm d} r \, r K_0
       \left( \frac{r}{\lambda}  \right)
        J_0 (qr) =   \frac{\phi_0 \, q_0^2}{q^2 + q_0^2},
\label{nusup-fluxa-Bq}
\end{eqnarray}
where $J_0(x)$ is a Bessel function and $q_0 =1/\lambda$.
Then, using cylindrical gauge, we have
${\bf A}({\bf q})  =$
$- i {\bf e}_A \, B(q)/q$,
${\bf e}_A = ({\bf B \times q})/(Bq)$ being
a unit vector.
Accordingly, in Eq.\,(\ref{nusup-fluxa-Q1})
one gets $A^i A^{j\ast} =$
$[B(q)/q]^2 \, e_A^i e_A^j $,
where $i,j$ = 1 or 2, for nonvanishing components.

Equations (\ref{nusup-fluxa-Q1}) and (\ref{nusup-fluxa-Jgeneral}) determine
the neutrino emissivity for any degree
of electron degeneracy and relativism.
We are interested in the case
of ultra-relativistic, strongly degenerate electrons.
In the relativistic limit Kaminker et al.\ (1997)
obtained
$
e_A^i e_A^j \, J_{i j} \approx  C_+^2 J_+
$,
where $J_+$ is given by their cumbersome Eqs.\ (16)--(17),
and $C_+^2$ is defined in our Eq.\ (\ref{nucrust-annih-Q1}).

Using  Eqs.\ (\ref{nusup-fluxa-Q1})
and returning to standard physical units we get
\begin{eqnarray}
    Q_{\rm flux}  &  =  & \frac{ G_{\rm F}^2 e^2 \phi_0^2 C_+^2}
                          {2268 \, \hbar^9 c^8} \,
                          (k_{\rm B} T)^6 q_0
                          {\cal N}_{\rm f} \, L
\nonumber \\
             & \approx &  2.66  \times 10^{16} \, \bar{B}_{13} \, T_9^6 \,
                  \left( { n_p \, m_p \over n_0  \,
                   m_p^\ast }\right)^{1/2} \, \alpha_T
             L \quad {\rm erg~s^{-1}~cm^{-3}}.
\label{nusup-fluxa-Qthrl}
\end{eqnarray}
Here,
we introduced the dimensionless quantity $L$, defined by
\begin{eqnarray}
     L & = & \frac{189}{(2\pi)^9 T^6 q_0 } \int \, {\rm d} {\bf p}  \,
         {\rm d} {\bf p}' \, {\rm d} {\bf k} \,
         \delta(p_z - p'_z - k_z)
         \left( {B(q) \over q \phi_0} \right)^2  \,
         \frac{\omega}{\epsilon \epsilon'} \,
         J_+ \, f (1 - f').
\label{nusup-fluxa-Lgeneral}
\end{eqnarray}
Actually Eq.\ (\ref{nusup-fluxa-Lgeneral})
is valid for any axially symmetric
distribution of fluxoid magnetic field $B(r)$ although we
will use the specific distribution,
Eqs.\,(\ref{nusup-fluxa-Br}) and (\ref{nusup-fluxa-Bq}).
An analogous quantity $L$
was introduced in the case of neutrino bremsstrahlung
in electron-nucleus scattering
(Sect.\ \ref{sect-nucrust-ebrems}).

The evaluation of $L$ is greatly simplified by the
strong electron degeneracy. In addition, we imply the inequality
$q_0 \ll p_{\rm Fe}$ typical
for neutron star cores.
It is convenient to replace
the integration over ${\bf p}'$ in Eq.\,(\ref{nusup-fluxa-Lgeneral})
by the integration over ${\bf q}$.
Since the main contribution
into the integral comes from the values $q \ll p_{\rm Fe}$
it is sufficient to use the approximation
of small-angle scattering (small momentum transfer $q$
and small neutrino momentum $k \sim T$).
According to Kaminker et al.\ (1997),
the leading term in the expression for $J_+$ 
under formulated conditions is 
$
            J_+  =  8 (\omega^2 - {\bf k}^2) (q/q_x)^2,
$
where the $x$ axis is chosen to lie in the $z{\bf p}$ plane.

Energy and momentum restrictions in the small-angle
approximation yield that
the kinematic condition $\omega^2 \geq {\bf k}^2$ requires  $q_r >0$,
$\omega > q_r/2$ and $k_t^2 <  k_0^2 = q_r(2 \omega - q_r)$,
where the subscript $r$ denotes the vector component along ${\bf p}$
($q_r=q_x \, \sin \theta$, $\theta$ being the electron pitch angle)
and the subscript $t$ denotes the component transverse to ${\bf p}$.
Then Eq.\ (\ref{nusup-fluxa-Lgeneral})  can be
presented in the form 
%
\begin{eqnarray}
               L & = & {189 \over 8 \pi^7 \, T^6 \, q_0}
               \int_0^\pi {\rm d} \theta \,
               \sin^3  \theta  \int_0^\infty {\rm d} q_x
               \int_0^\infty
               {\rm d} q_y \,
\nonumber       \\
               &  \times  &
               \left( { B(q)  \over \phi_0} \right)^2
               \,  \int_{q_r/2}^{\infty} {\rm d} \omega \,
               \left( \omega - {q_r \over 2} \right)^2
               \, { \omega^2 \over {\rm e}^{\omega / T}-1},
\label{nusup-fluxa-LRTfinal}
\end{eqnarray}
where  
$B(q)$ is given by
Eq.\ (\ref{nusup-fluxa-Bq}).
It can be shown that $L$ depends on 
dimensionless parameter
$
      t_0=k_{\rm B}T/(\hbar c q_0)=  T/T_0
       \approx
      0.00786 \, T_9
      (m_p^\ast/m_p)^{1/2}(n_0/n_p)^{1/2}/\alpha_T,
$
where $T_0 = T_p \alpha_T$, and
$T_p = \hbar \omega_p / k_{\rm B}$
is the proton plasma temperature corresponding to the
proton plasma frequency $\omega_p$.
Typically, $T_p \sim 1$ MeV in the neutron star cores.

In the low-temperature regime, $T \ll T_0$,
Eq.\ (\ref{nusup-fluxa-LRTfinal}) gives
$L = \pi /4$.
Neutrino emission in this regime
is very similar to neutrino bremsstrahlung
due to quasielastic scattering of electrons by atomic nuclei
(Sect. \ref{sect-nucrust-ebrems}). In particular,
the emissivity $Q_{\rm flux}$
is proportional to $T^6$. Therefore, we will call this
regime the {\it bremsstrahlung regime}.

In the high-temperature regime, $T \gg T_0$,
Eq.\ (\ref{nusup-fluxa-LRTfinal}) reduces to the
asymptotic equation
$L \approx  (189 /\pi^6) \, \zeta(5)/t_0$,
where $\zeta(5) \approx 1.037$.
The temperature dependence ($Q_{\rm flux} \propto T^5$) is the same as
for the synchrotron emission of neutrinos by
electrons (Sect.\ \ref{sect-nucrust-syn}).
Thus we will refer to this regime
as the {\it synchrotron regime}.

Let us compare the neutrino emissivity due to
$ef$-scattering
in the synchrotron regime
with the emissivity $Q_{\rm syn}$
of the ``purely synchrotron process''
given by Eq.\ (\ref{nucrust-syn-QB}).
For this purpose let us
treat $B$ in $Q_{\rm syn}$ as a local magnetic field of a fluxoid and
average $Q_{\rm syn} \propto B^2$
over an ensemble of fluxoids with the magnetic field
(\ref{nusup-fluxa-Br}).
In this way we have
$\bar{Q}_{\rm syn} = (2/3) \, Q_{\rm flux}$, i.e.,
the emissivities differ exactly by a factor of $2/3$.
Let us emphasize, that this emissivity $\bar{Q}_{\rm syn}$,
{\it averaged over a nonuniform magnetic field}, can differ from the
synchrotron emissivity $Q_{\rm syn}$ in {\it the initial uniform
field} $\bar{B}$ by a large factor of
$\overline{ B^2} / \bar{B}^2 = \phi_0 q_0^2 / (4 \pi \bar{B}) \sim$
$(d_{\rm f}/\lambda)^2$, where $d_{\rm f}$  is
the inter-fluxoid distance defined above.
This increase of $\bar{Q}_{\rm syn}$
is caused by the  magnetic field enhancement within
the fluxoids due to magnetic flux conservation.
The enhancement is much stronger than the reduction of the
neutrino-emission volume, i.e., the volume  occupied by the fluxoids.

An analysis shows, that
the difference by a factor of 2/3 comes from momentum space
available for neutrino-pair momentum {\bf k}. The
space is different in the
case of synchrotron radiation in a {\it uniform magnetic field}
and in the case of $ef$ scattering
{\it off magnetic inhomogeneities}  (in our case, fluxoids).
The detailed explanation is given by Kaminker et al.\ (1997).

Combining the above results,
Kaminker et al.\ (1997) proposed
the following fit expression for the quantity
$L$ in Eq.\,(\ref{nusup-fluxa-Qthrl}):
\begin{eqnarray}
    L & = & L_0 \, U \, V, \quad
       L_0  =  {\pi \over 4} \,
       {0.260 \, t_0 + 0.0133  \over
       t_0^2 + 0.25 \, t_0 + 0.0133}, \quad
       U = {2 \gamma+1  \over 3 \gamma+1 },
\label{nusup-fluxa-F1F2} \\
       \gamma & = & \left({ \hbar \omega_B^\ast  \over \mu_e} \right)^2
       { t_0 \over y_0^2}
       \approx \, 8.38 \times 10^{-12} \, T_9
       \bar{B}_{13}^2 \,
       \left( { n_0 \over n_e } \right)^{13/6} \,
       \left( {m_p^\ast \over m_p } \right)^{3/2} \,
       { 1 \over \alpha_T^3},
\label{nusup-fluxa-2/3} \\
       V & = & 1 + {4 \pi \bar{B} \lambda^2 \over \phi_0}
       \approx  1 + 0.00210 \, \bar{B}_{13} \,
       {m_p^\ast \, n_0 \over m_p\, n_p\, \alpha_T^2}.
\label{nusup-fluxa-unif}
\end{eqnarray}
Here, $y_0 = \hbar q_0/(2 p_{{\rm F}e})=\hbar /(2 \lambda p_{{\rm F}e})$
(assumed to be a small parameter, $y_0 \ll 1$),
$\omega_B^\ast=|e| Bc/\mu_e$ is the electron gyrofrequency,
and $L_0$ is given by the analytic expression
which fits the results of numerical calculations
of $L$ from Eq.\ (\ref{nusup-fluxa-LRTfinal})
with the error $<$ 1\% for any $t_0$.
The quantity $U$ ensures, somewhat arbitrarily,
the difference by a factor of 2/3 between the cases of
weakly and strongly nonuniform magnetic fields in the synchrotron regime.
The factor $V$ provides a
smooth transition from the $ef$ scattering to the pure
synchrotron emission in the case in which the fluxoid radius
$\lambda$ becomes
very large ($T \to T_{cp}$), the neighboring fluxoids overlap and the
overall magnetic field is nearly uniform.

Using these results, we can  follow evolution of $Q_{\rm flux}$
in the course of superconductivity onset. If $T \geq T_{cp}$,
the emissivity $Q_{\rm flux}$
is given by Eq.\ (\ref{nucrust-syn-QB}) since it
is essentially the same
as the synchrotron emissivity
$Q^{(0)}_{\rm syn}=Q_{\rm syn}(\bar{B})$ in a locally uniform
`primordial' magnetic field $B=\bar{B}$.
After $T$ falls only slightly below $T_{cp}$,
factor $U$ transforms from $U=2/3$ (`pure synchrotron')
to $U=1$ ($ef$ scattering in the synchrotron regime).
The fluxoid structure is still not very pronounced, i.e.,
$V \approx 4 \pi \bar{B} \lambda^2 / 
\phi_0 \approx \lambda^2 / d_{\rm f}^2 \gg 1 $,
and $Q_{\rm flux} \sim Q_{\rm syn}^{(0)}$.
When $T$ decreases to about $0.8 \, T_{cp}$,
we have the fluxoid radius $\lambda \approx\lambda_0$, and
$V \simeq 1$. Accordingly, $Q_{\rm flux} /Q^{(0)}_{\rm syn}$ grows
by a factor of $(d_{\rm f}/\lambda_0)^2$ due to the magnetic field
confinement within the fluxoids. Simultaneously,
$t_0$ decreases from very large values at  $T \to T_{cp}$
to  $t_0\simeq T_{cp}/T_p$.

Let us first consider the case in which the proton
plasma temperature is $T_p \ll T_{cp}$.
Then the emissivity
$Q_{\rm flux}$ is enhanced with respect to $Q^{(0)}_{\rm syn}$
by a factor of $(d_{\rm f}/\lambda_0)^2$ at $T \approx 0.8 T_{cp}$,
and this enhancement remains nearly constant over a wide temperature
range down to $T_p$. Within this range, the $ef$ scattering
operates in the synchrotron regime.
At lower temperatures, $T \lesssim T_p$,
the synchrotron regime transforms into the bremsstrahlung regime
and we have $Q_{\rm flux} \sim Q^{(0)}_{\rm syn}
(d_{\rm f}/\lambda_0)^2 \, (T/T_p)$.
Thus, for $T\lesssim T_p$ the emissivity $Q_{\rm flux}$
decreases with respect to $Q^{(0)}_{\rm syn}$ and
may become lower than $Q^{(0)}_{\rm syn}$.

In the opposite case, $T_p \gtrsim T_{cp}$, the $ef$ scattering operates
in the synchrotron regime only in a narrow
temperature range below $T_{cp}$ and transforms  into
the bremsstrahlung regime, 
$Q_{\rm flux} \sim $
$Q^{(0)}_{\rm syn}(d_{\rm f}/\lambda)^2 (T/T_p)$, at lower $T$.

\begin{figure}[!t]                         
\begin{center}
\leavevmode
\epsfysize=8.0cm
\epsfbox[65 35 375 375]{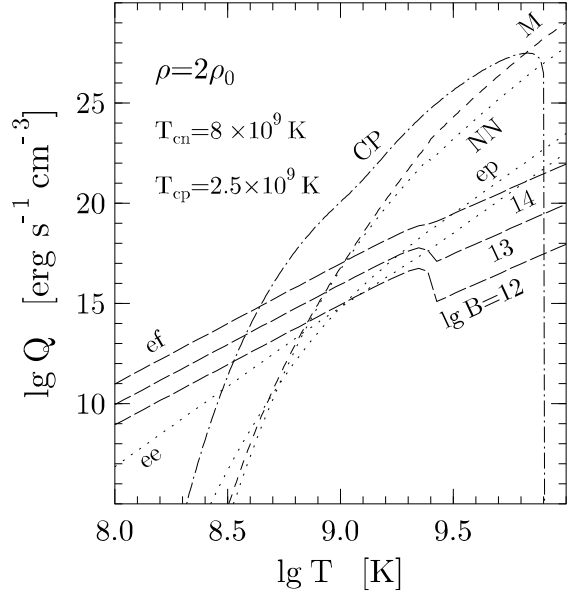}
\end{center}
\caption[]{\footnotesize
        Temperature profile of the neutrino emissivities
        in different reactions in superfluid magnetized
        neutron star matter
        at $\rho = 2 \, \rho_0$,
        $T_{cn}=8 \times 10^9$ K, $T_{cp}=2.5 \times 10^9$ K and
        $\bar{B}=10^{12}$, $10^{13}$, $10^{14}$ G.
        Curve $M$ corresponds to the modified Urca process,
        curve $CP$ to neutrino emission due to Cooper
        pairing, 
        curve $N\!N$ shows nucleon-nucleon bremsstrahlung,
        curves $ee$ and $ep$ show Coulomb bremsstrahlung,
        while $ef$ refers to synchrotron-fluxoid scattering.
        }
\label{fig-nusup-fluxa}
\end{figure}

The results are illustrated in Fig.\ \ref{fig-nusup-fluxa}
which shows temperature dependence of various neutrino
emissivities at $\rho= 2 \, \rho_0$,
$T_{cn}=8 \times 10^9$ K and $T_{cp}=2.5 \times 10^9$ K.
We present the emissivities of modified Urca process
(sum of the proton and neutron branches),
Cooper pairing processes (sum of neutron and proton reactions),
nucleon-nucleon bremsstrahlung (sum of $nn$, $np$ and $pp$
processes), $ep$ and $ee$ bremsstrahlungs, and
synchrotron-fluxoid emission. The emissivities
of all the processes but the latter one
are independent of $B$ as long as $B \lesssim 10^{18}$ G.
The synchrotron neutrino emissivity
is displayed for
three values of the magnetic field
$\bar{B}=10^{12}$, $10^{13}$ and $10^{14}$ G.
The straight lines for $T>T_{cp}$ correspond
to purely synchrotron emission in a locally uniform
magnetic field. If the proton superfluidity were
absent, these lines would be extended to lower $T$,
but neutrino bremsstrahlung in $pp$ collisions
would not be suppressed by the superfluidity
and would be dominant at lower $T$ displayed.
In our example
the superfluidity of protons and neutrons at low $T$
reduces all
neutrino processes which involve nucleons.
If the magnetic field were absent, the $ee$ scattering
would be the main neutrino mechanism at
$T \lesssim 3 \times 10^8$ K. However,  the $ef$
scattering in the presence of $\bar{B} \gtrsim
10^{12}$ G also becomes important and dominant.
Thus both processes, $ee$ scattering and
synchro-fluxoid neutrino emission, may contribute to
the residual neutrino emission of highly superfluid
neutron stars at low temperatures.

After the superfluidity onset ($T < T_{cp}$),
the magnetic field splits into fluxoids
and the synchrotron radiation transforms into the
radiation due to the $ef$ scattering.
This enhances $Q_{\rm flux}$ owing
to the field amplification
within the fluxoids. The enhancement
is more pronounced at lower $\bar{B} \sim 10^{12}$ G
at which the field amplification is stronger. This is
a rare case,
similar to Cooper-pairing neutrino emission,
in which the neutrino emissivity {\it increases} with
decreasing $T$. For the conditions displayed in Fig.\ \ref{fig-nusup-fluxa}
the proton plasma temperature $T_p =5.47 \times 10^{10}$ K
is much higher than the superfluid critical temperature $T_{cp}$.
Therefore, the synchrotron regime in the $ef$
scattering operates only at $T \sim T_{cp}$, during the 
fluxoid formation phase.
Very soon after $T$ falls below $T_{cp}$, the synchrotron
regime transforms into the bremsstrahlung regime
which operates further with decreasing $T$.

Although we have considered $npe$ matter and
superconductivity of protons, the presented results can be
extended to superconductivity
of other charged baryons. Similar synchrotron-fluxoid
neutrino emission can be produced by relativistic muons.
Analogous emission can be produced also by non-relativistic
charged baryons or muons but it requires additional
study. Nevertheless it is unlikely that
this latter emission can be much stronger than the
neutrino emission by electrons analyzed above.

In superconducting $npe$ matter,
the neutrinos can also be generated
in $pp$ collisions
of {\it normal} protons in the non-superfluid cores of the
fluxoids as well as
in  $en$ collisions. However, the non-superfluid fluxoid cores 
are very thin, the
emission volume is minor, and  the first process is
inefficient. Neutrino bremsstrahlung due
to $en$ scattering is also inefficient
since it occurs through electromagnetic interaction
involving neutron magnetic
moment (e.g., Baym et al.\ 1969) and since it can be suppressed
%
%
by the neutron superfluidity.

\newpage

\section{Neutrino emission and neutron star cooling}
\label{chapt-cool}

In this chapter we illustrate the interplay of various neutrino
processes in the cooling neutron stars.  Simulations of neutron star
cooling have a long history described, for instance, by Tsuruta (1998)
and Yakovlev et al.\ (1999b).  However, the physical input of the models
is being constantly updated as new, more accurate calculations become
available.  Instead of a historical review, we present 
(Sects.\  \ref{sect-cool-code} -- \ref{sect-cool-super}) the results
of the latest cooling simulations 
which are partly published elsewhere
(Gnedin et al.\ 2000).  We use neutron star models which
include both the core and the crust neutrino processes with the proper
treatment of possible superfluid effects.  We discuss how thermal
relaxation establishes in the crust and relates the internal temperature
to the observable effective surface temperature.  We emphasize the
importance of various neutrino emission processes in different layers of
neutron stars (Chapts.\ \ref{chapt-nucrust}--\ref{chapt-nusup}) for the
cooling theory and the interpretation of observations of isolated
neutron stars. In addition, in Sect.\ \ref{sect-cool-durmag}
we describe the results of recent cooling calculations
(Baiko and Yakovlev 1999) of neutron stars with strong internal
magnetic fields. They illustrate
the magnetic broadening of the threshold of direct Urca process
analyzed in Sect.\ \ref{sect-nusup-durmag}.

\subsection{Cooling equations and cooling code}
\label{sect-cool-code}

{\bf (a) Equations of thermal evolution}

Neutron stars are born very hot in supernova explosions, with the
internal temperature $T \sim 10^{11}$ K, but gradually cool down.  About
twenty seconds after the birth, the stars become fully transparent for
the neutrinos generated in numerous reactions in stellar interiors.  We
will consider cooling in the following neutrino-transparent stage.  The
cooling is realized via two channels, by neutrino emission from the
entire stellar body and by heat transport from the internal layers
to the surface resulting in thermal emission of photons.  Some
processes (e.g., frictional dissipation of the rotational energy or
Ohmic decay of the internal magnetic field,
see Sect. \ref{sect-cool-other}) may reheat stellar interior
thus delaying the cooling, especially at late stages, but we neglect
this for simplicity.  Since the internal neutron star structure may
be regarded as temperature-independent, the equations of stellar
structure are decoupled from the equations of thermal evolution.

The equations of neutron star structure and evolution are essentially
relativistic.  For a spherical star it is customary to adopt a
stationary, spherically symmetric space--time metric of the form
\begin{equation}
    {\rm d}s^2=c^2 \, {\rm d}t^2 \,{\rm e}^{2 \Phi}
         - \left( 1 - {2 G m \over r c^2} \right)^{-1} \,
         {\rm d}r^2 - r^2 \,
         ({\rm d} \theta^2 + \sin^2 \theta \, {\rm d} \phi^2),
\label{cool-metrics}
\end{equation}
where $G$ is the gravitational constant,
$t$ is a time-like coordinate, $r$ is a radial
coordinate,
$\theta$ and $\phi$ are the polar and azimuthal
angles, respectively; $\Phi=\Phi(r)$ is the metric function
which determines gravitational redshift, and
$m(r)$ represents the {\it gravitational mass}
enclosed within a sphere of radius $r$.
Stellar structure is described (e.g., Shapiro and Teukolsky 1983)
by the well known Oppenheimer-Tolman-Volkoff equation supplemented
by the equation for the metric function $\Phi(r)$ and the equation of state
of stellar matter. The boundary conditions at the stellar
surface $r=R$ are $P(R)=0$ and $m(R)=M$,
where $M$ is the total gravitational mass.
Outside the star the metric reduces to the Schwarzschild
form: $m(r)=M$ and $\exp \Phi(r)=(1-r_g/r)^{1/2}$, with
$r_g=2GM/c^2$ being the Schwarzschild radius.

The general relativistic equations of thermal evolution of
a spherically symmetric star were
derived by Thorne (1977).
They include, basically, two equations: {\it the thermal balance
equation} and {\it the thermal transport equation}.

The thermal balance equation can be written as
\begin{equation}
    { 1 \over 4 \pi r^2 {\rm e}^{2 \Phi}} \,
    \sqrt{1 - {2 G m \over c^2 r}} \,
    { \partial  \over \partial  r}
    \left( {\rm e}^{2 \Phi} L_r \right)
    = -Q_\nu - {C_v \over {\rm e}^\Phi} \,
     {\partial T \over \partial t},
\label{cool-therm-balance}
\end{equation}
where $Q_\nu$ is the
neutrino emissivity
[erg~cm$^{-3}$~s$^{-1}$],
$C_v$ is the specific heat capacity at constant volume [erg~cm$^{-3}$~K$^{-1}$],
and $L_r$ is the ``local luminosity" defined as the
non-neutrino heat flux [erg s$^{-1}$] transported through a sphere of 
radius $r$.

If the heat is carried by thermal conduction,
the equation of heat transport can be written as
\begin{equation}
    {L_r \over 4 \pi \kappa r^2} =
    - \sqrt{1 - {2Gm \over c^2 r}} \; {\rm e}^{-\Phi}
    {\partial \over \partial r} \left( T {\rm e}^\Phi \right),
\label{cool-Fourier}
\end{equation}
where $\kappa$ is the coefficient of thermal conductivity.
One needs to solve Eqs.\ (\ref{cool-therm-balance})
and (\ref{cool-Fourier}) simultaneously
to determine $L_r(r,t)$ and $T(r,t)$.

To facilitate the simulation of neutron
star cooling one usually subdivides the
problem {\it artificially} by analyzing 
separately heat transport
in the outer heat-blanketing envelope 
($R_b \leq r \leq R$) and in the interior
($r < R_b$). We follow this standard
prescription and take the blanketing envelope
extending to the density
$\rho_b=10^{10}$ g cm$^{-3}$ at $r=R_b$
($\sim 100$ meters under the surface).
The blanketing envelope is relatively thin
and contains negligibly small mass;
it includes no large sources of energy sink or generation;
it serves as a good thermal insulator of
the internal region; and its thermal relaxation time
is shorter than the time-scales of temperature variation
in the internal region. Accordingly, the bulk of neutrino
generation occurs in the internal region, which also contains most of the 
heat capacity. The thermal structure of the blanketing
envelope is studied separately by solving
Eqs.\ (\ref{cool-therm-balance}) and (\ref{cool-Fourier})
in the stationary, plane-parallel
approximation. The solution of these equations enables one to
relate the effective
surface temperature $T_s$ to the temperature $T_b$
at the inner boundary of the envelope. We use the
$T_s$--$T_b$ relation obtained by Potekhin et al.\ (1997)
for the envelope composed mostly of iron.

The effective temperature determines the photon luminosity,
$
  L_\gamma = L_r(R,t)=4 \pi \sigma R^2 T^4_s(t).
$
The quantities $L_\gamma$ and $T_s$ refer to the
locally-flat reference frame at the neutron star surface.
One often defines the ``apparent" luminosity $L_\gamma^\infty$,
``apparent" effective surface temperature $T_s^\infty$
and ``apparent" radius $R_\infty$ as would be registered by
a distant observer:
\begin{eqnarray}
&&   L_\gamma^\infty = L_\gamma  \, (1 - r_g/R) =
     4 \pi \sigma (T_s^\infty)^4 R_\infty^2,
\label{cool-L_gamma_infty}\\
&&   T_s^\infty = T_s \, \sqrt{1 - r_g/R}, \quad
     R_\infty = R/ \sqrt{1 - r_g/R}.
\label{cool-T_s_infty}
\end{eqnarray}
%

The main goal of the cooling theory is to calculate
the {\it cooling curve}, the dependence of
the local effective temperature $T_s$ or the apparent
temperature $T_s^\infty$ (or, equivalently, of
$L_\gamma^\infty$) on stellar age $t$. With the above simplifications,
the problem reduces to solving Eqs.\ (\ref{cool-therm-balance})
and (\ref{cool-Fourier}) in the 
neutron star interior, $r \leq R_b$.

Thermal history of an isolated neutron
star can be divided into
three stages: (a) {\it internal relaxation stage}
($t \lesssim 10$--1000 yr),
(b) {\it neutrino cooling stage} ($t \lesssim 10^5$ yr),
and (c) {\it photon
cooling stage} ($t \gtrsim 10^5$ yr).
As we discuss later, the transition from one stage to another may vary
depending on the stellar model.
After the thermal relaxation is over, the stellar
interior becomes isothermal. With 
the effects of general relativity, the quantity
\begin{equation}
      \widetilde{T}(t)= T(r,t)\; {\rm e}^{\Phi(r)}
\label{therm-Tconst}
\end{equation}
becomes constant throughout the internal stellar region.
This quantity can be called
the redshifted internal temperature.
Only the outermost stellar layer (within the heat blanketing
envelope) remains non-isothermal.
The approximation of isothermal interior greatly
simplifies the cooling equations (e.g., Glen and Sutherland 1980). 
%
%
The star
cools mainly via the neutrino emission from its core
at the neutrino cooling stage and via the photon emission from its
surface at the photon cooling stage.\\

{\bf (b) Numerical code and physics input}
 
In Sect.\ \ref{sect-cool-code}--\ref{sect-cool-super}
we present the results of
new cooling simulations
performed recently 
%
%
with the new evolutionary code developed by
one of the authors (OYG).
Using the Henyey-type scheme (Kippenhahn et al.\ 1967)
%
%
the code solves the system of
partial differential equations
(\ref{cool-therm-balance}) and (\ref{cool-Fourier}) on a
grid of spherical shells.  The hydrostatic model of the
neutron star with a given equation of state is calculated separately and
is fixed throughout the evolutionary calculation.  Full details of the new
code are available elsewhere
(http://www.ast.cam.ac.uk/$\sim$ognedin/ns/ns.html).

For simplicity, we consider the neutron star cores composed
of neutrons, protons and electrons.  The three most important
physical ingredients for the study of the thermal
evolution are the neutrino emissivity, specific heat capacity, and
thermal conductivity. The code includes all relevant sources of neutrino
emission: the direct and modified Urca processes, $nn$, $pp$, and $np$
bremsstrahlung in the core (Chapt.\ \ref{chapt-nucore}); 
as well as plasmon decay,
electron-nucleus ($eZ$) and $nn$ bremsstrahlung, $e^-e^+$ pair
annihilation in the crust (Chapt.\ \ref{chapt-nucrust}).  The
reduction of the neutrino reactions by superfluid nucleons
is properly included (Chapts.\ \ref{chapt-nusup} and
\ref{chapt-nucrust}). In the core, we consider the singlet-state pairing
of protons and the triplet-state pairing of neutrons (with zero
projection of the total angular 
momentum of the pair onto quantization axis, case B in
notations of Sect.\ \ref{sect-sf-gaps}).
In the crust, we consider the singlet-state
superfluidity of free neutrons at densities higher than the neutron drip
density. The adopted superfluid models are 
described in more detail in Sect. \ref{sect-cool-super}. 
We also include an additional neutrino emission due to Cooper
pairing of superfluid nucleons.
The effective masses of nucleons in the core and free
neutrons in the crust are set to be constant and equal to 0.7 of their
bare masses.

The heat capacity is contributed by neutrons, protons and electrons
in the core; and by electrons,
free neutrons, and atomic nuclei (ions) in the crust.
The effects of nucleon superfluidity on
the heat capacity of the nucleons in the core
and of the free neutrons in the crust
are incorporated according to Levenfish and Yakovlev (1994a).
The thermal conductivity in the crust is assumed to be due to
the electrons which scatter mainly off atomic nuclei. 
We use the subroutine kindly provided by
A.\ Potekhin which includes the 
form factor of proton charge distribution
for finite-size nuclei (Potekhin et al.\ 1999).
Relative to the approximation of point-like nuclei,
the form factor increases
the electron thermal conductivity in deep layers
of the crust ($\rho \sim 10^{14}$ g cm$^{-3}$) by almost a factor of 2--5.
The thermal conductivity
in the core is taken as a sum of the conductivities of
the electrons and neutrons. The electron
contribution is
calculated according to Gnedin and Yakovlev (1995)
and the neutron contribution according to
Baiko et al.\ (2000).

In the initial configuration the star has a constant redshifted
temperature throughout the interior,
$\widetilde{T} = 10^{10}$ K, and no heat flux, $L_r=0$. 
The calculated cooling curves are insensitive to the initial
temperature profile.  Also, to
improve numerical convergence the thermal conductivity in the core is
busted for the initial epoch $t < 10^{-2}$ yr.  Since the crust is
thermally detached from the core at such small age, this correction has
no effect on the cooling curves.

We use the equation of state of Negele and Vautherin (1973)
in the stellar crust with the smooth composition model
of ground-state matter to describe the properties of atomic
nuclei (Sect.\ \ref{sect-nucrust-introduc}).
We assume that the nuclei are spherical throughout the entire crust.
We use the three simple phenomenological models
proposed by Prakash et al.\ (1988) for the uniform matter
in the stellar core. The core--crust interface is placed at
$\rho_{cc}=1.5 \times 10^{14}$ g cm$^{-3}$. The equations of state
in the core correspond to the three choices
of the compression modulus of 
symmetric nuclear matter at saturation density:
$K_0=$ 120, 180 and 240 MeV. We will refer to these
examples as the
{\it soft}, {\it moderate}, and {\it stiff} equations of
state, for brevity, although in fact the last one
is intermediate
between the moderate and the stiff.
They give different neutron star models. 
For all three cases we use
the simplified
form of the symmetry energy proposed by Page and Applegate (1992),
in agreement with our previous work
(Yakovlev et al.\ 1999b and references therein).

\begin{table*}[!t]
\caption{Neutron star models}
\begin{center}
\begin{tabular}{||llllllll||}
\hline\hline
EOS & $M$  & $R$  & $\rho_c$ ($10^{14}$
    & $M_{\rm crust}$ & $\Delta R_{\rm crust}$$^c$ 
    & $M_{\rm D}$ & $R_{\rm D}$  \\
    & ($M_\odot$) & (km) & g cm$^{-3}$)   & ($M_\odot$)
    & (km)  & ($M_\odot$) & (km) \\
\hline
moderate & 1.1      & 12.20 &  8.50 & 0.050 & 1.66 & \ldots & \ldots \\
\ldots   & 1.2      & 12.04 &  9.52 & 0.044 & 1.45 & \ldots & \ldots \\
\ldots   & 1.3      & 11.86 & 10.70 & 0.039 & 1.26 & \ldots & \ldots \\
\ldots   & 1.4      & 11.65 & 12.20 & 0.033 & 1.09 & \ldots & \ldots \\
\ldots   & 1.44$^a$ & 11.54 & 12.98 & 0.031 & 1.02 & 0.000  & 0.00 \\
\ldots   & 1.5      & 11.38 & 14.20 & 0.028 & 0.93 & 0.065  & 2.84 \\
\ldots   & 1.6      & 11.01 & 17.20 & 0.022 & 0.77 & 0.301  & 4.61 \\
\ldots   & 1.7      & 10.37 & 23.50 & 0.016 & 0.59 & 0.685  & 5.79 \\
\ldots   & 1.73$^b$ &  9.71 & 32.50 & 0.011 & 0.47 & 0.966  & 6.18 \\
\hline
soft     & 1.0      & 11.61 & 10.36 & 0.046 & 1.69 & \ldots & \ldots \\
\ldots   & 1.1      & 11.31 & 12.17 & 0.039 & 1.42 & \ldots & \ldots \\
\ldots   & 1.12$^a$ & 11.24 & 12.69 & 0.037 & 1.36 & 0.000  & 0.00 \\
\ldots   & 1.2      & 10.98 & 14.54 & 0.032 & 1.18 & 0.087  & 3.14 \\
\ldots   & 1.3      & 10.56 & 17.98 & 0.025 & 0.97 & 0.302  & 4.61 \\
\ldots   & 1.4      &  9.95 & 24.18 & 0.018 & 0.74 & 0.606  & 5.55 \\
\ldots   & 1.46$^b$ &  8.91 & 40.40 & 0.010 & 0.51 & 0.955  & 5.96 \\
\hline
stiff    & 1.2      & 12.58 &  7.56 & 0.048 & 1.55 & \ldots & \ldots \\
\ldots   & 1.3      & 12.50 &  8.25 & 0.044 & 1.38 & \ldots & \ldots \\
\ldots   & 1.4      & 12.40 &  9.07 & 0.039 & 1.23 & \ldots & \ldots \\
\ldots   & 1.5      & 12.28 & 10.01 & 0.035 & 1.09 & \ldots & \ldots \\
\ldots   & 1.6      & 12.12 & 11.19 & 0.031 & 0.96 & \ldots & \ldots \\
\ldots   & 1.7      & 11.91 & 12.74 & 0.026 & 0.84 & \ldots & \ldots \\
\ldots   & 1.73$^a$ & 11.83 & 13.31 & 0.025 & 0.80 & 0.000  & 0.00 \\
\ldots   & 1.8      & 11.62 & 14.99 & 0.022 & 0.71 & 0.110  & 3.35 \\
\ldots   & 1.9      & 11.11 & 19.35 & 0.016 & 0.57 & 0.472  & 5.24 \\
\ldots   & 1.94$^b$ & 10.41 & 27.30 & 0.011 & 0.45 & 0.865  & 6.10 \\
\hline
\end{tabular}
\begin{tabular}{l}
  $^a$  threshold configuration for switching on direct Urca process \\
  $^b$  configuration with maximum allowable mass\\
  $^c$  $\Delta R_{\rm crust}$ is defined as $R - R_{\rm core}$
\end{tabular}
\label{tab-cool-model}
\end{center}
\end{table*}

The parameters of the models are summarized in Table \ref{tab-cool-model}.
We present stellar masses, radii, central densities,
crust masses, and crust thicknesses
for a number of models with each of the three equations of state
in the core, including
the configurations with the maximum allowable mass.
We define the crust thickness as 
$\Delta R_{\rm crust}=R - R_{\rm core}$ while
the proper geometrical thickness is
$\Delta R_{\rm crust}/\sqrt{1 - r_g/R}$.
The maximum masses of
the stable neutron stars are 1.46 $M_\odot$, 1.73 $M_\odot$
and 1.94 $M_\odot$ for the soft, moderate, and stiff
equations of state, respectively. If we fix the equation of state
and increase $M$ (or the central density $\rho_c$), 
the radii and crust masses
of the stable configurations get smaller, i.e. the stars
become more compact.  For a fixed stellar mass,
the star with stiffer equation of state
has larger radius and more massive crust.

The most important effect for neutron star cooling is
the operation of the powerful
direct Urca process (Sect.\ \ref{sect-nucore-Durca}).
For our simplified equations of state ($npe$ matter),
the direct Urca process is open if the ratio
of the proton to nucleon number densities
exceeds 1/9.
This condition is satisfied for the stellar models
in which the central density $\rho_c$ exceeds
the direct Urca threshold density.
This threshold is $\rho_{\rm crit}=1.269 \times 10^{15}$,
$1.298 \times 10^{15}$ and $1.331 \times 10^{15}$
g cm$^{-3}$, for the soft, moderate, and stiff
equations of state, respectively.
If $\rho_c > \rho_{\rm crit}$,
the stellar core has a central kernel, where the direct
Urca process leads to fast cooling.
The masses and radii of these kernels, $M_{\rm D}$ and
$R_{\rm D}$, are also given in Table \ref{tab-cool-model}.
In addition, we give the threshold stellar configurations
for which $\rho_c = \rho_{\rm crit}$.
They separate the sequence of stellar models into the low-mass
models, where the direct Urca process is forbidden,  
and high-mass models, where the direct Urca
is allowed. The mass of the central kernel increases
rapidly with $M$ in the latter models.

\subsection{Cooling of non-superfluid neutron stars}
\label{sect-cool-nonsup}

First, we investigate neutron star cooling without the effects of
superfluidity.  In this section we outline
the main features of the non-superfluid cooling,
which are generally well known from
previous calculations (e.g., Page and Applegate 1992,
Lattimer et al.\ 1994, Page 1998a, 1998b, and references
therein). We illustrate them with the cooling curves
obtained with our thermal evolution code.

\begin{figure}[th!]
\begin{center}
\leavevmode
\epsfysize=9cm \epsfbox{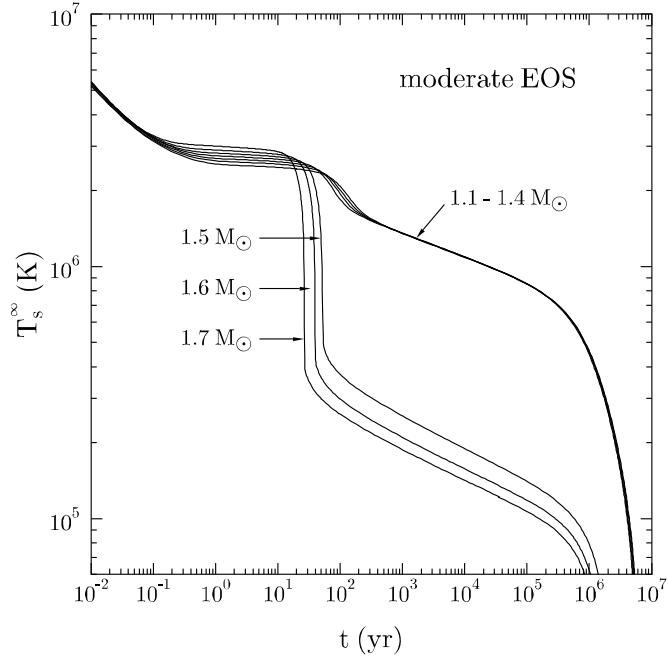}
\end{center}
\caption[ ]{ 
  Cooling curves for the neutron star models with 1.1, 1.2
  \ldots, 1.7 $M_\odot$ with the moderate equation of state
  and no superfluid effects (Gnedin et al.\ 2000).
}
\label{fig:c_nosf_main}
\end{figure}

\begin{figure}[ht!]
\begin{center}
\leavevmode
\epsfysize=9cm \epsfbox{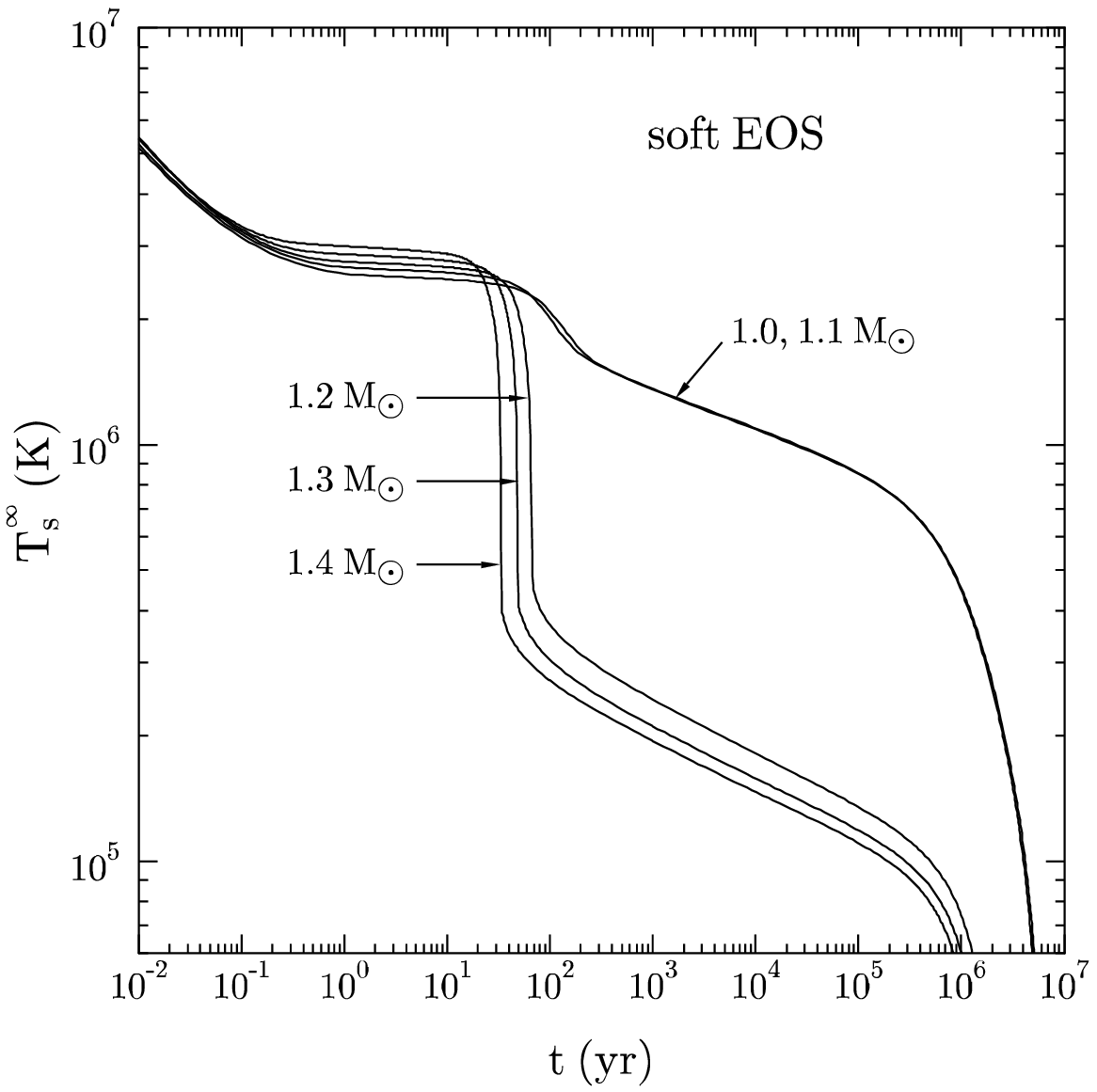}
\end{center}
\caption[ ]{
  Cooling curves for the non-superfluid neutron star models with
  1.0, 1.1, \ldots, 1.4 $M_\odot$ with the soft equation of state.
}
\label{fig:c_nosf_soft}
\end{figure}

\begin{figure}[ht!]
\begin{center}
\leavevmode
\epsfysize=9cm \epsfbox{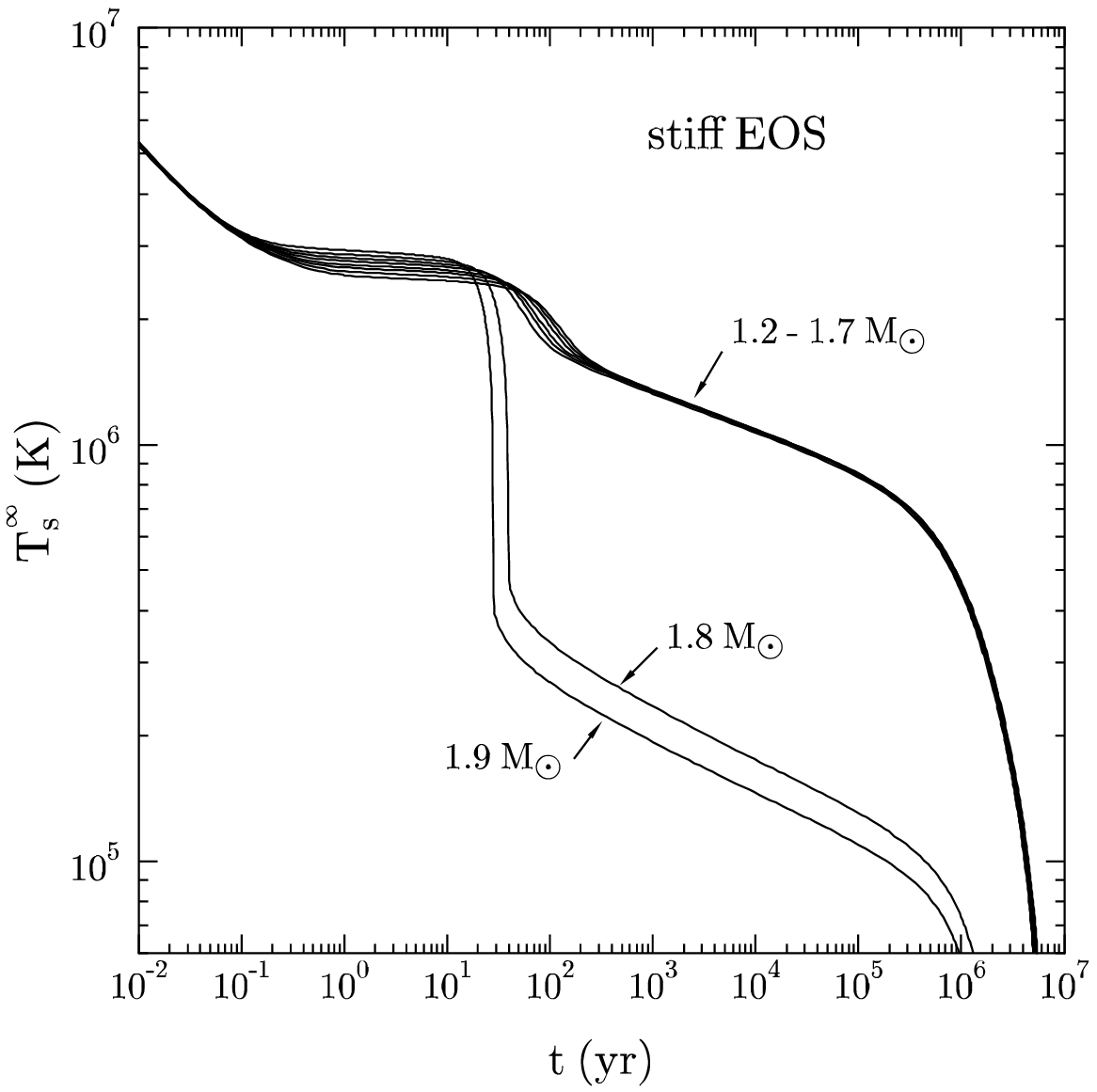}
\end{center}
\caption[ ]{
  Cooling curves for the non-superfluid neutron star models with
  1.2, 1.3, \ldots, 1.9 $M_\odot$ with the stiff equation of state.
}
\label{fig:c_nosf_stiff}
\end{figure}

\begin{figure}[ht!]
\begin{center}
\leavevmode
\epsfysize=9cm \epsfbox{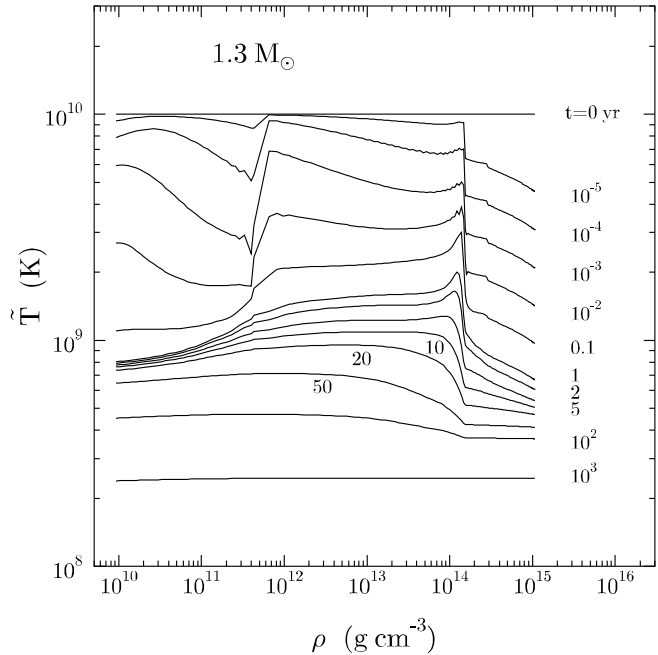}
\end{center}
\caption[ ]{
  Temperature profiles 
  in the interior  
  of the 1.3 $M_\odot$ model
  with the moderate equation of state and no superfluid effects 
  (Gnedin et al.\ 2000).
  Numbers next to curves show stellar age.
  Contours are at 0, $10^{-5}$, $10^{-4}$, $10^{-3}$, $10^{-2}$, $10^{-1}$,
  1, 2, 5, 10, 20, 50, 100, and 1000 yr.  After 1000 yr the
  stellar interior is isothermal.
}
\label{fig:config3_nosf}
\end{figure}

\begin{figure}[ht!]
\begin{center}
\leavevmode
\epsfysize=9cm \epsfbox{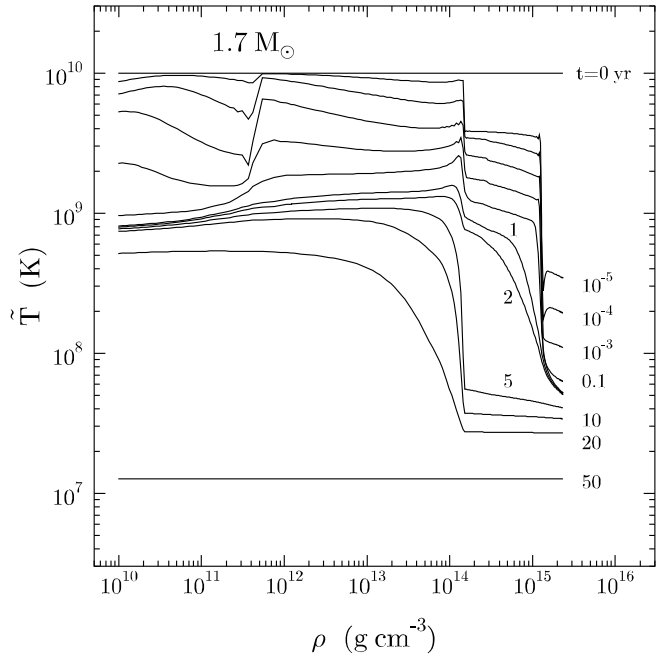}
\end{center}
\caption[ ]{
  Temperature profiles 
  in the interior
  of the 1.7 $M_\odot$ model with
  the moderate equation of state and no superfluid effects 
  (Gnedin et al.\ 2000).
  Contours are at 0, $10^{-5}$, $10^{-4}$, $10^{-3}$, $10^{-2}$, $10^{-1}$,
  1, 2, 5, 10, 20, and 50 yr.  After 50 yr the
  stellar interior is isothermal.
}
\label{fig:config7_nosf}
\end{figure}

\begin{figure}[ht!]
\begin{center}
\leavevmode
\epsfysize=9cm \epsfbox{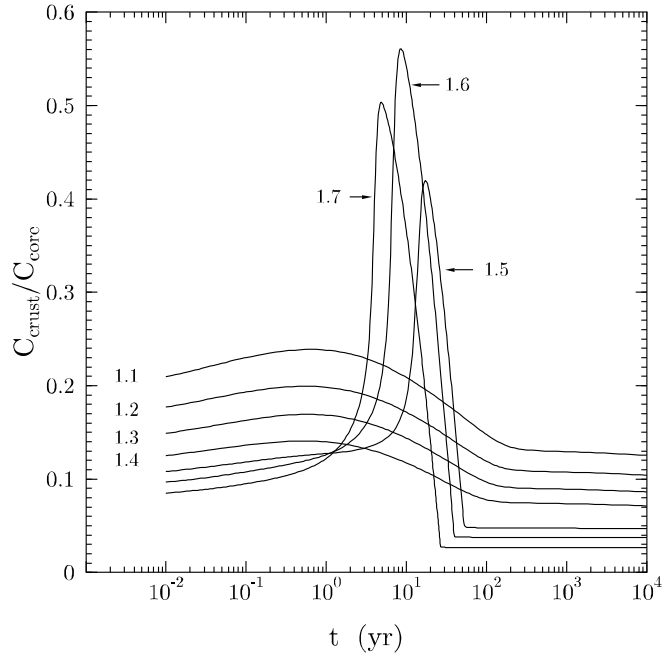}
\end{center}
\caption[ ]{
  Ratio of the integrated heat capacities in the crust and the core for
  the neutron star models of masses
  1.1, 1.2, \ldots, 1.7 $M_\odot$
  with the moderate equation of state (Gnedin et al.\ 2000).
}
\label{fig:heat}
\end{figure}

\begin{figure}[ht!]
\begin{center}
\leavevmode
\epsfysize=9cm \epsfbox{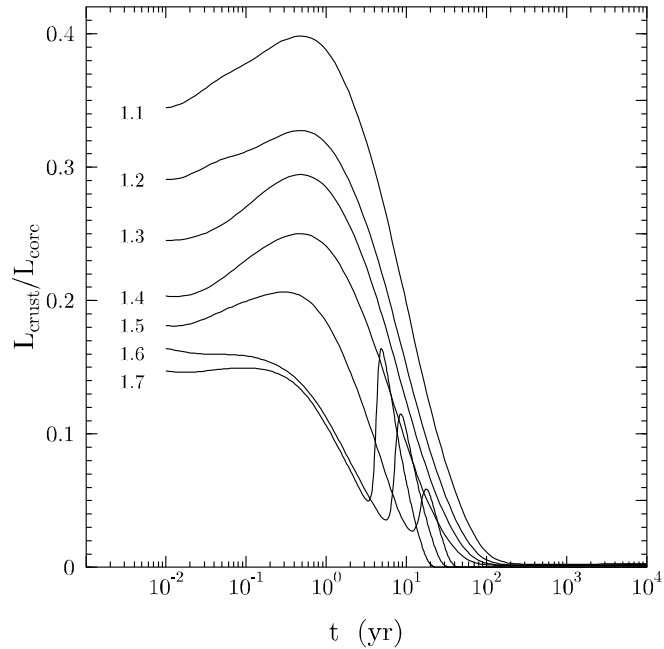}
\end{center}
\caption[ ]{
  Ratio of the neutrino luminosities in the crust and the core for
  the neutron star models of masses
  1.1, 1.2, \ldots, 1.7 $M_\odot$
  with the moderate equation of state (Gnedin et al.\ 2000).
}
\label{fig:lum}
\end{figure}

\begin{figure}[!th]
\begin{center}
\leavevmode
\epsfysize=9cm \epsfbox{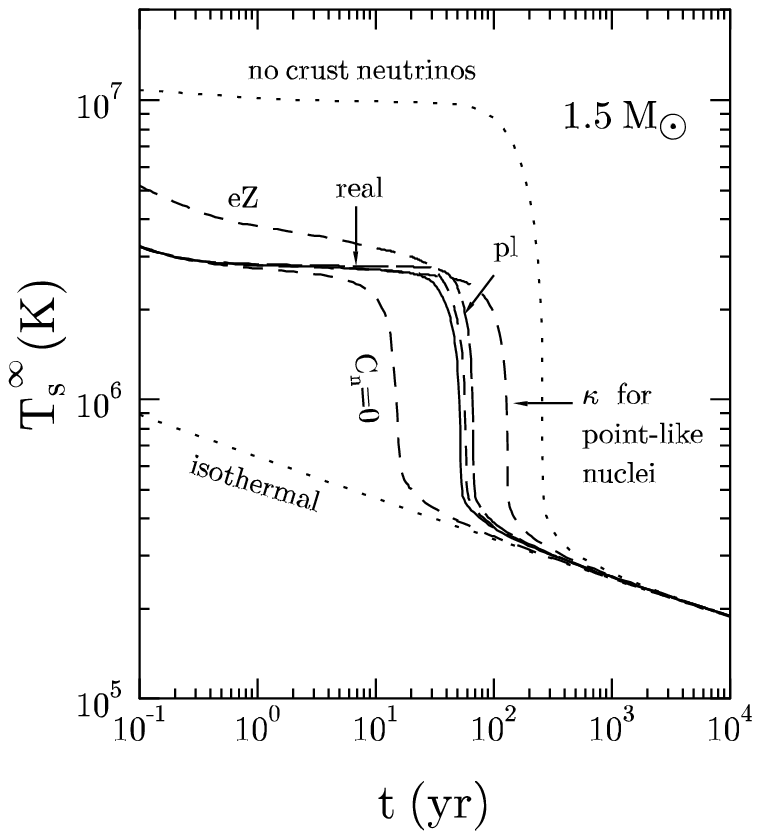}
\end{center}
\caption[ ]{
  Thermal relaxation epoch for the 1.5 $M_\odot$ model with the moderate
  equation of state and no superfluid effects (Gnedin et al.\ 2000).  
  Solid line is the real cooling
  curve, while dotted lines show the effect of omitting the neutrino emission
  from the crust (upper) or assuming infinite thermal
  conductivity in stellar interiors, at $\rho > 10^{10}$
  g cm$^{-3}$ (lower). The dashed curve $C_n=0$ is obtained
  neglecting neutron heat capacity in the crust. Another
  dashed curve is calculated using electron thermal
  conductivity $\kappa$ in the crust for pointlike neutrons.
  Two other dashed lines show
  cooling curves obtained by removing all neutrino
  mechanisms in the crust except one leading mechanism:
  plasmon decay (pl) or electron-nucleus
  bremsstrahlung ($eZ$).
  }
\label{fig:c5_cr}
\end{figure}

Figure \ref{fig:c_nosf_main} shows several 
cooling models of neutron stars with
different masses but the same moderate equation of state.  In the
low-mass models, $M < 1.44\, M_\odot$, the central density is not high
enough to switch on the direct Urca process.  These stars follow the
standard cooling scenario, with the cooling curves almost independent of
the stellar mass. The high-mass models go through the fast
cooling scenario and demonstrate a spectacular drop of the surface
temperature at the end of the thermal relaxation epoch,
$ t \sim 50$ yr, due to
the emergence of the cooling wave on the surface.
If observed, this effect can give
valuable information 
on the properties
of matter at sub-nuclear densities at the bottom of the 
neutron star crust (Lattimer et al.\ 1994).  
We analyze this in more detail in subsequent
sections.  The same, although much
less pronounced, effect takes place in the case of slow cooling.
Let us emphasize that no young neutron stars have been
detected so far to realize this method in practice,
but they can be observed with Chandra, XMM, and future X-ray missions.

Since the neutrino emissivity of the direct Urca process is
several orders of magnitude larger than that of
the modified Urca processes, the fast cooling regime
in high-mass stars ($M > 1.44\, M_\odot$)
can be established even if the central
kernel, where the direct Urca process is allowed, occupies a small
fraction of the stellar core (Page and Applegate 1992).
The cooling curves, again, depend
weakly on the mass.  This weak mass dependence
of both slow- and fast cooling curves at the neutrino cooling stage
has a simple explanation. The cooling rate is determined by
the ratio of the neutrino luminosity and the total (integrated)
heat capacity. Both quantities depend on $M$ in a similar way,
so that their ratio is almost independent of $M$.

Figures \ref{fig:c_nosf_soft} and \ref{fig:c_nosf_stiff}
show the cooling curves for the soft and stiff equations
of state, respectively.
The cooling curves for all three equations of state
are qualitatively similar, although fast cooling switches on at
different masses.
The change of the slope of the cooling curves at $t \sim 10^5$--$10^6$ yr
manifests the transition from the neutrino to the photon cooling stage.

It is remarkable that the surface temperature of a star at the
initial cooling stage (the first 50 years) is rather
independent of the equation of state, stellar mass, or the core neutrino
luminosity. The surface
temperature is mainly determined by the 
thermal energy content, neutrino losses, and transport
processes 
in the crust. The core and the crust are thermally
decoupled, and the effective surface temperature does not reflect
the thermal state of the stellar core.

In contrast, the evolution of the central temperature, $T(0,t)$, is
drastically different for the slow and fast cooling scenarios at all
times. 
In the low-mass models, the dependence of $T(0,t)$ on
time is identical, with a small offset in normalization.  This is due to
a simple temperature dependence of the dominant neutrino emissivity and
the heat capacity. In the absence of superfluidity, $C_v \propto T$.
If the neutrino emission is dominated by the modified Urca
($Q_\nu \propto T^8$),  then
$T(0,t) \propto t^{-1/6}$ at all times up to $t \sim
10^5$ yr.  Afterwards, photon emission from the surface comes into
play and changes the scaling law throughout the isothermal core.

In the models with fast cooling, where the dominant neutrino process is
the direct Urca ($Q_\nu \propto T^6$), the central temperature follows
the scaling relation $T(0,t) \propto t^{-1/4}$ 
for $t \lesssim 10^{-2}$ yr.  But
then until $t \lesssim 10$ yr, it stays almost constant at about $10^8$ K
as the heat flows from the warmer outer core, in which direct Urca process
is prohibited, into the inner core. 
During the thermal relaxation epoch, $10 \lesssim t \lesssim 100$ yr,
the central temperature declines again by a factor of several. After
the thermalization, $T(0,t)$
once again follows approximately the $t^{-1/4}$
law until the photon emission overtakes in the late stage of the evolution.
   
\subsection{Thermal relaxation in the crust}
\label{sect-cool-relax}

In this section we show the initial stage of thermal relaxation
in a neutron star, using models with
the moderate equation of state as an example.
Figures \ref{fig:config3_nosf} and \ref{fig:config7_nosf} illustrate the
effect of thermal relaxation on the internal temperature profiles for
the slow and fast cooling scenarios, respectively.  Until the age of
about 1 yr, the neutron star core, the inner and the outer crusts form
almost independent thermal reservoirs.  The region around $4\times
10^{11}$ g cm$^{-3}$, where free neutrons appear in the crust, seems
to be the most effective at cooling owing to the
powerful neutrino emission (see below).  The outer crust cools to $10^9$ K
in less than a month, while the inner parts remain much hotter.  The
core also cools independently but is unable to affect the inner crust
layers due to the slow thermal conduction. During the first years
the central kernel of the 1.7 $M_\odot$ model in Fig.\
\ref{fig:config7_nosf} remains much colder than the outer core.
This is because the kernel is cooled by the powerful
direct Urca process and thermal conduction is still
unable to establish thermal relaxation throughout the core.
Almost full core relaxation is achieved in 10 years.

After the first year, the crust
temperature profiles of the slow and fast
cooling scenarios start to differ. In the former, the temperature
gradient between the core and the crust is slowly eroded, as the
cooling
wave from the center reaches the surface. In the latter, the
temperature gradient continues to grow until it reaches a maximum at $t
\sim 10$ yr.  Then a huge amount of heat releases from the crust and
leads to a spectacular drop of the surface temperature by an order of
magnitude (which corresponds to lowering the photon
luminosity by four orders of magnitude !).  
At $t=50$ yr, the entire star is already isothermal.  Note,
that despite larger temperature gradients, thermal relaxation proceeds
overall quicker in the fast cooling scenario.

Prior to thermal relaxation, the contributions of the neutron star crust
to the integrated heat capacity and neutrino luminosity are significant
(Figs.\ \ref{fig:heat} and \ref{fig:lum}).
For the slow-cooling models with the moderate equation of state, the
heat capacity in the
crust ranges from 10\% to 20\% of that in the core, with the larger
fraction in the low-mass models (where crusts occupy larger
fraction of the volume). In the fast-cooling
models, the ratio of the crust to core heat capacities
reaches a maximum of 55\% at $t \sim 10$ yr before
dropping to under 10\% after the relaxation.  Similarly, the integrated
neutrino luminosity of the crust is about 15\%--40\% of that of the core at $t
\sim 1$ yr, and then drops to a tiny fraction at later times.

The crust of the neutron star is responsible for the delayed reaction of
the surface to the cooling of the core.
To investigate the most important mechanism in this process we have run
several test models switching on and off various ingredients.

Figure \ref{fig:c5_cr} shows these test models for a rapidly cooling
1.5 $M_\odot$ neutron star.
Following Lattimer et al.\ (1994) we define the relaxation
time $t_w$ of a rapidly cooling young star as the moment
of the steepest fall of the surface temperature $T_s^\infty(t)$.
The relaxation time of the real model is $t_w=51$ yr.
Switching off the neutrino emission from
the crust, while keeping the heat capacity, 
slows down the thermalization epoch by a factor of five,
to 260 yrs.  
The surface temperature prior to that also
stays much higher, at $10^7$ K.  The effect is similar, but less
pronounced, in the standard cooling scenario.

The importance of the individual neutrino mechanisms for the crust cooling
varies at different epochs.  
First, for $t \lesssim 10^{-2}$ yr in the
fast cooling scenario or for 
$t \lesssim 3\times 10^{-3}$ yr in the slow
cooling scenario, the $e^-e^+$ pair emission dominates.  As the
temperature drops below $5\times 10^9$ K, this process quickly fades
away.  The next
epoch is controlled by plasmon decay; it
dominates for $10^{-2} \lesssim t \lesssim 10$ yr (fast cooling) or $3\times
10^{-3} \lesssim t \lesssim 10$ yr (slow cooling).  Figure \ref{fig:c5_cr}
demonstrates that if plasmon decay is the only neutrino process in the
crust, the resulting cooling curve of the 1.5 $M_\odot$ star is not very
different from that with all other processes included.  In this case the
epoch of thermalization is delayed until $t_w \approx 68$ yr.

The last epoch of thermal relaxation lasts for the period $10 \lesssim t
< 100$ yr (fast cooling) or $10 \lesssim t < 1000$ yr (slow cooling),
when either electron-nucleus
or neutron-neutron bremsstrahlung
is important.  In fact, both neutrino processes
give almost identical cooling curves in the absence of superfluidity.
However, free neutrons in the crust are thought to be in a
superfluid state (see Sect.\
\ref{sect-cool-super}) which strongly suppresses $nn$
bremsstrahlung.
Therefore, electron-nucleus bremsstrahlung
is likely to be the dominant neutrino mechanism
in this last epoch.  Plasmon decay and
electron-nucleus bremsstrahlung together would reproduce accurately
the full 
cooling curve shown in Figure \ref{fig:c5_cr}.

The rate of relaxation is sensitive to the heat capacity of free
neutrons in the crust, $C_n$.  If this heat capacity is suppressed by
strong neutron superfluidity (discussed in more detail in Sect.\
\ref{sect-cool-super}), relaxation proceeds much faster.  To imitate
this effect, we set $C_n = 0$ and obtain $t_w \approx 15$ yr.


Finally, the relaxation epoch depends on
the thermal conductivity of the inner crust.
For instance, a neglect of finite sizes of atomic nuclei 
in the electron--nucleus scattering rate
would lower the electron thermal conductivity
at the crust base ($\rho \gtrsim 10^{13}$ 
g cm$^{-3}$) by a factor of 2--5 (Gnedin et al.\ 2000).
If we were to use this less realistic thermal conductivity,
we would have calculated much longer relaxation time, about 130 yr
(Fig.\ \ref{fig:c5_cr}).
On the other hand, if the thermal conductivity
in the stellar interior ($\rho > \rho_b$) were infinite, we would have obtained
an isothermal cooling scenario.
A sharp drop of the surface temperature associated with
the relaxation would disappear.

\subsection{The relaxation time}
\label{sect-reltime}

In this section we focus on the duration of the thermal relaxation stage.
The problem has been studied
in a number of papers, with the most 
detailed and thorough work by
Lattimer et al.\ (1994; also see references therein).
%
%
According to Lattimer et al.\ (1994) the relaxation 
time of rapidly cooling neutron stars with various masses
is determined mainly by the crust
thickness $\Delta R_{\rm crust}$ and scales as
\begin{equation}
  t_w \approx t_1 \, 
    \left( \Delta R_{\rm crust} \over 1~{\rm km} \right)^2
    \, {1 \over (1-r_g/R)^{3/2}},
\label{cool-tw}
\end{equation} 
where the normalized relaxation time $t_1$ depends on microscopic
properties of matter such as the thermal conductivity and
heat capacity. 
If neutrons are superfluid in the crust,
$t_1$ is sensitive to the magnitude and
density dependence of the critical temperature
of the superfluidity, as we discuss later.  
In slowly cooling neutron stars thermal relaxation has weaker effect
on the cooling curve, although the relaxation time appears to be similar.

The dependence of $t_w$ 
on the thermal conductivity $\kappa$ and heat capacity
$C_v$ follows from a simple estimate
of the thermal relaxation time 
in a uniform slab of width $l$:
\begin{equation}
  t_w \sim C_v l^2/\kappa.
\label{eq:slab}
\end{equation} 
In a thin crust ($\Delta R_{\rm crust} \ll R$) with
the effects of general relativity, the proper width is
$l= \Delta R_{\rm crust}/\sqrt{1-r_g/R}$,
which gives $t_w \propto 1/(1-r_g/R)$ in Eq.\ (\ref{cool-tw}).
An additional factor $1/\sqrt{1-r_g/R}$ in $t_w$
accounts for the gravitational dilation of time intervals.

For the non-superfluid stars
with the core--crust interface placed at $\rho_0/2$,
which is close to our value $\rho_{cc} = 1.5 \times 10^{14}$ g cm$^{-3}$, 
Lattimer et al.\ (1994)
obtained $t_1 \approx 26$ yr.
Our models show similar scaling, $t_1 = 28.5 \pm 1$ yr.
There is a small variation ($\pm 1$ yr) with the equation of state:
the soft equation of state leads 
to slightly lower $t_1$, while the
stiff one leads to slightly higher $t_1$, for the same value of
$\Delta R^2_{\rm crust} (1-r_g/R)^{-3/2}$.

In order to clarify the dependence of $t_1$ on physical
properties of the crust, we have run a number of test
cooling models with $M = 1.5 \,
M_\odot$ varying the heat capacity and thermal conductivity 
within the crust (at $\rho_b \leq \rho \leq \rho_{cc}$)
by a fixed factor 1/8, 1/4, 1/2, 2, 4, 8.  
Our test models show that the relaxation time $t_w$ is indeed
quite sensitive to these variations,
in agreement with the qualitative
estimate,
Eq.\ (\ref{eq:slab}), and the results of Lattimer et al.\ (1994).
It is important that the
variations of $\kappa$ and $C_v$
do not invalidate the scaling relation for the
relaxation time, Eq.\ (\ref{cool-tw}), but only affect
the normalization $t_1$. Moreover,
if $C_v$ is increased and $\kappa$ is decreased,
the dependence of $t_1$ on the values of heat capacity and thermal
conductivity is described by
a simple scaling relation,
$t_1 \propto C_v/\kappa^{0.8}$.

Let us emphasize that the crust relaxation time is sensitive
to the values of $\kappa$ and $C_v$ at 
sub-nuclear densities, at which
the properties of matter are very model-dependent.
For instance, the nuclei at $\rho \sim 10^{14}$ g cm$^{-3}$
may be strongly non-spherical
(rods, plates, etc.; Sect.\ \ref{sect-nucrust-introduc})
which is not taken into account in our calculations.
Moreover,
the thermal conductivity has not been calculated so far
for the phase of non-spherical nuclei.
Although we have varied $\kappa$ and $C_v$ within
the broad density range ($\rho_b \leq \rho \leq \rho_{cc}$)
in our test models,
we expect that the relaxation time is most sensitive
to the values of $\kappa$ and $C_v$ in the density range 
$\rho_{cc}/10 \lesssim \rho \leq \rho_{cc}$ and in the temperature range
$10^8 \lesssim T \lesssim 10^9$ K. 

However, if $C_v$ were noticeably lower 
or $\kappa$ noticeably higher than in our basic
nonsuperfluid models, the relaxation time
would have saturated at $t_w \approx 12$ yr. 
This is the time it takes the inner
core with the direct Urca emission 
to equilibrate thermally with the outer core 
(cf Figure \ref{fig:config7_nosf}).  
More generally, this is the core relaxation time $t_{\rm core}$,
which is independent of the parameters of the crust.
It can be estimated using the same formalism of heat diffusion, 
Eq.\  (\ref{eq:slab}), through a slab of material between the
direct-Urca-allowed kernel and the boundary of the core,
with $l = R_{\rm core}-R_{\rm D}$. 
Thus, the fast cooling models may have two distinct 
relaxation times, in the
core and in the crust, and the latter is typically longer,
at least for non-superfluid models.

\begin{figure}[!t]
\begin{center}
\leavevmode
\epsfysize=7cm 
\epsfbox[75 270 560 600]{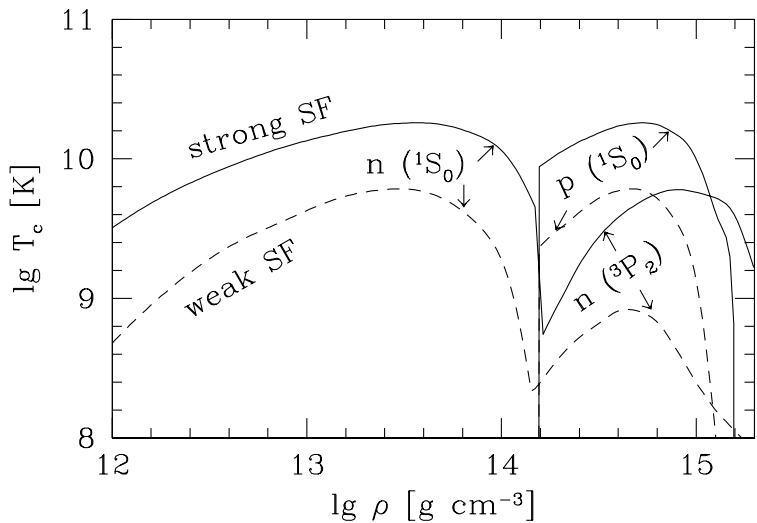} 
\end{center}
\caption[ ]{
  Density dependence of the critical temperatures
  of superfluidity (SF) of free neutrons 
  in the inner crust,
  and neutrons and protons in the core
  for the strong (solid lines) and
  weak (dashed lines) superfluid models (see text for details).    
}
\label{fig-cool-tc}
\end{figure}

\begin{figure}[ht!]
\begin{center}
\leavevmode
\epsfysize=9cm \epsfbox{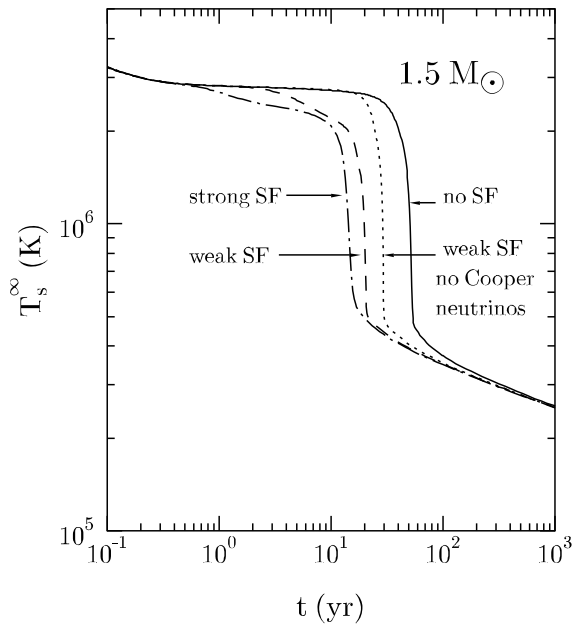}
\end{center}
\caption[ ]{
  Superfluid effects in the crust of the 1.5 $M_\odot$ neutron star
  model with the non-superfluid core and
  moderate equation of state (Gnedin et al.\ 2000).  
  Dashed line is for the case of
  weak neutron superfluidity (see text), while 
  dashed and dot line is for the
  case of strong superfluidity. Dotted line is obtained
  neglecting the Cooper-pair emission for the weak-superfluid model.
  Solid line is the cooling curve for the non-superfluid crust.
}
\label{fig:c5_crsf}
\end{figure}

\begin{figure}[!th]
\begin{center}
\leavevmode
\epsfysize=9cm \epsfbox{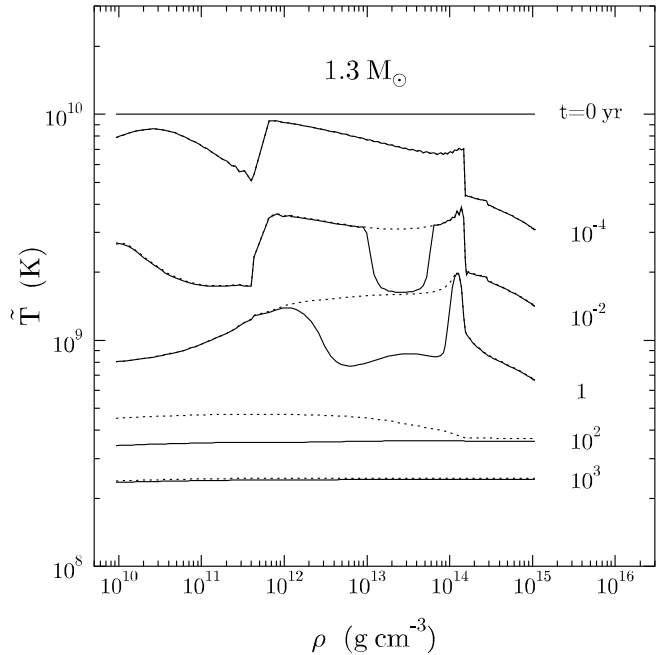}
\end{center}
\caption[ ]{
  Temperature profiles 
  in the interior of the 1.3 $M_\odot$ model
  (non-superfluid core with the moderate equation of state)
  with (solid lines) and
  without (dots) weak crust superfluidity of free neutrons
  (Gnedin et al.\ 2000).  
  Numbers next
  to curves show the stellar age.
  The contours are at 0, $10^{-4}$, $10^{-2}$, 1,
  100, and 1000 yr.
}
\label{fig:config3_crsf2}
\end{figure}

\begin{figure}[!th]
\begin{center}
\leavevmode
\epsfysize=9cm \epsfbox{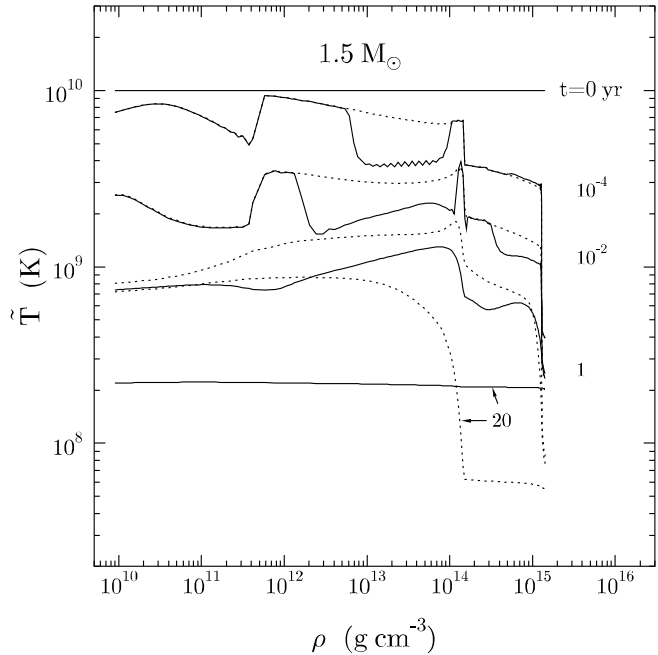}
\end{center}
\caption[ ]{
  Temperature profiles (solid lines)
  in the interior of the 1.5 $M_\odot$ model with
  strong superfluidity both
  in the crust and the core (Gnedin et al.\ 2000). Numbers next to
  curves show the stellar age. The contours are at 0, $10^{-4}$,
  $10^{-2}$, 1, and 20 yr. Dotted lines show the temperature profiles
  of the non-superfluid star.
}
\label{fig:config5_sf1}
\end{figure}

\begin{figure}[ht!]
\begin{center}
\leavevmode
\epsfysize=9cm \epsfbox{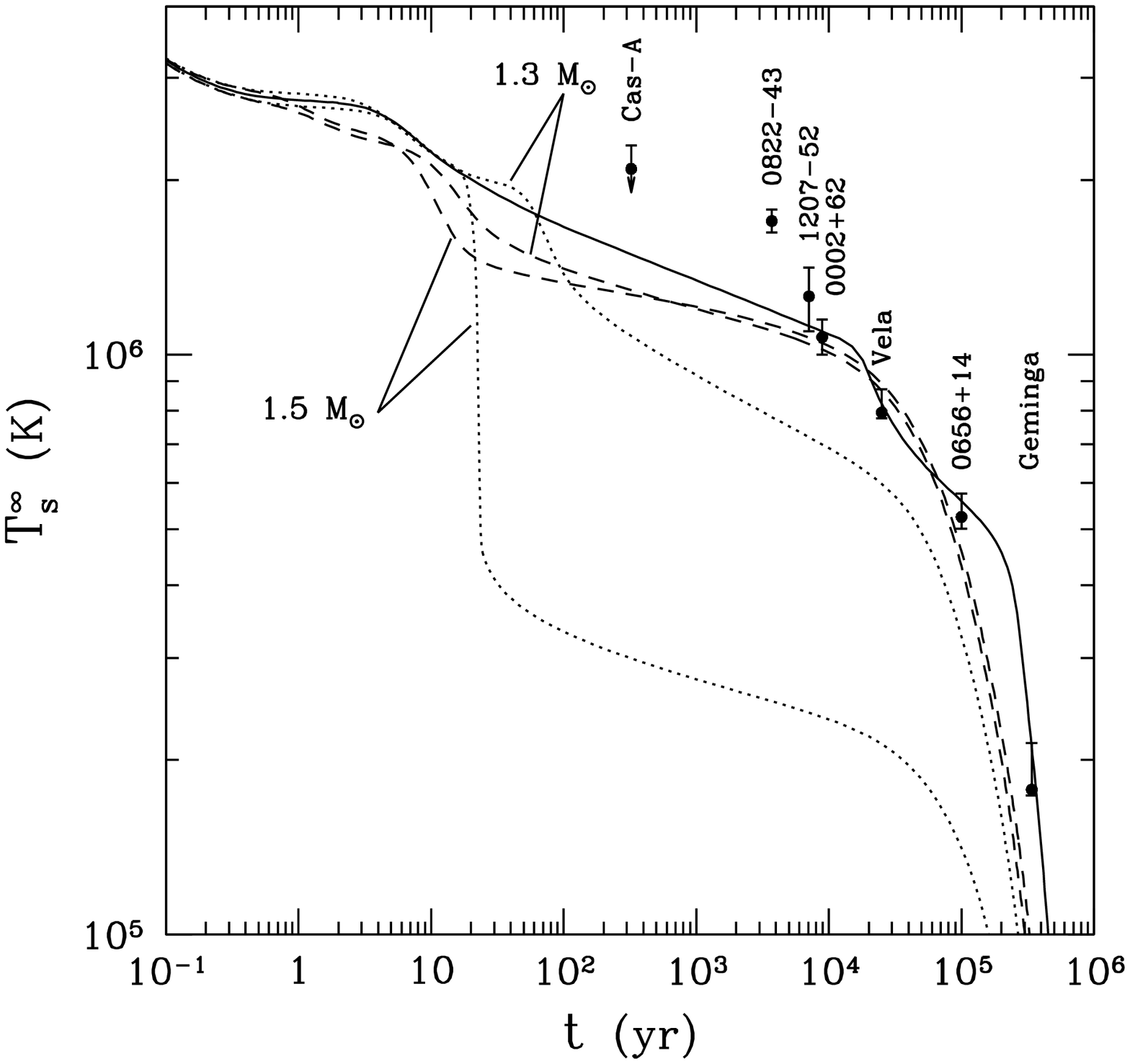}
\end{center}
\caption[ ]{
  Cooling curves of the superfluid models 
  with the moderate equation of state.
  Dots and dashes are for the cases of weak and strong
  superfluidity, respectively, for the 1.3 and 1.5 $M_\odot$ models.
  Solid line is for the 1.4 $M_\odot$ model with weak neutron
  superfluidity in the crust and constant critical temperatures of the
  baryon superfluidity in the core, $T_{cn} = 2\times 10^8$ K, $T_{cp} =
  1.3\times 10^8$ K.  Data points with error bars show the estimates of the
  surface temperature of several isolated neutron stars
  (Table \protect{\ref{tab-cool-data}}).
}
\label{fig:c_sf}
\end{figure}

\subsection{Cooling of superfluid neutron stars}
\label{sect-cool-super}

Free neutrons in the crust and both neutrons and protons in the core of
a neutron star are likely to be in the superfluid (SF) state 
(Sect.\ \ref{sect-overview-struct}).
We assume the singlet-state pairing of the protons,
and either singlet-state or triplet-state pairing of
the neutrons. In a uniform, low-density
matter (near the core-crust interface)
the neutron pairing is known to be of the singlet type,
but it switches to the triplet-state type at higher densities.
At any place in the star we adopt the neutron
pairing type that corresponds to the larger energy gap.
Various microscopic
theories predict a large scatter of the critical temperatures
of the neutron and proton superfluids, $T_{cn}$ and $T_{cp}$,
depending on the nucleon-nucleon potential model and the many-body theory
employed in calculations (see Yakovlev et al.\ 1999b
for references).  However, we can calculate
the cooling curves under different model assumptions
on $T_{cn}$ and $T_{cp}$, and try to constrain the critical
temperatures by comparing the theoretical cooling curves with
observations.

As an example we will use two models, a {\it weak} and a {\it strong}
superfluidity. They are the extensions of the models
used in Sects.\ \ref{sect-nucrust-nn} and \ref{sect-nucrust-overlook}
to describe the superfluidity in neutron star crusts.
The density dependence of the critical temperatures
in both models is shown in Fig.\ \ref{fig-cool-tc}.
The model of strong superfluidity corresponds to the higher critical
temperatures $T_c$.
It is based on the rather large energy gaps
calculated by Elgar{\o}y et al.\ (1996) for the singlet-state pairing
(with the maximum gap of about 2.5 MeV)
and by Hoffberg et al.\ (1970)
for the triplet-state pairing. The weak superfluid model
makes use of the sufficiently small superfluid gaps
derived by Wambach et al.\ (1993)
(with the maximum gap of about 1 MeV) for the singlet-state
superfluid and by Amundsen and {\O}stgaard (1985)
for the triplet-state neutron superfluid.
For simplicity, we use the same function $T_c(n)$
to describe the singlet pairing of free neutrons in the crust
($n=n_n$) and of the protons in the core ($n=n_p$).
In addition, we will use the {\it synthetic}
superfluid models adopting either weak or strong superfluidity
in the crust and constant density-independent values $T_{cn}$ and
$T_{cp}$ in the core.

It turns out that {\it superfluidity in the crust} affects
the cooling curves at the {\it initial thermal relaxation stage},
while {\it superfluidity in the core} affects cooling at
{\it later stages}.

First, consider the effects of neutron superfluidity in the crust.
If the temperature $T$ falls much below the
critical temperature $T_{cn}$, the superfluidity
reduces strongly the neutron heat capacity and $nn$
neutrino bremsstrahlung.
While the latter is compensated by
electron-nucleus bremsstrahlung,
the former effect leads to a faster thermal relaxation.  In
addition, a new neutrino emission process is allowed in the
superfluid state, the neutrino emission due to
Cooper pairing of free neutrons.  This process further
accelerates the cooling and thermal relaxation of the crust.

Figure \ref{fig:c5_crsf} demonstrates the aforementioned effects on
the fast cooling of the 1.5 $M_\odot$ neutron star.
The stellar core is assumed to be non-superfluid while
the crust superfluidity is taken to be either weak or strong.
The thermal
relaxation stage occurs about 2.5 times earlier in the case of weak
superfluidity and about 3.6 times earlier in the case of strong superfluidity,
compared to the non-superfluid crust.  And also,
while the inclusion of the neutrino emission
due to Cooper pairing leads to faster cooling
for $t < 20$ yr, most of the accelerating effect is due to the reduction
of the heat capacity. Therefore, the {\it duration of the thermal
relaxation stage is greatly reduced by the effect of superfluidity
on heat capacity of free neutrons in the stellar crust}.
As 
in the preceding section,
the relaxation time
$t_w$ of the superfluid crust
satisfies the same scaling relation, Eq.\ (\ref{cool-tw}), 
as for non-superfluid crust
but with a different scaling parameter $t_1$.
We have $t_1=11$ years for the weak crustal superfluidity
and $t_1=8$ years for the strong one.

The shortening of the thermal relaxation phase in a rapidly cooling star,
due to the superfluid reduction
of the crustal heat capacity, was emphasized by Lattimer et al.\ (1994).
These authors find that 
the relaxation time becomes three times as short,
in qualitative agreement with our results.
Our calculations indicate that
the effect is sensitive to the model of neutron superfluidity
in the crust and therefore, it can be used to test such models.

Figure \ref{fig:config3_crsf2} shows how the weak neutron superfluidity
in the crust
carves out the temperature profiles in the standard cooling
scenario.  For the first $10^{-4}$ yr, when the temperature is above
$T_{cn}$, the thermal structure is
identical to that of the non-superfluid model.  Later, the region
where the critical temperatures are the highest cools much faster than
the neighboring layers.  The acceleration of cooling
is again mainly due to the
reduction of the heat capacity and switching on the Cooper--pairing
neutrino emission. 
A sequence of points in Fig.\ \ref{fig:config3_crsf2}
in which solid lines start to deviate from the dotted ones
reproduces the density profile of $T_{cn}$ shown 
in Fig.\ \ref{fig-cool-tc}.
As the temperature falls further,
wider density regions become affected, producing shells of cool matter
surrounded by hotter layers on both sides.
After the cooling wave from
the core reaches the outer crust, the star settles into almost the same
isothermal state as the non-superfluid model, but faster.

The effects are much stronger if the superfluidity is allowed for in the
stellar core.  Figure \ref{fig:config5_sf1} shows the combined effect of
the strong core and crust superfluids.  Both neutrons
and protons are superfluid in the core of this 1.5 $M_\odot$ neutron
star. In addition to the trough in the crust layers, the core develops
a complex thermal structure.  
All the sources of
neutrino emission and the nucleon heat capacity in the core
are affected by the superfluidity, while the electron
heat capacity is not (and becomes dominant in superfluid cores).
As soon as the temperature drops significantly below
the critical temperatures $T_{cn}$ or $T_{cp}$, a new
powerful Cooper-pairing neutrino emission mechanism comes into play.
It starts in the inner part of the core and drives the temperature down.
At $t \sim 10^{-2}$ yr that region is even cooler than in the
non-superfluid model, while the other parts of the core are slightly hotter.
By the age of 1 yr, this cool region
includes all of the core except the inner kernel. 
As a result, thermal relaxation proceeds on a shorter timescale,
and at $t \lesssim 100$ yr the stellar interior is isothermal
(as seen also from Fig. \ref{fig:c_sf}).
Thus, for the large assumed values
of $T_{cn}$ and $T_{cp}$, the neutrino
emission due to Cooper pairing becomes so strong
that, instead of slowing down, the presence of the core superfluidity
accelerates the cooling.

When thermal relaxation is over and the isothermal state
is established throughout the star, the cooling is mainly regulated
by the neutrino luminosity and heat capacity of the stellar
core. The neutrino processes and heat capacity of the crust
cease to play a significant role except for the very low-mass
stars with large crusts, which we do not discuss
here.  We find, however, that for some superfluid models
the neutrino luminosity
of the crust may affect the cooling for a short period of time
during the transition from the neutrino cooling era to the photon
era.

The cooling of neutron stars after thermal relaxation can be
considered in the isothermal approximation. The superfluid effects
in the core on the
isothermal cooling have been discussed in detail by Yakovlev et al.\
(1999b), assuming the critical temperatures
$T_{cn}$ and $T_{cp}$ to be density-independent.
Thus, we
only summarize the results of Yakovlev et al.\
(1999b) and present some illustrative examples.

\newcommand{\rrr}{\rule{0cm}{0.4cm}}
\newcommand{\hh}{\rule{0.5cm}{0cm}}
\newcommand{\hb}{\rule{0.4cm}{0cm}}
 
\begin{table*}[!t]   
\caption[]{Surface temperatures of several neutron stars
inferred from observations using hydrogen atmosphere models}
\label{tab-cool-data}
\begin{center}
\begin{tabular}{|| l | l | l | c | l ||}
\hline
\hline
 Source       & lg~t &  lg~$T_s^\infty$ &  Confid. & References   \\
              & [yr] &  [K]             &  level   &           \\ 
\hline
\hline
NS candidate  in Cas A    & 2.51 & $<6.32^{+0.04 \rrr}_{-0.04}$ & 99\%  &
Pavlov et al.\ (2000) \\
RX$\,$J0822-43 & 3.57 & $ ~~~6.23^{+0.02 \rrr}_{-0.02} $ & 95.5\% &
Zavlin et al.\ (1999b) \\
1E$\,$1207-52  & 3.85 & $ ~~~6.10^{+0.05 \rrr}_{-0.06} $ & 90\% &
Zavlin et al.\ (1998) \\
RX$\,$J0002+62 & $3.95^{a)}$ & $ ~~~6.03^{+0.03 \rrr}_{-0.03} $ & 95.5\%&
Zavlin et al.\ (1999a) \\
PSR~0833-45 (Vela) & $4.4^{b)}$ & $ ~~~5.90^{+0.04 \rrr}_{-0.01}$ & 90\%  &
Page et al.\ (1996)\\
PSR~0656+14 & 5.00 & $ ~~~5.72^{+0.04\rrr}_{-0.02} $ & $^{c)}$ &
Anderson et al.\ (1993) \\
PSR~0630+178 (Geminga) & 5.53 & $ ~~~5.25^{+0.08\rrr}_{-0.01} $ & 90\% &
Meyer et al.\ (1994) \\
\hline
\hline
\end{tabular}
\begin{tabular}{l}
  $^{a)}\,$\rrr{\footnotesize The mean age taken according
 to Craig et al.\ (1997).~~~~~~~~~~~~~~~~~~~~~~~~~~~~~~~~~~~~~~~~~~~~~~~~~~~~~~
                    }\\[-0.7ex]
  $^{b )}\,$\rrr{\footnotesize According to
        Lyne et al.\ (1996).}\\[-0.7ex]
  $^{c)}\,$\rrr{\footnotesize Confidence level is not indicated
                in the cited reference.}\\[0.7ex] 
\end{tabular}
\end{center}
\end{table*}

The main conclusions on the cooling of neutron stars
with superfluid cores after thermal relaxation are as
follows :\\
(a) The nucleon superfluidity in the core may be the strongest
regulator of the cooling.\\
(b) The superfluidity may greatly delay or accelerate
the cooling in the neutrino cooling era, depending on
the values of the critical temperatures $T_{cn}$ and
$T_{cp}$.\\
(c) The superfluidity reduces
the large difference between the fast and slow cooling scenarios.
Under certain conditions, it makes the slow cooling look like
the fast and vice versa.\\
(d) The superfluidity accelerates the
cooling in the photon cooling era by reducing the heat capacity
of the stellar core.

Some superfluid effects on the thermal evolution of 1.5 $M_\odot$
(`fast' cooling) and
1.3 $M_\odot$ (`slow' cooling)
neutron stars are demonstrated in Fig.\ \ref{fig:c_sf}.
The dotted lines show the cooling curves of the models with weak
superfluidity.  In these models, the core superfluidity
is indeed too weak to eliminate the
great difference between the neutrino
luminosities provided by the direct and modified
Urca processes. Accordingly, the cooling curve of
the 1.5 $M_\odot$ star gives a typical example of the fast cooling,
while the cooling curve of the 1.3 $M_\odot$ star
is typical for slow cooling. The dashed curves show cooling
of the stars in the strong superfluid regime.
The strong superfluidity is seen to be sufficient to
reduce the difference between the fast and slow cooling scenarios. Both
curves look alike and are much closer to the
family of the slow-cooling curves
than to the family of the fast-cooling ones
(cf Fig. \ref{fig:c_nosf_main}).

In addition, Fig.\ \ref{fig:c_sf} shows some observational
data.  In the past few years, a great progress has been made
in detecting thermal emission from several isolated neutron
stars. In spite of many observational difficulties
(described, e.g., by  Pavlov and Zavlin 1998 and
Yakovlev et al.\ 1999b), some of the detections
are thought to be quite
reliable.  Such are
the three
closest middle--age radio pulsars (Vela, Geminga,
PSR~0656+14) and the three sufficiently young
radio-silent neutron stars
in supernova remnants (1E~1207--52 in the remnant PKS~1209--51/52,
RX~J0002+62 near CTB-1, and RX~J0822--4300 in Puppis A).  We also use
the results of recent observations of the
neutron star candidate in the supernova
remnant Cas A.
The objects are listed
in Table \ref{tab-cool-data}, ordered by their age.
The characteristic age is determined either from the
neutron star spin-down rate or from the morphology of the supernova remnant.
The effective surface temperature $T_s^\infty$
can be determined by fitting
the observed spectra either with the blackbody spectrum
or with the model spectra of the neutron star atmospheres, and
the results appear to be different. The nature of the difference
and the problems associated with the theoretical interpretation
of observations 
are discussed, for instance, by Yakovlev et al.\ (1999b).
For illustration, in Table \ref{tab-cool-data}
we present the surface temperatures
obtained with the hydrogen
atmosphere models.
For the source in Cas A, only the upper limit on
$T_s^\infty$ was obtained due to the low
photon statistics (Pavlov et al.\ 2000).
%
%
More discussion of the
values of $T_s^\infty$ for other sources 
in Table \ref{tab-cool-data} is given
by Yakovlev et al.\ (1999b).
We assume that the neutron star surfaces are
covered by thin layers of hydrogen or helium,
with the mass $\lesssim 10^{-13} \, M_\odot$.
This amount of light elements
is too small to affect the thermal insulation of the envelope
and stellar cooling but it is high enough to 
form the `atmospheric' spectrum of thermal radiation.

Now we can compare the observational data with the theoretical cooling
curves in Fig.\ \ref{fig:c_sf}. The source RX$\,$J0822-43
is too hot to fit our present cooling models
although it can be explained (Yakovlev et al.\ 1999b)
by a superfluid neutron star model
with the same equation of state with an addition of the outer
shell of light elements.
The range of $T_s^\infty$ allowed by the error-bars of other sources lies
higher than the cooling curves of the
weakly superfluid models but close to the
cooling curves of the strongly superfluid models.
Therefore, strong superfluidity is more suitable for the
interpretation of observations, although it does not
fit all the data.

Our aim here is to illustrate the method of probing
the superfluid state of neutron stars by studying their
cooling history. We do not intend to give a detailed comparison of
the theoretical
cooling curves for various superfluid models.  Instead,
in Fig.\ \ref{fig:c_sf} by the solid line we show an additional cooling
curve for a 1.4 $M_\odot$ neutron star
with the weak superfluidity in the crust
and the rather low fixed critical temperatures
of baryon superfluidity in the core: 
$T_{cn} = 2\times 10^8$ K and $T_{cp}
= 1.3 \times 10^8$ K.  This model fits the observational
data for the five neutron stars at once and
lies below the upper limit for the Cas A source. Thus, it explains
the observations of these six sources by a cooling 
curve
of one neutron star with the indicated constant values of $T_{cn}$ and $T_{cp}$
in the core. This demonstrates once again that
the superfluidity of nucleons in neutron star cores
is a strong regulator of the cooling and can
enable one, in principle, to find the agreement between
the theory and observations. On the other hand, 
the cooling curves are {\it sensitive
to the density dependence} of the critical
temperatures $T_{cn}(\rho)$ and $T_{cp}(\rho)$.
Although the accurate determination of $T_{cn}$ and $T_{cp}$ as
functions of density is ambiguous, the 
high sensitivity of the cooling curves 
to the theoretical assumptions should
enable one to constrain the critical temperatures by this method.

\subsection{Cooling of neutron stars with strong internal magnetic fields}
\label{sect-cool-durmag}

In this section we discuss some effects of strong
internal magnetic fields on the cooling of isolated
neutron stars. An internal magnetic field can affect
the neutron star evolution in many ways depending on
the superfluid (superconducting) state of the stellar
interiors. We focus here
on the magnetic broadening of
the direct Urca threshold studied in Sect.\ \ref{sect-nusup-durmag}.
We follow the analysis by Baiko and Yakovlev (1999).
%
%

Consider
a set of neutron star models with the moderate equation
of state, varying the central density and the
magnetic field strength. In the absence of the internal magnetic field
the chosen equation of state opens the direct Urca process
at densities $\rho \geq  \rho_{\rm crit} = 12.976 \times 10^{14}$ g cm$^{-3}$,
or for stellar masses $M \geq M_c = 1.442 \, M_\odot$
(Table \ref{tab-cool-model}).

As pointed out in Sect.\ \ref{sect-cool-nonsup}, the cooling history
of a non-magnetized neutron star is
extremely sensitive to the stellar
mass: if the mass exceeds $M_c$
the star cools rapidly via the direct Urca process, while
the cooling of the low--mass star is mainly due to
the modified Urca process and, therefore, is slower.
The effect of the magnetic field would be to speed up
the cooling of the star with the mass below
$M_c$, because the strong
field opens direct Urca process
even if it is forbidden at $B=0$
(Sect.\ \ref{sect-nusup-durmag}).

The magnetic field in the neutron star core may evolve
on timescales $\sim 10^6$--$10^7$ yr. This happens if
the electric currents supporting the magnetic field are located
in those regions of the core where the protons,
as well as the neutrons, are non-superfluid. If so,
the electric currents transverse to the field
undergo accelerated ohmic decay due to the strong magnetization
of charged particles (e.g., Haensel et al.\ 1990).
The consequences would be
twofold. Firstly, if the strong field occupies a
large fraction of the core, the decay produces an additional source of heating
which would delay the cooling of the core.
Secondly, the magnetic field decay reduces
the magnetic broadening of the direct Urca threshold and, thus,
the neutrino emission from 
the layers where the direct Urca is forbidden at $B=0$
(by decreasing the factor $R_B$ in Eq.\ (\ref{nusup-durmag-Qnu3}),
also see Fig.\ \ref{fig-nusup-durmag}).
If, however, the neutrons
are strongly superfluid, there is no acceleration of the field decay
(Haensel et al.\ 1990; {\O}stgaard and Yakovlev 1992)
%
%
and the ohmic decay time of the internal
magnetic field is longer than
the age of the Universe (Baym et al.\ 1969).
%
%
On the other hand, the microscopic calculations
of the energy gaps of neutron superfluidity
suggest that the critical
temperature $T_{cn}$ can be rather high at densities
$\lesssim 10^{15}$ g cm$^{-3}$, but may decrease
at higher densities (see Fig.\ \ref{fig-cool-tc}).
Thus the electric currents could persist in the outer
core, where neutron superfluid
is available and the enhanced field--decay mechanism
does not work,
while the direct Urca process operates in the inner
core and is not suppressed by the superfluidity.
We adopt this latter scenario.  We assume the presence
of the magnetic field $B$ in the stellar kernel,
where the direct Urca process is allowed, and
assume that only neutron superfluidity may be available
(protons are non-superfluid).
Thus, 
the entire stellar core is not superconducting and 
the magnetic field does not vary over the timescales
$t \lesssim 10^7$ yr of our interest, being frozen into the outer core.
For simplicity, the density dependence of
the neutron critical temperature
(the triplet-state pairing, case B)
is taken as a step function:
$T_{cn}=10^{10}$ K
at $\rho<7.5 \times 10^{14}$ g cm$^{-3}$, and
$T_{cn}=0$ at higher $\rho$. As in the previous sections,
all main neutrino emission processes in the stellar core
have been taken into account
suppressed properly by the neutron superfluidity
in the outer core. The emissivities of all the processes
but the direct Urca are unaffected by the magnetic field.  

To emphasize the effect of the internal magnetic field on
the cooling, we neglect the presence of
the crustal magnetic field and use
the relationship between the surface and internal
temperatures for $B=0$ (see Sect.\ \ref{sect-cool-code}).

\begin{figure}[!ht]
\vspace{-0.5cm}
\begin{center}
\leavevmode
\epsfysize=8.8cm
\epsfbox{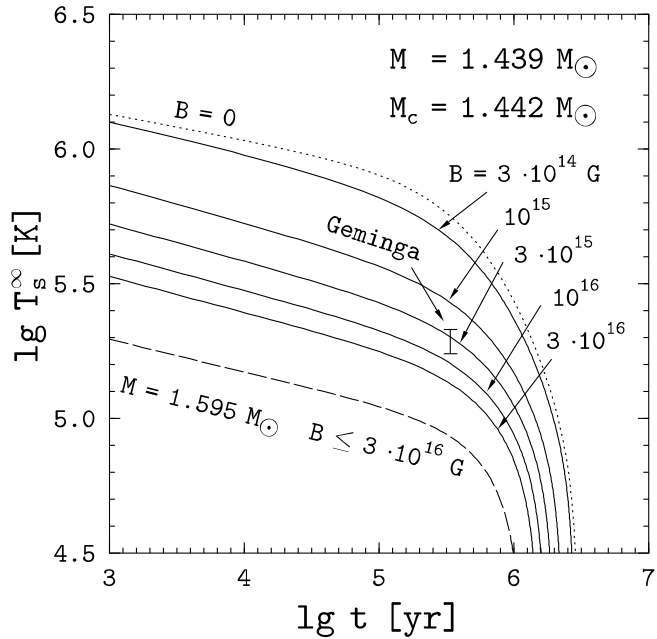}   
\end{center}
\vspace{-1cm}
\caption[ ]{Surface temperature as seen
by a distant observer as a function of age
of a neutron star with the moderate equation of state
(Baiko and Yakovlev 1999).
The dashed curve is for the star of mass
$M=1.595 M_\odot$, well above the threshold mass $M_c$,
with magnetic field $0 \leq B \leq 3 \times 10^{16}$ G.
The dotted and solid curves are for the
$1.439 M_\odot$ star.  The vertical bar shows the observed
range of $T_s^\infty$ for the Geminga pulsar.
}
\label{fig-cool-durmag1}
\end{figure}

The typical cooling curves are shown in
Fig.\ \ref{fig-cool-durmag1}. 
The dashed line illustrates fast
cooling of the massive $1.595 \, M_\odot$ model. Its mass
is well above the threshold mass $1.442 \, M_\odot$,
i.e., the direct
Urca process is already allowed in a large
portion of the core at $B=0$ (Table \ref{tab-cool-model}).
Even a very high internal field $B \lesssim 3 \times 10^{16}$ G
does not affect the cooling of this model.
New regions
of the core, where the field opens the direct Urca process,
add a negligible fraction of the neutrino luminosity
to the already large field-free luminosity.

%
%
%

The upper dotted curve
is calculated for the star with mass
1.439 $M_\odot$ 
at $B=0$.
It represents slow
cooling via the standard neutrino reactions
with the direct Urca being forbidden.
The solid curves illustrate the effect of the
magnetic field on the stars of the same mass.
If the stellar mass is slightly (by 0.2\%) below $M_c$ 
the cooling curve starts to deviate
from the standard one for not too high
fields, $B=3 \times 10^{14}$ G. Stronger fields,
$(1$--$3) \times 10^{15}$ G, produce the cooling scenarios
intermediate between the standard and fast ones, while still
higher fields $B \gtrsim 10^{16}$ G open
the direct Urca in a large fraction of the inner core
(Fig.\ \ref{fig-nusup-durmag})
and initiate the enhanced cooling. If, however, the stellar mass
is below the threshold by about 8\%, only
a very strong field $B=3 \times 10^{16}$ G can keep
the direct Urca process
open to speed up the cooling.

%
%
%

These results indicate that the magnetic field
in the inner core can indeed
enhance the cooling provided the stellar mass
is close to the threshold value $M_c$.
If $B = 3 \times 10^{16}$ G, the effect is significant
in a mass range
$(M - M_c) \lesssim 0.1 \, M_c$.
For lower fields
the range becomes smaller. If, for instance,
$B = 3 \times 10^{15}$ G, the mass range is as narrow
as $(M - M_c) \lesssim 0.015 \, M_c$.

%
\begin{figure}[!ht]
\vspace{-0.6cm}
\begin{center}
\leavevmode
\epsfysize=8.8cm
\epsfbox[20 20 250 280]{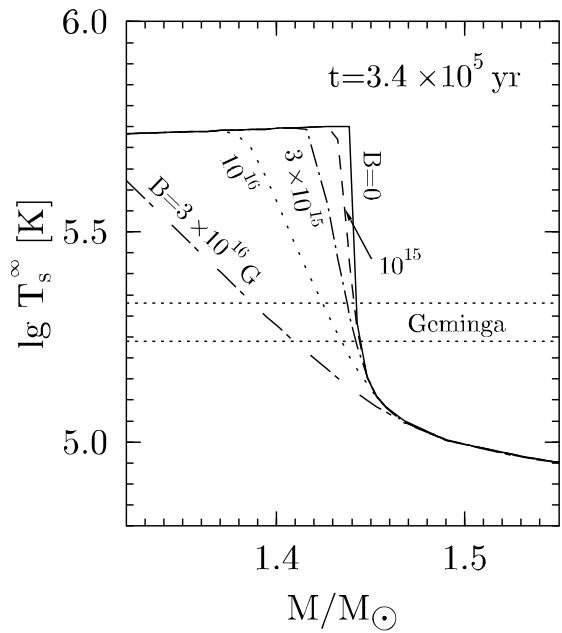} 
\end{center}
\vspace{-0.4cm}
\caption[ ]{Surface temperature of the neutron star
of Geminga's age versus stellar mass
for several values of the internal magnetic field
$B=0$, $10^{15}$,
$3 \times 10^{15}$, $10^{16}$, and $3 \times 10^{16}$ G.
Horizontal dotted lines show the confidence interval
of the surface temperature of the Geminga pulsar.
}
\label{fig-cool-durmag3}
\end{figure}
%
%

These findings may help in the interpretation of the
observational data. For illustration,
consider observations of the thermal radiation
from the Geminga pulsar
(Table \ref{tab-cool-data}).
Adopting the dynamical
age $t = 3.4 \times 10^5$ yr, we can place
Geminga's error-bar in
Fig.\ \ref{fig-cool-durmag1}.
Let us consider the above cooling model with the possible strong
magnetic field $B$ near the stellar center, which is unrelated to the
much weaker surface field.
If $B=0$, Geminga lies
between the lines of standard ($M \leq M_c$) and
fast ($M > M_c$) cooling.
It is clear that by tuning the mass slightly
above $M_c$ we can force the cooling curve to cross
the error bar. This effect
is demonstrated in Fig.\ \ref{fig-cool-durmag3}.
The figure shows the effective surface temperature
versus mass of the neutron star of Geminga's age
for several values of the central magnetic field.
If the mass $M$ exceeds $M_c$ for $B=0$,
the surface temperature decreases sharply
indicating that the transition from the standard cooling
to the fast cooling occurs in a very narrow mass range
just above $M_c$.
The smallness of the acceptable mass range
makes it fairly improbable that Geminga's mass lies
in this range. Accordingly the suggested interpretation
of Geminga's cooling with $B=0$ is unlikely.

The situation becomes strikingly different in the presence
of the strong magnetic field.
It broadens the range of masses
transient between the slow and fast cooling and shifts it
below $M_c$.
The acceptable mass range is now much wider.
At $B=10^{16}$ G it is about 0.01 $M_\odot$,
while at $B = 3 \times 10^{16}$ G it is about 0.02 $M_\odot$.
Correspondingly, the chances that
Geminga's mass falls into this range are better.

\subsection{Other aspects of neutron star cooling}
\label{sect-cool-other}

We have described the effects of various neutrino reactions
on the cooling of neutron stars.
For completeness, let us mention some other issues related
to the cooling problem.\\

{\bf (a) Cooling of stars with neutron superfluidity containing gaps
with nodes.} It is a triplet-state neutron superfluidity with the
$|m_J|=2$ projection of the total 
angular momentum of the Cooper pair
onto the quantization axis
(case C, in notations of Sect.\ \ref{sect-sf-gaps}).  The present microscopic
theories of neutron superfluidity are very model-dependent, and one
cannot exclude the appearance of neutron superfluidity of this
kind.  Since such superfluidity reduces the emissivity of the neutrino
reactions less strongly
than the nodeless superfluidity of type A or B (Chapt.\ \ref{chapt-nusup}),
it should be less efficient in regulating neutron star cooling.  The first
simulations of the standard and fast cooling with this type of
superfluidity were done by
Schaab et al.\ (1998b) using the simplified reduction factors
of the neutrino reactions.
These authors relied on the model calculations
of neutron pairing in a strong magnetic field
of Muzikar et al.\ (1980), who showed that the field
$B \gtrsim 10^{16}$ G made the pairing of type
C energetically more favorable than that of type B.\\

{\bf (b) Cooling of stars with hyperons and muons.}
The effects of hyperons in neutron star cores
have been reviewed by Balberg et al.\ (1999).
%
%
The first models of cooling neutron stars involving hyperons
were calculated by Haensel and Gnedin (1994).
%
%
Schaab et al.\ (1998a) included the possible effects of hyperon
superfluidity
%
%
using the approximate reduction factors of the neutrino reactions
and the approximate expressions for the neutrino emissivity
of the Cooper pairing process.
Cooling of superfluid hyperon stars was also analyzed by
Page (1998b). \\
%
%

{\bf (c) Cooling of stars with exotic cores.}
The exotic composition of the cores, such as kaon or pion
condensates or quarks, affects the neutrino emission
(Sect.\ \ref{sect-nucore-exotica}) and hence the
cooling.
These effects have been investigated in a number of papers,
see, e.g., Schaab et al.\ (1996, 1997a) and references therein.\\
%
%
%

{\bf (d) Cooling of stars with surface magnetic fields.}
The surface magnetic field affects the thermal
conductivity of the heat-blanketing envelope
(Sect.\ \ref{sect-cool-code}) and changes
the relation between the temperature $T_b$
at the bottom of the envelope
and the photon luminosity from the surface.
A detailed study of this relation
for the magnetic
field normal to the surface was carried out by
Van Riper (1988).
He also analyzed in detail
(Van Riper 1991)
the effects of such magnetic fields on neutron star cooling.
%
%
%
Page (1995) and
Shibanov and Yakovlev (1996)
%
%
%
reconsidered the same problem for
the dipole configuration of the magnetic field
and showed that it affected the cooling less strongly
and in a qualitatively different way.

Let us also mention a recent series of papers by Heyl, Hernquist and
collaborators
(see, e.g., Heyl and Hernquist 1998;
Heyl and Kulkarni 1998, and references therein)
devoted to the cooling of magnetars, the neutron stars with
superstrong magnetic fields $10^{14}$--$10^{16}$~G.
%
%
%
These fields may reduce the thermal insulation
of the heat-blanketing layers, making magnetar's surface
much hotter at the early cooling stage than the
surface of an ``ordinary'' neutron star.   Note, that the microscopic
properties of matter in superstrong magnetic fields
(equation of state, thermal conductivity) are poorly known,
so still much work is required to solve the problem completely.\\

{\bf (e) Cooling of stars with outer shells
composed of light elements.}
The thermal evolution can also be affected
by the presence of a thin
(mass $\lesssim 10^{-8}\, M_\odot$) shell of light elements
(H, He) at the surface of a neutron star
%
(Chabrier et al.\ 1997;
Potekhin et al.\ 1997;
Page, 1997, 1998a, 1998b).
%
%
%
%
%
%
Owing to the higher electron thermal conductivity
of the light-element plasma, the surface of a non-magnetized star
appears to be noticeably warmer in the neutrino cooling era.\\

{\bf (f) Reheating mechanisms.}
Cooling of neutron stars can be affected by
possible reheating of their interiors
by several reheating mechanisms.
The realistic mechanisms
turn out to be important at late
cooling stages, $t \gtrsim 10^4$ yr.
The most popular is the
viscous dissipation of the rotational
energy inside the star (see, e.g., Umeda et al.\ 1994,
Van Riper et al.\ 1995, Page 1998a, and references therein).
%
%
%

Another reheating source can be provided
by the energy release associated with a weak deviation
from beta-equilibrium in the stellar core
(Reisenegger 1995, Sect.\ \ref{sect-nucore-noneq}).
%
%
Reheating of the non-superfluid core
can also be produced by the ohmic dissipation
of the internal magnetic field
due to the enhancement of electric resistance
across the field
(Haensel et al.\ 1990;
Yakovlev 1993;
Shalybkov 1994;
Urpin and Shalybkov 1995; also see Yakovlev et al.\ 1999b
for critical discussion of other works).
%
%
%
%
%
Finally, reheating of an old neutron star
($t \gtrsim 10^7$ yr) can be provided by the ohmic decay
of the magnetic field in the crust
(Miralles et al.\ 1998). \\
%

To summarize,
the neutrino emission and
cooling of isolated neutron stars can be affected by many processes.
However, we hope we know the main cooling regulators.
For instance, the thermal history of the young
neutron stars, $t \lesssim 10$--1000 yr,
is determined mainly by the properties of their
inner crusts (Sect.\ \ref{sect-cool-relax}).
Cooling of the older, middle-aged
stars ($t \lesssim 10^5$--$10^6$ yr)
is most sensitive to the neutrino emission
from their cores, which depends on the composition
of dense matter and its superfluid properties.
Using a simple model with the standard $npe$ composition
and varying the critical temperatures of the neutron and
proton superfluidity, we can fit the observed
thermal radiation from a number of neutron
stars (Sect.\ \ref{sect-cool-super}).
Cooling of the old isolated neutron stars ($t \gtrsim 10^6$ yr)
does not depend on the neutrino emission.  However, the thermal
history of the old accreting neutron stars is
determined by various processes in the crusts, including
the non-equilibrium beta processes accompanied by neutrino emission
(Sect.\ \ref{sect-nucrust-beta}).

\section{Conclusions}

We described a variety of neutrino reactions
in neutron star crusts (Chapt.\ \ref{chapt-nucrust})
and cores (Chapts.\ \ref{chapt-nucore} and \ref{chapt-nusup})
and their effect on the cooling (Chapt.\ \ref{chapt-cool}).
We mainly focused on the non-exotic stars,
although we considered briefly the cores composed of exotic matter
(Sect.\ \ref{sect-nucore-exotica}).
In particular, we discussed the
following neutrino reactions:
electron-positron pair annihilation, plasmon
decay, electron-nucleus bremsstrahlung,
direct and modified Urca processes, nucleon-nucleon
bremsstrahlung, as well as their modifications and
analogs.
In addition, we considered some reactions
which are less known in the literature, for instance,
electron-electron bremsstrahlung and the related
lepton processes. We described the
reduction of the reaction rates by superfluidity
of the particles involved 
and also analyzed a specific neutrino emission
produced by Cooper pairing of the particles (Sect.\ \ref{chapt-nusup}).
We discussed in some detail
the effects of strong magnetic fields
on the direct Urca process (Sect.\ \ref{sect-nusup-durmag})
and analyzed the specific reactions
opened by the magnetic field, such as the
neutrino synchrotron emission by electrons
(Sect.\ \ref{sect-nucrust-syn}) and
electron scattering off fluxoids (Sect.\ \ref{sect-nusup-fluxa}).

In Chapt.\ \ref{chapt-cool} we illustrated the effects of the neutrino
reactions on neutron star cooling.  This gives one of the methods to
study the internal structure of neutron stars by comparing cooling
theory with observations.  Our selected examples confirm the
importance of the neutrino reactions in the cooling, especially for the
young and middle-aged neutron stars, $t \lesssim 10^5-10^6$ yr. For
instance, cooling of the middle-aged stars is very sensitive to
the superfluidity of baryons in their cores.  This allows one
to constrain the parameters of
nucleon superfluidity and the equation of state in the
cores, and thus shed light on the main `mystery' of neutron stars (Sect.\
\ref{sect-overview-introduct}).

It is important to emphasize that much work is required to complete the
study of neutrino reactions in neutron stars.  Some unsolved problems
were mentioned in Chapts.\ \ref{chapt-nucrust}--\ref{chapt-nusup}.  For
example, it would be useful to reconsider the modified Urca process
and baryon-baryon bremsstrahlung using modern models of
strong interactions. Another interesting problem
is to calculate exactly the superfluid reduction of the modified Urca
process and nucleon-nucleon bremsstrahlung under the joint effect of
superfluidity of the reacting particles, without using the
approximate similarity criteria (Sect.\ \ref{sect-nusup-similar}).  It
would also be important to analyze accurately the neutrino emission in
various reactions involving nucleons (free neutrons and nucleons bound
in nuclei) in the crust, such as the neutron-nucleus
bremsstrahlung (Sect.\ \ref{sect-nucrust-nn}).
In spite of these unsolved problems, we think
that we understand correctly the
main features of the neutrino emission in neutron stars.

Finally, we are hopeful that the cooling theory
combined with other astrophysical
studies of neutron stars (Sect.\ \ref{sect-overview-introduct}),
and supplied with the new exciting observational results,
will soon enable one to reach deeper understanding of the nature
of superdense matter.

\bigskip

{\bf Acknowledgments.}  
We are grateful to D.A.\ Baiko, V.G.\ Bezchastnov, A.B.\ Koptsevich,
K.P.\ Levenfish, G.G.\ Pavlov, A.Y.\ Potekhin, 
A.M.\ Shelvakh, Yu.A.\ Shibanov, 
D.N.\ Voskresensky, V.E.\ Zavlin
for discussions of the problems included in this review.  
The work was
supported in part by RFBR (grant No.\ 99-02-18099), INTAS (grant No.\
96-0542), KBN (grant No.\ 2 P03D 014 13), 
PAST professorship of French MENRT, PPARC, and NSF (grant No. PHY94-07194).

\end{document}